\documentclass[10pt,final,journal,twocolumn,singlespaced]{IEEEtran}
\usepackage{cite}

\usepackage{times}
\usepackage{epsfig}
\usepackage{graphicx}
\usepackage{amsmath}
\usepackage{amssymb}
\DeclareGraphicsExtensions{-eps-converted-to.pdf,.jpeg,.png}
\usepackage{epstopdf}
\usepackage{multirow}
\usepackage{makecell}

\usepackage[pagebackref=true,breaklinks=true,letterpaper=true,colorlinks,bookmarks=false]{hyperref}
\newcommand{\etal}{\textit{et al}. }
\newcommand{\ie}{\textit{i}.\textit{e}. }
\newcommand{\eg}{\textit{e}.\textit{g}. }
\begin{document}
%

\title{Hyperspectral Image Super-resolution via Deep Spatio-spectral Convolutional Neural Networks}

\author{Jin-Fan Hu, Ting-Zhu Huang,~\IEEEmembership{Member,~IEEE}, Liang-Jian Deng,~\IEEEmembership{Member,~IEEE}, Tai-Xiang Jiang,~\IEEEmembership{Member,~IEEE}, Gemine Vivone,~\IEEEmembership{Senior Member,~IEEE}, and Jocelyn~Chanussot,~\IEEEmembership{Fellow,~IEEE}
	\thanks{The work is supported by National Natural Science Foundation of China grants 61702083, 61772003 and 61876203, and the Fundamental Research Funds for the Central Universities JBK2001011.}
	\thanks{J. -F. Hu, T. -Z. Huang and L. -J. Deng is with the School of Mathematical Sciences, University of Electronic Science and Technology of China, Chengdu, Sichuan, 611731, China (e-mail: hujf0206@163.com; tingzhuhuang@126.com; liangjian.deng@uestc.edu.cn).}
	\thanks{T.-X. Jiang is with the FinTech Innovation Center, Financial Intelligence and Financial Engineering Research Key Laboratory of Sichuan province, School of Economic Information Engineering, Southwestern University of Finance and Economics, Chengdu, Sichuan, 610074, China (e-mail: taixiangjiang@gmail.com).}
	\thanks{G. Vivone is with the Department of Information Engineering, Electrical Engineering and Applied Mathematics, University of Salerno, 84084 Fisciano, Italy and with the Institute of Methodologies for Environmental Analysis, CNR-IMAA, 85050 Tito Scalo, Italy (e-mails: gvivone@unisa.it; gemine.vivone@imaa.cnr.it).}
	\thanks{J. Chanussot is with Univ. Grenoble Alpes, Inria, CNRS, Grenoble INP, LJK, Grenoble, 38000, France (e-mail: jocelyn.chanussot@gipsa-lab.grenoble-inp.fr).}
}

\maketitle

\begin{abstract}
Hyperspectral images are of crucial importance in order to better understand features of different materials. To reach this goal, they leverage on a high number of spectral bands. However, this interesting characteristic is often paid by a reduced spatial resolution compared with traditional multispectral image systems.	In order to alleviate this issue, in this work, we propose a simple and efficient architecture for deep convolutional neural networks to fuse a low-resolution hyperspectral image (LR-HSI) and a high-resolution multispectral image (HR-MSI), yielding a high-resolution hyperspectral image (HR-HSI). The network is designed to preserve both spatial and spectral information thanks to an architecture from two folds: one is to utilize the HR-HSI at a different scale to get an output with a satisfied spectral preservation; another one is to apply concepts of multi-resolution analysis to extract high-frequency information, aiming to output high quality spatial details. Finally, a plain mean squared error loss function is used to measure the performance during the training. Extensive experiments demonstrate that the proposed network architecture achieves best performance (both qualitatively and quantitatively) compared with recent state-of-the-art hyperspectral image super-resolution approaches. Moreover, other significant advantages can be pointed out by the use of the proposed approach, such as, a better network generalization ability, a limited computational burden, and a robustness with respect to the number of training samples.
\end{abstract}


\section{Introduction}
\label{sec:intro}
Traditional multispectral images (MSIs, \eg RGB images) usually contain a reduced number of spectral bands providing a limited spectral information. It is well-known that, the more spectral bands we have, the better we would understand the latent spectral structure. Since hyperspectral imaging can obtain more spectral bands, it has become a non-negligible technology that is able to capture the intrinsic properties of different materials.
However, due to the physical limitation of imaging sensors, there is a trade-off between the spatial resolution and the spectral resolution in a hyperspectral image (HSI), therefore it is burdensome to obtain an HSI with a high spatial resolution.
In this condition, hyperspectral image super-resolution by fusing a low-resolution hyperspectral image (LR-HSI) with a high-resolution multispectral image (HR-MSI) is a promising way to address the problem.
\begin{figure}[!t]
	\begin{center}
		\begin{minipage}{ 0.98\linewidth}
			{\includegraphics[width=1\linewidth]{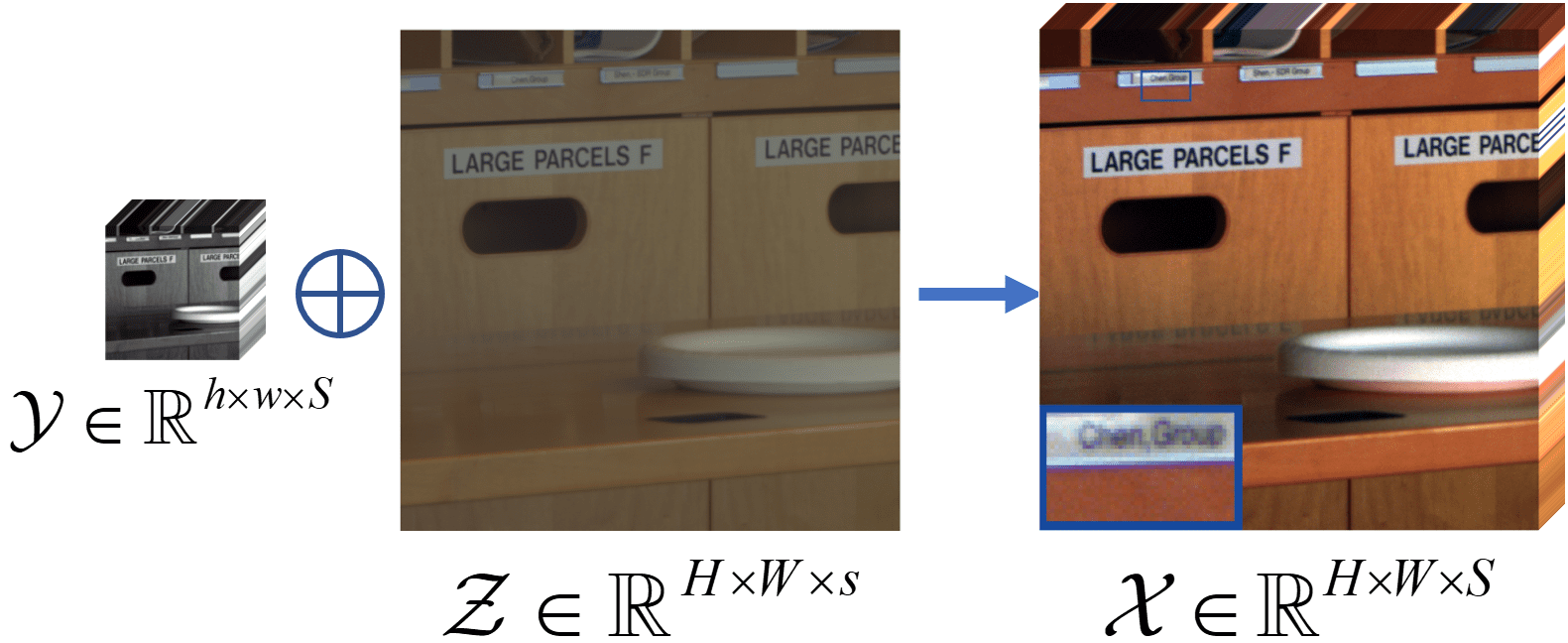}}
			\vspace{0.1pt}
		\end{minipage}
		
		\begin{minipage}{ 0.98\linewidth}
			\begin{minipage}{ 0.32\linewidth}
				{\includegraphics[width=1\linewidth]{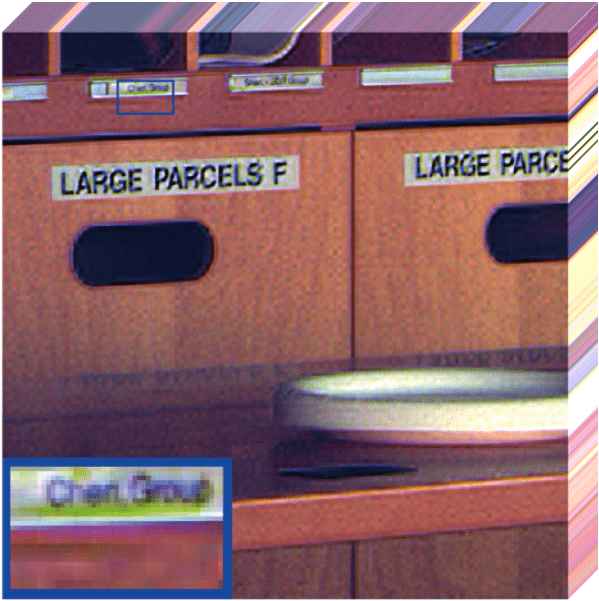}}
				\centering
				{(a) LTTR \cite{LTTR}}
			\end{minipage}
			\begin{minipage}{ 0.32\linewidth}
				{\includegraphics[width=1\linewidth]{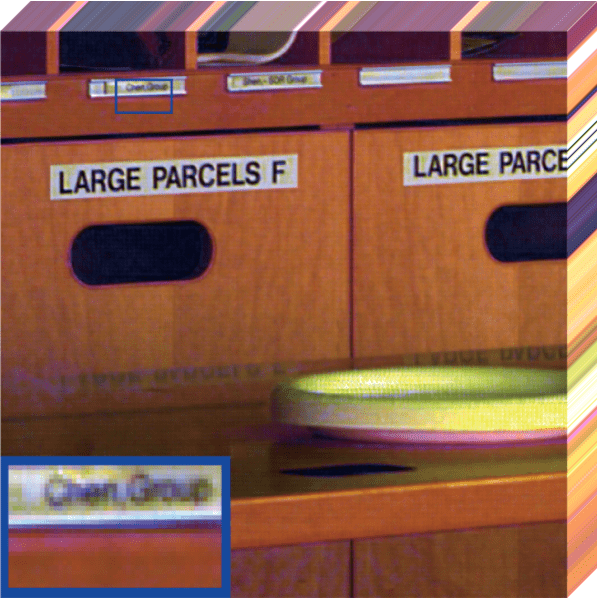}}
				\centering
				{(b) MHFnet \cite{xie2019multispectral}}
			\end{minipage}
			\begin{minipage}{ 0.32\linewidth}
				{\includegraphics[width=1\linewidth]{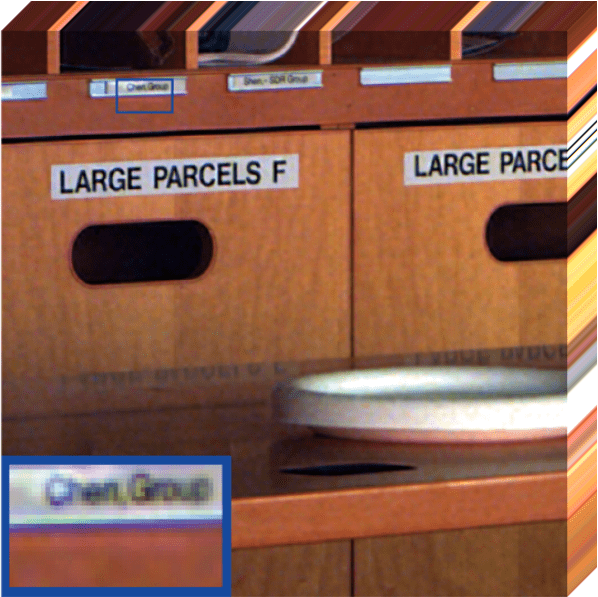}}
				\centering
				{(c) HSRnet}
			\end{minipage}
			\centering
		\end{minipage}
	\end{center}
	\caption{First row: the schematic diagram of hyperspectral image resolution on a test image from the Harvard dataset ($ h $ and $ w $ represent the height and width of LR-HSI, $ H $ and $ W $ denote the height and width of HR-MSI, $ s$ and $ S $ denote the spectral band number of HR-MSI and LR-HSI, respectively). The right image is the ground-truth HR-HSI, $ \mathcal{X}$. Second row: the results obtained by (a) LTTR (PSNR = 41.20dB), (b) MHFnet (PSNR = 38.70dB), and (c) the proposed HSRnet (PSNR = 43.93dB), where PSNR stands for the peak signal-to-noise ratio. Note that all the images are displayed with pseudo-color red, green, and blue (RGB) format using R = 28-th band, G = 12-th band, and B = 1-st band. Besides, MHFnet and HSRnet are both trained on the same CAVE dataset.}\label{fig:topimg}
\end{figure}

Many researchers have focused on hyperspectral image super-resolution to increase the spatial resolution of LR-HSI proposing several algorithms.
These latter are mainly based on the following models:
\begin{equation}\label{eq:re1}
\begin{aligned}
\mathbf{Y}  = \mathbf{XBS}, ~~
\mathbf{Z}  = \mathbf{RX},
\end{aligned}
\end{equation}
where $\mathbf{Y} \in \mathbb{R}^{S \times hw}$, $ \mathbf{Z} \in \mathbb{R}^{s \times HW}$ and $\mathbf{X} \in \mathbb{R}^{S \times HW}$ represent the mode-3 unfolding matrices of LR-HSI ($ \mathcal{Y} \in \mathcal{R}^{h \times w \times S}$), HR-MSI ($ \mathcal{Z} \in \mathcal{R}^{H \times W \times s}$) and the latent HR-HSI ($ \mathcal{X} \in \mathcal{R}^{H \times W \times S}$), respectively, $ h $ and $ w $ represent the height and width of LR-HSI, $ H $ and $ W $ denote the height and width of HR-MSI, $ s$ and $ S $ denote the spectral band number of HR-MSI and LR-HSI, respectively. Additionally, $ \mathbf{B} \in \mathbb{R}^{HW \times HW} $ is the blur matrix, $ \mathbf{S} \in \mathbb{R}^{HW \times hw} $ denotes the downsampling matrix, and $ \mathbf{R} \in \mathbb{R}^{s \times S} $ represents the spectral response matrix.
It is worth to be remarked that coherently with the notation adopted above, in this paper, we denote scalar, matrix, and tensor in non-bold case, bold upper case, and calligraphic upper case letters, respectively.

Based on the models in (\ref{eq:re1}), many related approaches have been proposed. Different prior knowledge or regularization terms are integrated in those methods.
However, the spectral response matrix $ \mathbf{R} $ is usually unknown, thus the traditional methods need to select or estimate the matrix $ \mathbf{R} $ and other involved parameters.
Additionally, the related regularization parameters used in these kinds of approaches are often image-dependent.

Recently, with the tremendous development of neural networks, deep learning has become a promising way to deal with the hyperspectral image super-resolution problem.
In \cite{dian2018deep}, Dian \etal mainly focus on the spatial detail recovery learning image priors via a convolutional neural network (CNN). These learned priors have been included into a traditional regularization model to improve the final outcomes getting better image features than traditional regularization model-based methods.
 In \cite{xie2019multispectral}, Xie \etal  propose a model-enlightened deep learning method for hyperspectral image super-resolution. This method has exhibited an ability to preserve the spectral information and spatial details, thus obtaining state-of-the-art hyperspectral image super-resolution results.

However, deep learning-based approaches for hyperspectral image super-resolution also encounter some challenges. First of all, these methods sometimes have \textit{complicated architectures} with millions of parameters to estimate. Second, due to the complicated architecture and large-scale training data, \textit{expensive computation and storage} are usually involved. Third, deep learning-based methods are data-dependent, which usually holds a \textit{weak network generalization}. Thus, the model trained on a specific dataset could poorly perform on a different kind of dataset. Instead, \textit{the proposed network architecture can easily handle the above-mentioned drawbacks}.


In this paper, the proposed network architecture (called HSRnet from hereon) can be decomposed into two parts. One part is to preserve the spectral information of HR-HSI by upsampling the LR-HSI. The other part is mainly to get the spatial details of HR-HSI by training a convolutional neural network with the high-frequency information of HR-MSI and LR-HSI as inputs. By imposing the similarity between the network output and the reference (ground-truth) image, we can efficiently estimate the parameters involved in the network. In summary, this paper mainly consists of the following contributions:
\begin{enumerate}
	\item The proposed network architecture is \textit{simple} and \textit{efficient}. As far as we know, it obtains better qualitative and quantitative performance than recent state-of-the-art hyperspectral image super-resolution methods. For example, our method shows significant improvements with respect to two state-of-the-art methods, one is deep learning-based \cite{xie2019multispectral} and the other one is regularization-based \cite{LTTR}, see also Fig. \ref{fig:topimg}. Besides, the proposed architecture involves fewer network parameters than other deep learning-based approaches thanks to our simple network design, more details are presented in Sec. \ref{vs}.
	\item The network architecture has a \textit{promising generalization} ability to yield competitive results for different datasets even though the network is trained only on a specific dataset. This is due to the use of high-pass filters to feed the network with high-frequency spatial information. Extensive experiments corroborate this conclusion, see Fig. \ref{F:harvard-1} and Tab. \ref{harvard10-ave}.
	\item Multi-scale information is integrated into our network architecture, which significantly improves the performance of the proposed method. The effectiveness of a multi-scale module has been proven in many computer vision works \cite{8767931,ms2,ren2016single,zhang2018multi,8931240,8049485} and further discussed in Sec. \ref{newstruct}.
	\item The network shows a good \textit{robustness to the number of training samples}, which indicates that our method could get very high performance with a different amount of training data. Furthermore, \textit{shorter training and testing times} compared with a state-of-the-art deep learning-based approach (see Tab. \ref{diffrernt-num-t}) have been remarked.
\end{enumerate}

The rest of the paper is outlined as follows. Section \ref{related} presents the related works about the hyperspectral super-resolution problem. Section \ref{main} introduces the proposed network architecture. In Section \ref{exp}, extensive experiments are conducted to assess the effectiveness of the proposed architecture. Furthermore, some discussions about the image spectral response, the network generalization, the computational burden, and the use of the multi-scale module are provided to the readers.


\begin{figure*}[t]
	\begin{center}
		\begin{minipage}{ 0.82\linewidth}
			{\includegraphics[width=1\linewidth]{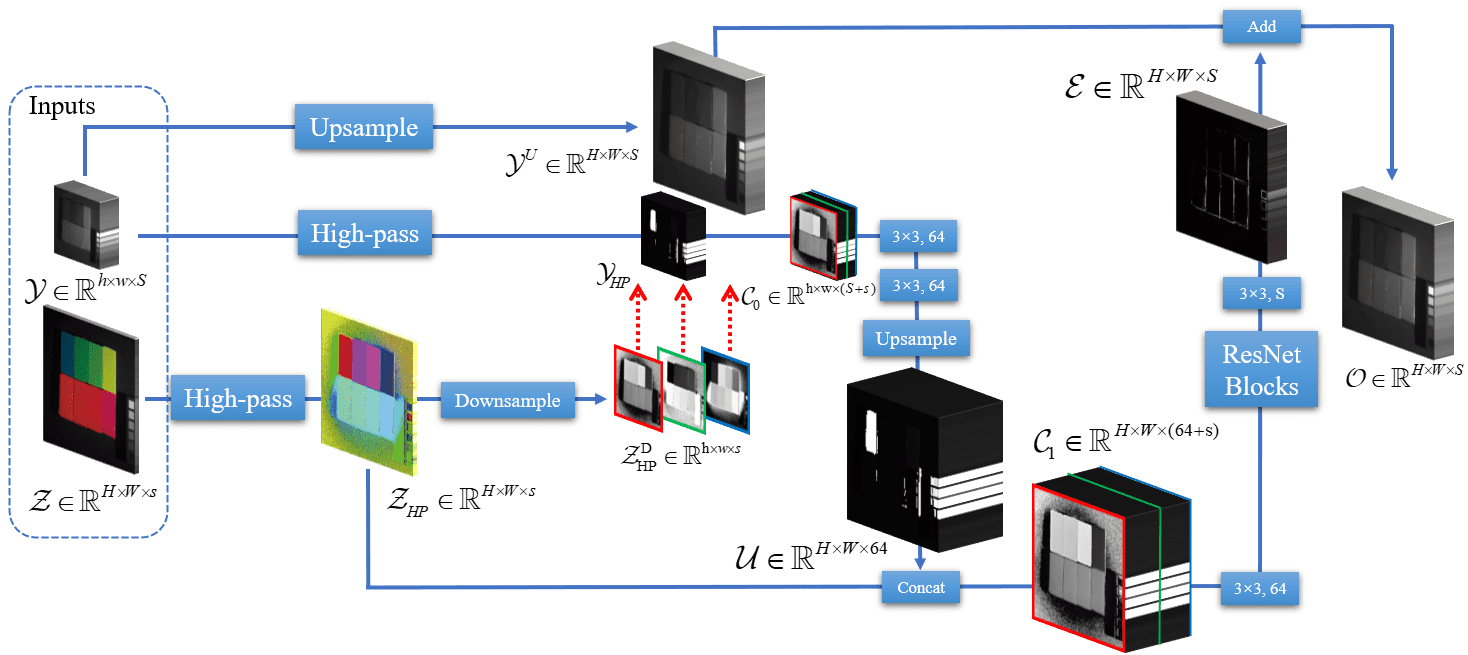}}
			\centering
			{(a)}
		\end{minipage}
		\begin{minipage}{ 0.14\linewidth}
			{\includegraphics[width=1\linewidth]{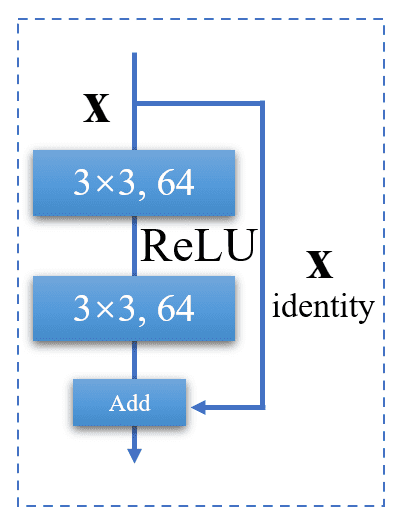}}
			\centering
			{(b)}
		\end{minipage}
	\end{center}
	\caption{The flowchart of the proposed network architecture (HSRnet). (a) Architecture of our HSRnet. LR-HSI $\mathcal{Y}$ and HR-MSI $\mathcal{Z}$ are the two inputs, and the $\mathcal{O}$ is the final output. (b) Illustration of one ResNet block with two layers and 64 kernels (size $3\times 3$) for each layer.
	}\label{structure}
\end{figure*}

\section{Related Works}\label{related}
Hyperspectral image super-resolution is a popular topic, which is receiving more and more attention. In particular, the combination of hyperspectral data with higher spatial resolution multispectral images is representing a fruitful scheme leading to satisfying results. Recent fusion or super-resolution approaches can be roughly categorized into two families: model-based approaches and deep learning-based methods.

Model-based approaches are classic solutions. Indeed, many works have been already published \cite{6502715,8444767,fu2019variational,kanatsoulis2018hyperspectral,zhang2018spatial,8768351,xing2018pansharpening,aiazzi2007improving,han2017hyperspectral,GLP-HS,1518950,CNMF,LTMR,LTTR,pan2019multispectral,zhang2016multispectral,yuan2018multiscale,zhang2019pan,dian2019multispectral,rong2014pansharpening,rong2012low,aly2014regularized,liu2017variational,7312998} for the super-resolution problem. For instance, Dian \etal \cite{dian2019multispectral} exploit the spectral correlations and the non-local similarities by clustering the HR-MSI in order to create clusters with similar structures. Low tensor-train rank prior is used in \cite{oseledets2011tensor}, the so-called \cite{LTTR} method. The tensor train (TT) rank consists of ranks of matrices formed by a well-balanced matricization scheme. The effectiveness of low TT rank (LTTR) prior has been utilized in \cite{bengua2017efficient}, which shows ability in image and video reconstruction. Compared to normal matrix ranks, the tensor rank keeps more abundant information about the data cube. Then, they regard the super-resolution as an optimization problem that, with the help of low tensor-train rank constraint, has a satisfying solution under the well-known alternating direction multipliers minimization (ADMM) \cite{bioucasdias2010alternating} framework.

Deep learning-based methods have recently showed exceptional performance in the field of image super-resolution, see \eg  \cite{sr1,sr2,sr3,xie2019multispectral,dian2018deep,Palsson2017Multispectral,inproceedings,yang2018hyperspectral,shao2018remote,li2017hyperspectral,yao2018pixel,vitale2019cnn,palsson2018sentinel,han2019hyperspectral,liu2018deep,huang2015new,rao2017residual,liu2018psgan}.
A powerful example is provided by the so-called PanNet developed in \cite{inproceedings}. Here, Yang \etal designed a new architecture training the deep-learning network with high-pass filtered details rather than original images. This is done in order to simultaneously preserve the spatial and spectral structures. Thanks to the use of high-pass filters, a greater generalization capability is observed. 
Another instance of deep learning-based methods for solving the hyperspectral image super-resolution issue is provided in \cite{xie2019multispectral}, where a model-based deep learning method is proposed. The method exhibits a great ability to preserve structures and details, as well as it obtains state-of-the-art results. Unlike other deep learning-based methods that mainly regard the image super-resolution issue as a simple regression problem, this approach is based on the generation mechanism of the HSI and the MSI to build a novel fusion model. It adopts the low rankness knowledge along with the spectral mode of the HR-HSI under analysis. Instead of solving the model by traditional alternating iterative algorithms, the authors design a deep network learning the proximal operators and model parameters by exploiting CNNs.

\section{The Proposed HSRnet}\label{main}

In this section, we introduce first the regularization-based model for the hyperspectral image super-resolution problem. Motivated by the above-mentioned model, we propose our network architecture that will be detailed in Sec. \ref{N_A}.

\subsection{Problem Formulation}
Estimating the HR-HSI from LR-HSI and HR-MSI is an ill-posed inverse problem. Thus, prior knowledge is introduced exploiting regularization terms under the maximum a posteriori (MAP) framework. Those methods can be formulated as:
\begin{equation}
\min_{\mathbf{X}} \mathcal{L} =  \lambda_{1}f_{1}(\mathbf{X}, \mathbf{Y})+  \lambda_{2}f_{2}(\mathbf{X}, \mathbf{Z})+ R(\mathbf{X}),
\end{equation}
where $\mathbf{X}, \mathbf{Y}, \mathbf{Z}$ are the mode-3 unfolding matrices of tensor HR-HSI, LR-HSI, and HR-MSI, respectively, which have been introduced in Sec. \ref{sec:intro}. $\lambda_{1}$ and $\lambda_{2}$ represent two regularization parameters,
$ f_{1} $ and $ f_{2} $ force the spatial and spectral consistency, respectively, and $ R$ stands for the regularization term depending on the prior knowledge.
In general, $ f_{1} $ and $ f_{2} $ are defined based on the relations in (\ref{eq:re1}), \ie,
\begin{equation}
\begin{aligned}
f_{1} (\mathbf{X},\mathbf{Y}) &= \|\mathbf{Y}-\mathbf{X B S}\|_{F}^{2},\\
f_{2} (\mathbf{X},\mathbf{Z}) &= \|{\mathbf{Z}}-\mathbf{R} \mathbf{X}\|_{F}^{2},
\end{aligned}
\end{equation}
where $ {\left\| {\bf{X}} \right\|_F}$ = $\sqrt {\sum {\sum {x_{ij}^2} } }$ is the Frobenius norm. In particular, the regularization term $R$ is crucial for regularization-based methods.

Deep learning can be viewed as an estimation problem of a function mapping input data with ground-truth (labeled) data. In our case, starting from the input images (\textit{i.e.}, LR-HSI and HR-MSI), we can estimate the mapping function $\mathit{f}$ by minimizing the following expression:
\begin{equation}\label{mapping}
\begin{aligned}
\min_{\mathbf{\Theta}} ~\mathcal{L}=\left\|f_\mathbf{\Theta}(\mathbf{Y},\mathbf{Z})-\mathbf{X}\right\|_{F}^{2},
\end{aligned}
\end{equation}
where $\mathbf{Y}$ and $\mathbf{Z}$ are the LR-HSI and the HR-MSI, respectively, and $\mathbf{X}$ is the reference (ground-truth) HR-HSI. The mapping function $\mathit{f}$ can be viewed as a deep convolutional neural network, thus $\mathbf{\Theta}$ represents the parameters of the network. Besides, the prior knowledge can be viewed as being implicitly expressed by the learned parameters. In the next subsection, we will present the network architecture recasting the problem as in (\ref{mapping}), where the function $\mathit{f}$ is estimated thanks to several examples provided to the network during the training phase.

\subsection{Network Architecture}\label{N_A}



Fig. \ref{structure} shows the proposed HSRnet for the hyperspectral image super-resolution problem. From the figure, it is easy to see that we decompose the network into two parts, such that the two parts can preserve the most crucial characteristics of a hyperspectral image, \textit{i.e.,} the spectral information and the spatial details.

\subsubsection{Spectral preservation} 
\begin{figure}[t]
	\begin{center}

		{\includegraphics[height=0.65\linewidth,width=0.9\linewidth]{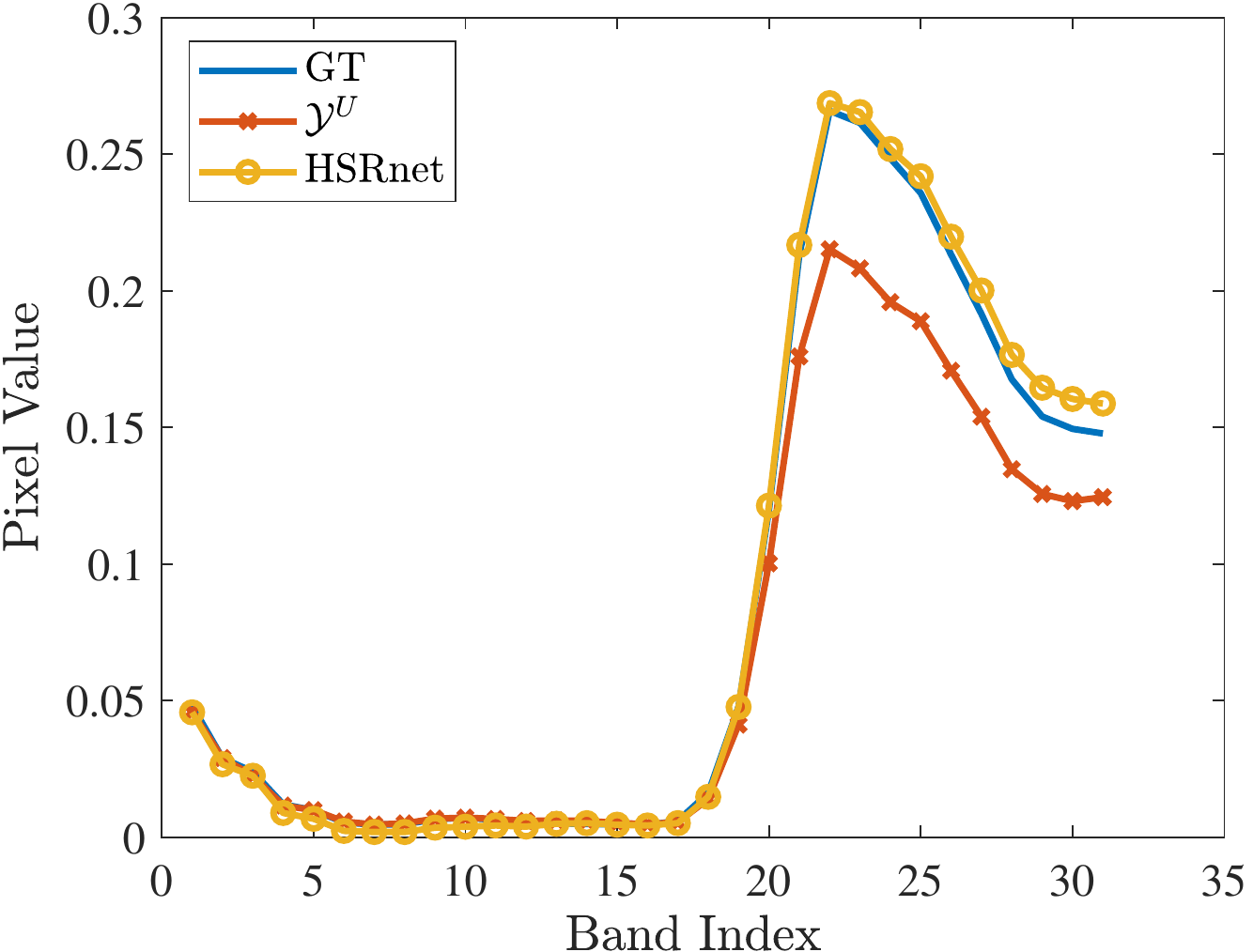}}
		
		\caption{Sampled spectral signatures for the object at pixel (175, 400) as obtained by the (ground-truth) HR-HSI, the upsampled LR-HSI $\mathcal{Y}^{U}$, and the estimated version of the high resolution HSI exploiting the proposed HSRnet.}\label{sp-gt-lms}
	\end{center}
\end{figure}

The LR-HSI $ \mathcal{Y} \in \mathbb{R}^{h \times w \times S}$ \footnote{We use three coordinates format to better represent the 3D hyperspectral image, \textit{i.e.},  ${h \times w \times S}$.} has the same spectral band number as the ground-truth HR-HSI $ \mathcal{X} \in \mathbb{R}^{H \times W \times S}$. Indeed, most of the spectral information of the HR-HSI is contained in the LR-HSI (the remaining part is due to the spectral information of the high resolution spatial details). In order to corroborate it, we plot the sampled spectral signatures obtained by the ground-truth HR-HSI $\mathcal{X}$ and by the corresponding upsampled LR-HSI $\mathcal{Y}^{U} \in \mathbb{R}^{H \times W \times S}$ in Fig. \ref{sp-gt-lms}. It is easy to be noted that the plots are very close to each other indicating that $\mathcal{Y}^{U}$ holds most of the spectral content of $\mathcal{X}$. Therefore, in order to guarantee a spectral preservation, we simply upsample $ \mathcal{Y}$ getting $\mathcal{Y}^{U}$ (as shown in the top part of Fig. \ref{structure}(a)). 

Admittedly, $\mathcal{Y}^{U}$ is able to preserve the spectral information, but many spatial details are lost (which can retain part of the spectral information). Instead, the proposed HSRnet can learn the spectral information of the HR-HSI, even preserving the spatial counterpart. As a result, the final outcome of the proposed HSRnet clearly shows an almost perfect spectral preservation, see Fig. \ref{sp-gt-lms}.

\subsubsection{Spatial preservation} Since the HR-MSI $\mathcal{Z} \in \mathbb{R}^{H \times W \times s}$ contains high spatial resolution information, we aim to use $\mathcal{Z}$ to extract spatial details injecting them into the final hyperspectral super-resolution image. Moreover, $\mathcal{Y}$ still contains some spatial details, thus we also consider employing $\mathcal{Y}$ to extract them. However, we do not simply concatenate $\mathcal{Z}$ and $\mathcal{Y}$ together taking them into the network because that will not lead to a satisfying detail preservation. Indeed, we calculate first the spatial details at the LR-HSI scale, called $\mathcal{Y}_{HP}$ in Fig. \ref{structure}. In particular, we simply take them from high-pass filtering the LR-HSI. Moreover, we add other details at the same scale by extracting them from the HR-MSI $\mathcal{Z}$. This is done by filtering and then downsampling the HR-MSI $\mathcal{Z}$ getting $\mathcal{Z}_{HP}^{D}$, see Fig. \ref{structure} again. This information has the advantage to occupy less memory and to require less computational burden to be processed compared to the original information in $\mathcal{Z}$. Finally, we concatenate this information, \textit{i.e.} $\mathcal{Y}_{HP}$ and $\mathcal{Z}_{HP}^{D}$, to get $\mathcal{C}_0 \in \mathbb{R}^{h \times w \times (S+s)}$.

In order to complete the multiresolution analysis, thus introducing a multi-scale module in our network, the details at the HR-MSI scale are also extracted. This is performed by simply filtering them using a properly designed high-pass filter. These details, denoted as $\mathcal{Z}_{HP}$, can be concatenated with $\mathcal{C}_0$ (\textit{i.e.}, the details at the lower scale) after this latter is properly convoluted and upsampled to the HR-MSI scale. Thus, $\mathcal{C}_1 \in \mathbb{R}^{H \times W \times (64+s)}$ indicates the concatenation of the details at two different scales (the LR-HSI one and the HR-MRI one). This represents the input of the ResNet implementing the well-known concept of multi-resolution analysis often considered in previously developed researches (\eg \cite{ms2,ren2016single,zhang2018multi,8931240,8049485}) either by designing diverse kernel sizes for convolution \cite{ms2,ren2016single} or extracting different spatial resolutions by filtering input data \cite{zhang2018multi,8931240,8049485}.

It is worth to be remarked that the high-pass filtering step is realized by the subtraction of the original image and its low-pass version, which is obtained by an average filter with a kernel size equal to $6 \times 6$. The upsampled operation is implemented by deconvolving with a kernel of size $6 \times 6$. Moreover, the concatenation operator is about adding the multispectral bands with high spatial resolution (3 bands, RGB image) into the hyperspectral bands (as shown in Fig. \ref{structure}). In this work, the red, the green, and the blue slices of $\mathcal{Z}_{HP}^{U}$ and $\mathcal{Z}_{HP}$ are inserted as the head, the middle, and the tail frontal slices to complement the spectral information of the hyperspectral image.

\begin{figure}[t]
	\begin{center}
		\begin{minipage}{ 0.4\linewidth}
			{\includegraphics[width=1\linewidth]{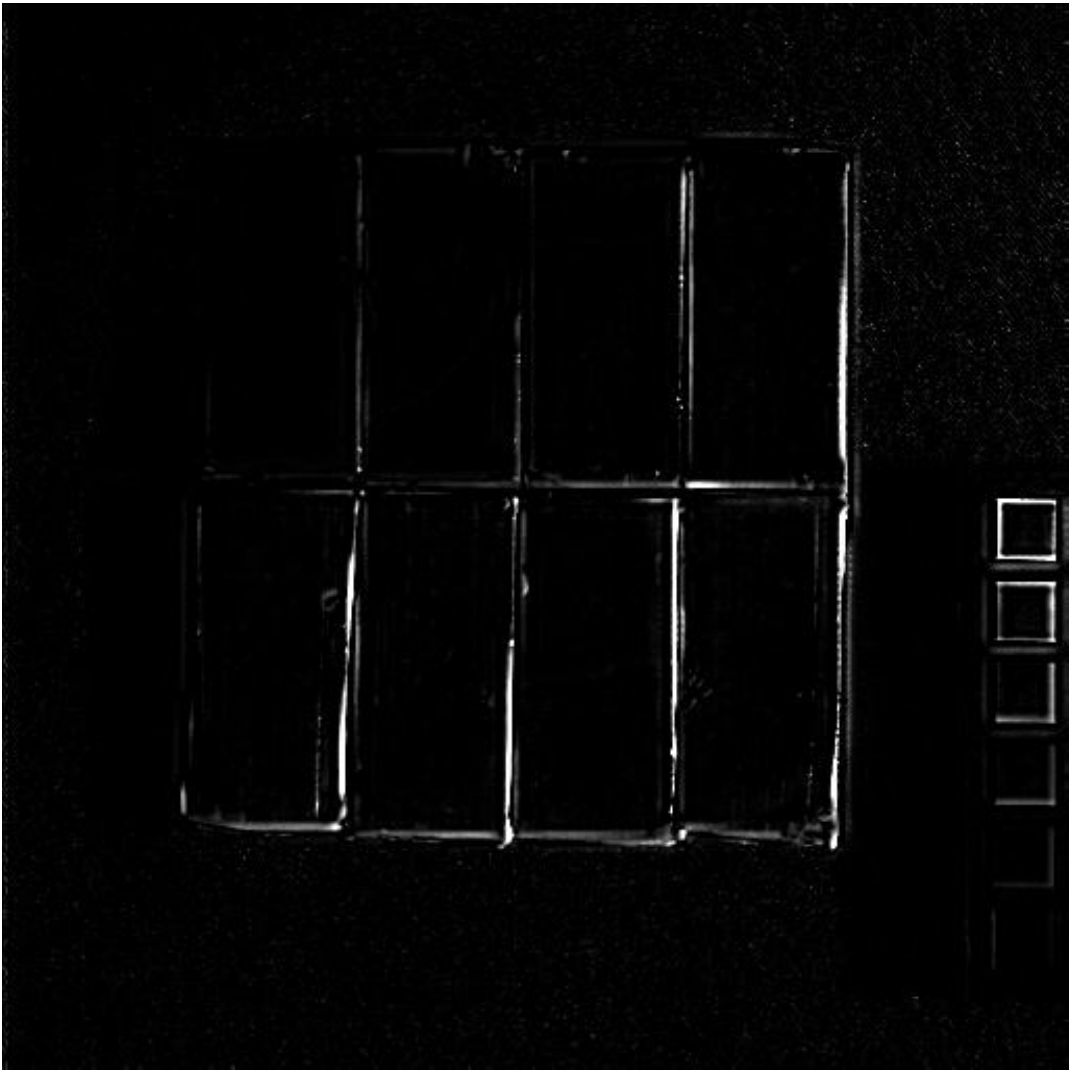}}
			\centering
			{(a)}
		\end{minipage}
		\begin{minipage}{ 0.4\linewidth}
			{\includegraphics[width=1\linewidth]{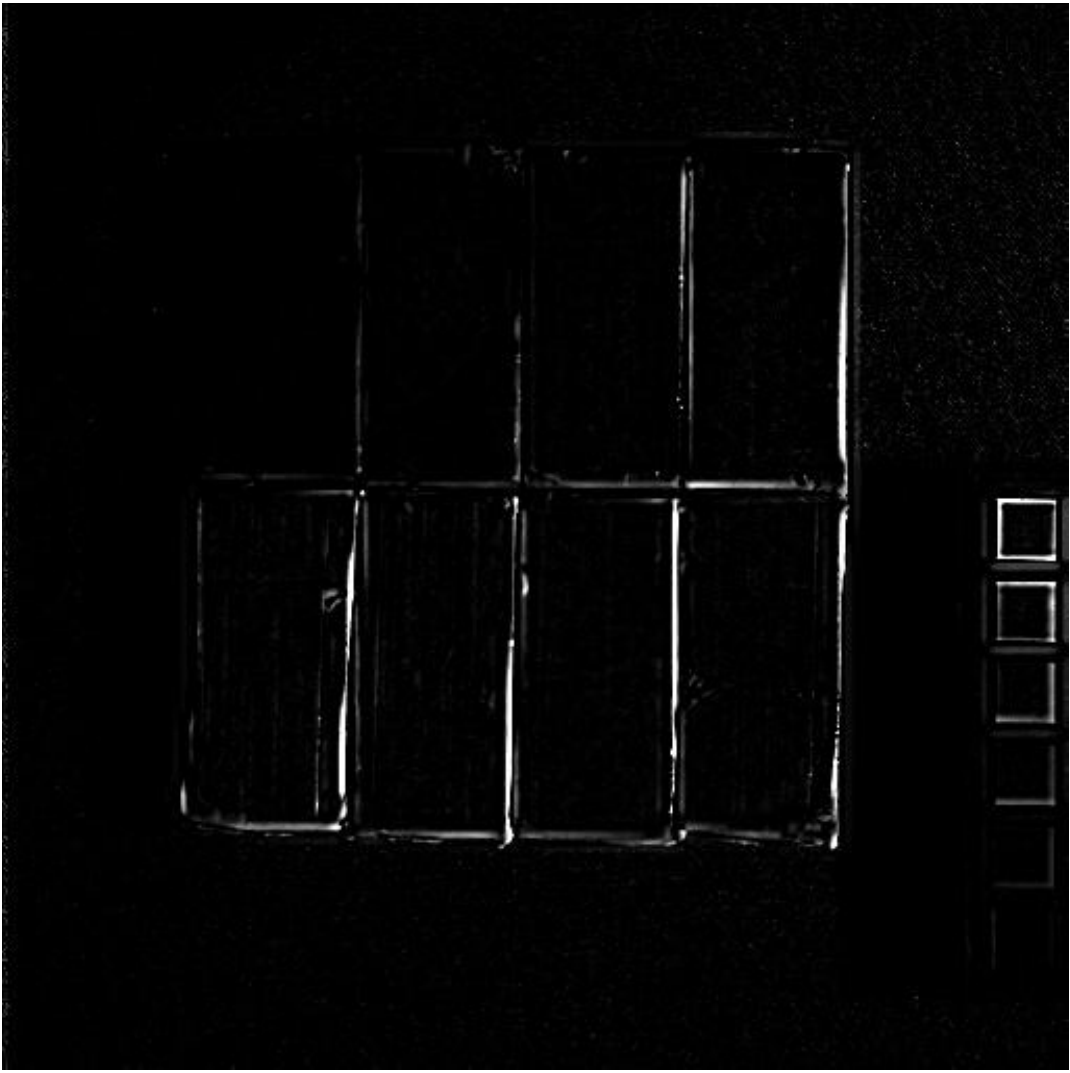}}
			\centering
			{(b)}
		\end{minipage}
		\begin{minipage}{14pt}
			\centering
			{\includegraphics[height=98pt]{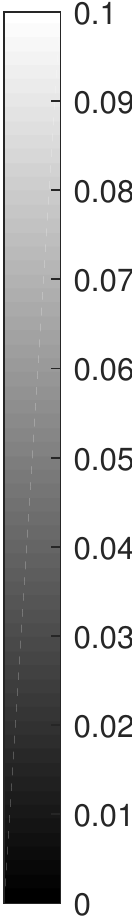}}
			\vspace{2.5pt}
		\end{minipage}
		\caption{The residual maps: (a) $\mathcal{E}= \mathcal{O} - \mathcal{Y}^{U}$ and (b) $\mathcal{E}_{gt}= \mathcal{X} - \mathcal{Y}^{U}$.}\label{resnet}
	\end{center}
\end{figure}

Fig. \ref{resnet} shows a comparison between $\mathcal{E}$ and $\mathcal{E}_{gt}$. From the figure, it is clear that $\mathcal{E}$ (\textit{i.e.}, the details extracted by the proposed approach) and $\mathcal{E}_{gt}$ (\textit{i.e.}, the details extracted by using the reference image) are very close to each other validating the effectiveness of the proposed network design. This result is only obtained thanks to the use of a multi-scale module combining details at two different scales guaranteeing a better spatial detail content in input of the ResNet.


\subsubsection{Loss function}
After obtaining the spectral preserved $\mathcal{Y}^{U}$ image and the spatial preserved $\mathcal{E}$ image from the ResNet fed by the image cube $\mathcal{C}_1$, we subsequently add the two outputs together to get the outcome. Thus, the loss function exploited during the training phase to drive the estimation of the function mapping in (\ref{mapping}) can be defined as

\begin{equation}\label{loss}
\begin{aligned}
\min_{\mathbf{\Theta}}\mathcal{L}=\left\|f_\mathbf{\Theta}(\mathcal{Y}_{HP},\mathcal{Z}_{HP}^{D}, \mathcal{Z}_{HP})+ \mathcal{Y}^{U}-\mathcal{X}\right\|_{F}^{2},
\end{aligned}
\end{equation}
where $f_\mathbf{\Theta}(\cdot)$ is the mapping function that has as input the details at the two different scales used to estimate the spatial preserved image $\mathcal{E}$ and the upsampled LR-HSI $\mathcal{Y}^{U}$. The loss function imposes the similarity between the network output $f_\mathbf{\Theta}(\mathcal{Y}_{HP}^{U},\mathcal{Z}_{HP}^{D},\mathcal{Z}_{HP}) + \mathcal{Y}^{U}$ and the reference (ground-truth) $\mathcal{X}$ image.

%
%
%
%

\subsection{Network Training}\label{Sec-Train}

\subsubsection{Training data} 
\begin{figure}[t]
	\begin{center}
		\begin{minipage}{ 0.155\linewidth}
			{\includegraphics[width=1\linewidth]{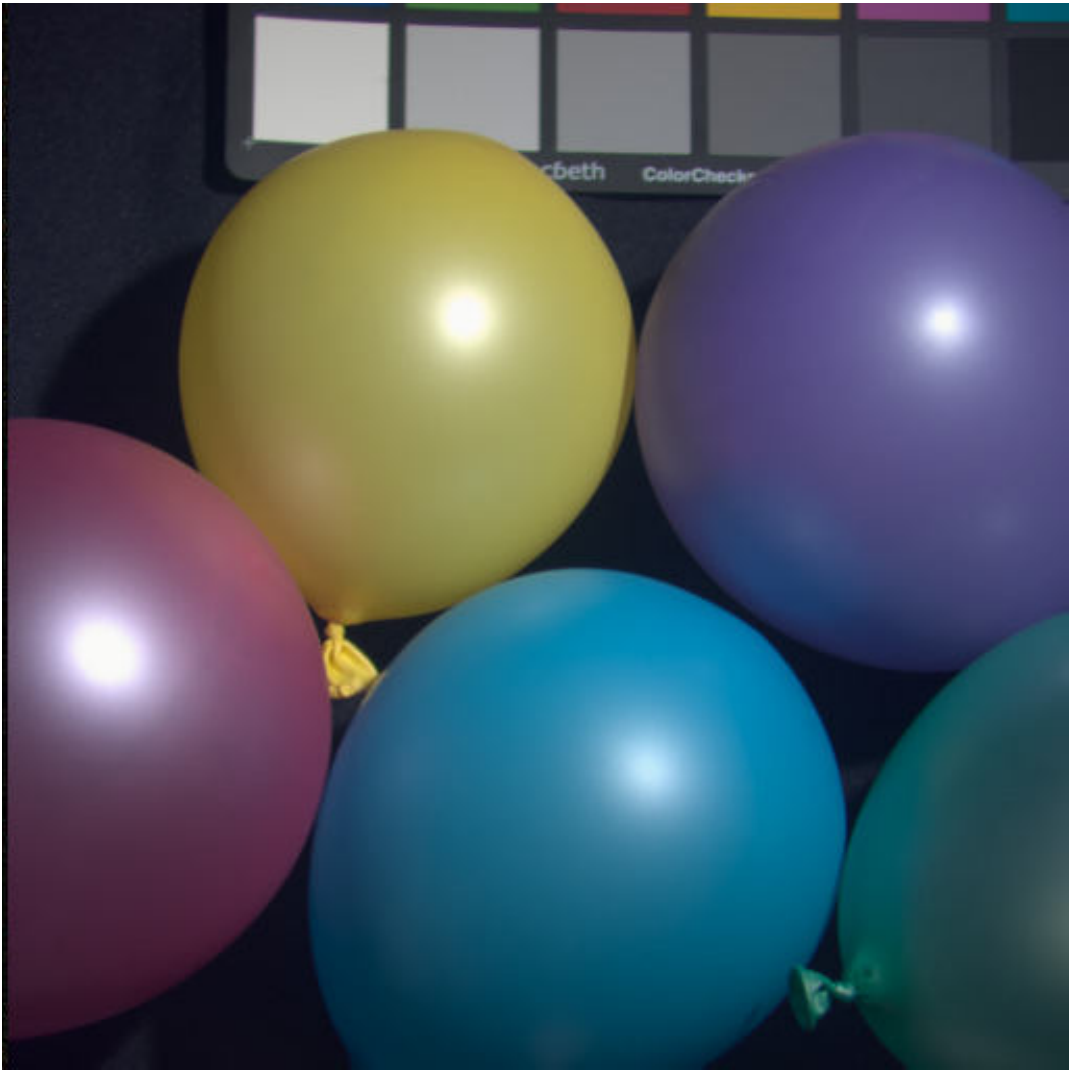}}
			\centering
			{(a)}
		\end{minipage}
		\begin{minipage}{ 0.155\linewidth}
			{\includegraphics[width=1\linewidth]{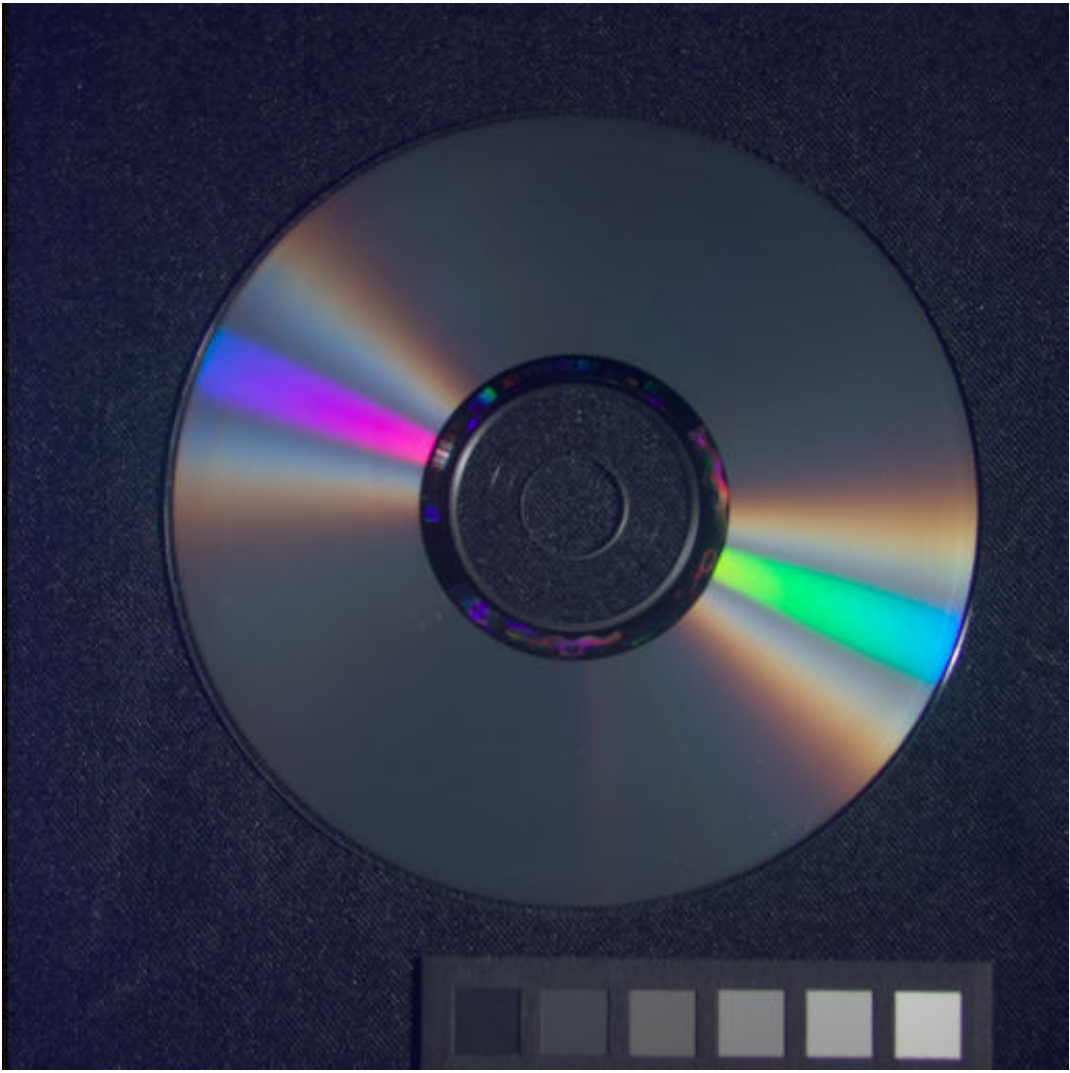}}
			\centering
			{(b)}
		\end{minipage}
		\begin{minipage}{ 0.155\linewidth}
			{\includegraphics[width=1\linewidth]{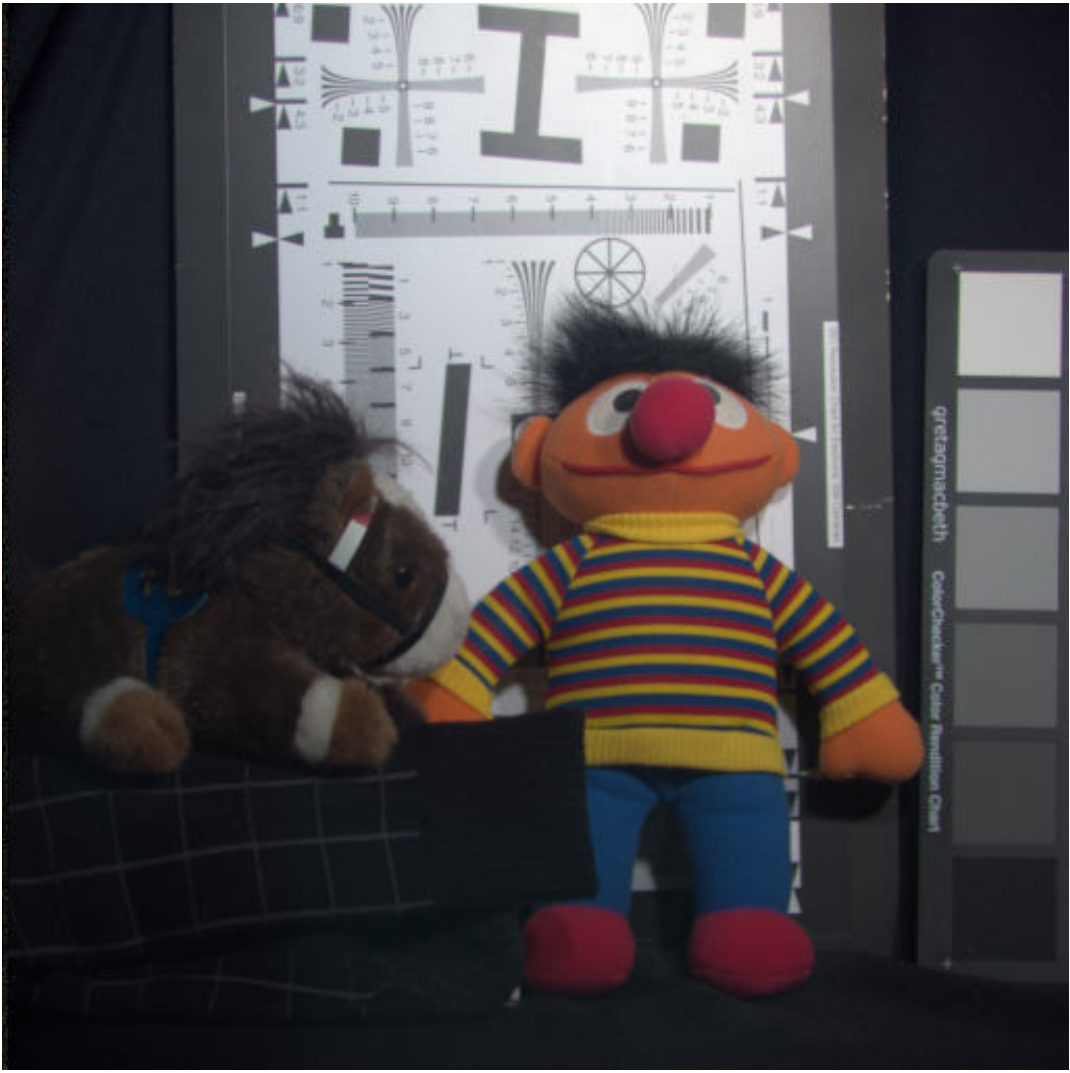}}
			\centering
			{(c)}
		\end{minipage}
		\begin{minipage}{ 0.155\linewidth}
			{\includegraphics[width=1\linewidth]{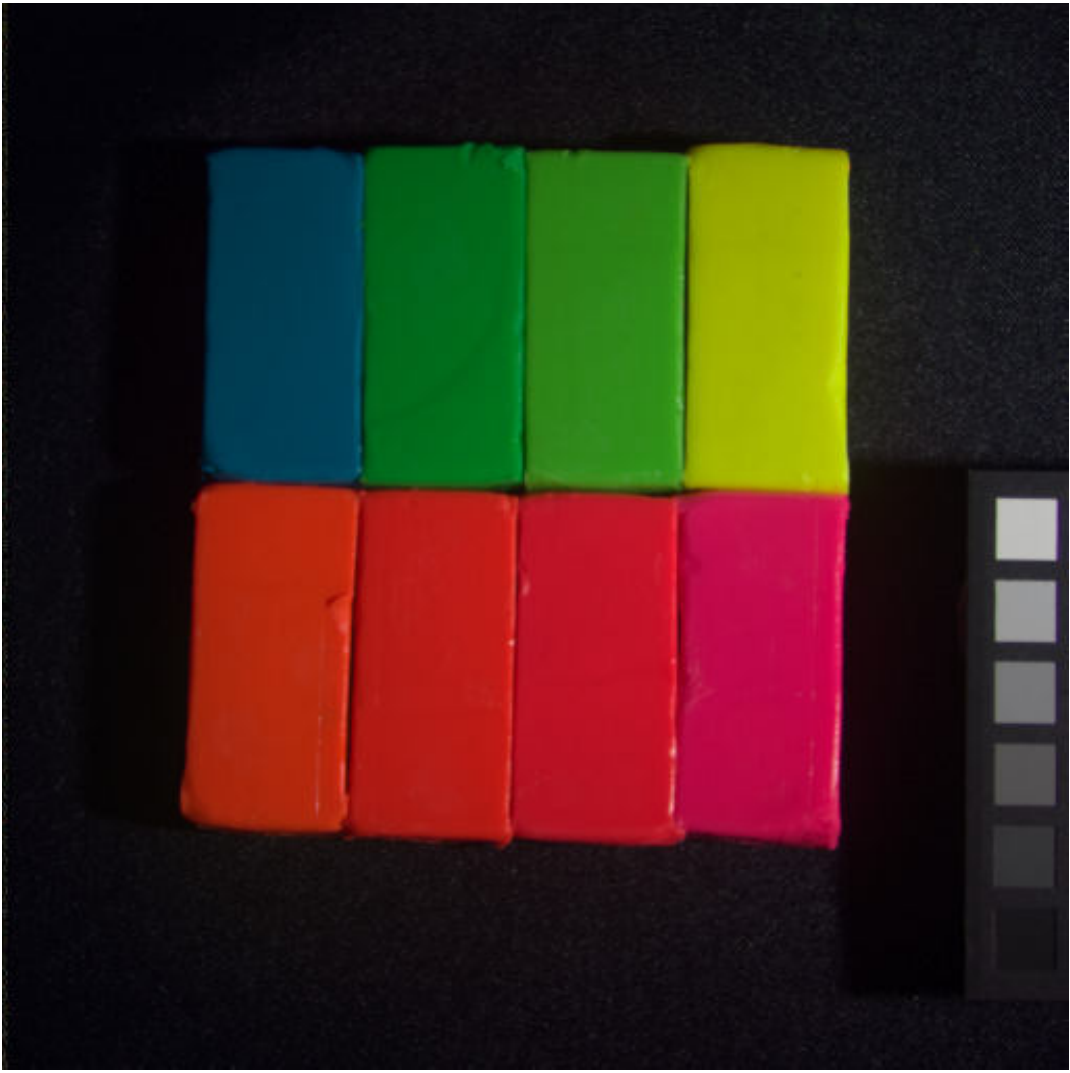}}
			\centering
			{(d)}
		\end{minipage}
		\begin{minipage}{ 0.155\linewidth}
			{\includegraphics[width=1\linewidth]{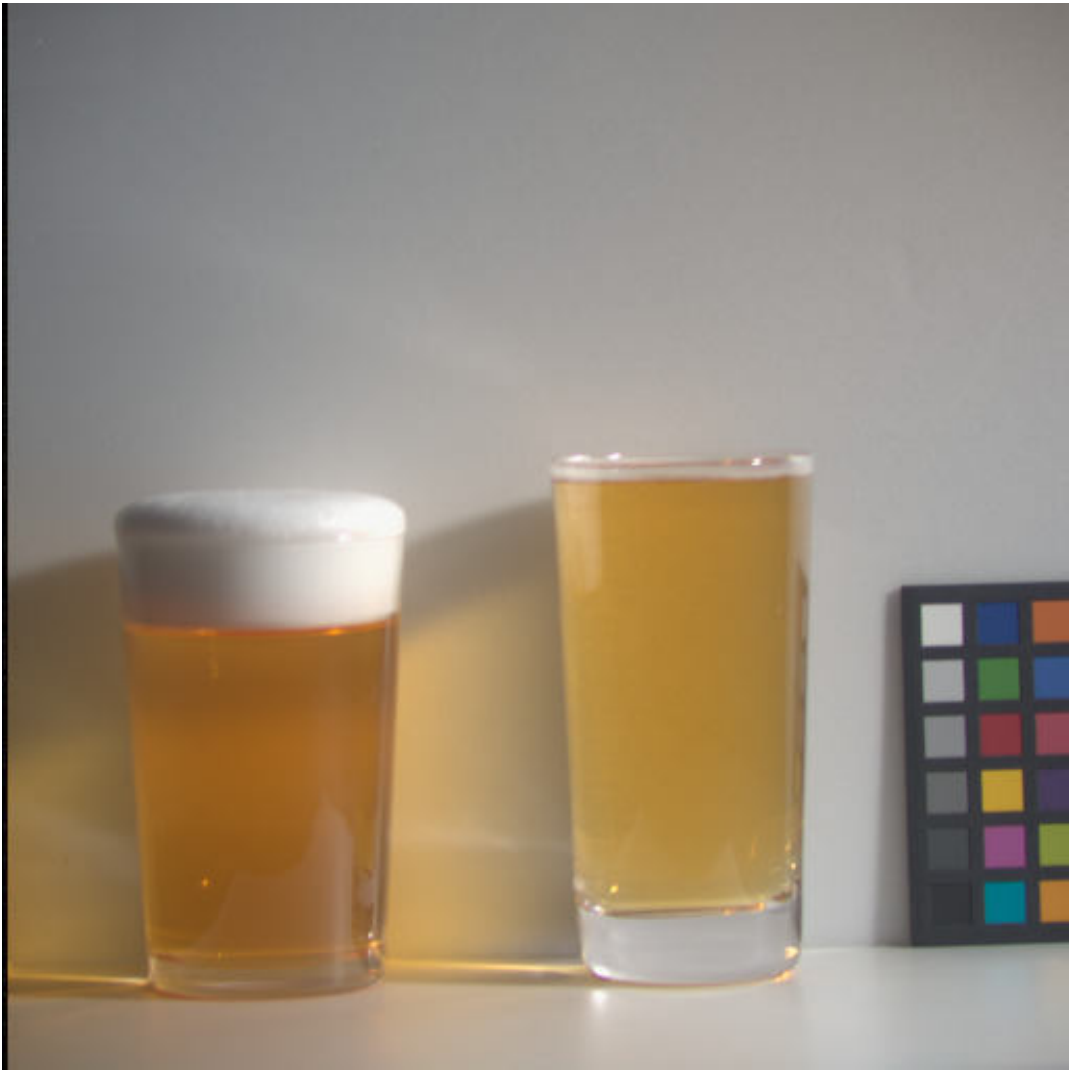}}
			\centering
			{(e)}
		\end{minipage}
		\begin{minipage}{ 0.155\linewidth}
			{\includegraphics[width=1\linewidth]{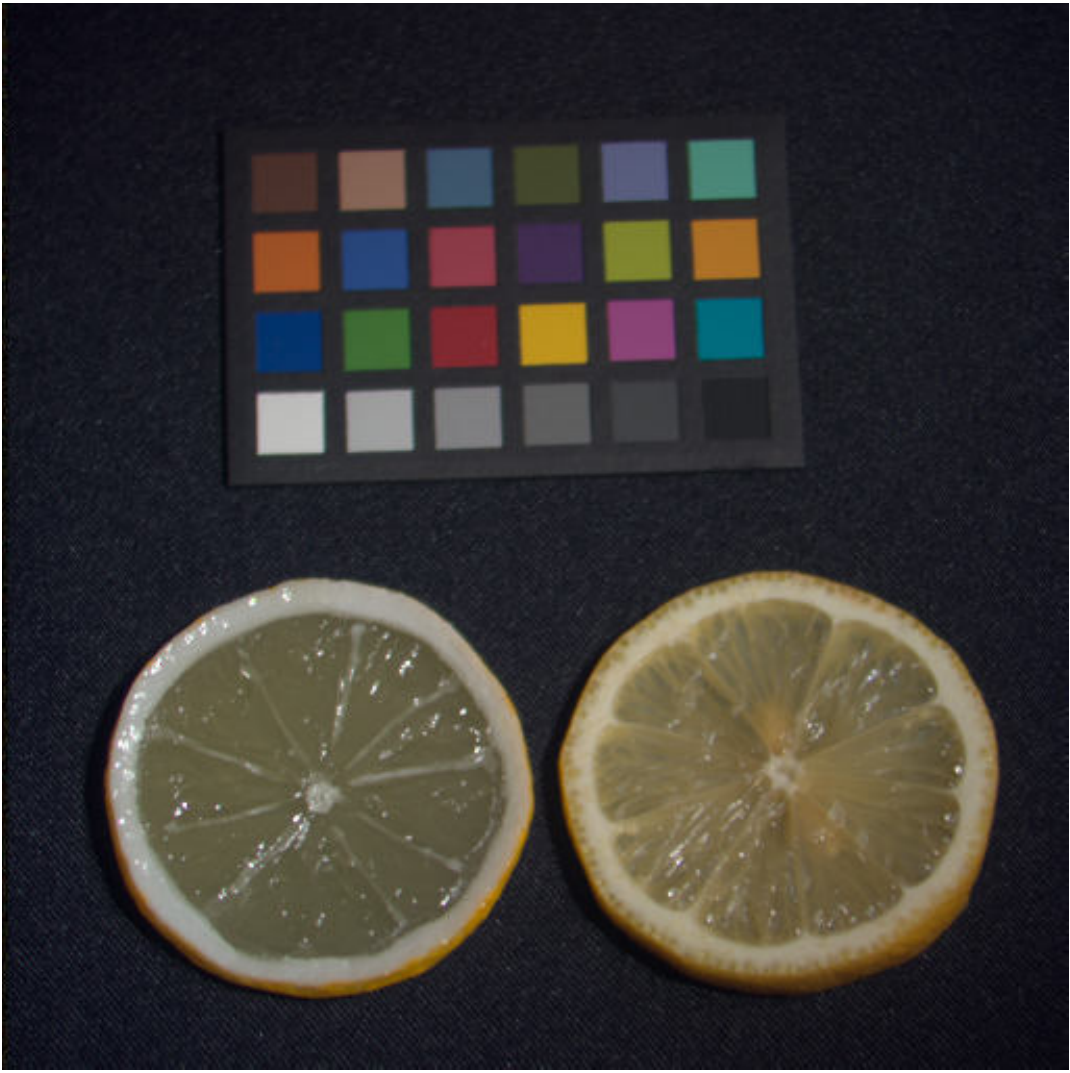}}
			\centering
			{(f)}
		\end{minipage}
		\centering
	\end{center}
	
	\begin{center}
		
		\begin{minipage}{ 0.155\linewidth}
			{\includegraphics[width=1\linewidth]{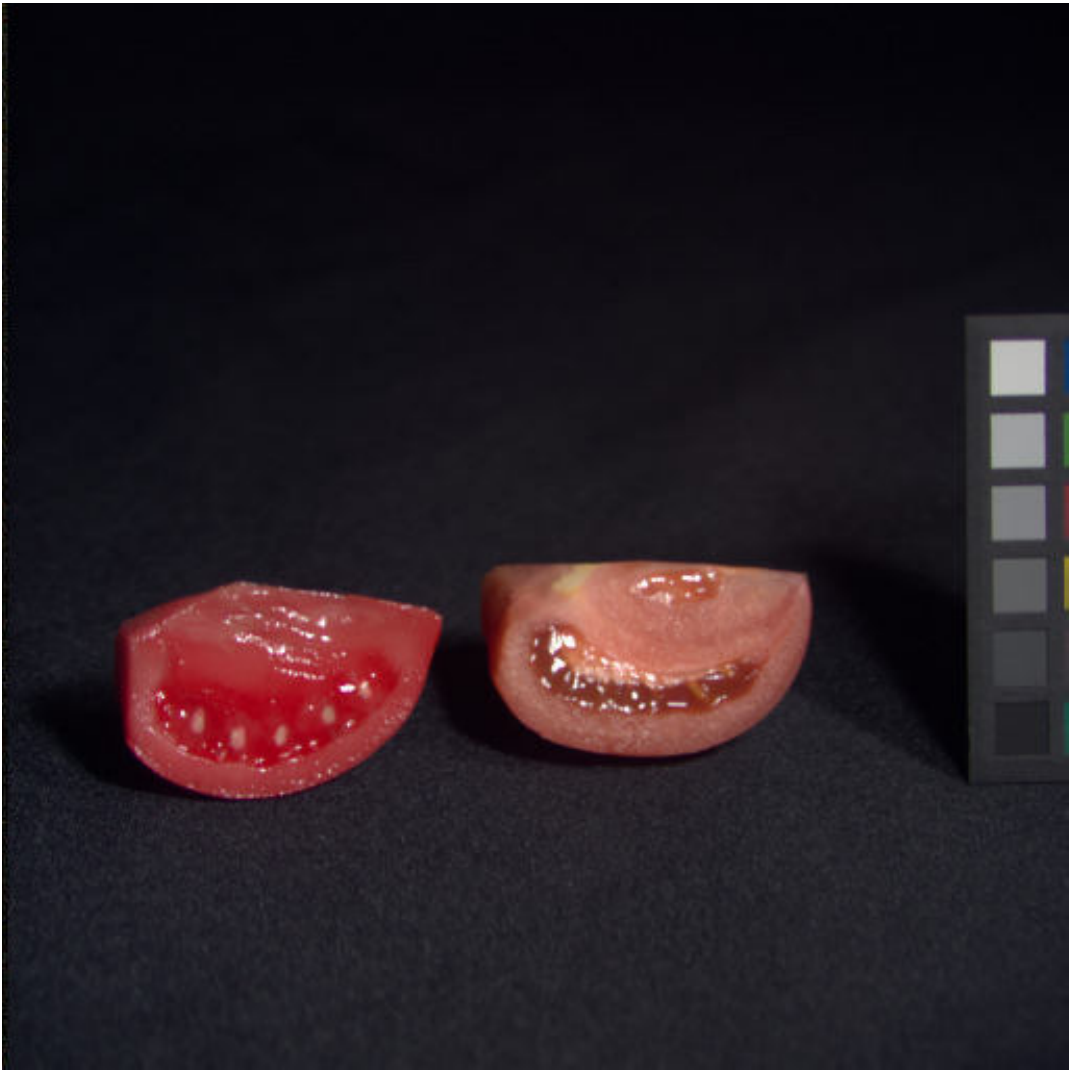}}
			\centering
			{(g)}
		\end{minipage}
		\begin{minipage}{ 0.155\linewidth}
			{\includegraphics[width=1\linewidth]{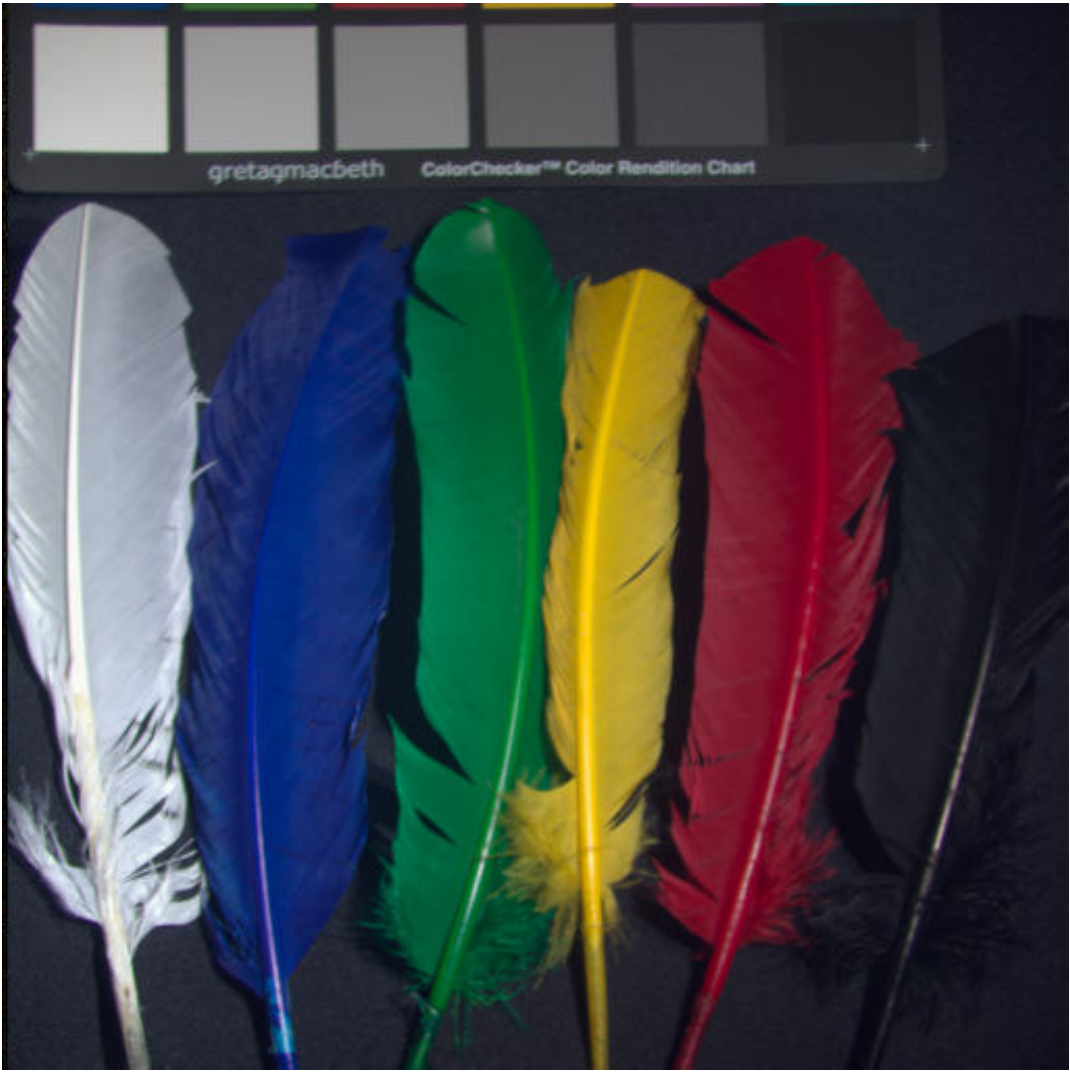}}
			\centering
			{(h)}
		\end{minipage}
		\begin{minipage}{ 0.155\linewidth}
			{\includegraphics[width=1\linewidth]{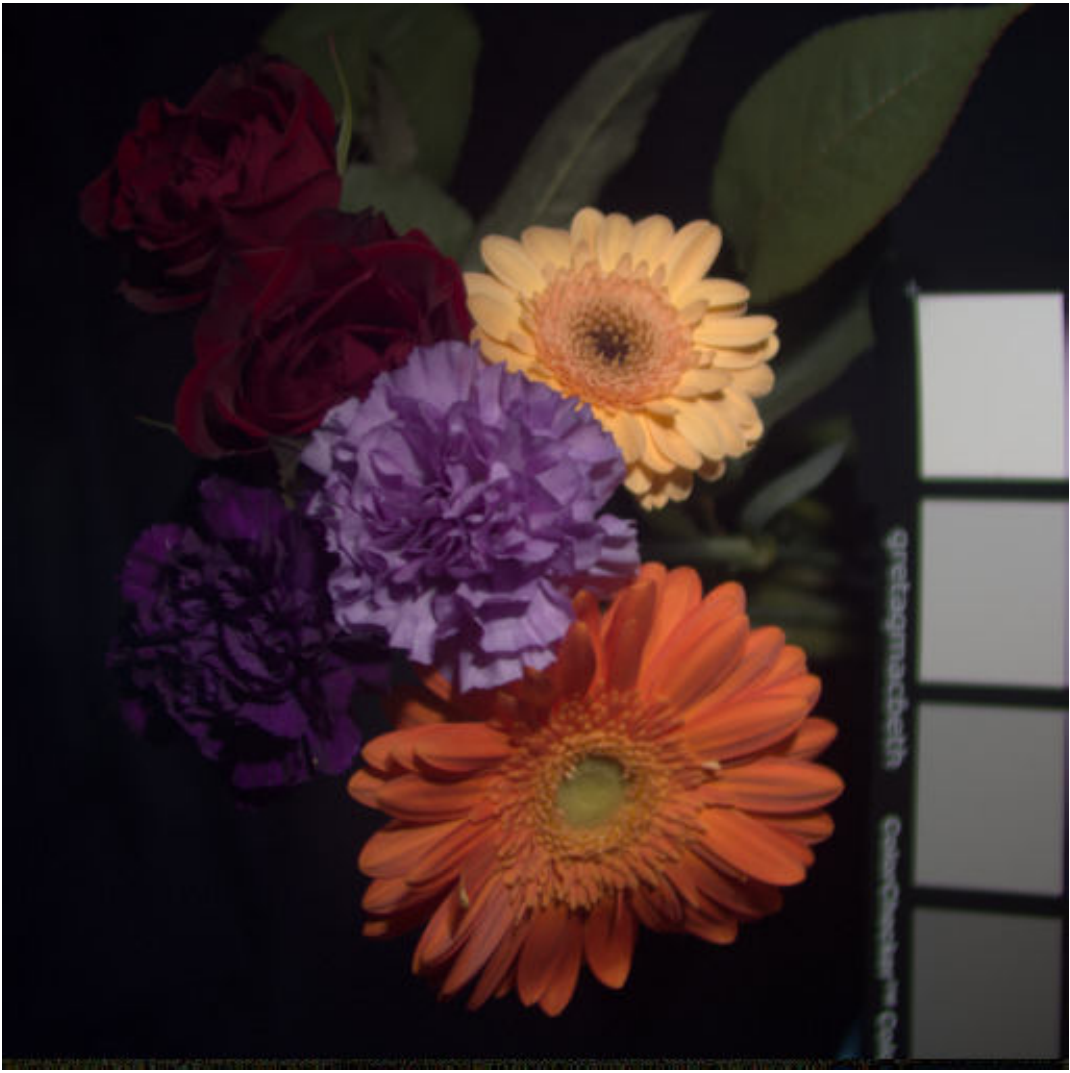}}
			\centering
			{(i)}
		\end{minipage}
		\begin{minipage}{ 0.155\linewidth}
			{\includegraphics[width=1\linewidth]{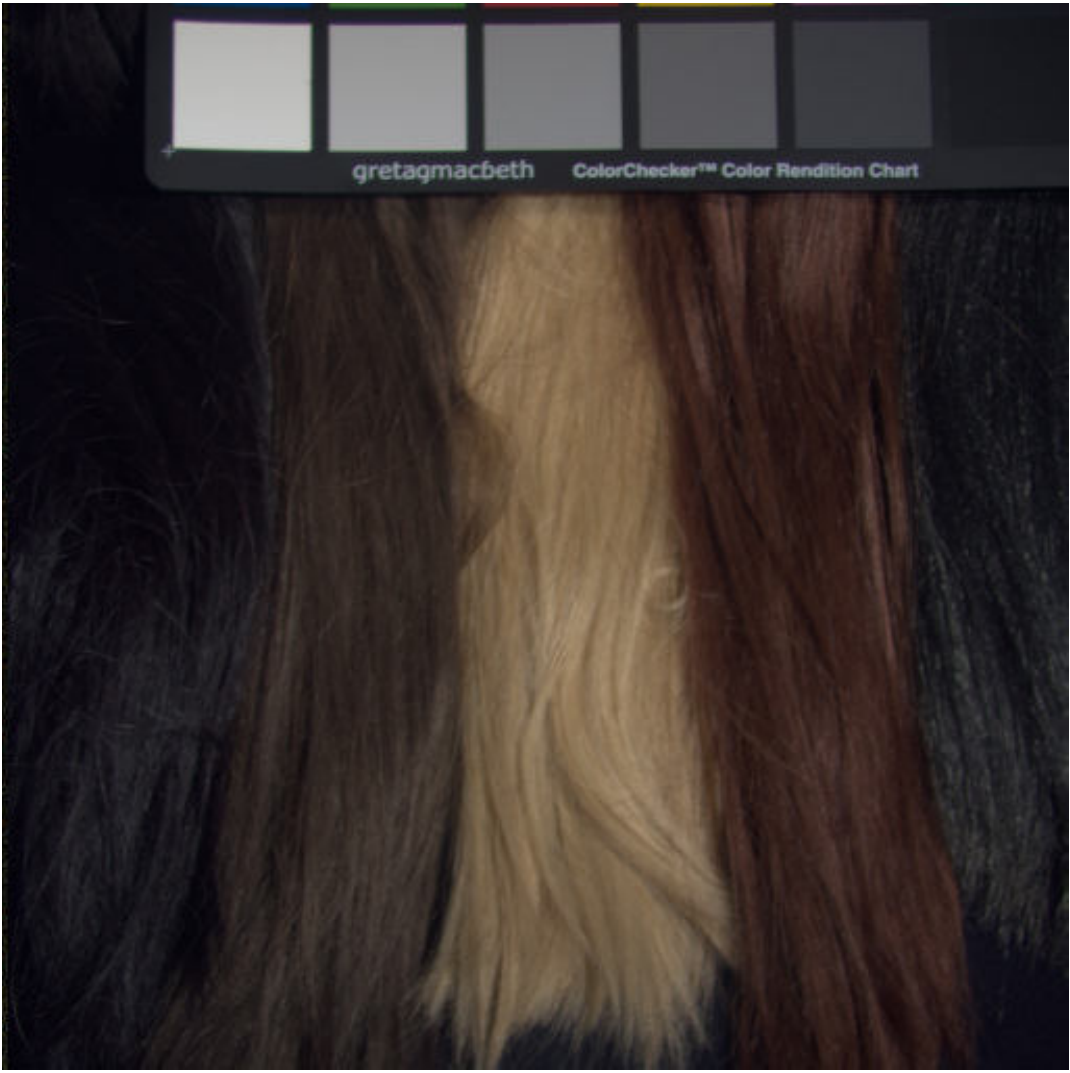}}
			\centering
			{(j)}
		\end{minipage}
		\begin{minipage}{ 0.155\linewidth}
			{\includegraphics[width=1\linewidth]{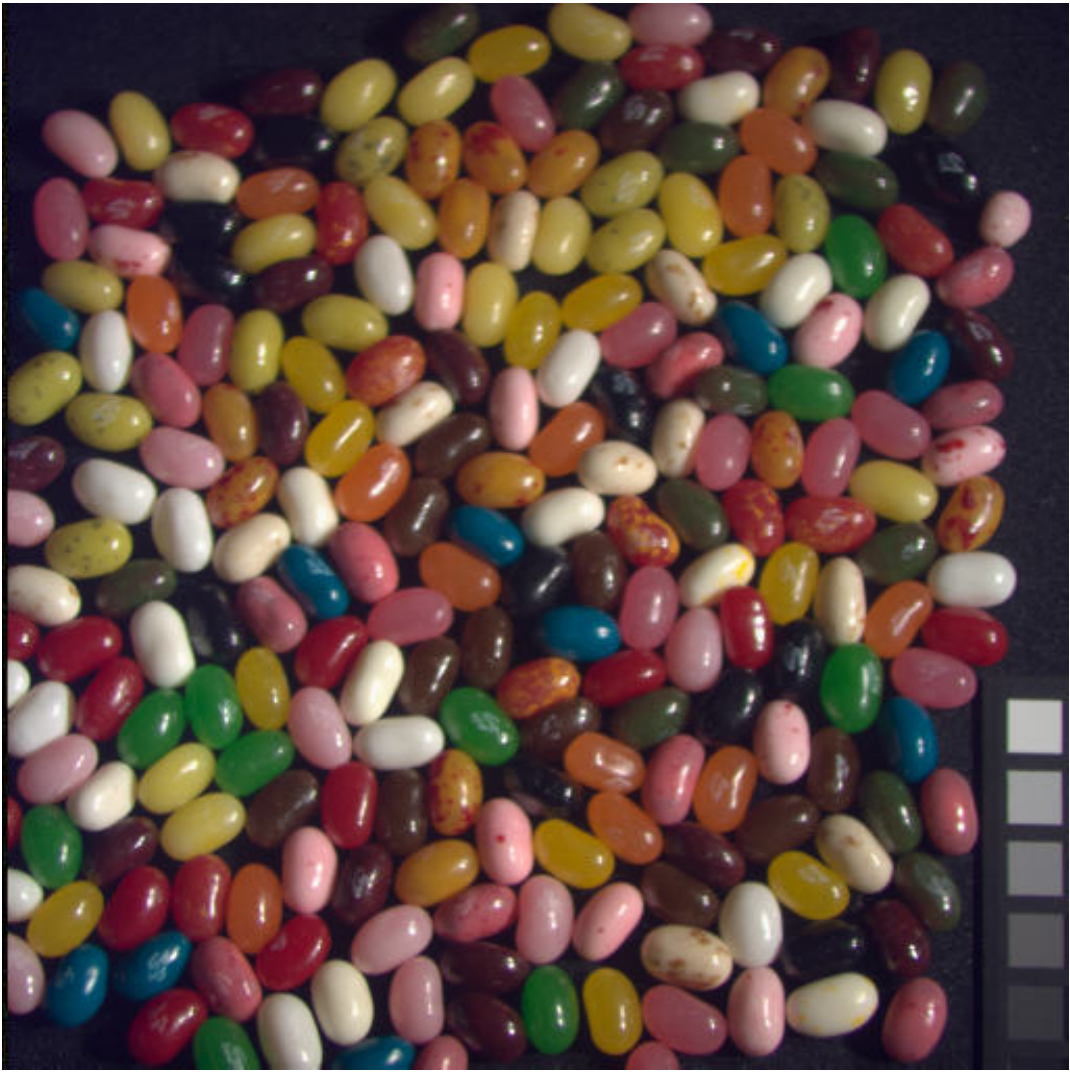}}
			\centering
			{(k)}
		\end{minipage}
		\centering
	\end{center}
	
	\caption{The $11$ testing images from the CAVE dataset. (a) \textit{balloons}, (b) \textit{cd}, (c)\textit{chart and stuffed toy}, (d) \textit{clay}, (e) \textit{fake and real beers}, (f) \textit{fake and real lemon slices}, (g) \textit{fake and real tomatoes},(h)\textit{feathers},(i)\textit{flowers},(j)\textit{hairs},(k)\textit{jelly beans}.} \label{cave_test}
\end{figure}

In the work, we mainly use the CAVE dataset \cite{yasuma2010generalized} for training the network. It contains 32 hyperspectral images with size $512 \times 512$ and $31$ spectral bands. Additionally, each hyperspectral image also has a corresponding RGB image with size $512 \times 512$ and $3$ spectral bands (\textit{i.e.}, the HR-MSI image). We selected $20$ images \footnote{We selected the same 20 images as for the training of the MHFnet.}for training the network, and the other $11$ images to be considered for testing\footnote{One image, \textit{i.e.}, ``Watercolors'', is discarded as it is unavailable for use.}, as done for the MHFnet in \cite{xie2019multispectral}. The CAVE test images are shown in Fig. \ref{cave_test}.

\subsubsection{Data simulation}
We extracted 3920 overlapped patches with a size of $64 \times 64 \times 31$ from the 20 images of the CAVE dataset used as ground-truth, thus forming the HR-HSI patches. Accordingly, the LR-HSI patches are generated starting from the HR-HSI by applying a Gaussian blur with kernel size equal to $3\times 3$ and standard deviation equal to 0.5 and then downsampling the blurred patches to the size of $16 \times 16$, \textit{i.e.,} with a downsampling factor of 4. Moreover, the HR-MSI patches (\textit{i.e.,}, the RGB patches) are generated similarly as for the HR-HSI patches, but using the corresponding (already available) RGB data. Thus, other 3920 patches of size of $64 \times 64 \times 3$ are available to represent the HR-MSI. Following these indications, the patches for the training phase are the $80\%$ of the whole dataset and the rest (\textit{i.e.}, the $20\%$) is used for the testing phase.

\subsubsection{Training platform and parameters setting}
\begin{figure}[t]
	\begin{center}
		
		{\includegraphics[width=0.9\linewidth]{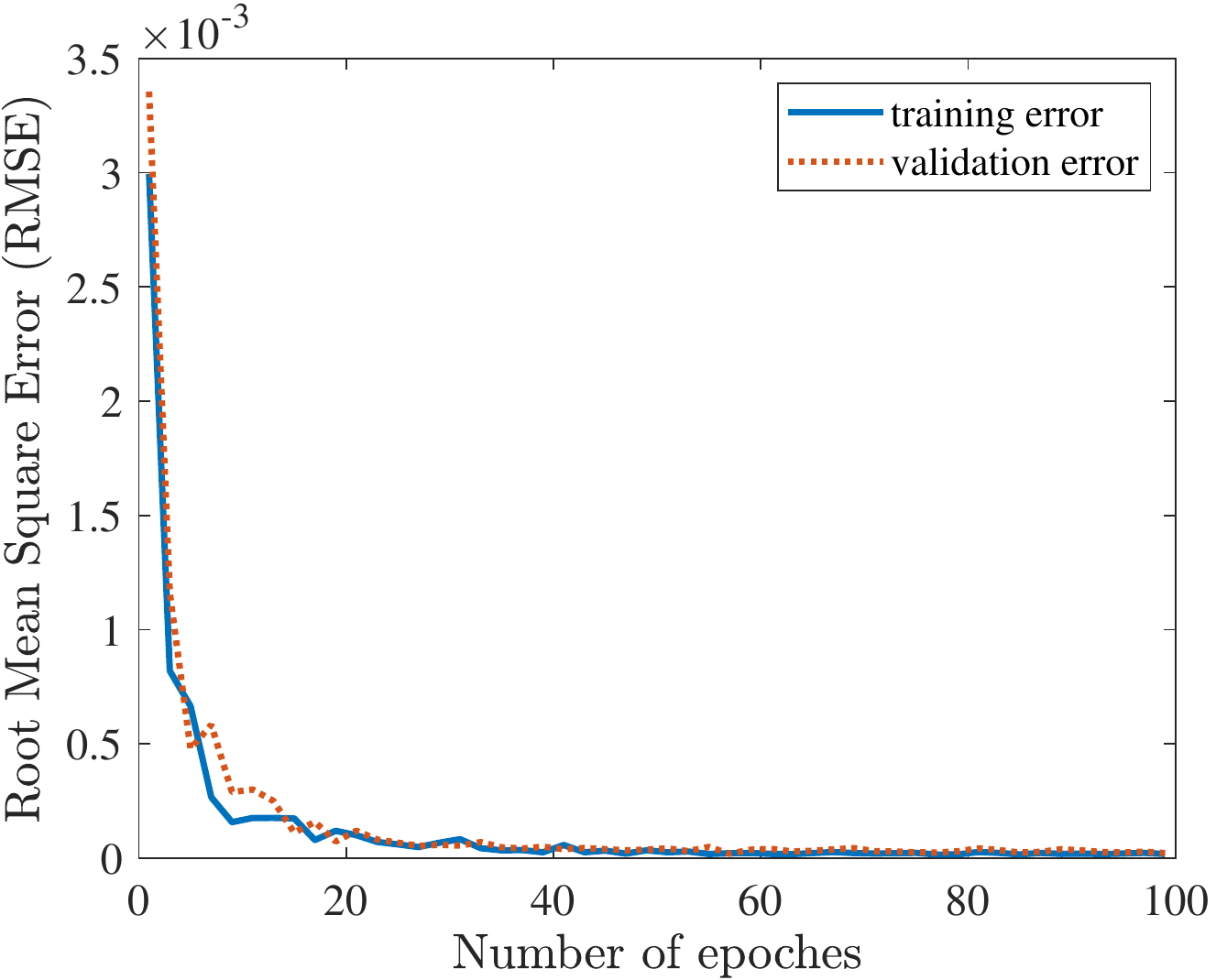}}
		\caption{Training and validation errors for the proposed HSRnet.}\label{error}
	\end{center}
\end{figure}

The proposed network is trained on Python 3.7.4 with Tensorflow 1.14.0 and Linux operating system with NVIDA GPU GeForce GTX 2080Ti. We use Adam optimizer with a learning rate equal to 0.0001 in order to minimize the loss function (\ref{loss}) by 100,000 iterations and 32 batches. The ResNet block in our network architecture is crucial. Indeed, we use 6 ResNet blocks (each one with two layers and 64 kernels of size $3\times 3$ for each layer. See Fig. \ref{structure}). Fig. \ref{error} shows the training and validation errors of the proposed HSRnet confirming the convergence of the proposed convolutional neural network using the above-mentioned parameters setting.

\section{Experimental Results}\label{exp}

In this section, we compare the proposed HSRnet with several state-of-the-art methods for the hyperspectral super-resolution problem. In particular, the benchmark consists of the CNMF method\footnote{http://naotoyokoya.com/Download.html} \cite{CNMF}, the FUSE approach\footnote{http://wei.perso.enseeiht.fr/publications.html} \cite{FUSE}, the GLP-HS method\footnote{http://openremotesensing.net/knowledgebase/hyperspectral-and-multispectral-data-fusion/}\cite{GLP-HS}, the LTTR technique\footnote{https://github.com/renweidian}\cite{LTTR}, the LTMR approach\footnote{https://github.com/renweidian}\cite{LTMR}, the MHFnet\footnote{https://github.com/XieQi2015/MHF-net} \cite{xie2019multispectral}, and the proposed HSRnet approach. For a fair comparison, the MHFnet is trained on the same training data as the proposed approach. Furthermore, the batch size and the training iterations of the MHFnet are set to 32 and 100,000, respectively, as for the proposed approach.

Two widely used benchmark datasets, \textit{i.e.,} the CAVE database\footnote{http://www.cs.columbia.edu/CAVE/databases/multispectral/} \cite{yasuma2010generalized} and the Harvard database\footnote{http://vision.seas.harvard.edu/hyperspec/download.html}\cite{chakrabarti2011statistics}, are selected.

For quantitative evaluation, we adopt four quality indexes (QIs), \textit{i.e.,} the peak signal-to-noise ratio (PSNR), the spectral angle mapper (SAM) \cite{yuhas1993determination}, the erreur relative globale adimensionnelle de synth\`{e}se (ERGAS) \cite{wald2002data}, and the structure similarity (SSIM) \cite{wang2004image}.
The SAM measures the average angle between the spectral vectors of the target and of the reference image. Instead, the ERGAS represents the fidelity of the image based on the weighted sum of mean squared errors. The ideal value in both the cases is zero. The lower the index, the better the quality. Finally, PSNR and SSIM are widely used to evaluate the similarity between the target and the reference image. The higher the index, the better the quality. The ideal value for SSIM is one.

\subsection{Results on CAVE Dataset}\label{caveexp}

\begin{table}[t]
	\centering\renewcommand\arraystretch{1}\setlength{\tabcolsep}{6pt}\footnotesize
	\caption{Average QIs and related standard deviations of the results on 100 patches extracted from the testing images on the CAVE dataset. The best values are highlighted in boldface.}
	\begin{tabular}{l|c|c|c|c}
		\Xhline{1.2pt}
		Method & PSNR & SAM & ERGAS & SSIM \\ \hline
		CNMF& 	31.4$\pm$3.3 & 5.95$\pm$4.0 & 8.19$\pm$19.2 	& 0.96$\pm$0.04 \\ 
		FUSE&	28.9$\pm$3.3 & 10.36$\pm$6.4& 7.23$\pm$6.0 		& 0.91$\pm$0.07 \\
		GLP-HS& 30.8$\pm$4.0 	 & 6.28$\pm$3.9 & 5.9$\pm$4.8 		& 0.94$\pm$0.05 \\ 
		LTTR& 	31.1$\pm$3.6 & 7.36$\pm$3.5 & 6.55$\pm$5.3 		& 0.94$\pm$0.04 \\
		LTMR& 	30.6$\pm$3.5 & 7.61$\pm$3.6 & 7.25$\pm$6.8 		& 0.93$\pm$0.04 \\ 
		MHFnet& 35.1$\pm$5.9 & 7.29$\pm$7.2 & 30.7$\pm$146.4 	& 0.96$\pm$0.03 \\ 
		HSRnet& \textbf{38.2}$\pm$5.3 & \textbf{2.94}$\pm$1.8 & \textbf{2.99}$\pm$3.6 		& \textbf{0.99}$\pm$0.01 \\ \hline
		Best value& +$ \infty $ & 0 & 0 & 1 \\ \Xhline{1.2pt}
	\end{tabular}
	\label{cave-ave}
\end{table}

In order to point out the effectiveness of all the methods on different kinds of scenarios, we divide first the remaining 11 testing images on the CAVE dataset into small patches of size $128 \times 128$. Then, 100 patches are randomly selected. We exhibit the average QIs and corresponding standard deviations of the results for the different methods on these patches in Table \ref{cave-ave}.
From Table \ref{cave-ave}, we can find that the proposed HSRnet significantly outperforms the compared methods.
In particular, the SAM value of our method is much lower than that of the compared approaches (about the half with respect to the best compared method). This is in agreement with our previously developed analysis, namely that the proposed HSRnet is able to preserve the spectral features of the acquired scene.

\begin{table}[t]
	\centering\renewcommand\arraystretch{1}\setlength{\tabcolsep}{6pt}\footnotesize
	\caption{Average QIs and related standard deviations of the results on 11 testing images on the CAVE datasets. The best values are highlighted in boldface.}
	\begin{tabular}{l|c|c|c|c}
		\Xhline{1.2pt}
		Method 	& PSNR & SAM & ERGAS & SSIM \\ \hline
		CNMF 	& 32.2$\pm$4.5 	& 14.96$\pm$5.2 	& 8.79$\pm$4.8 	& 0.911$\pm$0.04 \\ 
		FUSE 	& 31.5$\pm$2.5 	& 17.71$\pm$7.8 	& 9.07$\pm$6.2 	& 0.870$\pm$0.05 \\ 
		GLP-HS 	& 35.4$\pm$2.7	& 7.91$\pm$3.0 	& 5.61$\pm$3.6 	& 0.946$\pm$0.02 \\ 
		LTTR 	& 36.8$\pm$2.8 	& 6.65$\pm$2.5 	& 5.66$\pm$2.8 	& 0.957$\pm$0.03 \\ 
		LTMR 	& 36.2$\pm$2.7 	& 7.66$\pm$2.9 	& 5.70$\pm$2.7 	& 0.949$\pm$0.03 \\ 
		MHFnet 	& 43.3$\pm$2.8 	& 4.34$\pm$1.5 	& 2.33$\pm$1.4 	& 0.989$\pm$0.01 \\ 
		HSRnet 	& \textbf{44.0}$\pm$2.9 	& \textbf{3.09}$\pm$1.0	& \textbf{1.93}$\pm$1.0 	& \textbf{0.992}$\pm$0.00 \\ \hline
		Best value& +$ \infty $ & 0 & 0 & 1 \\ \Xhline{1.2pt}
	\end{tabular}
	\label{cave11-ave}
\end{table}

\begin{table}[t]
	\setlength{\tabcolsep}{2pt}
	\caption{QIs of the results by different methods and the running times on (a) \emph{balloons}, (d)  \emph{clay}, and (e) \emph{fake and real beers} on the CAVE dataset. G indicates that the method is running on the GPU device, while C denotes the use of the CPU. The best values are highlighted in boldface.}
	\centering{
		\begin{tabular}{l|ccccccc}
			\Xhline{1.2pt}
			\multicolumn{8}{c}{(a) 512 $\times$ 512} \\ \hline
			Method & CNMF & FUSE & GLPHS & LTTR & LTMR & MHFnet & HSRnet\\
			PSNR & 31.26 & 32.02 & 39.73 & 39.13 & 39.21 & 45.24 & \textbf{49.51} \\ 
			SAM & 9.89 & 10.56 & 3.29 & 3.29 & 4.15 & 2.91 & \textbf{1.64} \\ 
			ERGAS & 4.57 & 4.30 & 1.81 & 2.11 & 2.11 & 1.06 & \textbf{0.59} \\ 
			SSIM & 0.926 & 0.928 & 0.975 & 0.980 & 0.980 & 0.992 & \textbf{0.996} \\ 
			\Xhline{1.2pt}
			\multicolumn{8}{c}{(d) 512 $\times$ 512} \\ \hline
			Method & CNMF & FUSE & GLPHS & LTTR & LTMR & MHFnet & HSRnet\\
			PSNR & 31.35 & 32.18 & 37.59 & 37.09 & 37.06 & 43.09 & \textbf{45.06} \\ 
			SAM & 17.56 & 17.68 & 10.68 & 7.00 & 7.64 & 7.71 & \textbf{4.60} \\ 
			ERGAS & 7.19 & 9.25 & 4.78 & 5.20 & 5.23 & 2.94 & \textbf{2.06} \\ 
			SSIM & 0.926 & 0.900 & 0.963 & 0.976 & 0.973 & 0.986 & \textbf{0.993} \\ 
			\Xhline{1.2pt}
			\multicolumn{8}{c}{(e) 512 $\times$ 512} \\ \hline
			Method & CNMF & FUSE & GLPHS & LTTR & LTMR & MHFnet & HSRnet\\
			PSNR & 30.41 & 35.98 & 37.57 & 38.99 & 38.66 & 41.97 & \textbf{45.97} \\ 
			SAM & 4.81 & 3.97 & 1.25 & 1.97 & 2.18 & 1.62 & \textbf{0.94} \\ 
			ERGAS & 2.19 & 1.70 & 1.23 & 1.25 & 1.26 & 0.76 & \textbf{0.42} \\ 
			SSIM & 0.965 & 0.962 & 0.969 & 0.975 & 0.972 & 0.986 & \textbf{0.992} \\ \hline
			\makecell[c]{Average \\ time(s)}&\footnotesize{27.1(C)}&\footnotesize	{1.9(C)}&\footnotesize	{4.6(C)}&\footnotesize {767.8(C)}&\footnotesize {271.3(C)}&\footnotesize	{4.4(G)}&\footnotesize {\textbf{1.7}(G)}\\
			
			\Xhline{1.2pt}
	\end{tabular}}
	\label{qresult-4CAVE}
\end{table}

\begin{figure*}[htb]
	\centering
	\begin{minipage}[t]{0.94\linewidth}
		\begin{minipage}[t]{0.12\linewidth}
			{\includegraphics[width=1\linewidth]{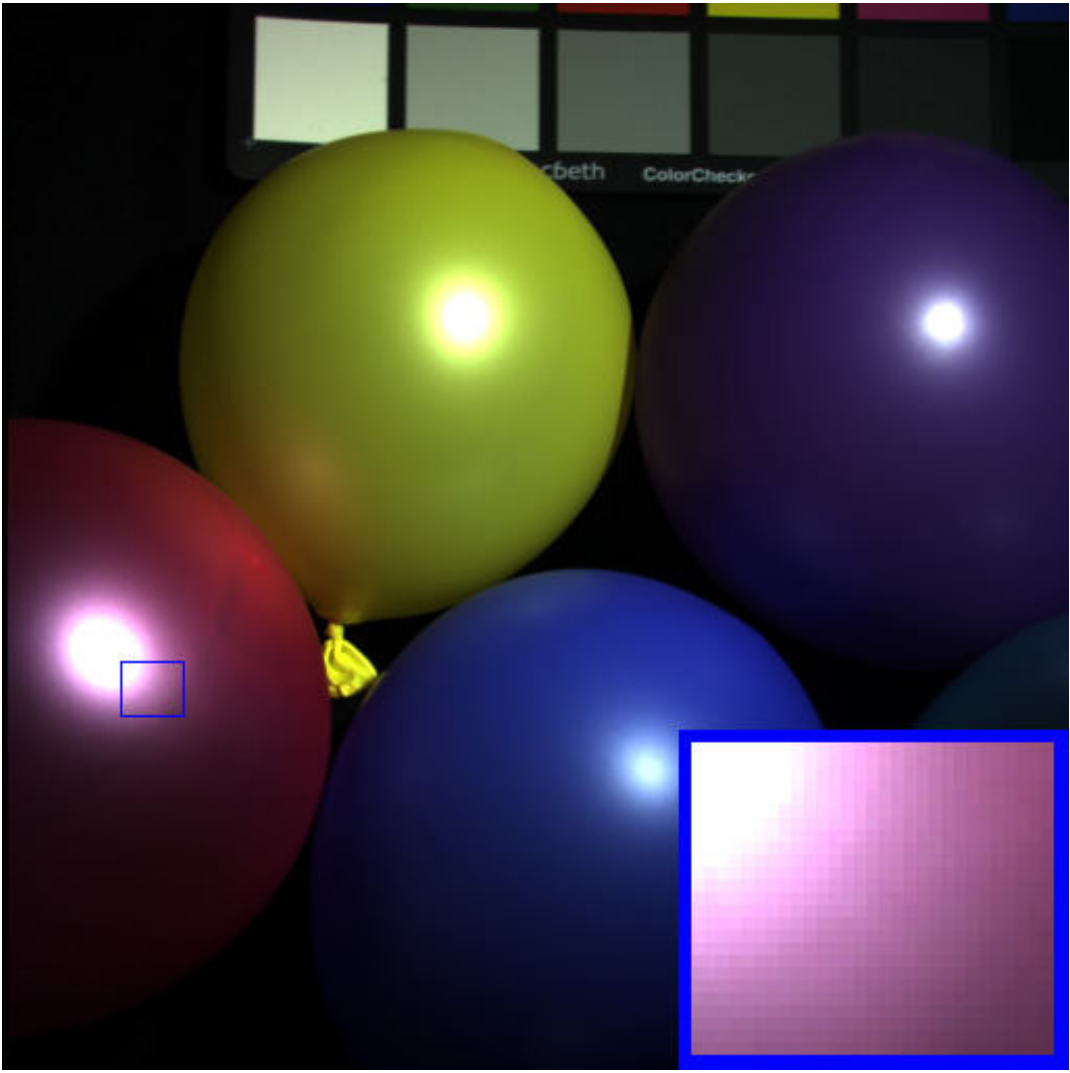}}
			{\includegraphics[width=1\linewidth]{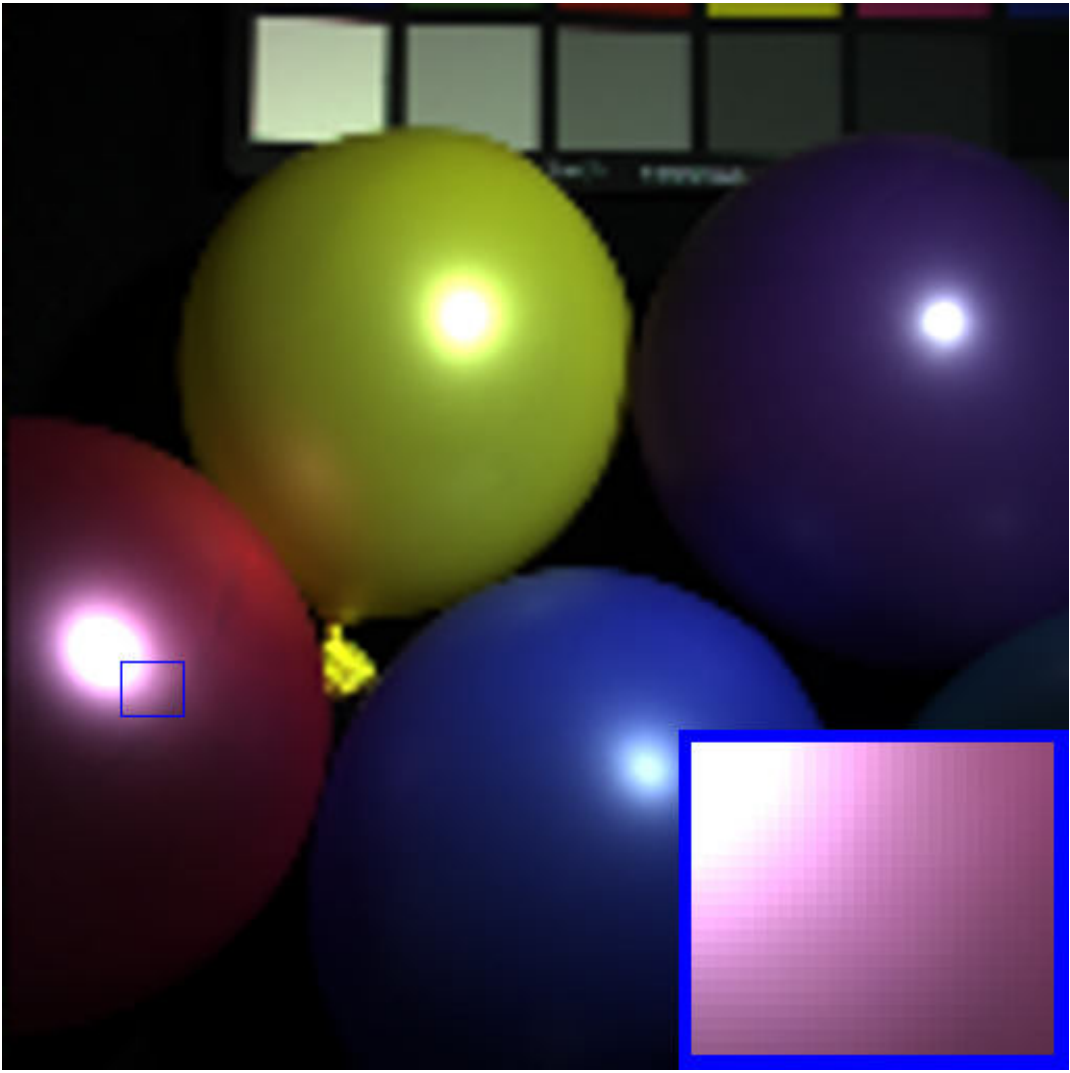}}
			\centering
			
		\end{minipage}
		\begin{minipage}[t]{0.12\linewidth}
			{\includegraphics[width=1\linewidth]{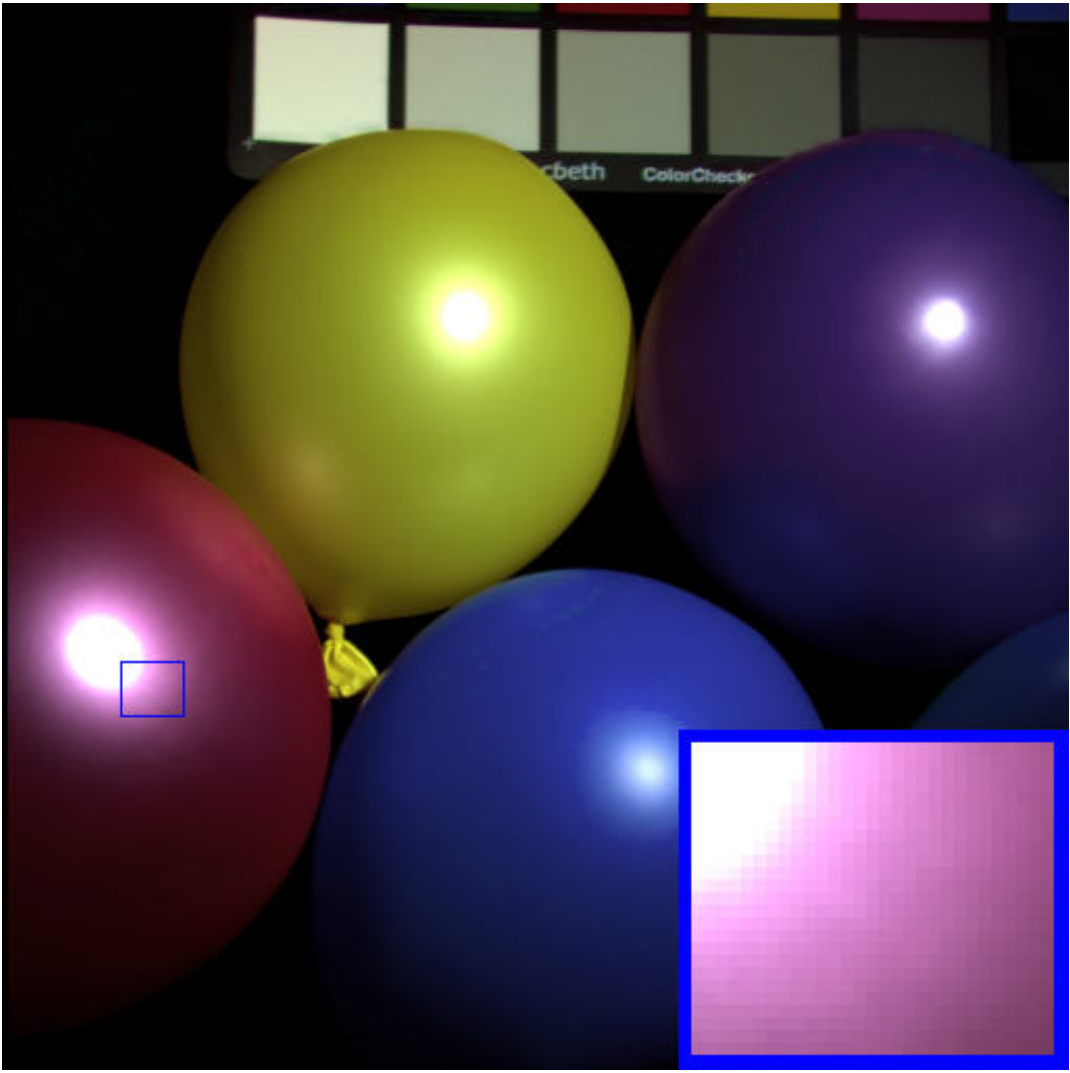}}
			{\includegraphics[width=1\linewidth]{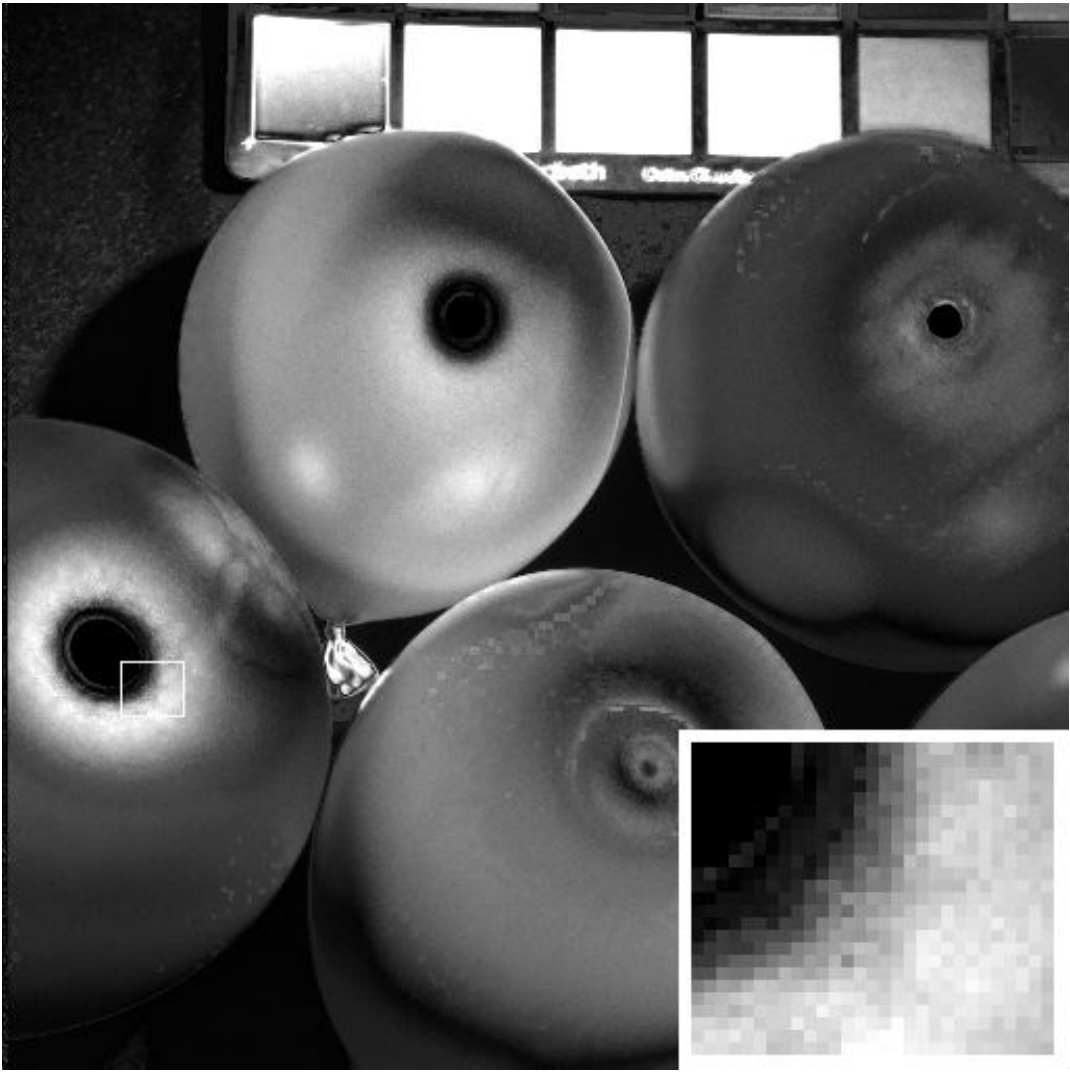}}
			\centering
			
		\end{minipage}
		\begin{minipage}[t]{0.12\linewidth}
			{\includegraphics[width=1\linewidth]{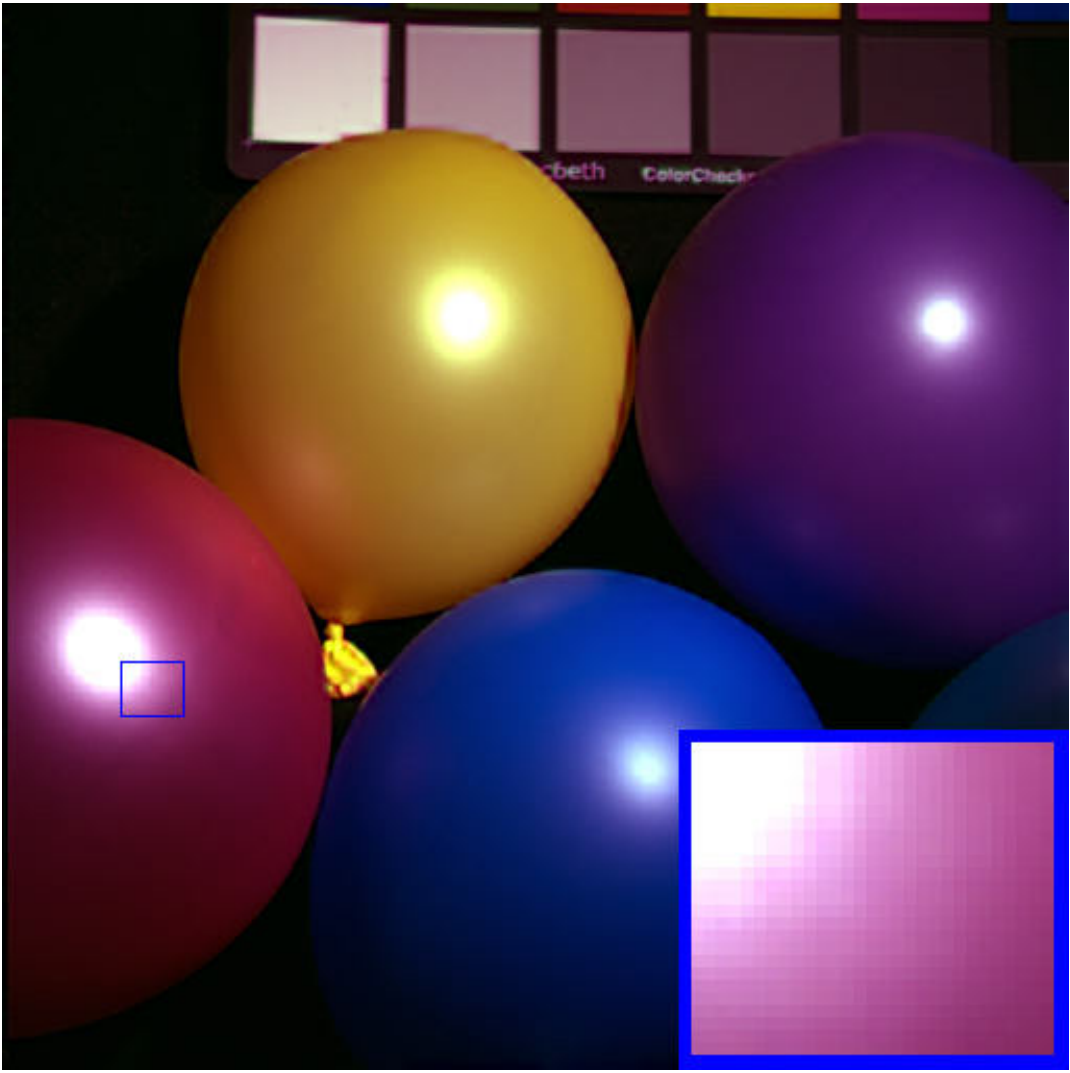}}
			{\includegraphics[width=1\linewidth]{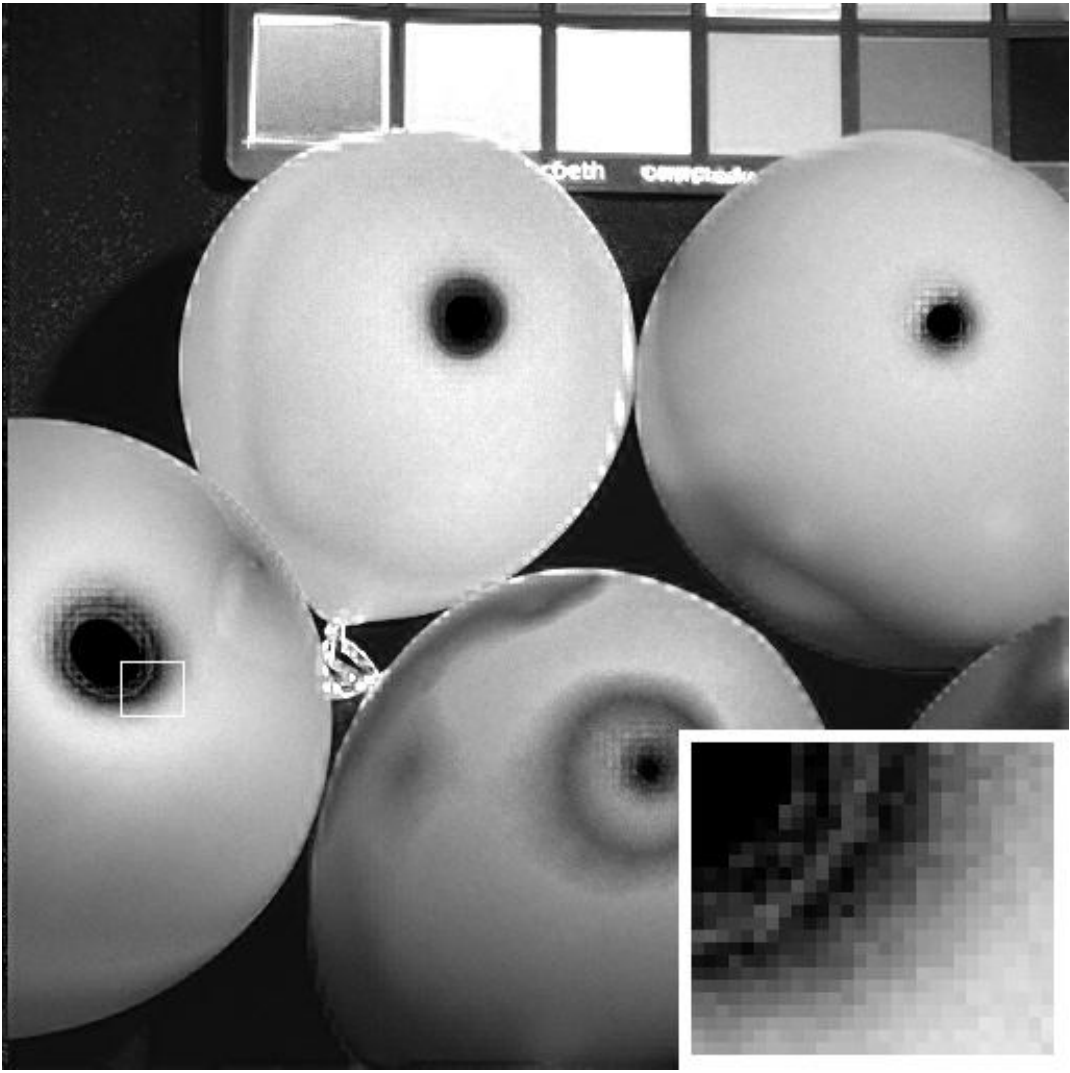}}
			\centering
			
		\end{minipage}
		\begin{minipage}[t]{0.12\linewidth}
			{\includegraphics[width=1\linewidth]{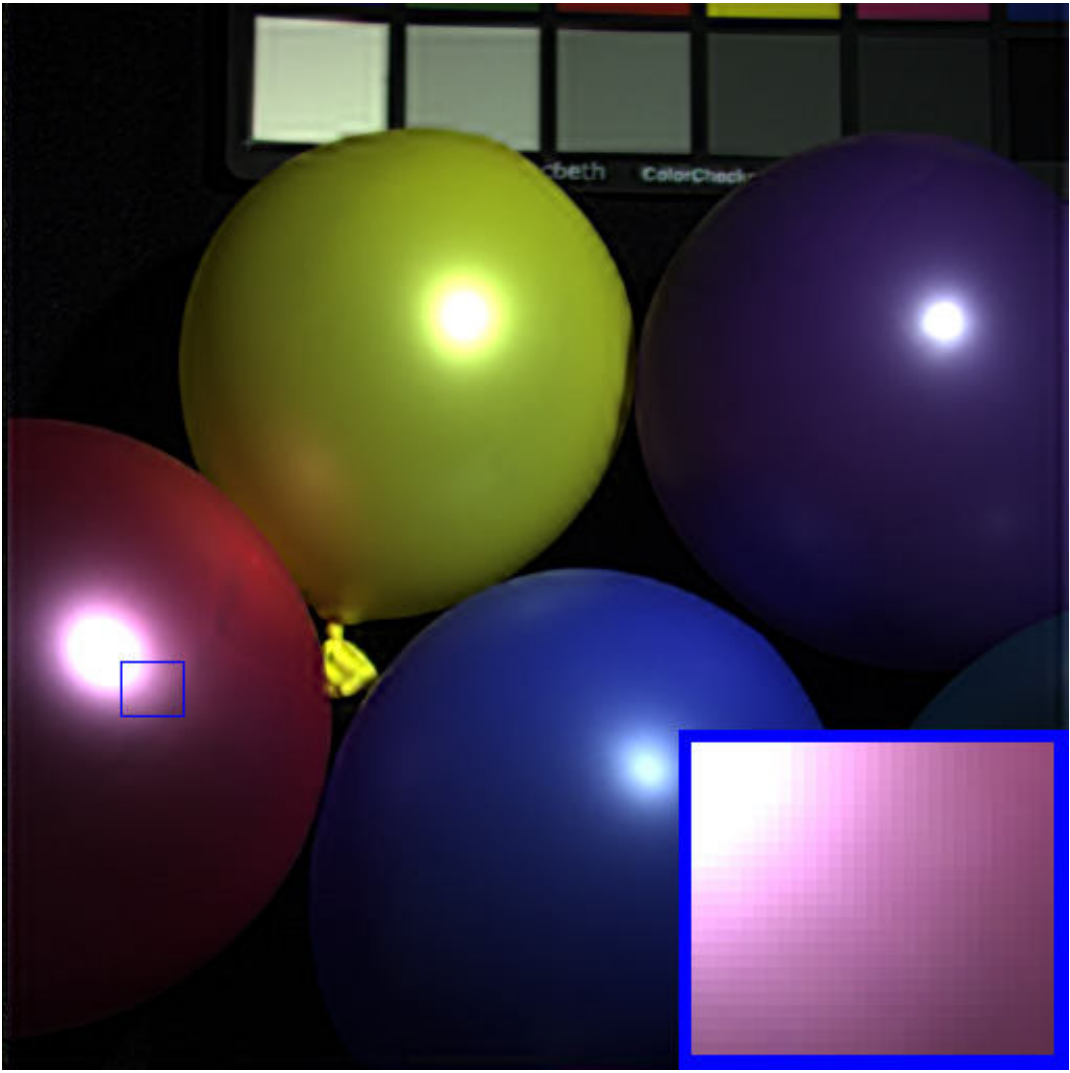}}
			{\includegraphics[width=1\linewidth]{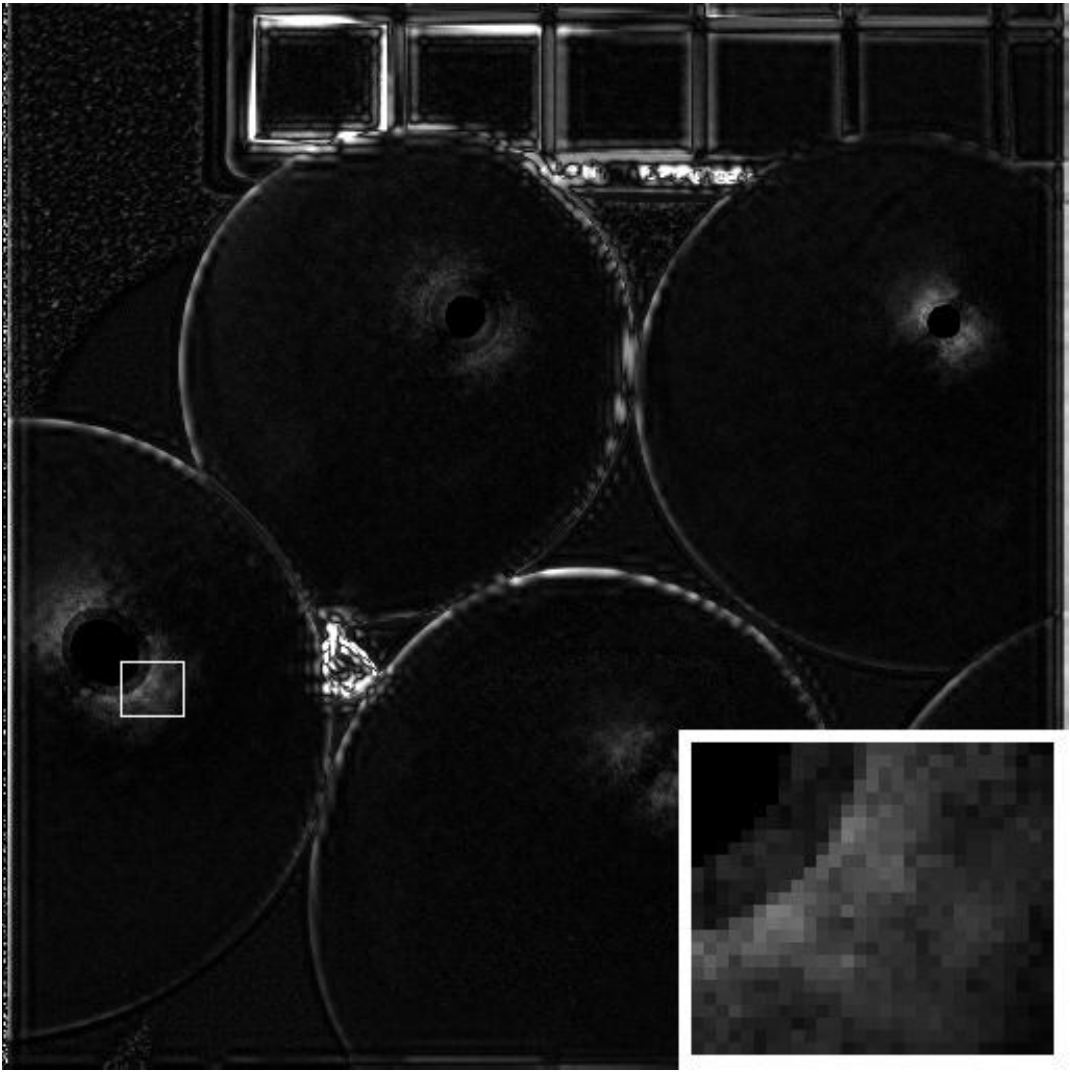}}
			\centering
			
		\end{minipage}
		\begin{minipage}[t]{0.12\linewidth}
			{\includegraphics[width=1\linewidth]{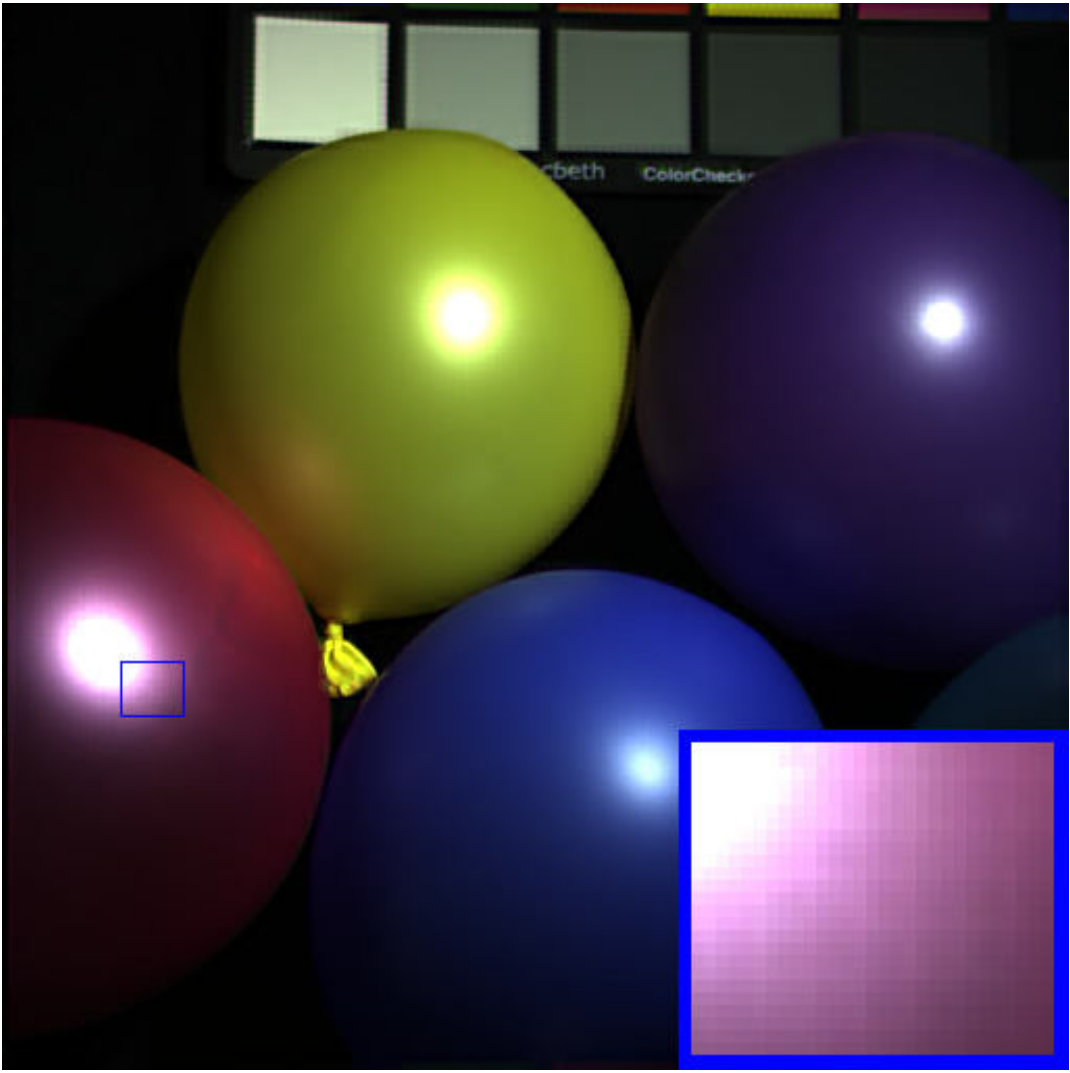}}
			{\includegraphics[width=1\linewidth]{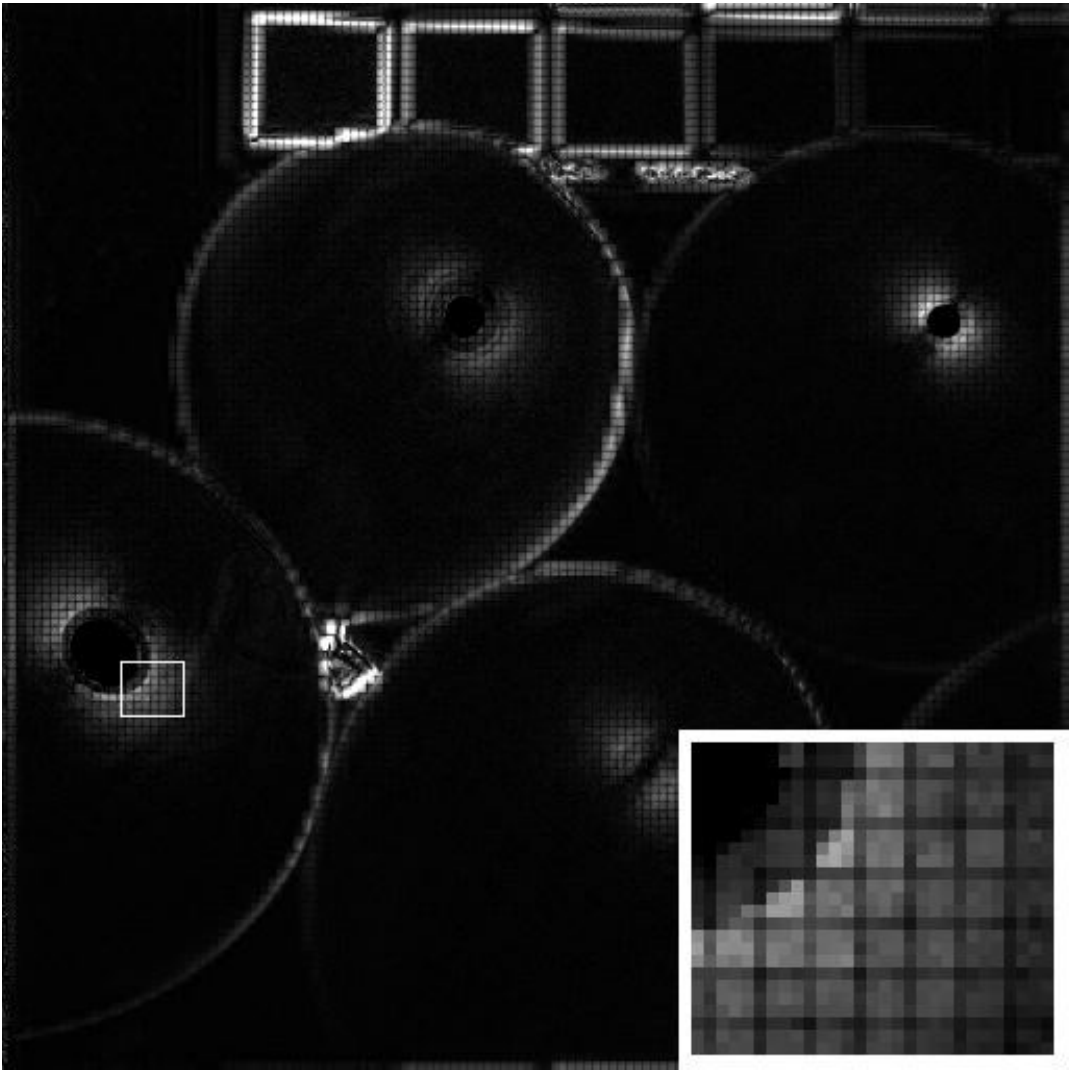}}
			\centering
			
		\end{minipage}
		\begin{minipage}[t]{0.12\linewidth}
			{\includegraphics[width=1\linewidth]{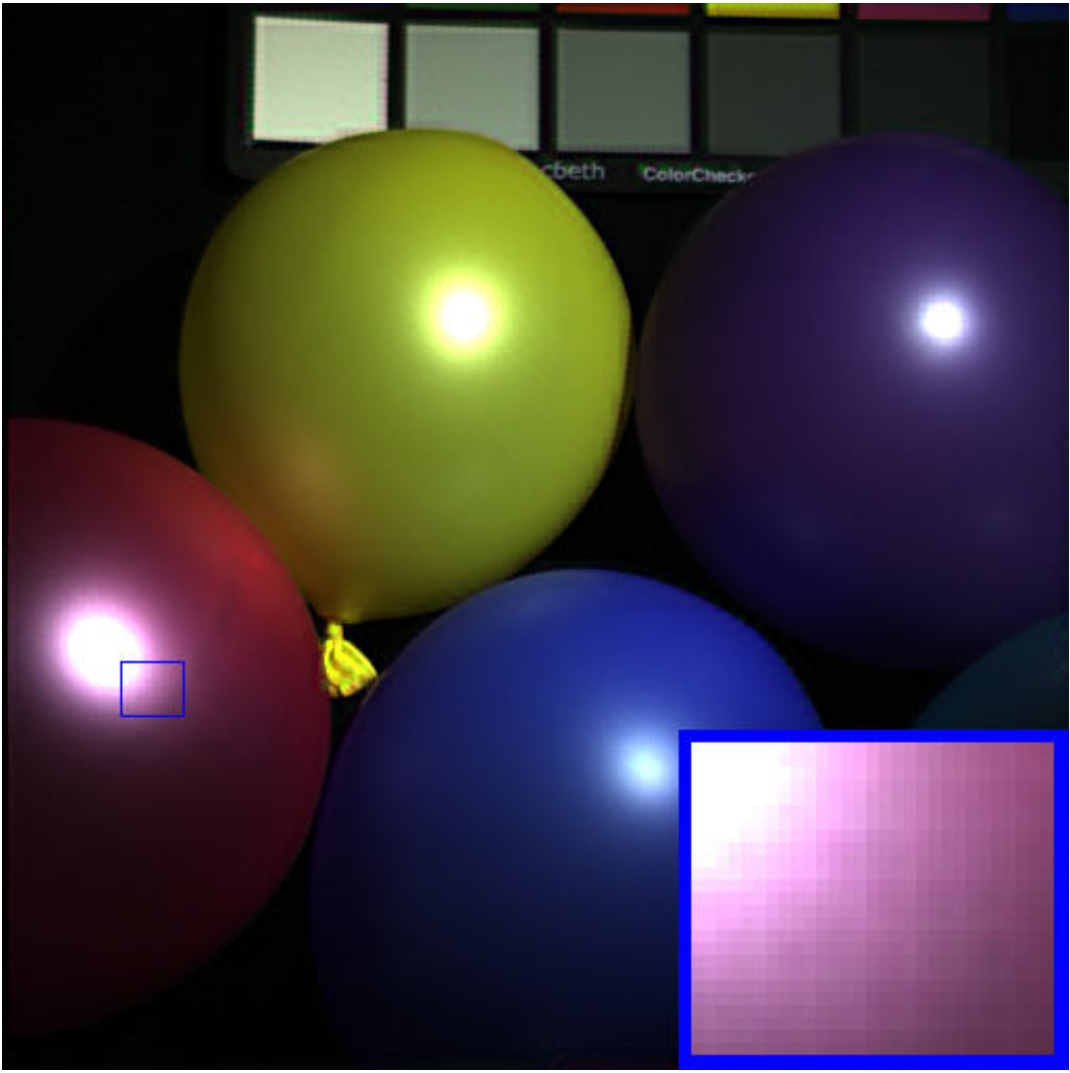}}
			{\includegraphics[width=1\linewidth]{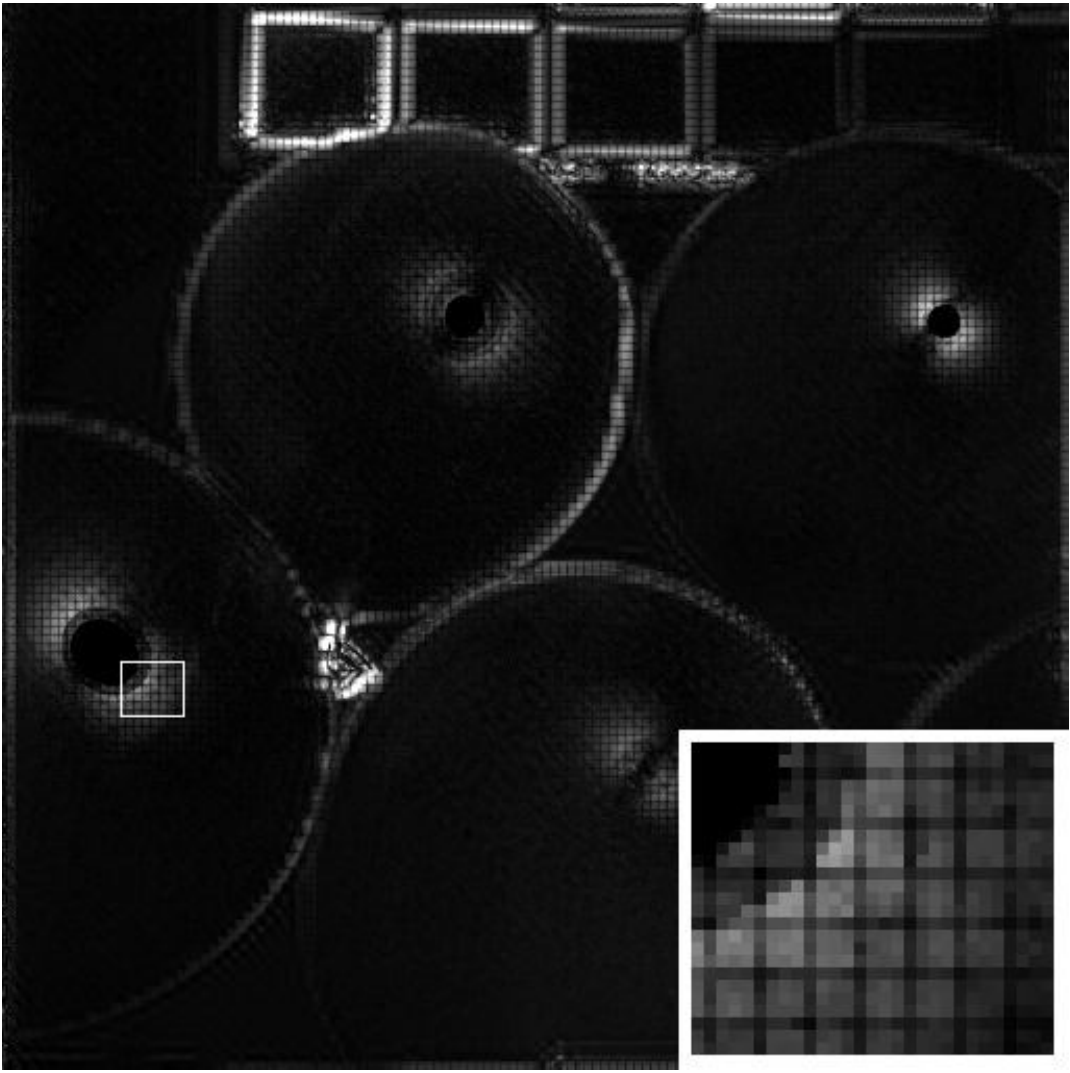}}
			\centering
			
		\end{minipage}
		\begin{minipage}[t]{0.12\linewidth}
			{\includegraphics[width=1\linewidth]{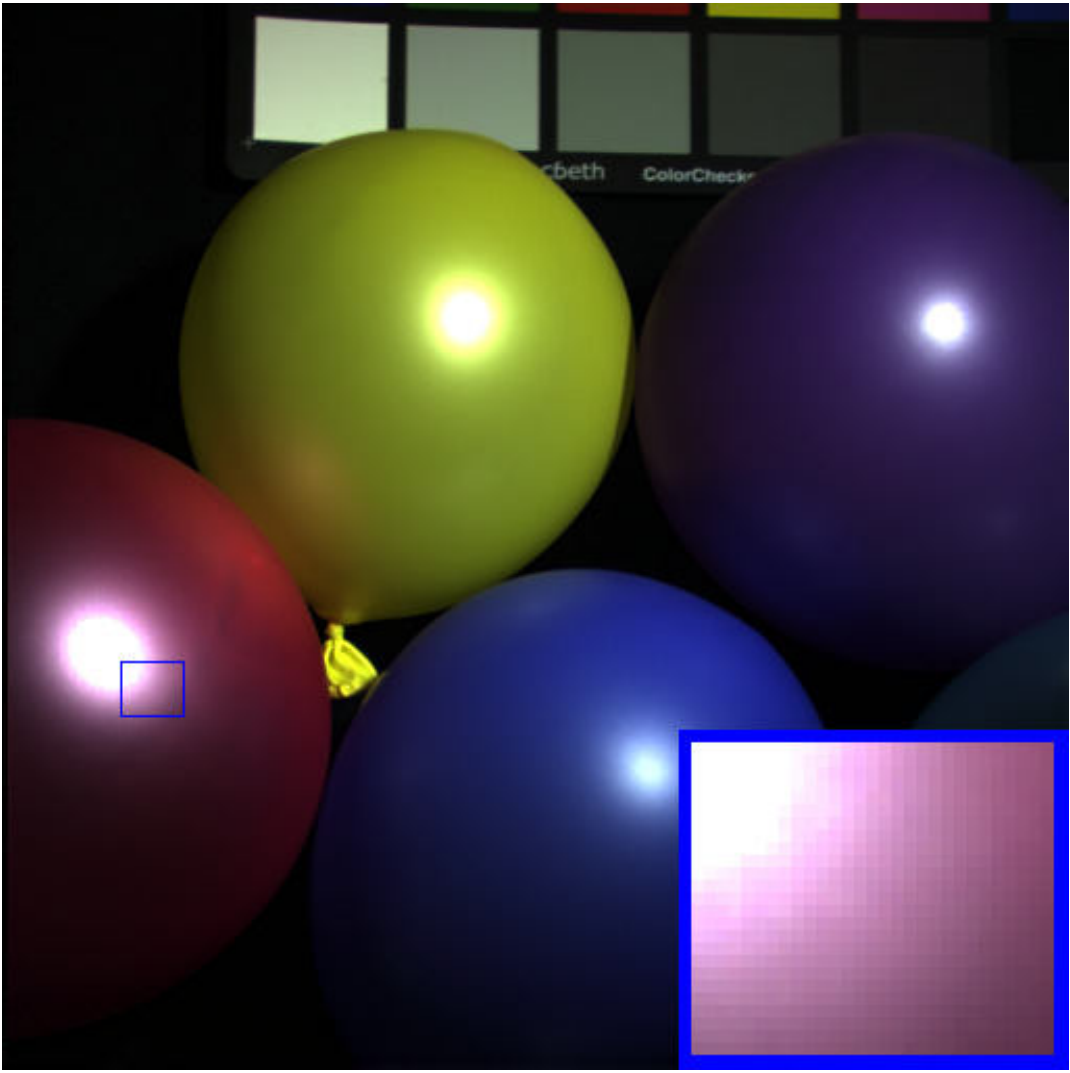}}
			{\includegraphics[width=1\linewidth]{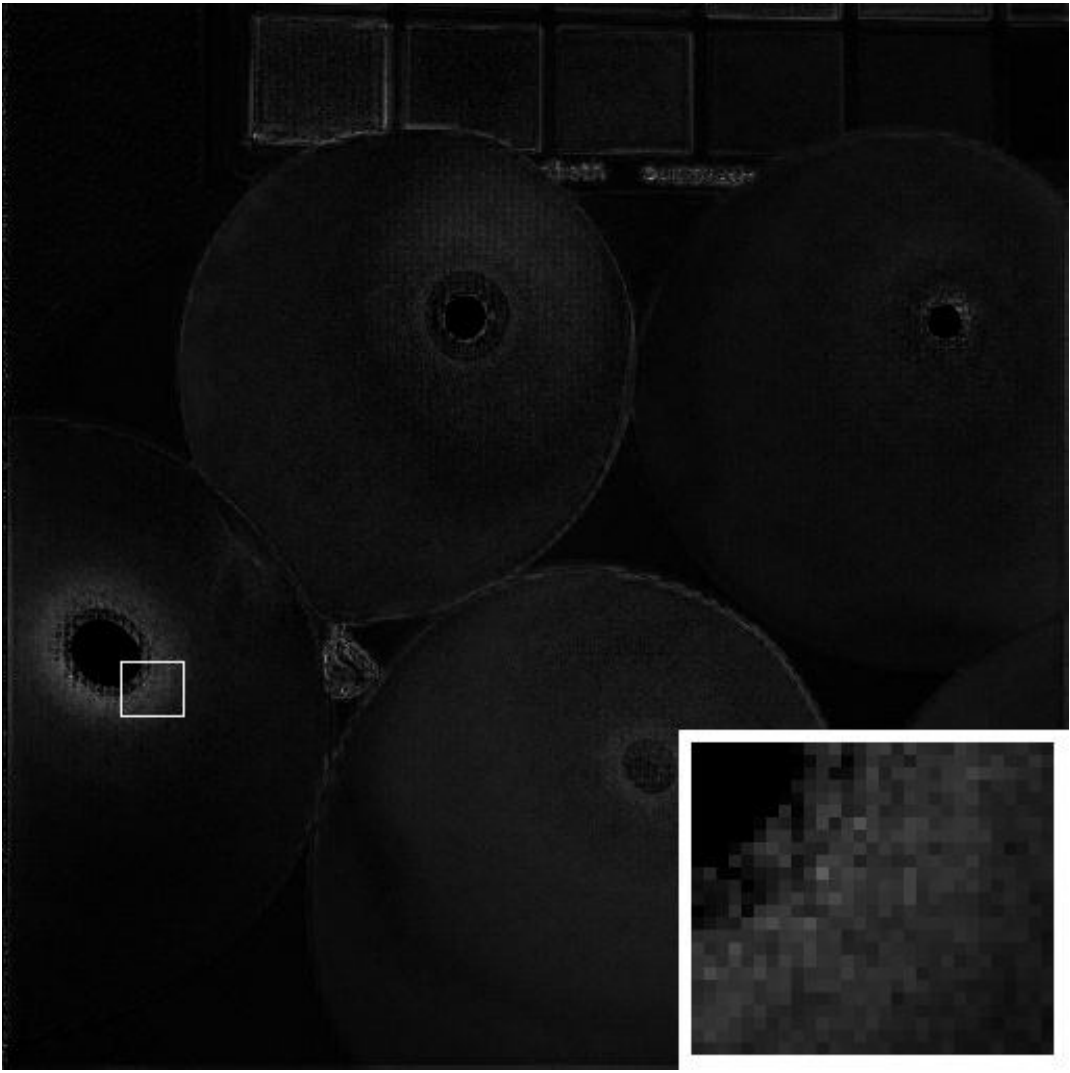}}
			\centering
			
		\end{minipage}
		\begin{minipage}[t]{0.12\linewidth}
			{\includegraphics[width=1\linewidth]{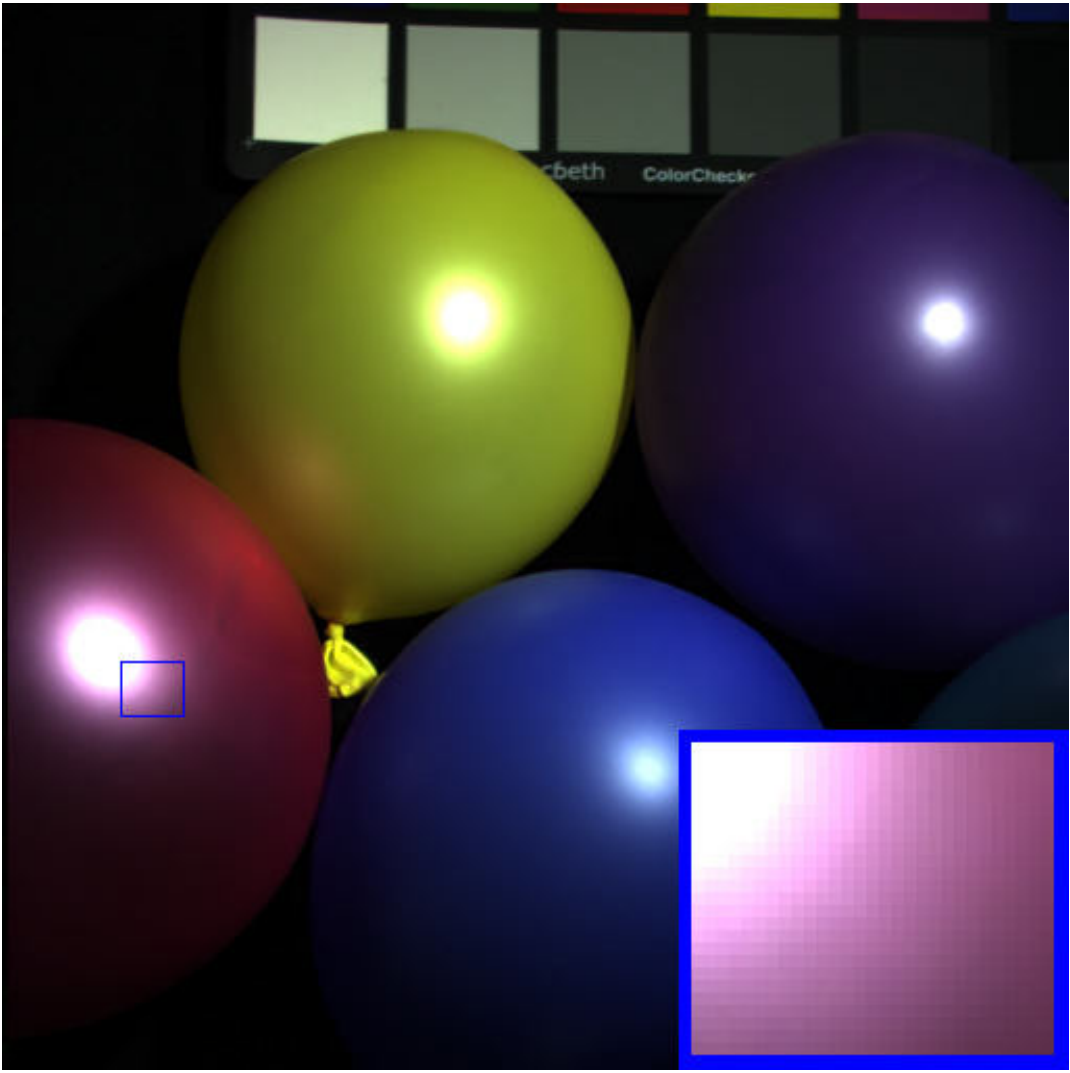}}
			{\includegraphics[width=1\linewidth]{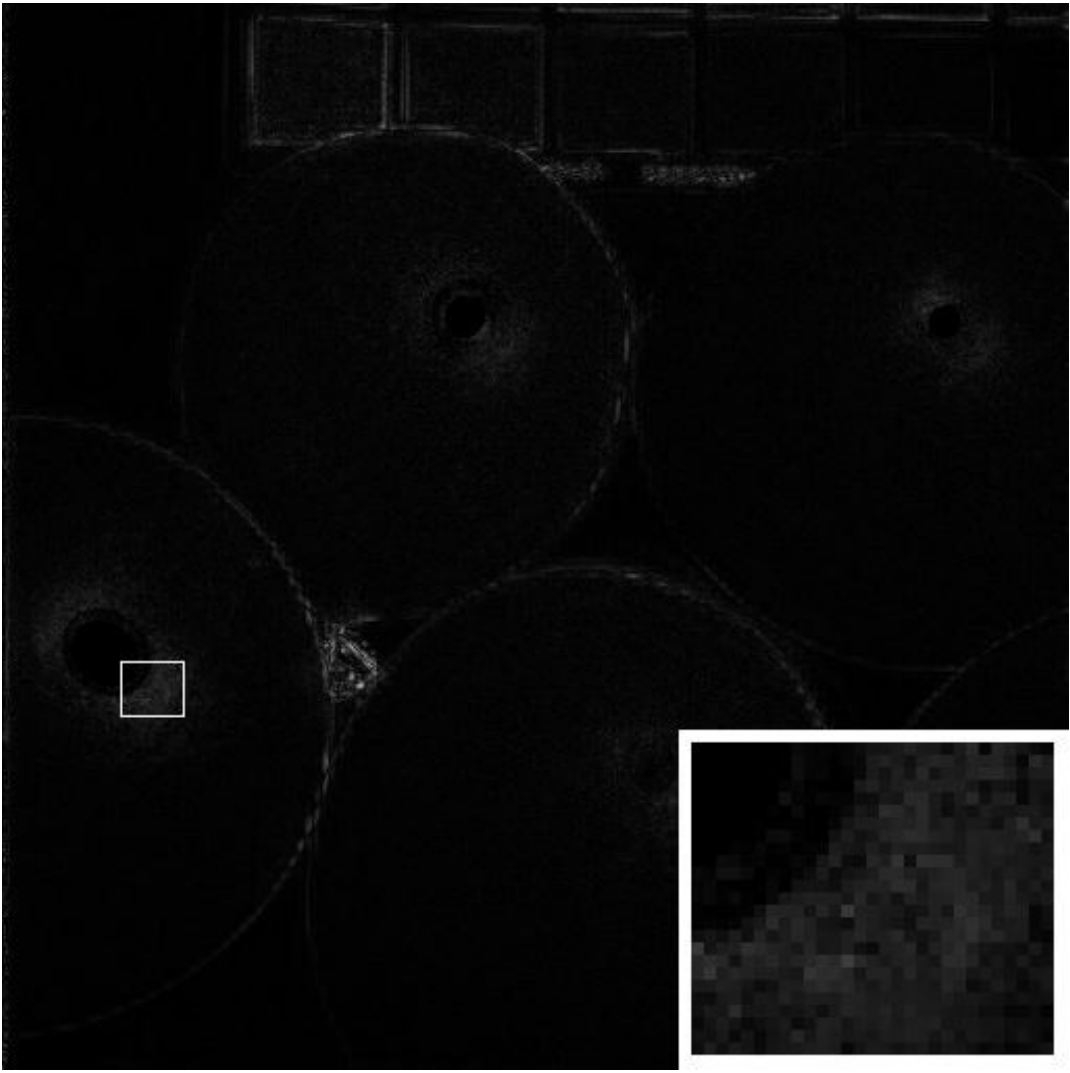}}
			\centering
		\end{minipage}
		
		\vspace{5pt}
		
		\begin{minipage}[t]{0.12\linewidth}
			{\includegraphics[width=1\linewidth]{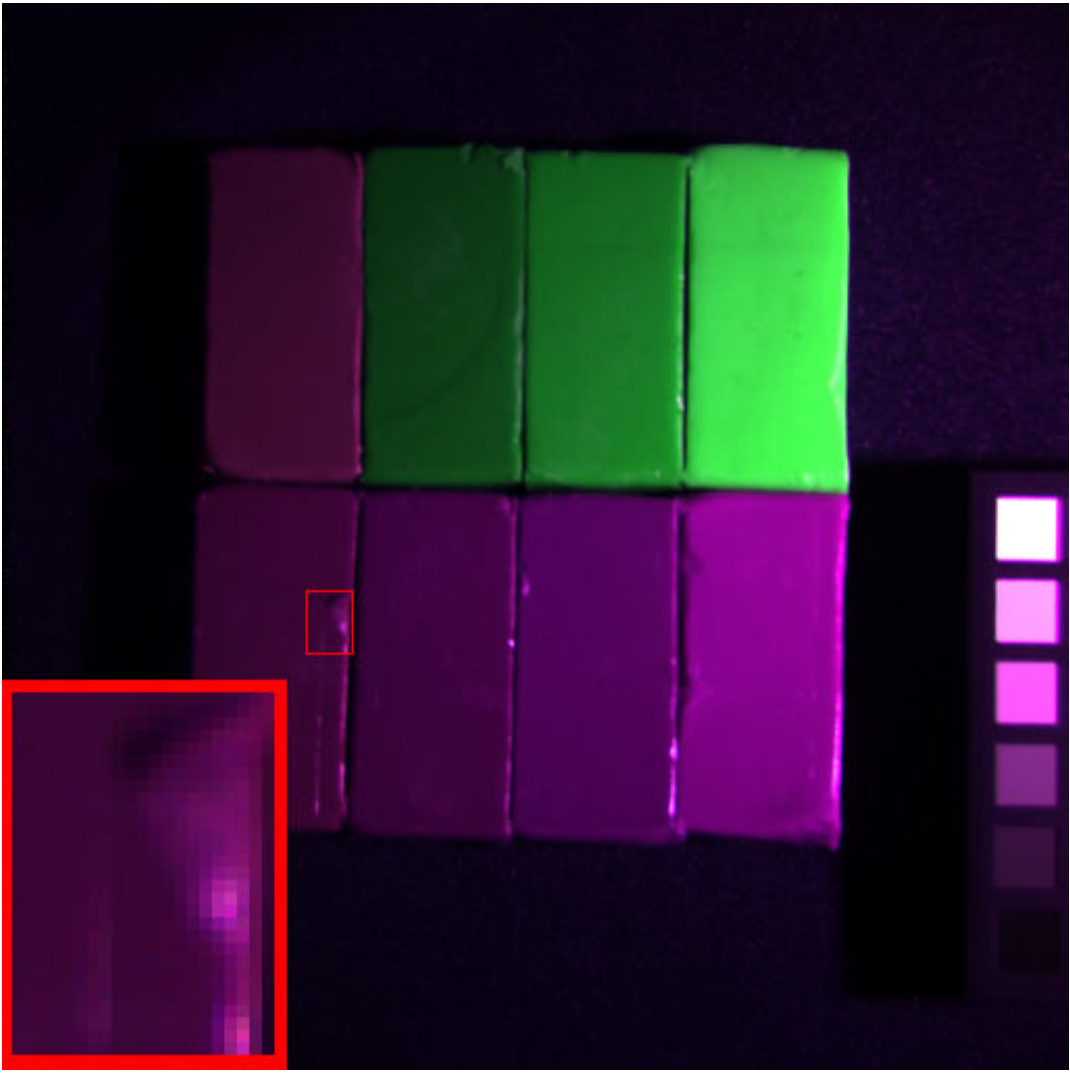}}
			{\includegraphics[width=1\linewidth]{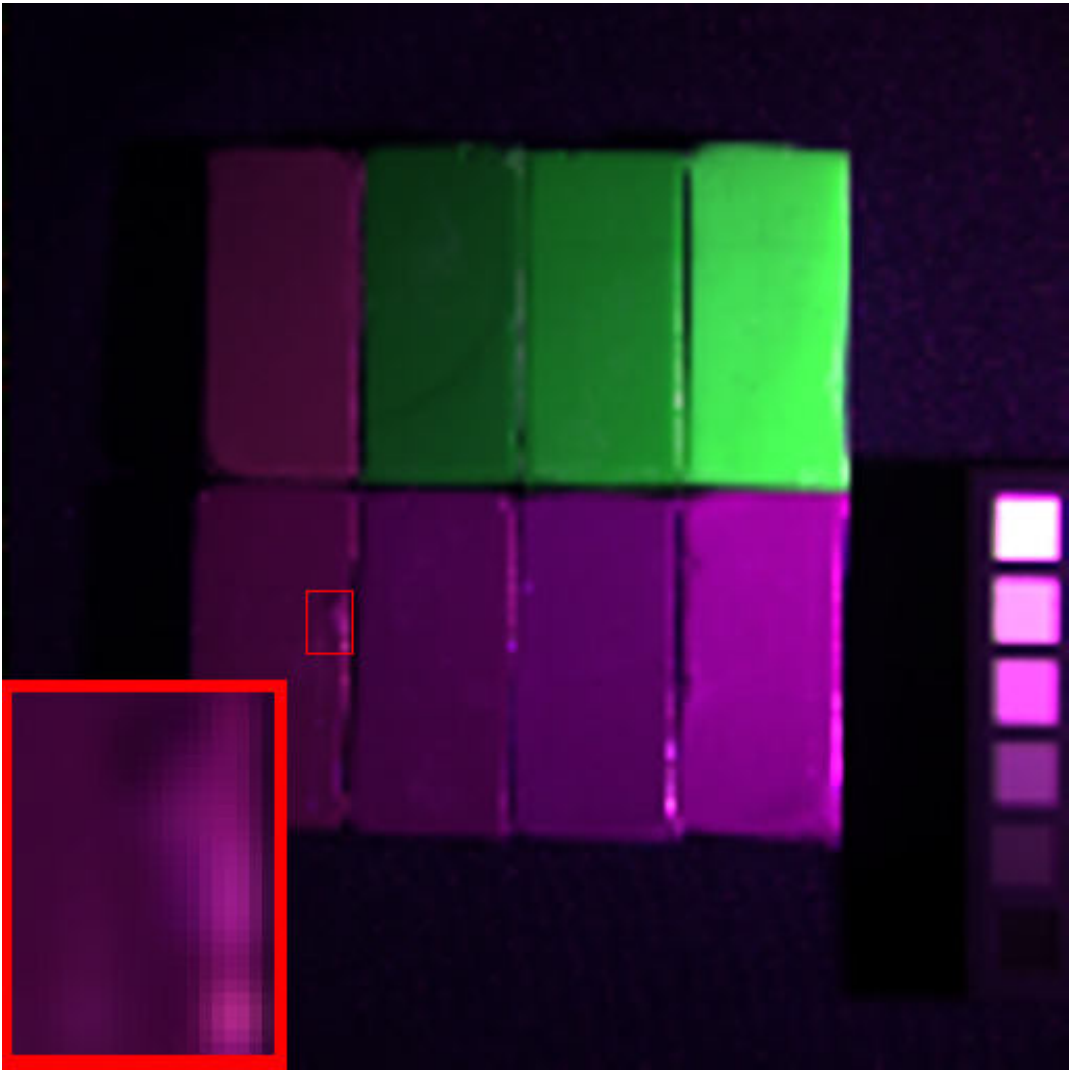}}
			\centering
			
		\end{minipage}
		\begin{minipage}[t]{0.12\linewidth}
			{\includegraphics[width=1\linewidth]{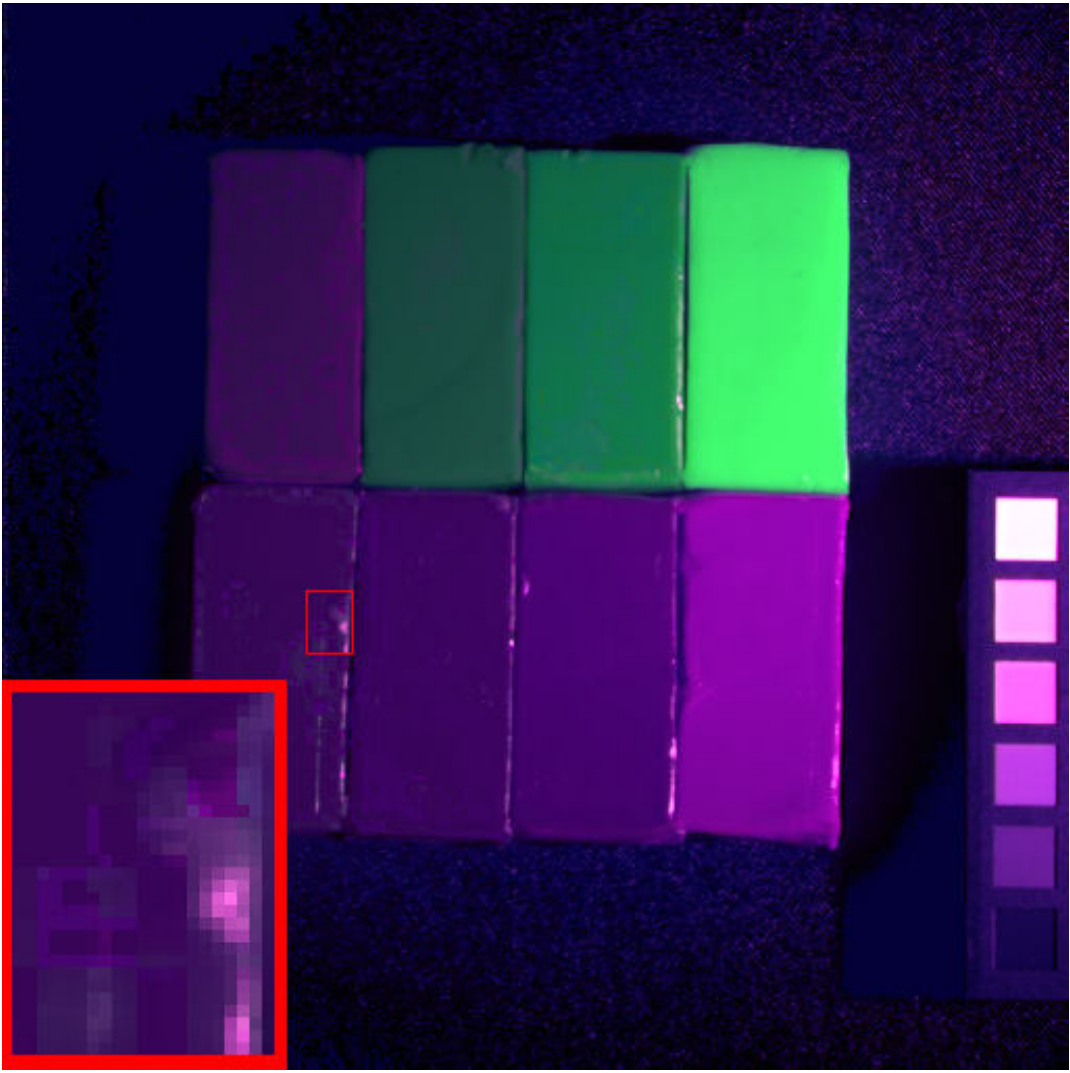}}
			{\includegraphics[width=1\linewidth]{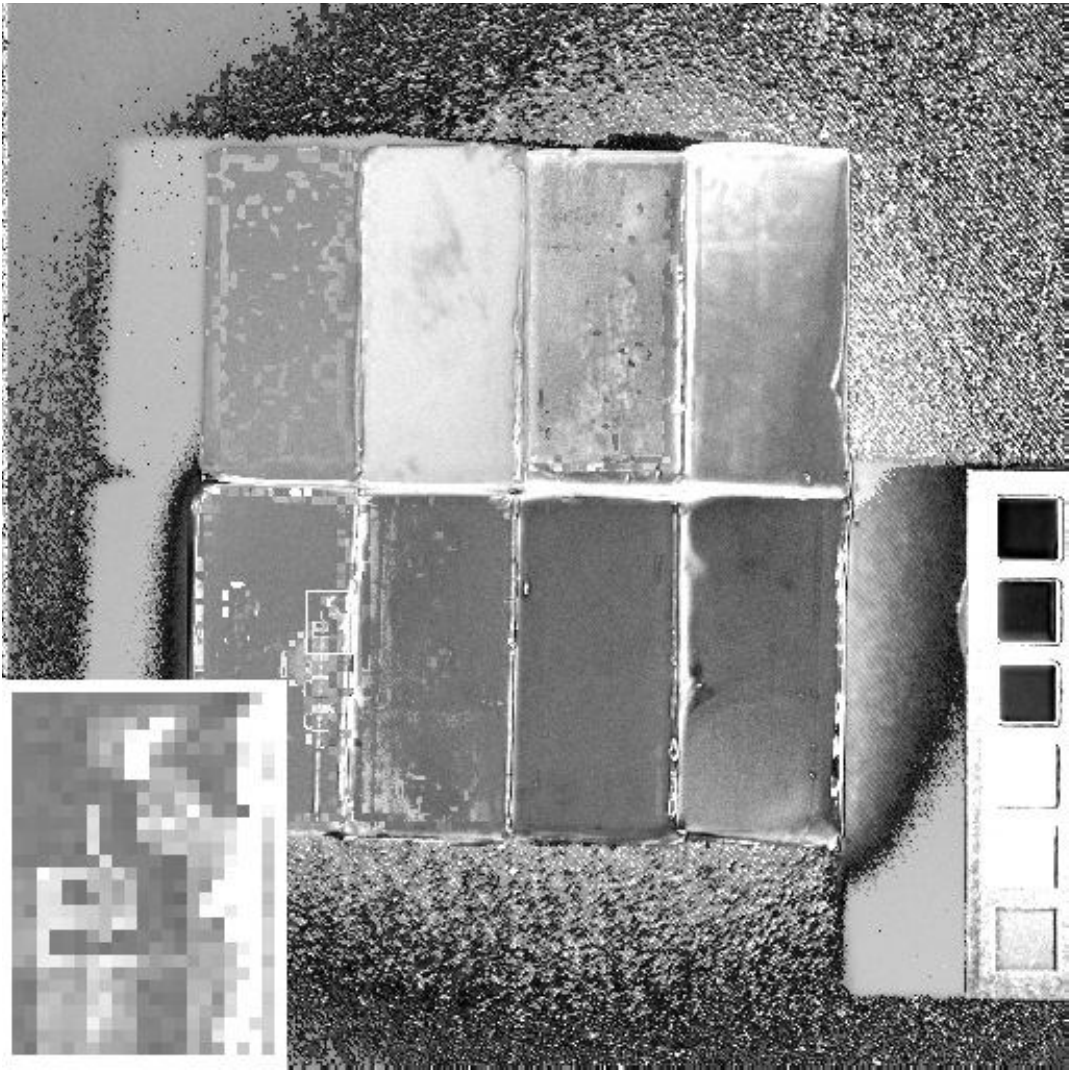}}
			\centering
			
		\end{minipage}
		\begin{minipage}[t]{0.12\linewidth}
			{\includegraphics[width=1\linewidth]{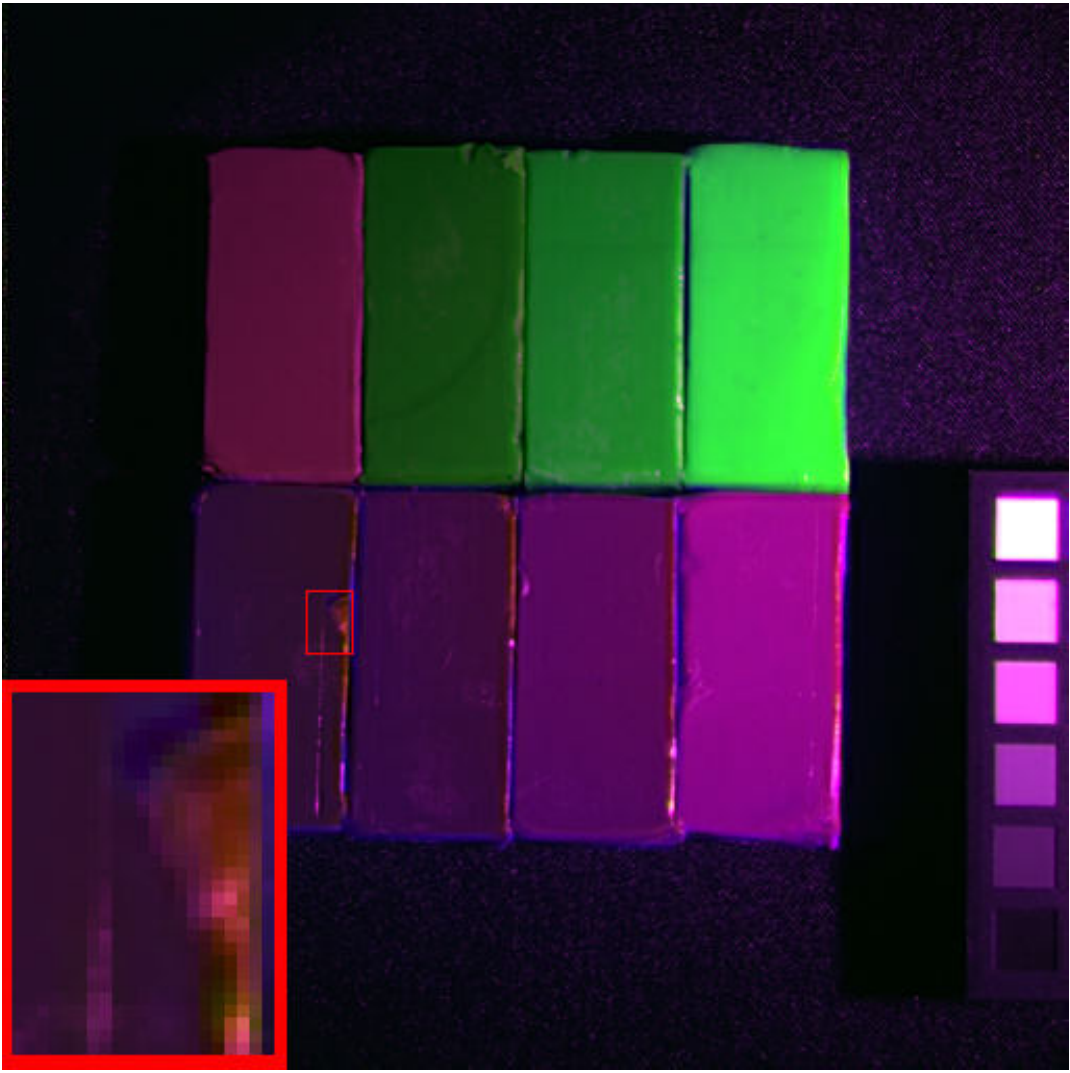}}
			{\includegraphics[width=1\linewidth]{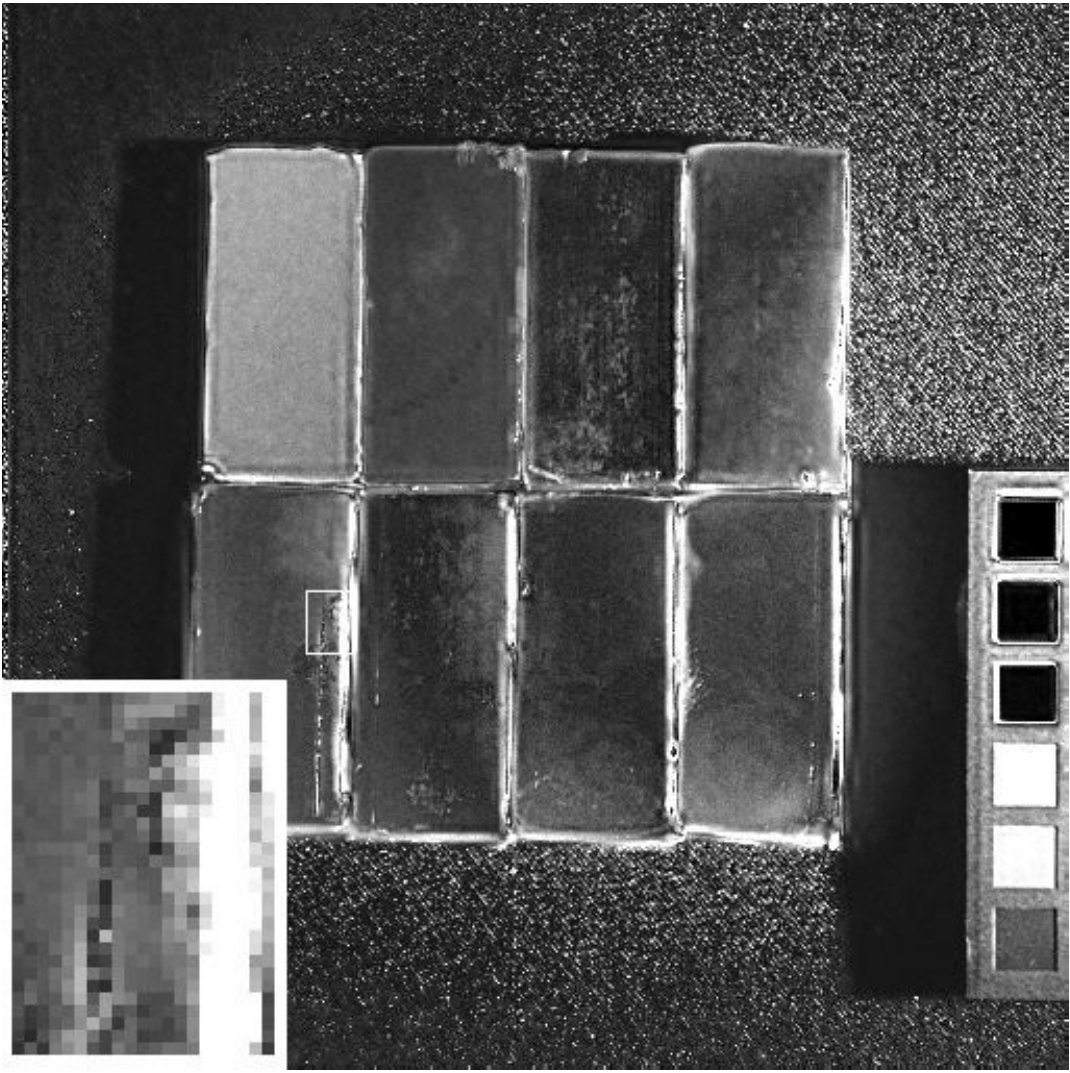}}
			\centering
			
		\end{minipage}
		\begin{minipage}[t]{0.12\linewidth}
			{\includegraphics[width=1\linewidth]{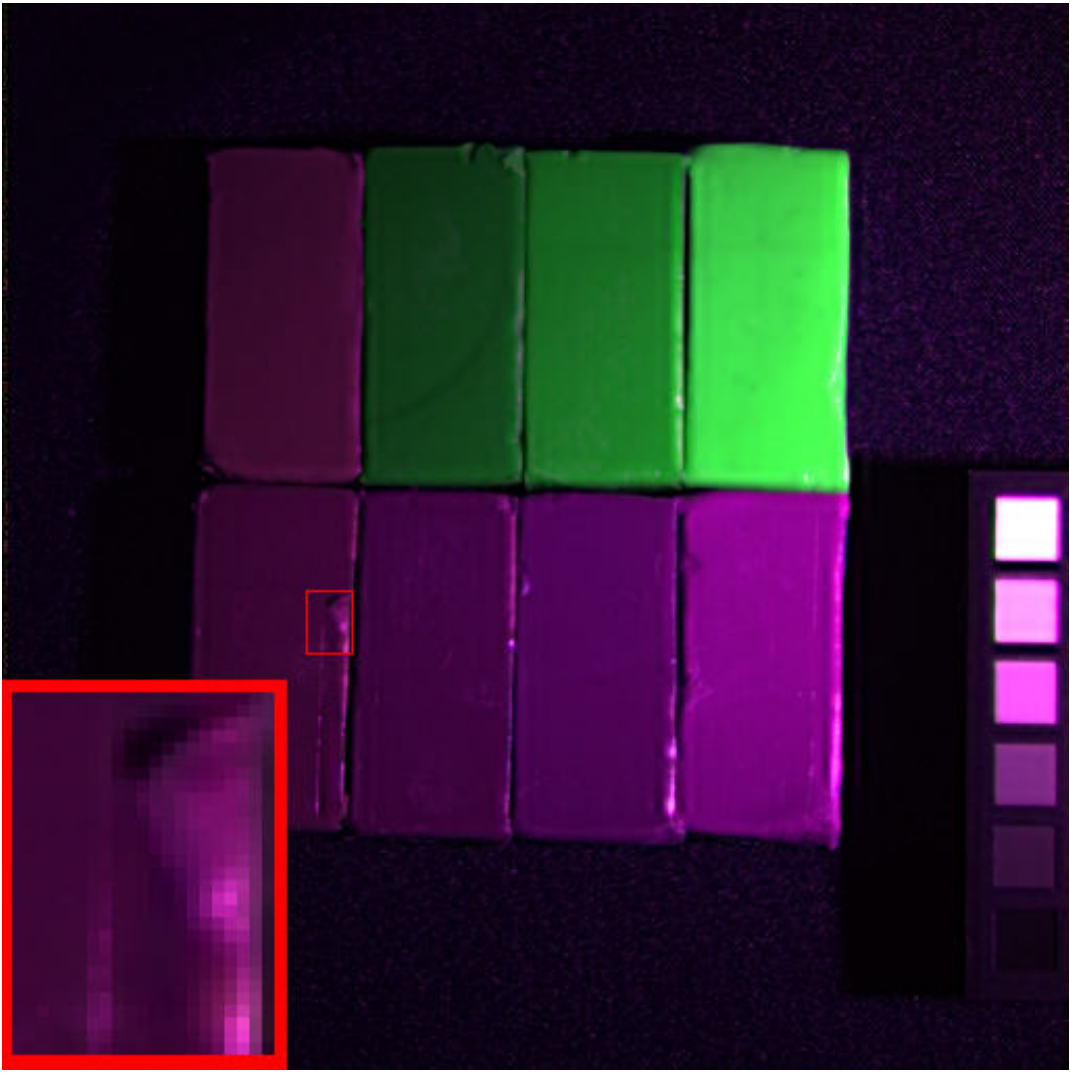}}
			{\includegraphics[width=1\linewidth]{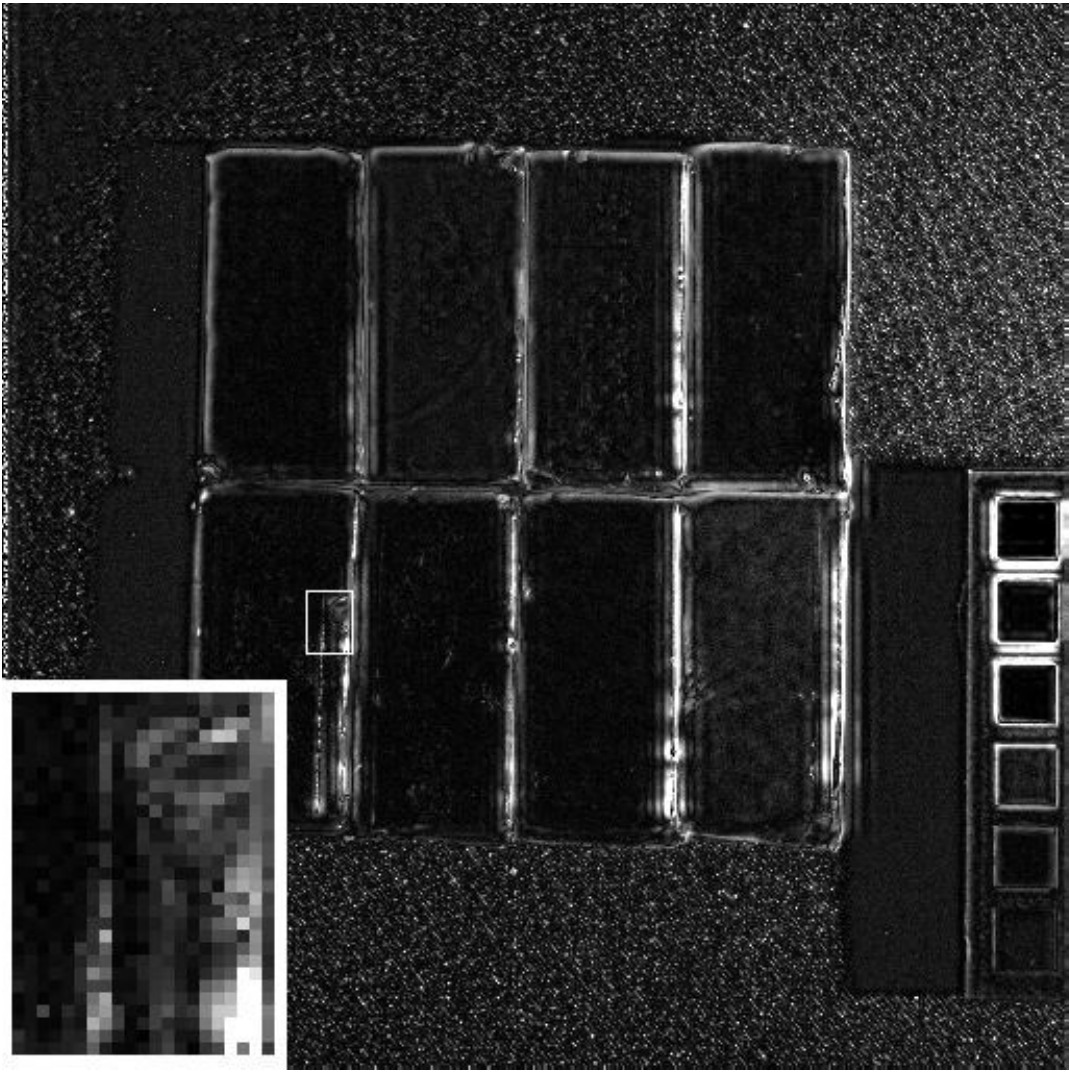}}
			\centering
			
		\end{minipage}
		\begin{minipage}[t]{0.12\linewidth}
			{\includegraphics[width=1\linewidth]{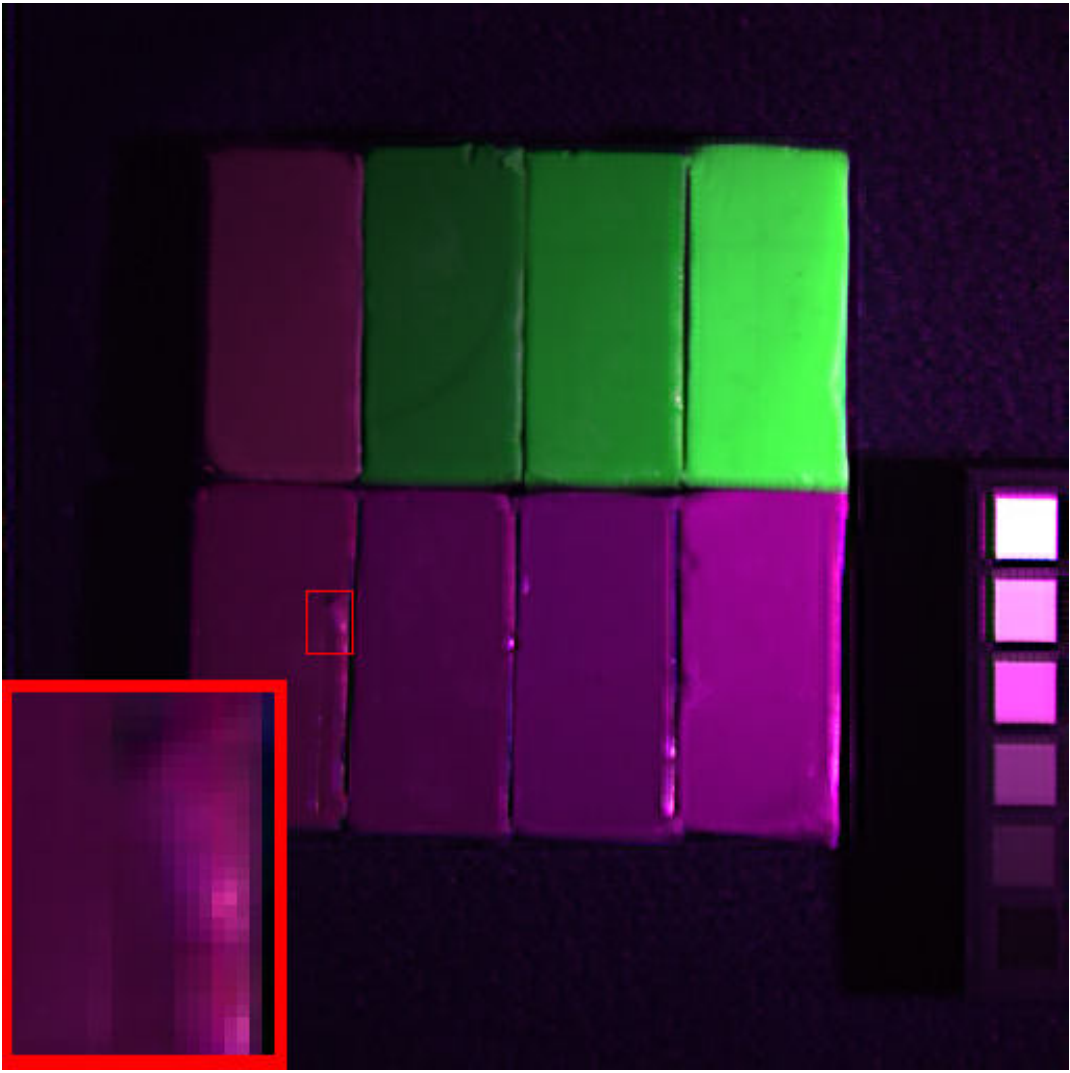}}
			{\includegraphics[width=1\linewidth]{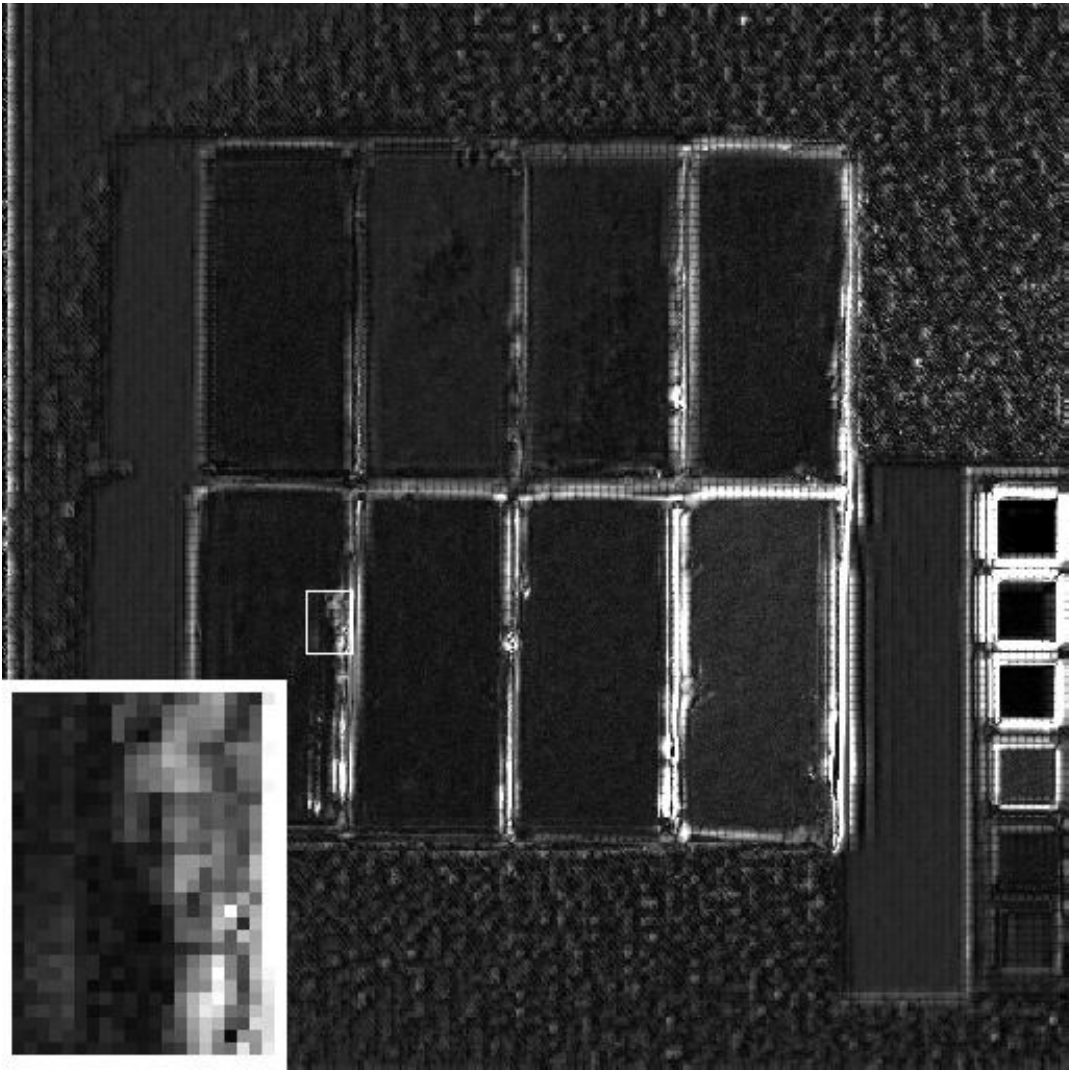}}
			\centering
			
		\end{minipage}
		\begin{minipage}[t]{0.12\linewidth}
			{\includegraphics[width=1\linewidth]{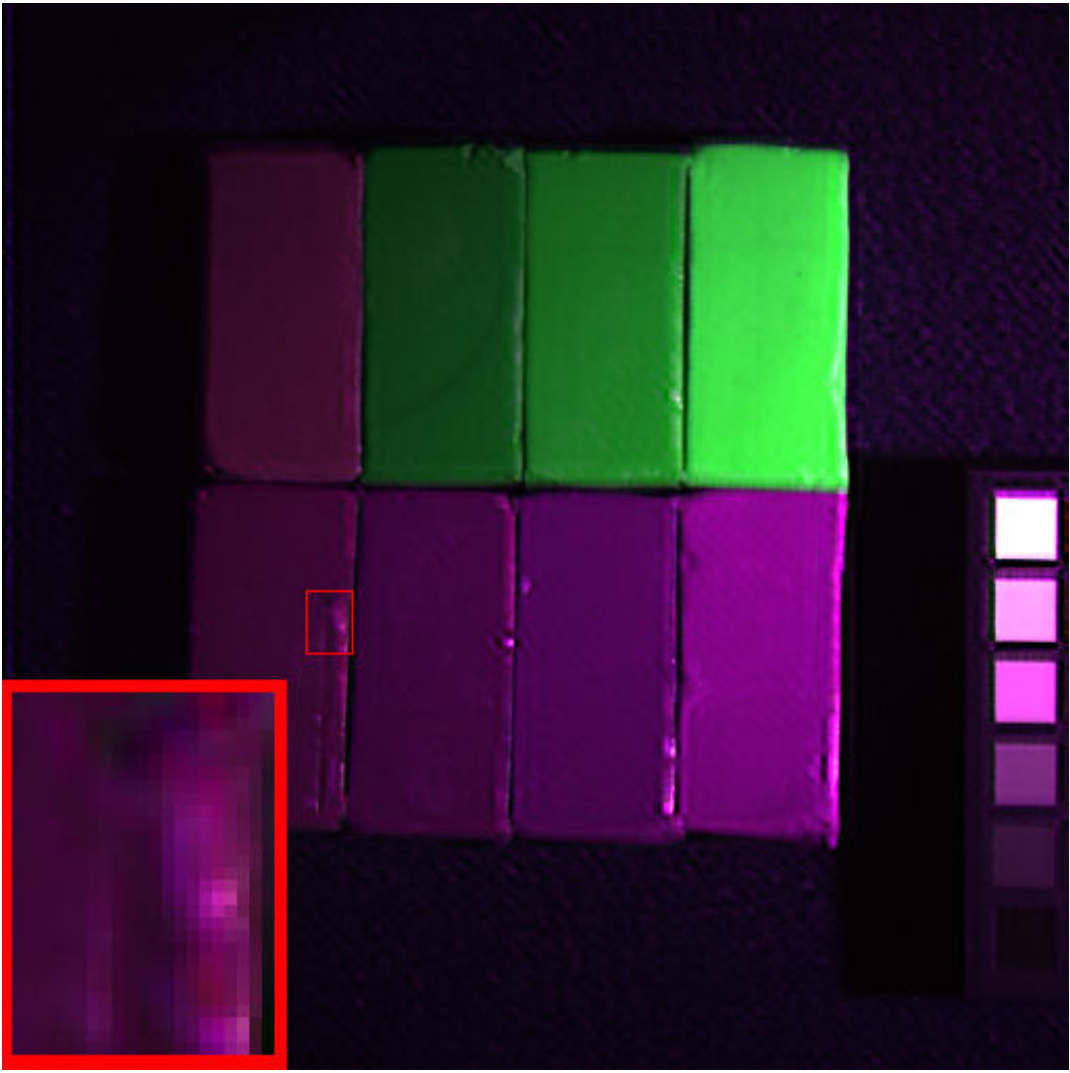}}
			{\includegraphics[width=1\linewidth]{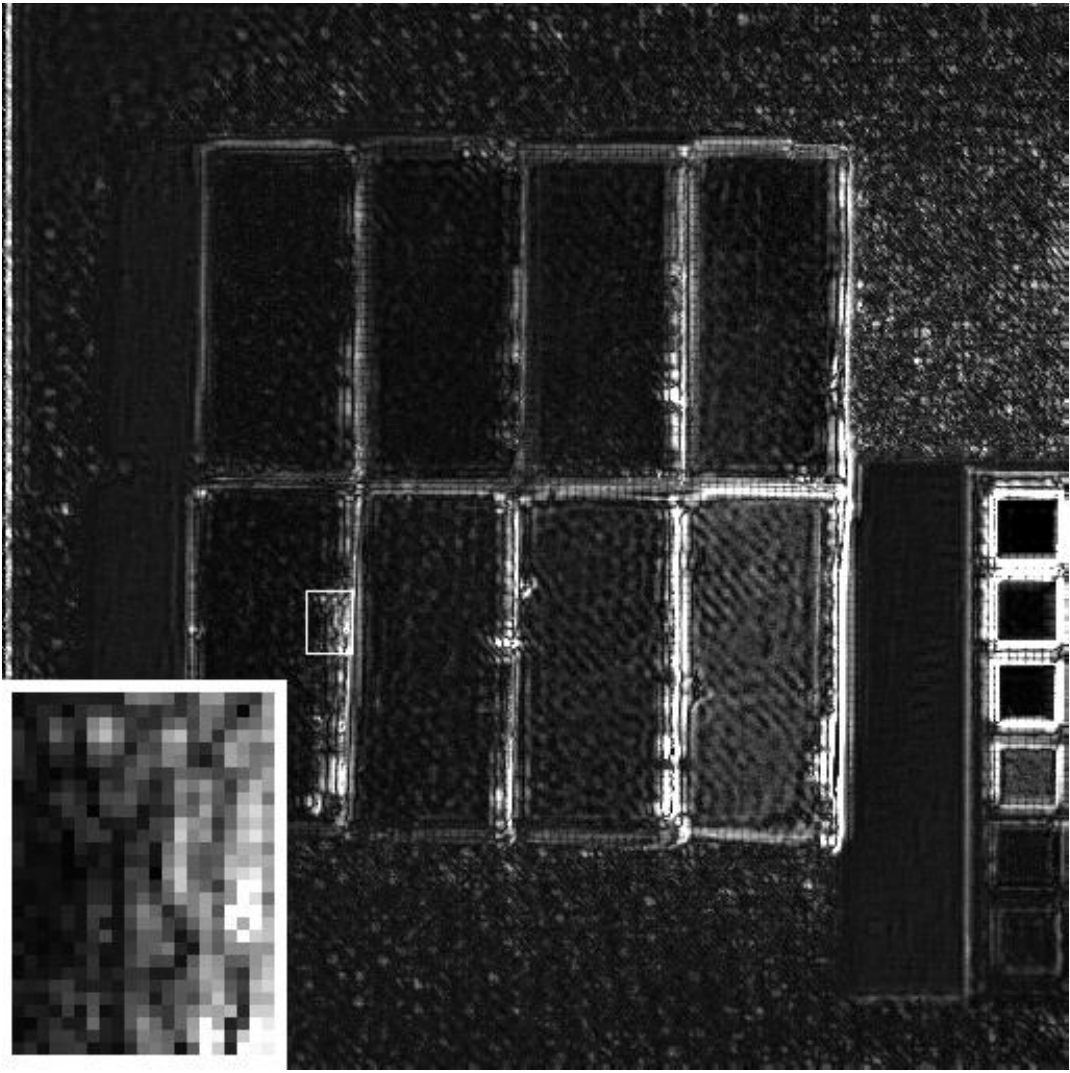}}
			\centering
			
		\end{minipage}
		\begin{minipage}[t]{0.12\linewidth}
			{\includegraphics[width=1\linewidth]{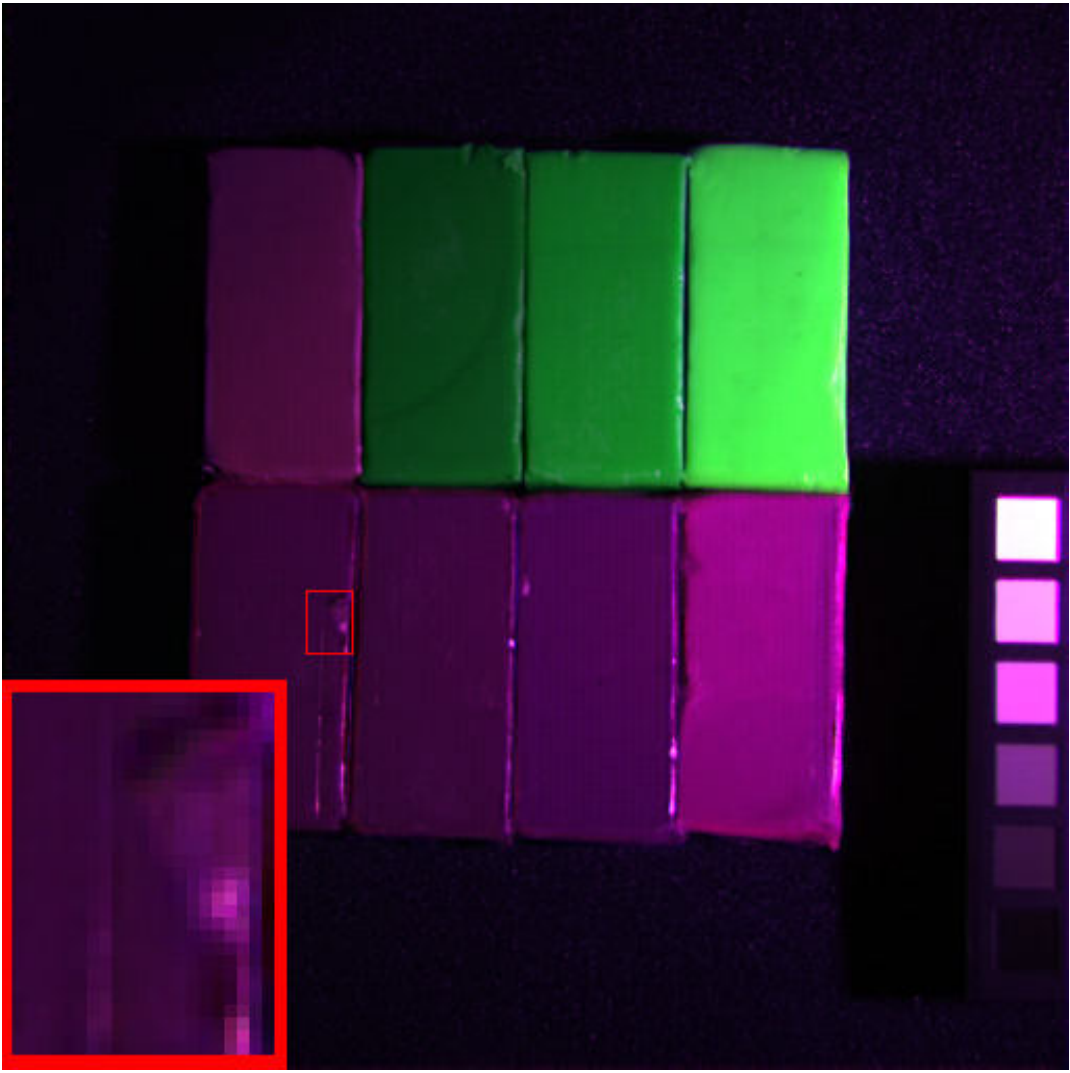}}
			{\includegraphics[width=1\linewidth]{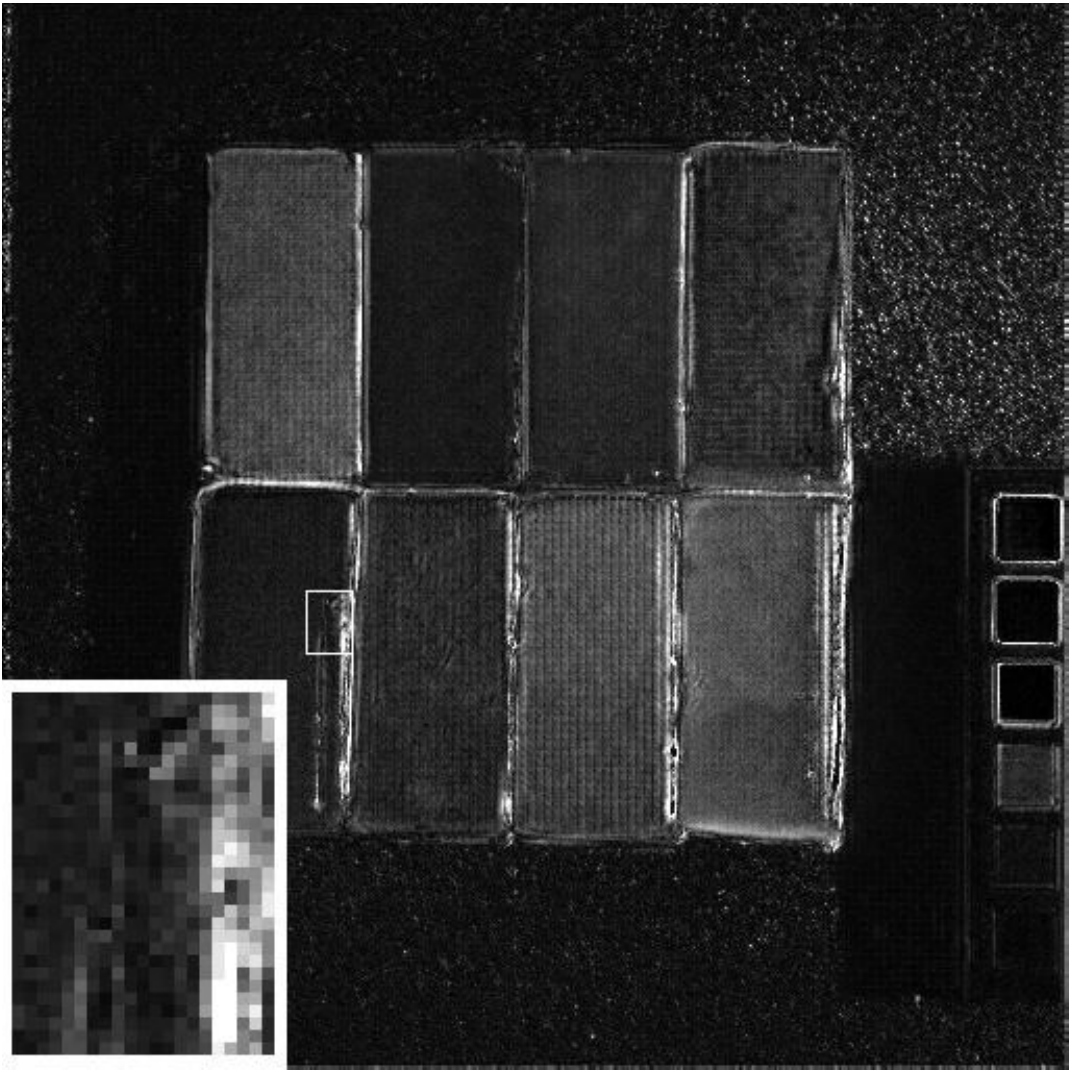}}
			\centering
			
		\end{minipage}
		\begin{minipage}[t]{0.12\linewidth}
			{\includegraphics[width=1\linewidth]{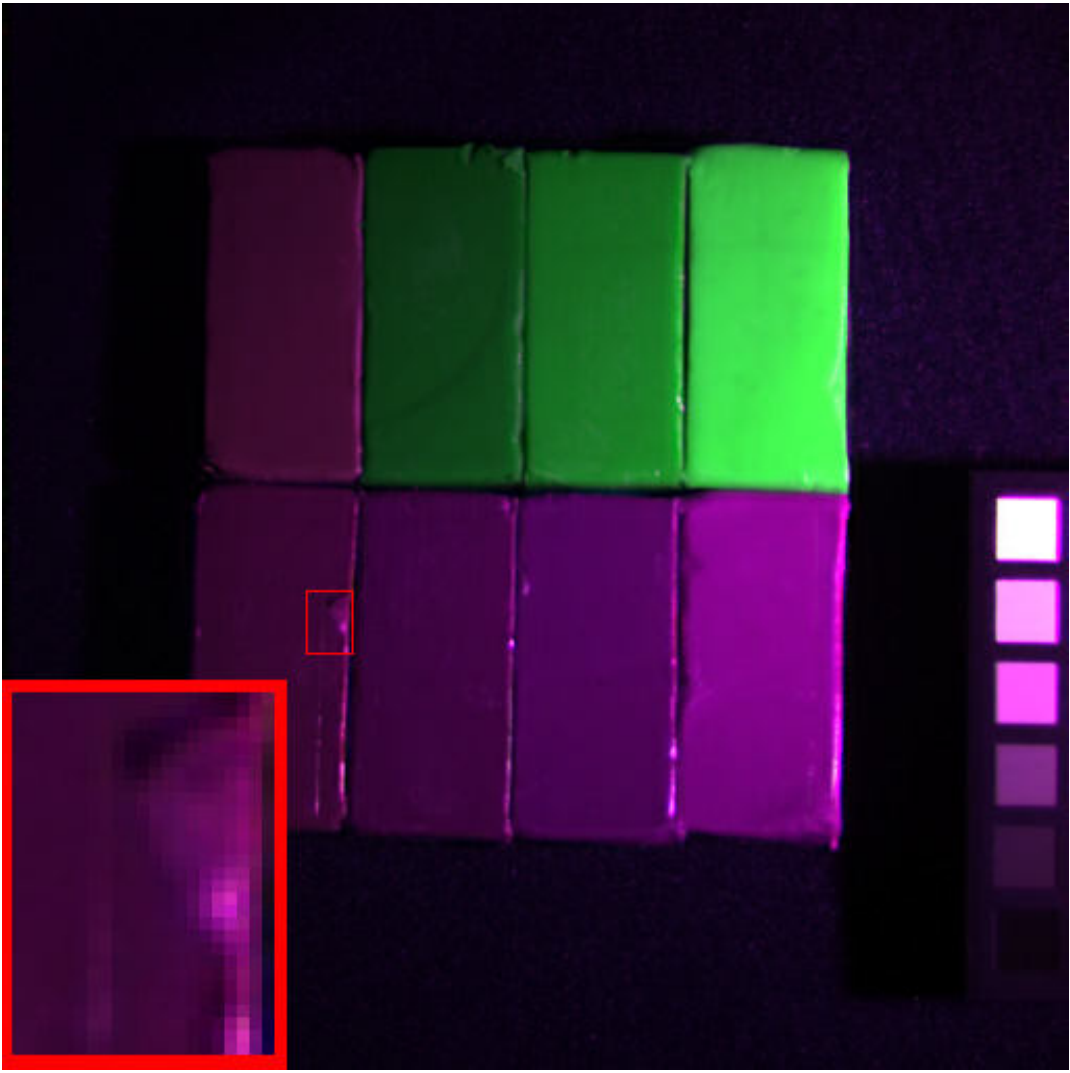}}
			{\includegraphics[width=1\linewidth]{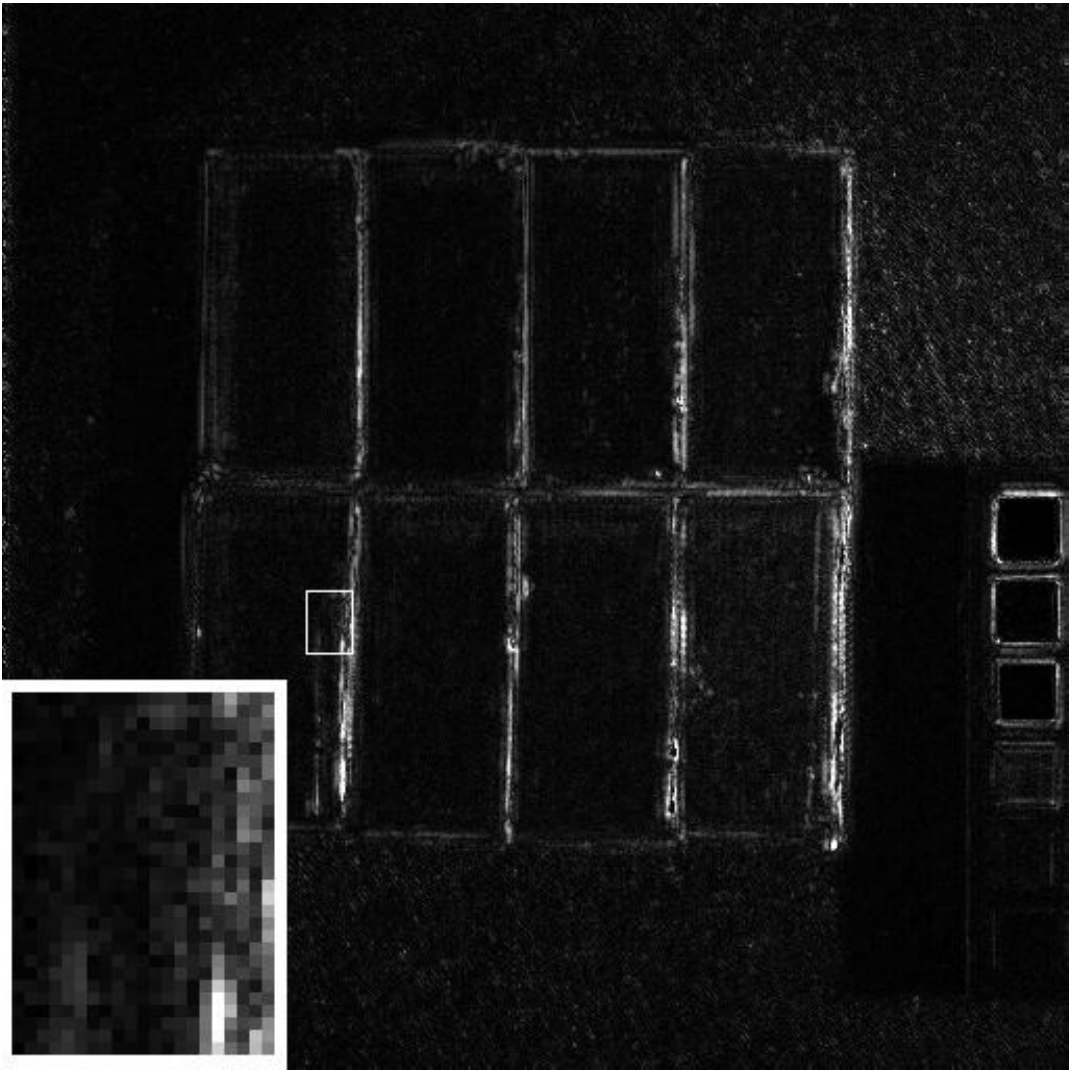}}
			\centering
		\end{minipage}
		
		\vspace{5pt}
		
		\begin{minipage}[t]{0.12\linewidth}
			{\includegraphics[width=1\linewidth]{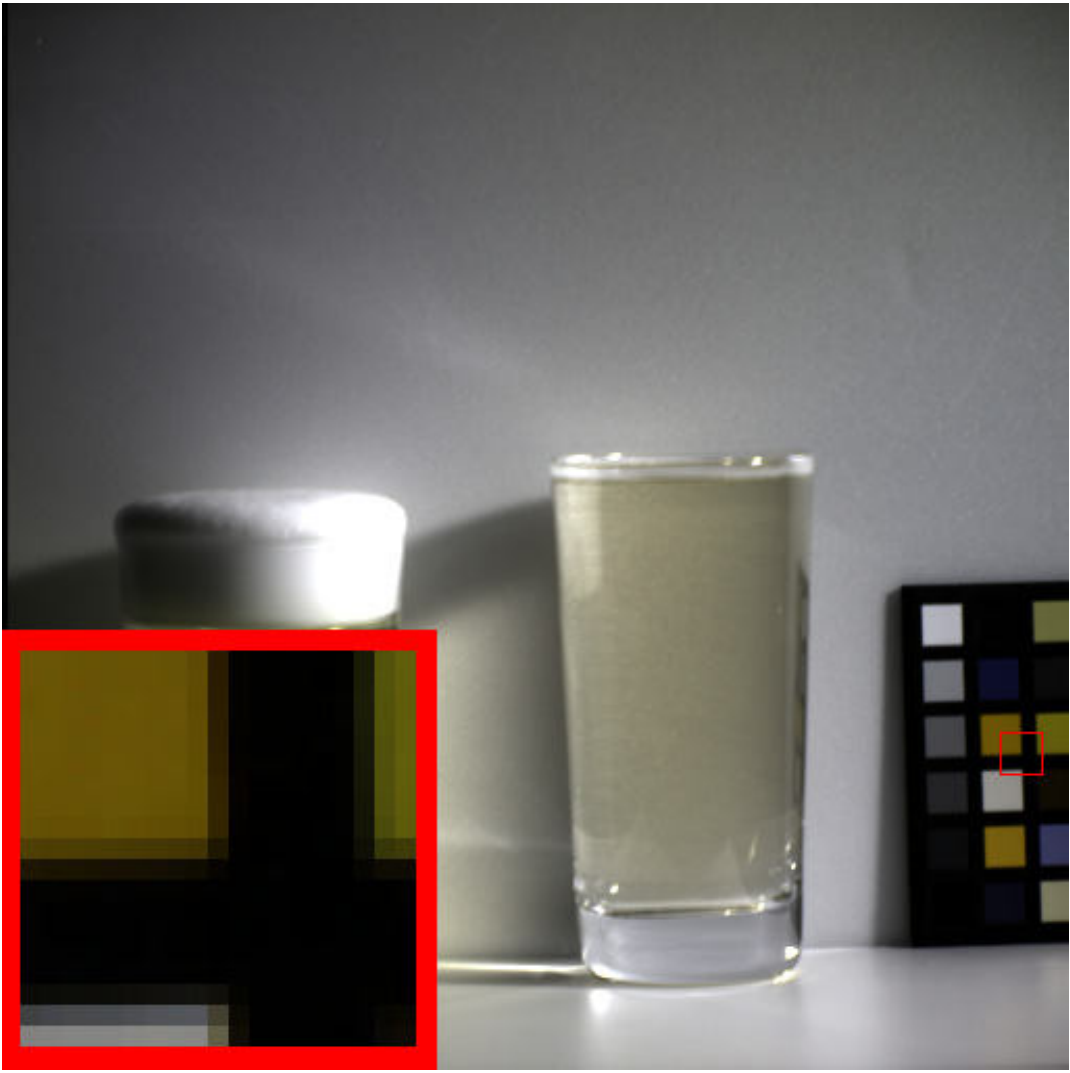}}
			{\includegraphics[width=1\linewidth]{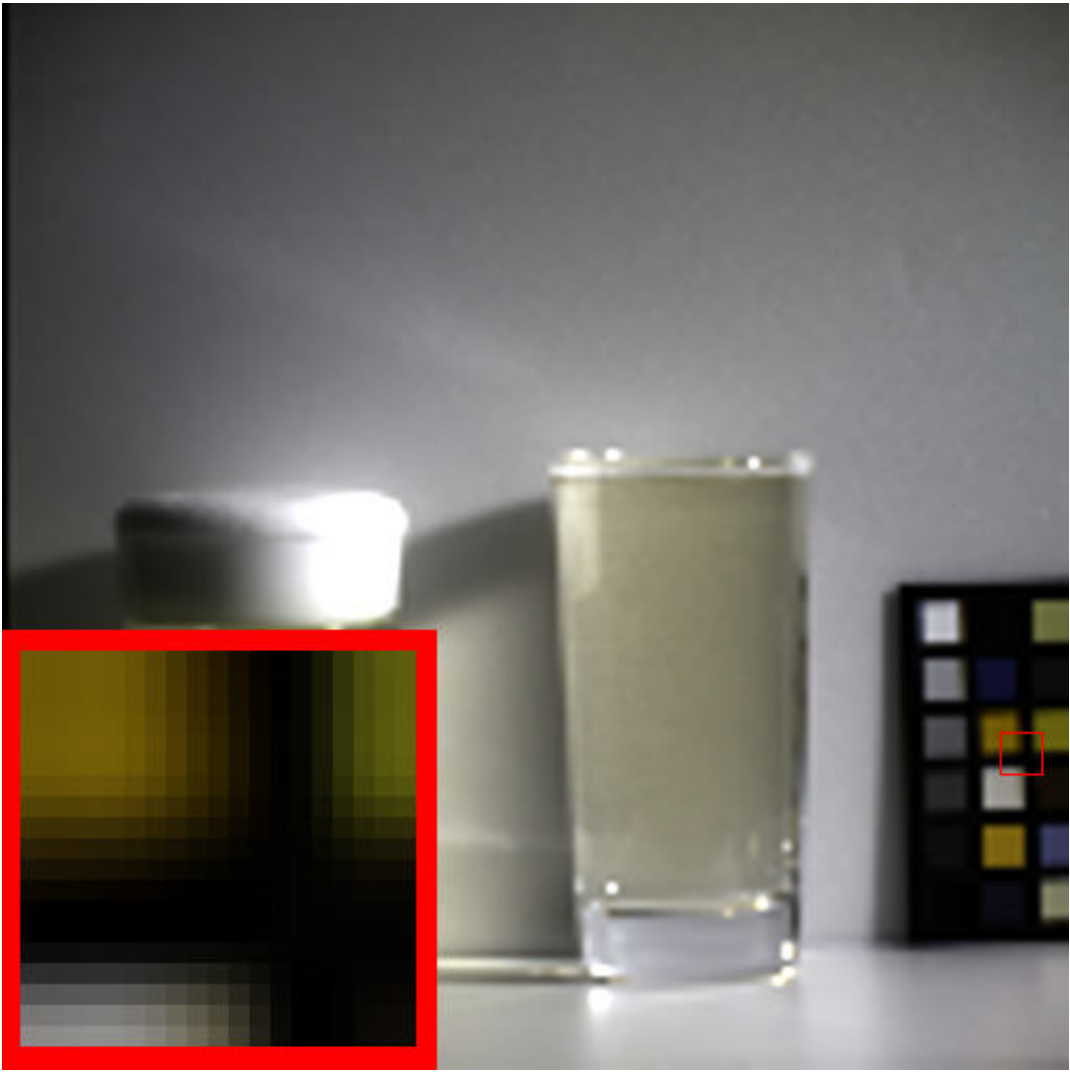}}
			\centering
			{ GT}
		\end{minipage}
		\begin{minipage}[t]{0.12\linewidth}
			{\includegraphics[width=1\linewidth]{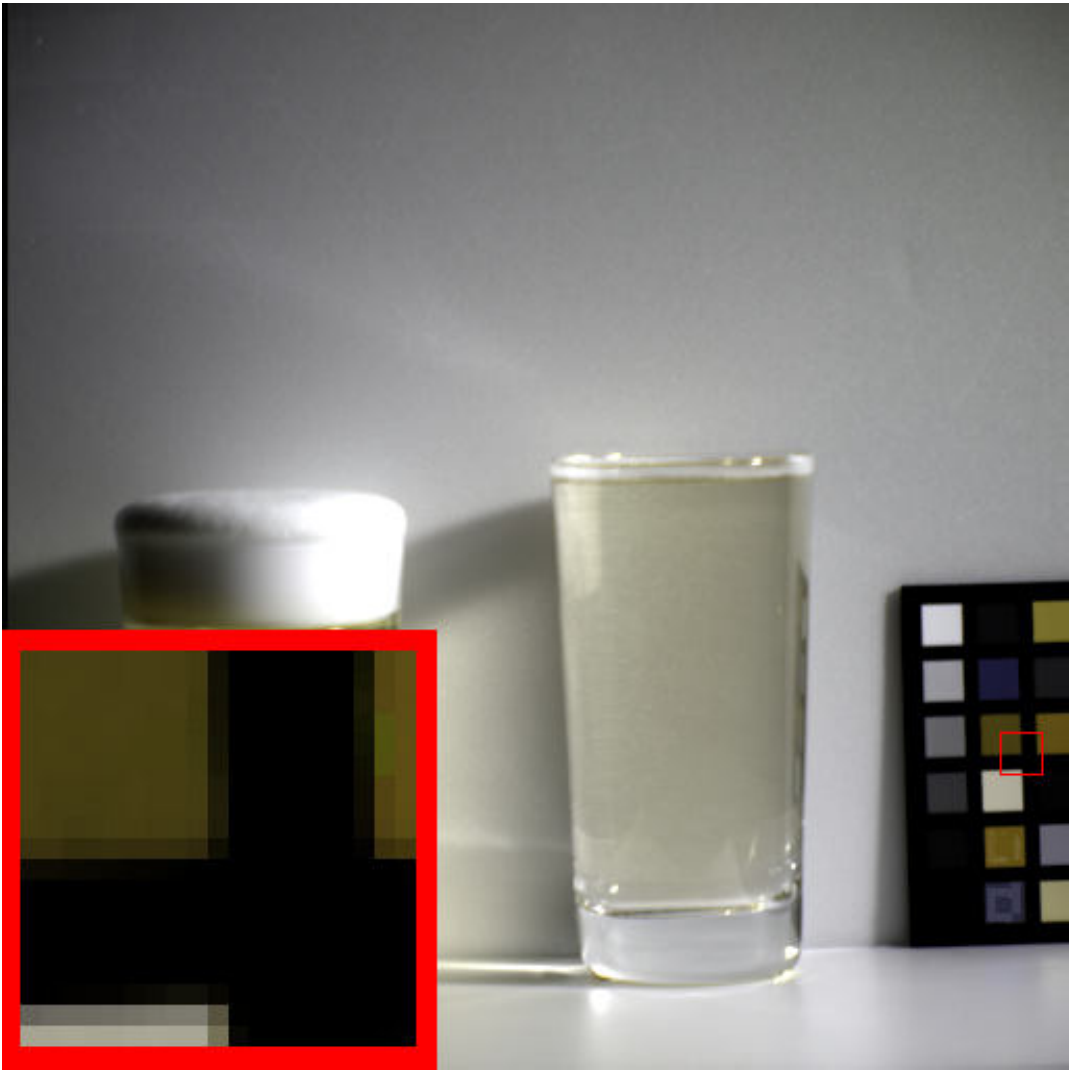}}
			{\includegraphics[width=1\linewidth]{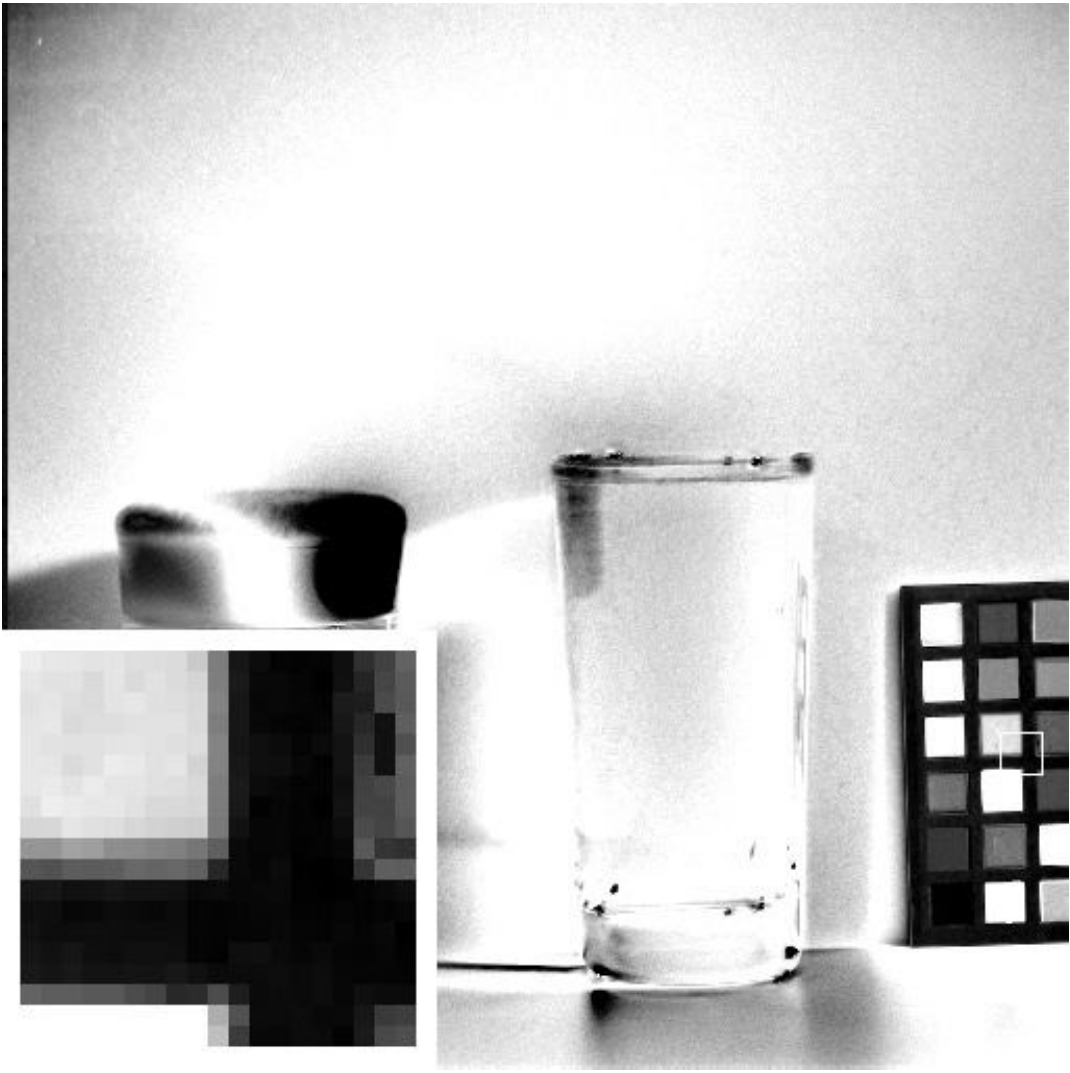}}
			\centering
			{CNMF\cite{CNMF}}
		\end{minipage}
		\begin{minipage}[t]{0.12\linewidth}
			{\includegraphics[width=1\linewidth]{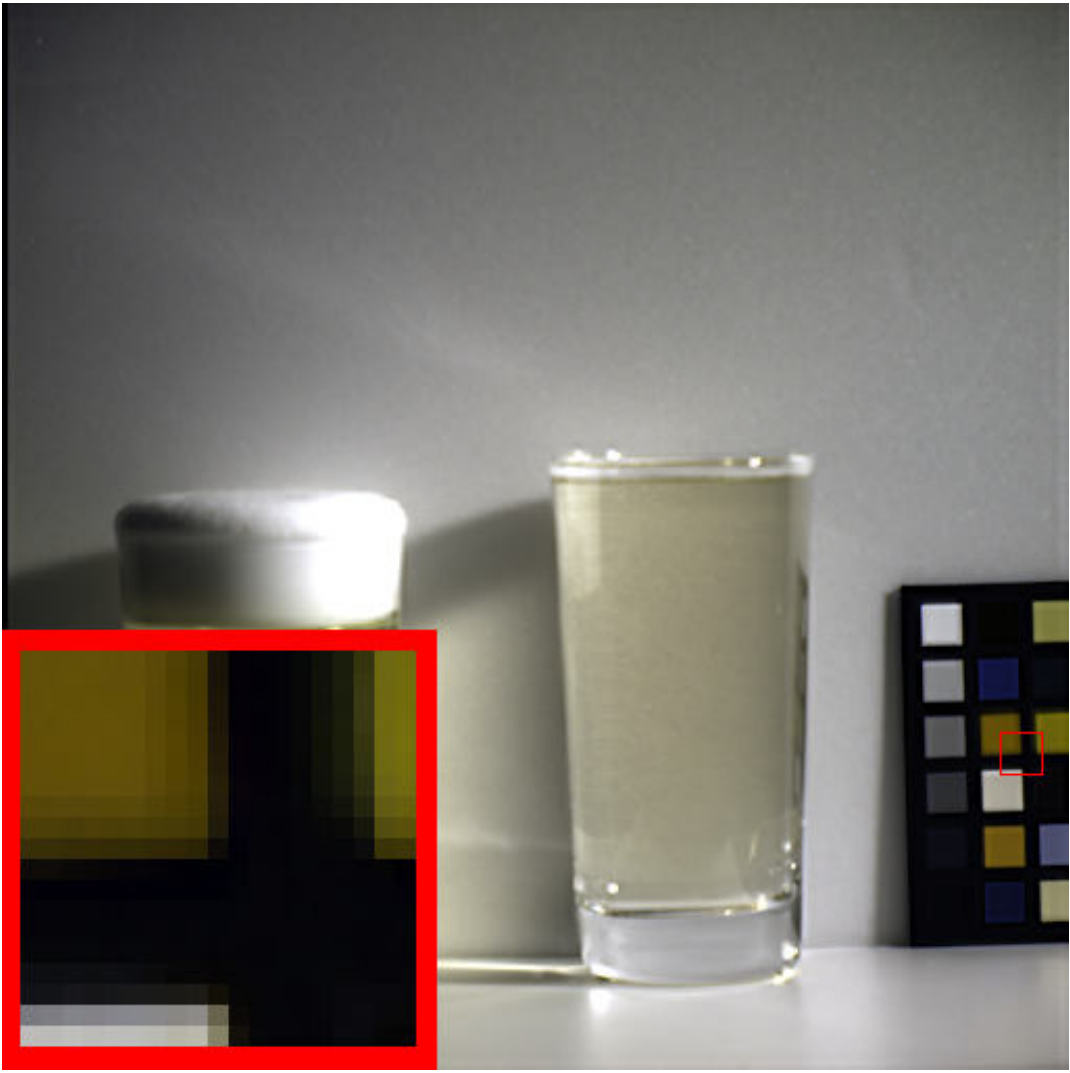}}
			{\includegraphics[width=1\linewidth]{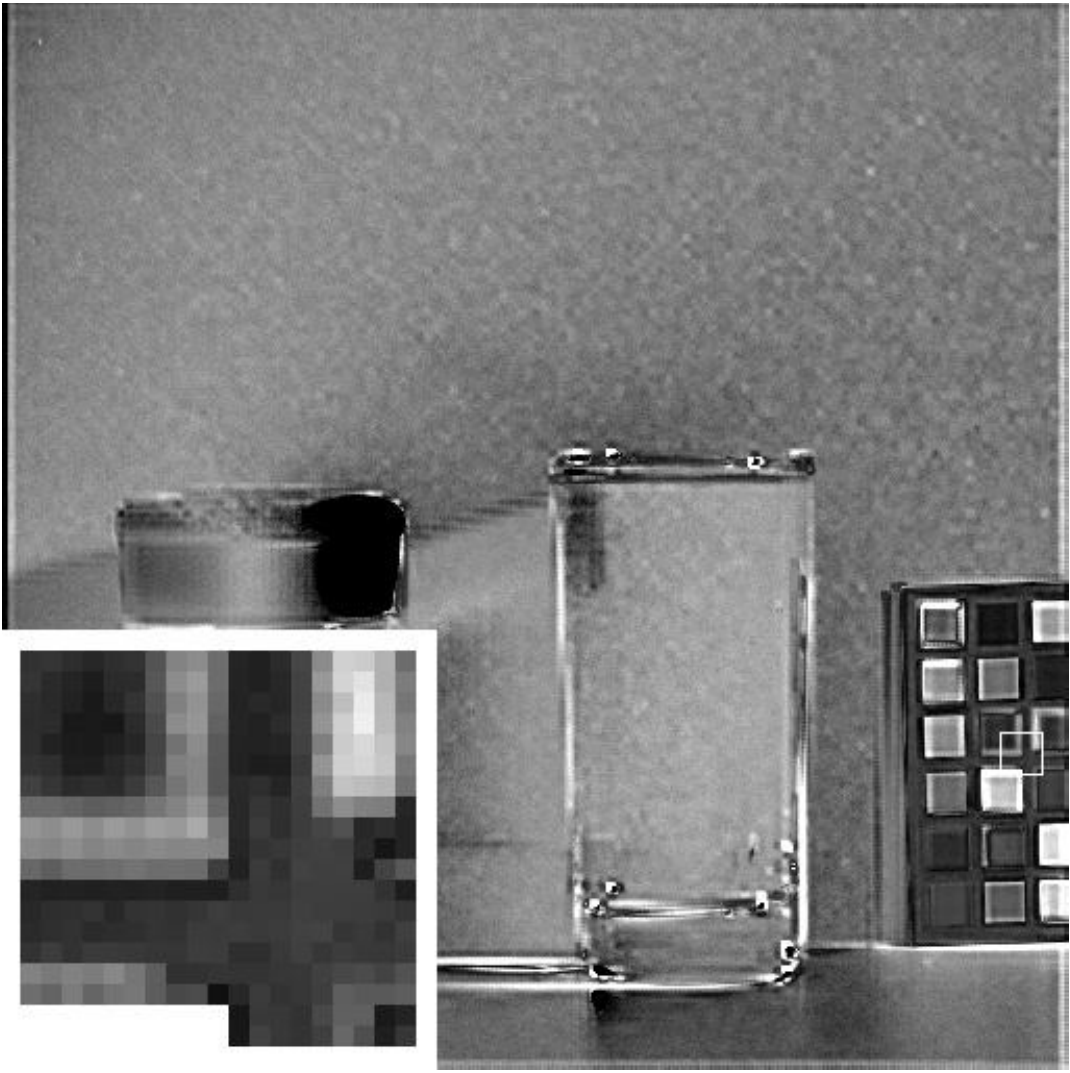}}
			\centering
			{ FUSE\cite{FUSE}}
		\end{minipage}
		\begin{minipage}[t]{0.12\linewidth}
			{\includegraphics[width=1\linewidth]{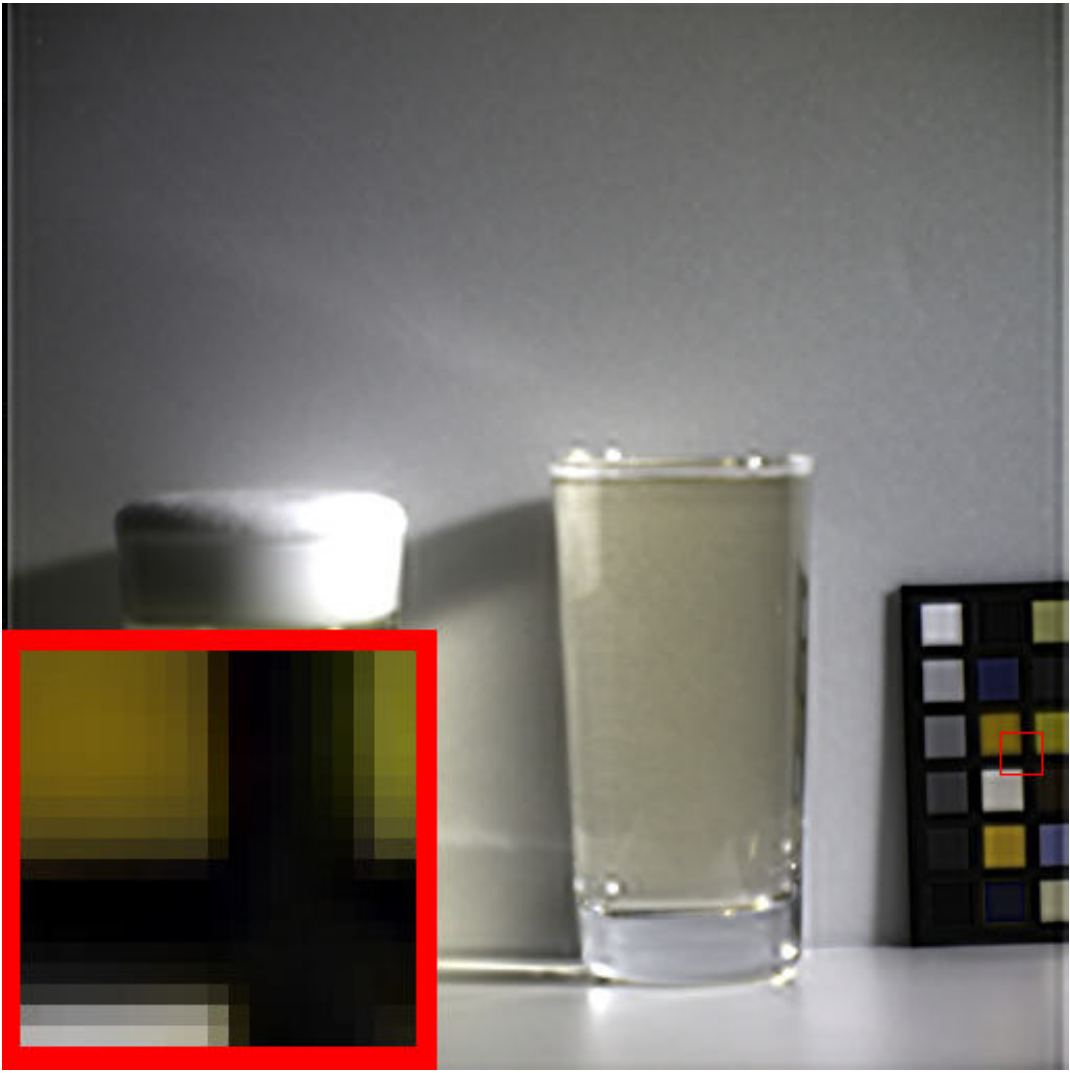}}
			{\includegraphics[width=1\linewidth]{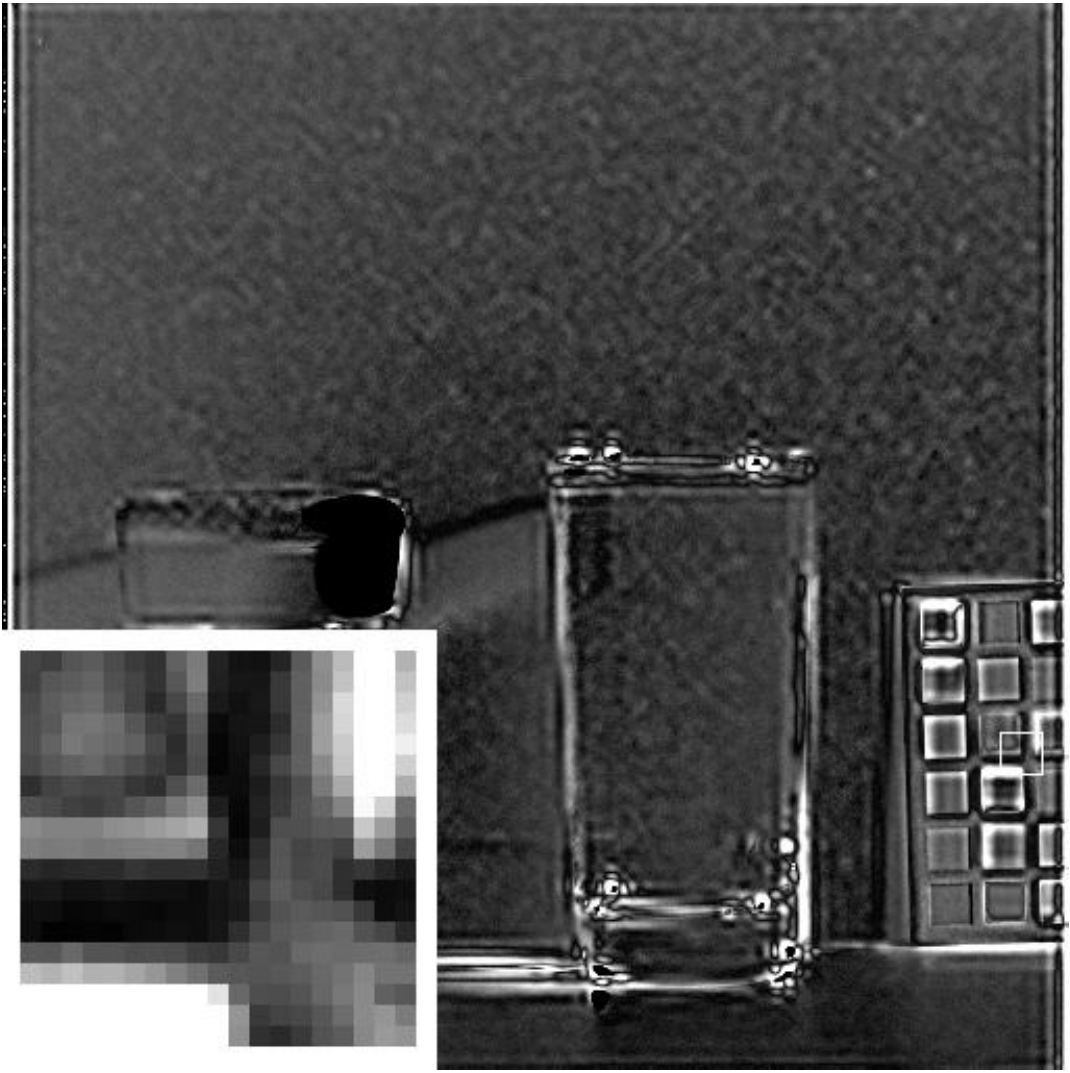}}
			\centering
			{GLP-HS\cite{GLP-HS}}
		\end{minipage}
		\begin{minipage}[t]{0.12\linewidth}
			{\includegraphics[width=1\linewidth]{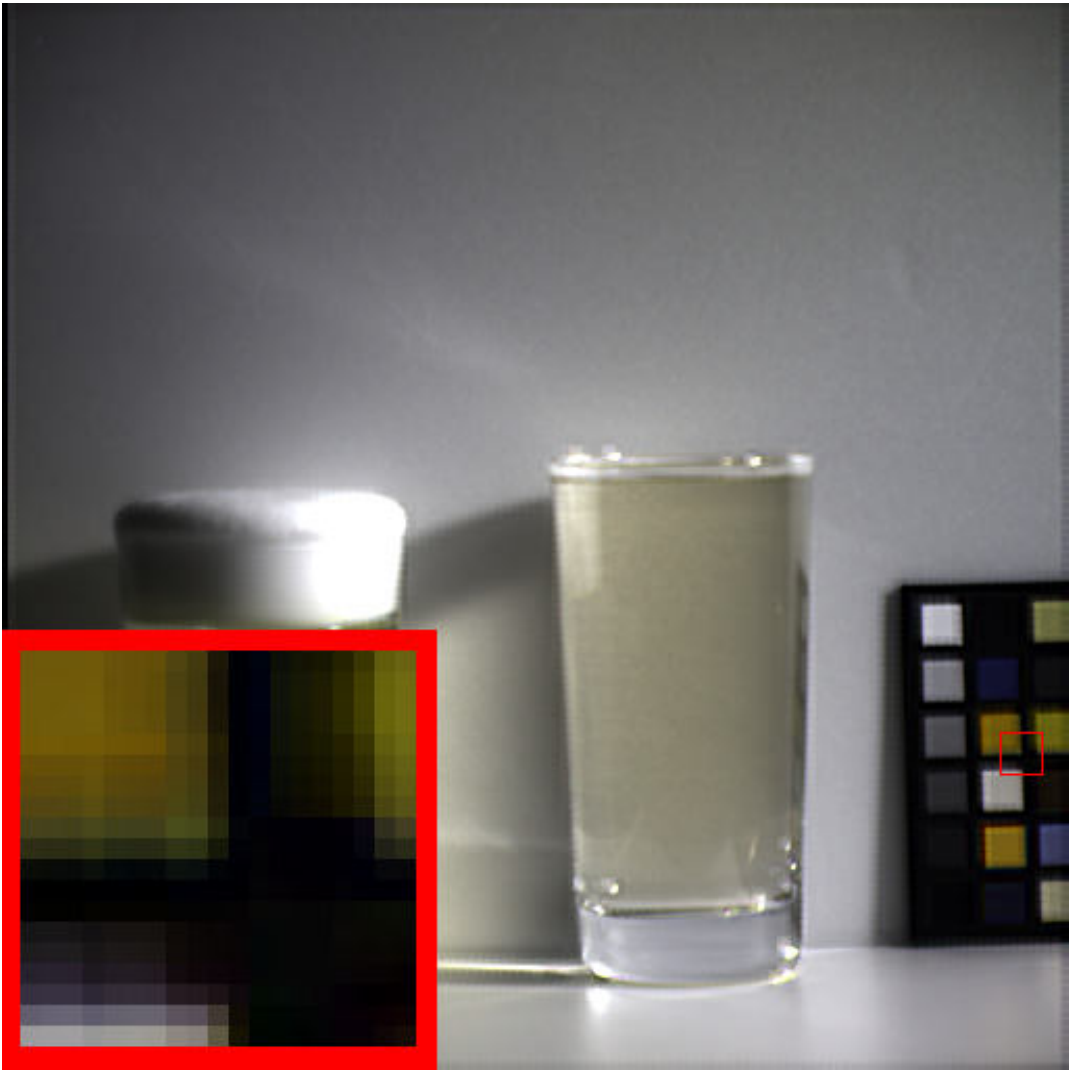}}
			{\includegraphics[width=1\linewidth]{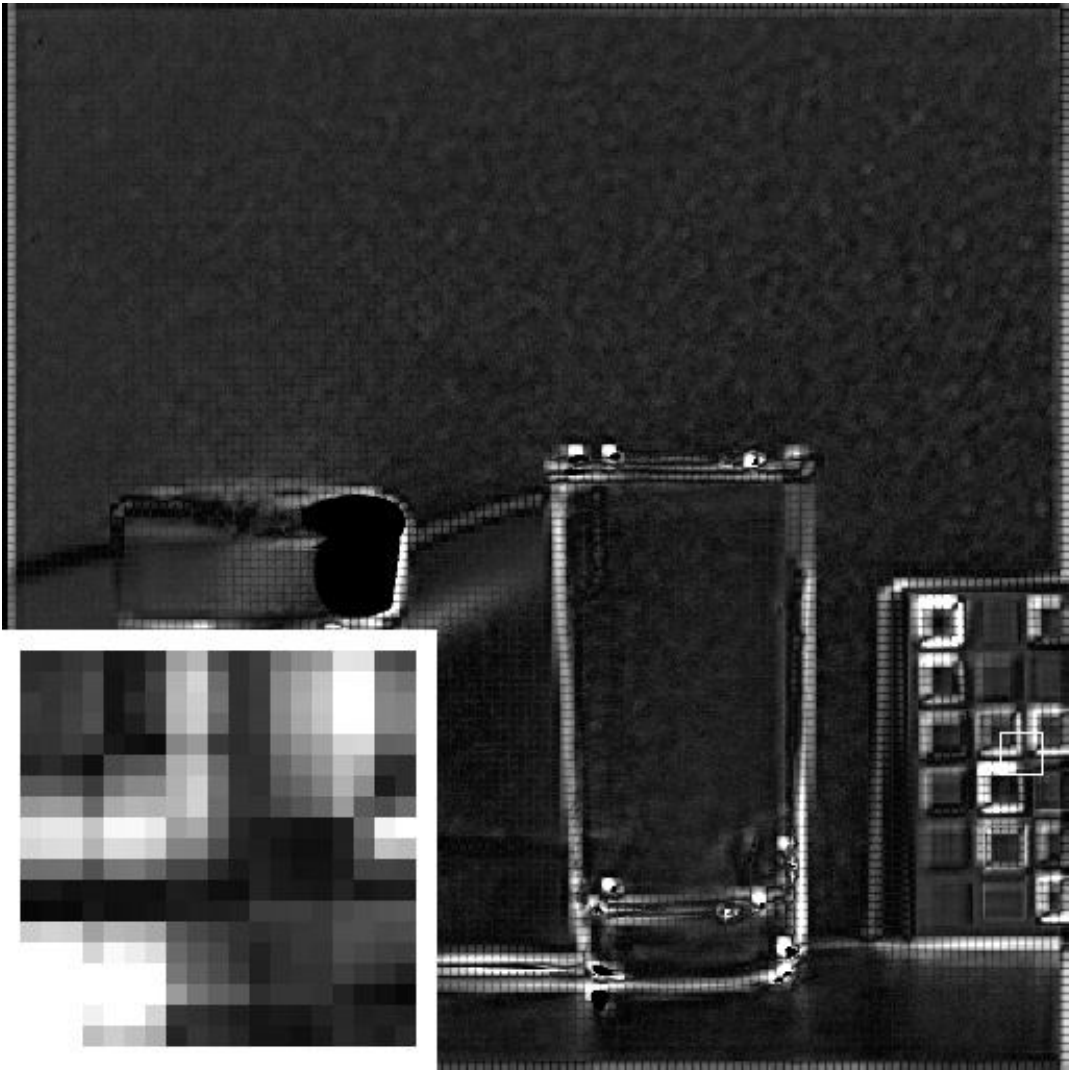}}
			\centering
			{LTTR\cite{LTTR}}
		\end{minipage}
		\begin{minipage}[t]{0.12\linewidth}
			{\includegraphics[width=1\linewidth]{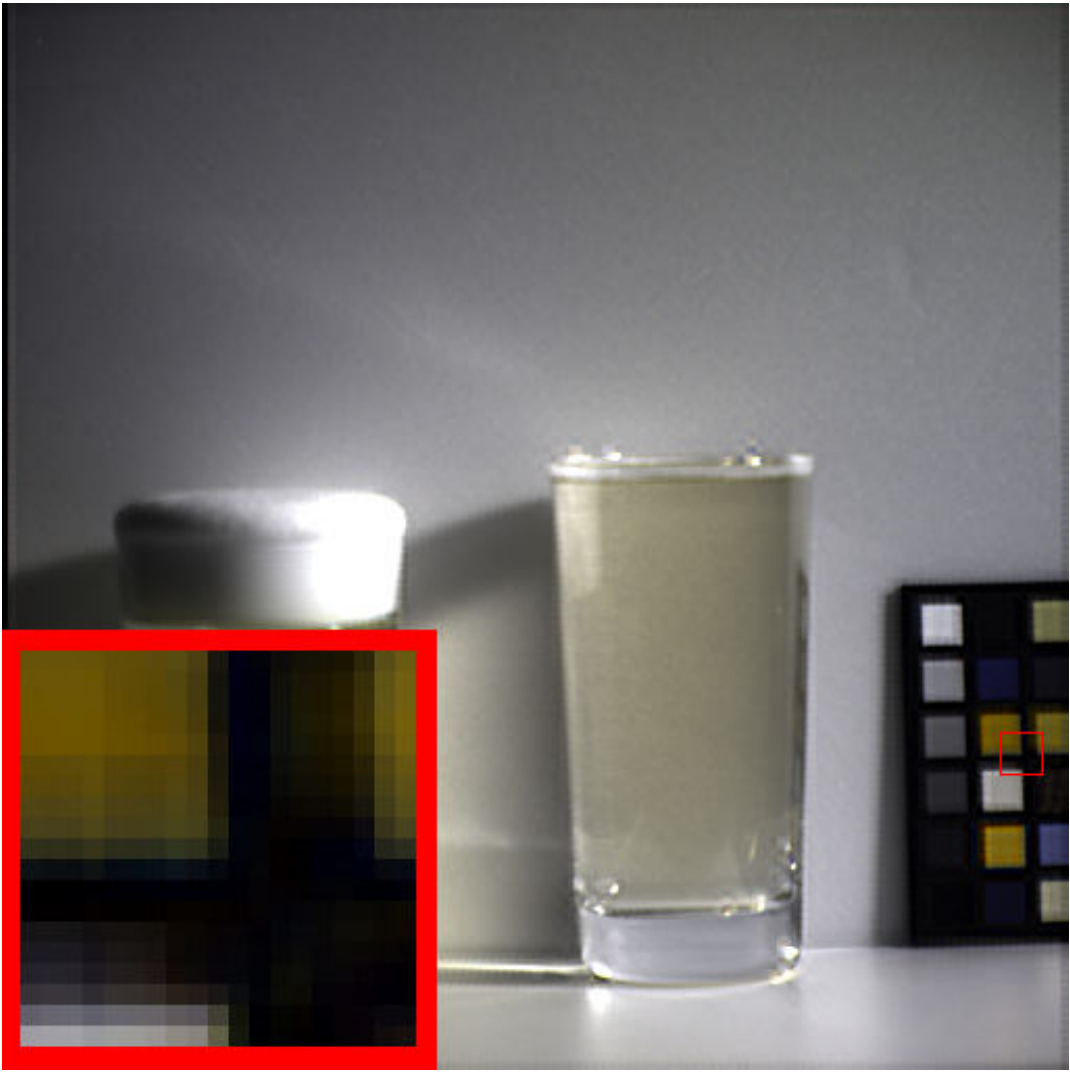}}
			{\includegraphics[width=1\linewidth]{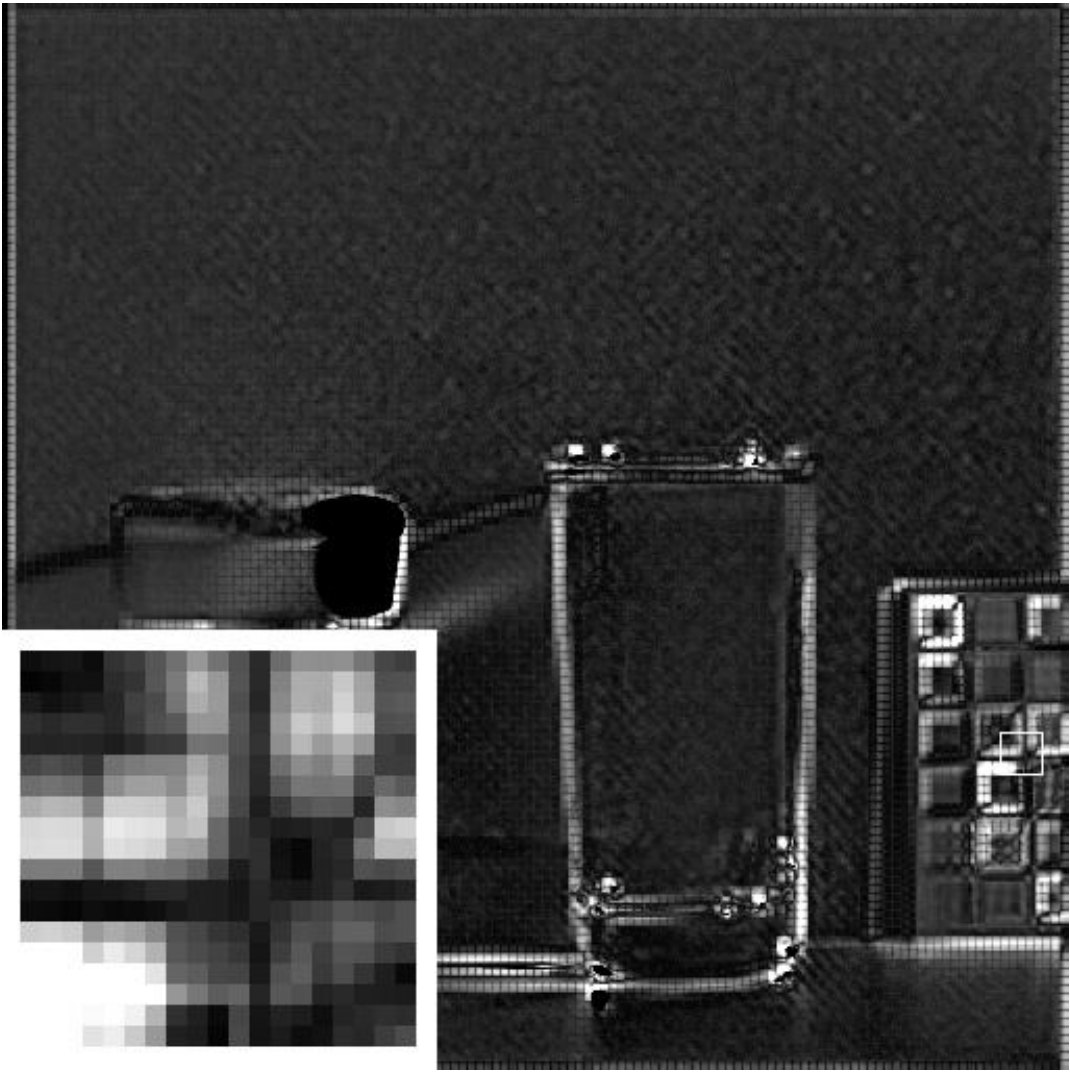}}
			\centering
			{LTMR\cite{LTMR}}
		\end{minipage}
		\begin{minipage}[t]{0.12\linewidth}
			{\includegraphics[width=1\linewidth]{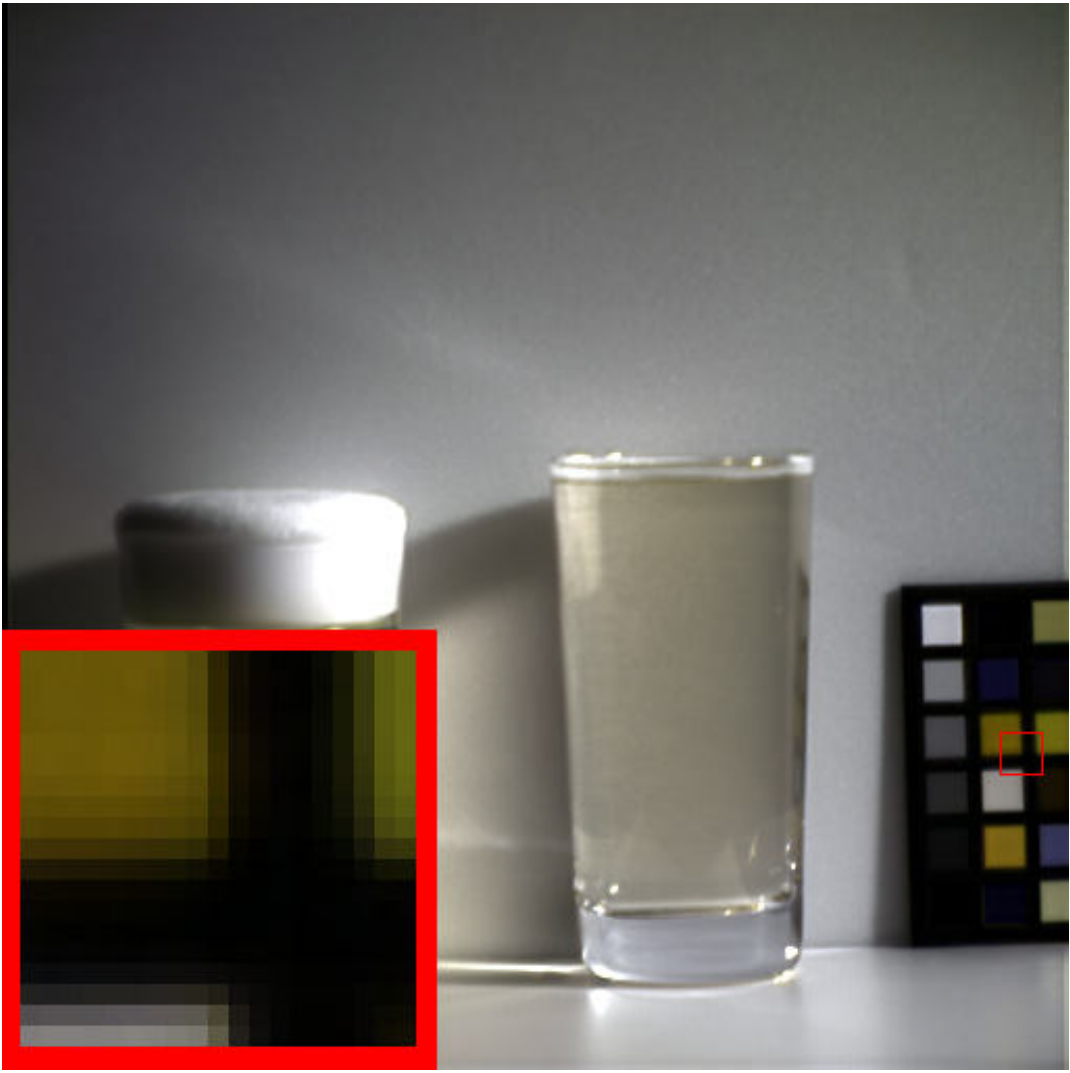}}
			{\includegraphics[width=1\linewidth]{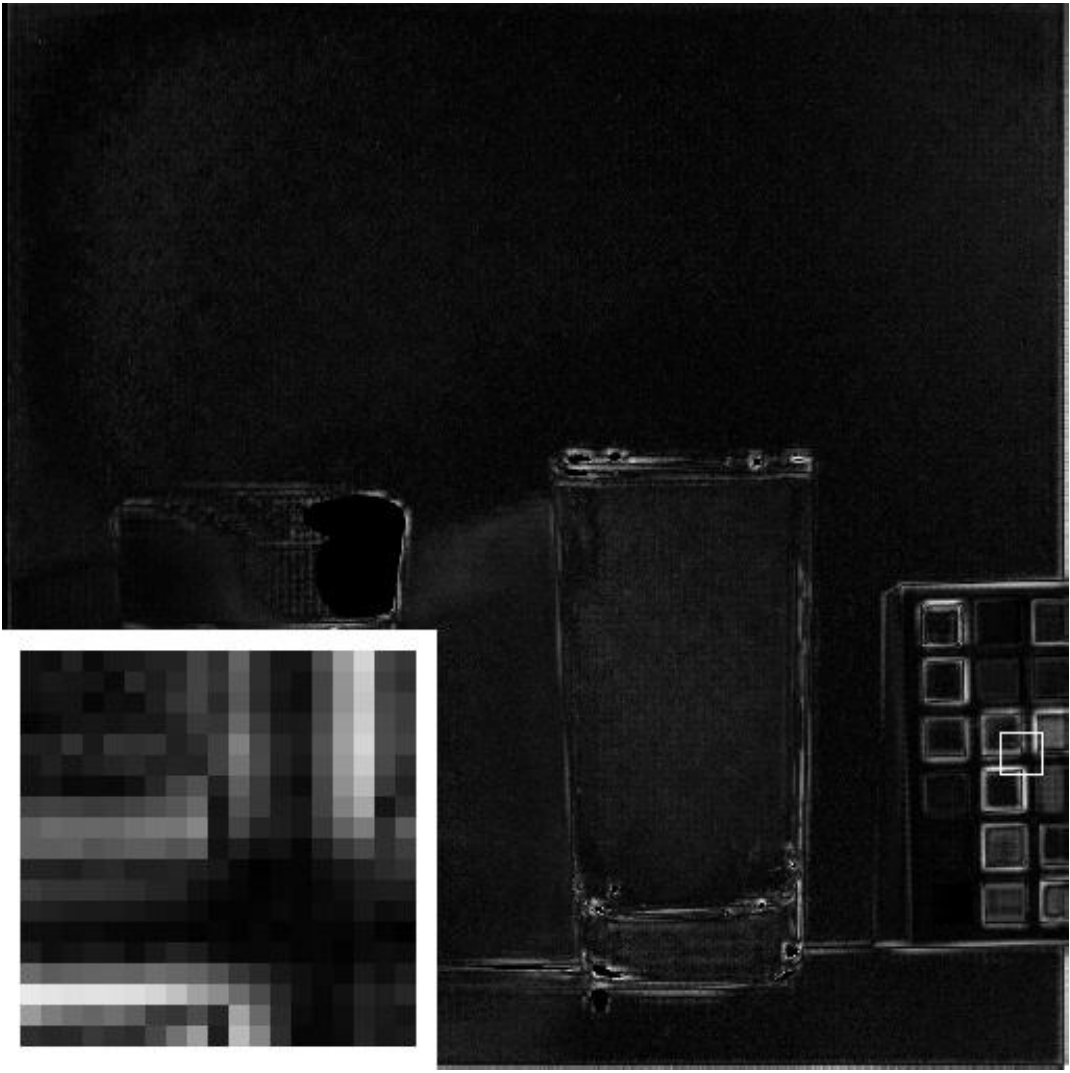}}
			\centering
			{MHFnet\cite{xie2019multispectral}}
		\end{minipage}
		\begin{minipage}[t]{0.12\linewidth}
			{\includegraphics[width=1\linewidth]{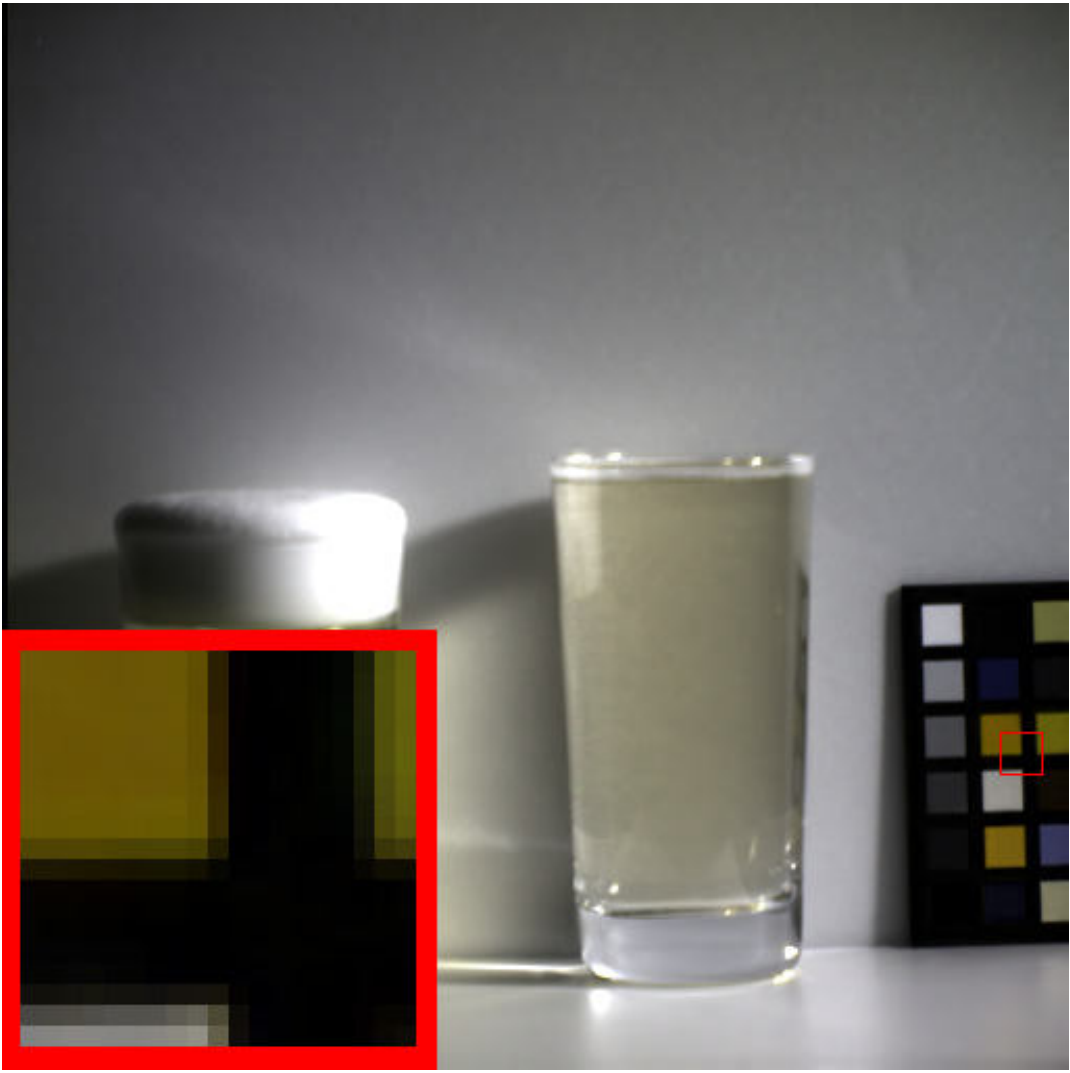}}
			{\includegraphics[width=1\linewidth]{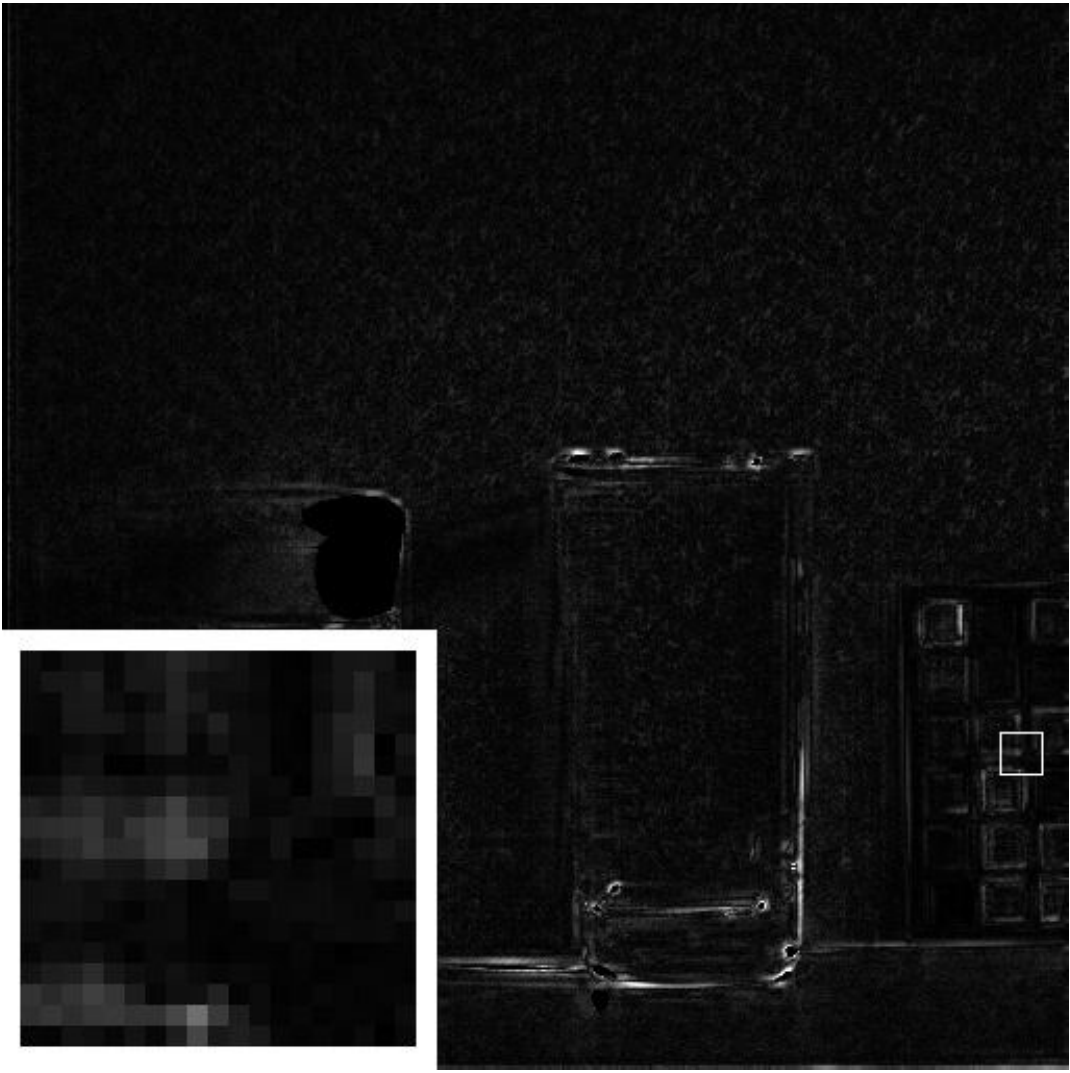}}
			\centering
			{HSRnet}
		\end{minipage}
	\end{minipage}
	\begin{minipage}[t]{0.04\linewidth}
		\vspace{0.7cm}
		{\includegraphics[height=10\linewidth,width=1\linewidth]{colobargray_01-eps-converted-to.pdf}}
	\end{minipage}
	\caption{The first column: the true pseudo-color images from the original CAVE dataset and the corresponding LR-HSI images of \textit{balloons} (R-23, G-18, B-7) (1st-2nd rows), \textit{clay} (R-3, G-16, B-2) (3rd-4th rows), and \textit{fake and real beers} (R-24, G-23, B-18) (5th-6th rows). 2nd-8th columns: the true pseudo-color fused products and the corresponding residuals for the different methods in the benchmark pointing out some close-ups to facilitate the visual analysis.}
	\label{F:cave}
\end{figure*}

\begin{figure}[t]
	\centering\renewcommand\arraystretch{0.65}\small
	\begin{tabular}{c}
		\includegraphics[height= 0.5\linewidth,width=0.85\linewidth]{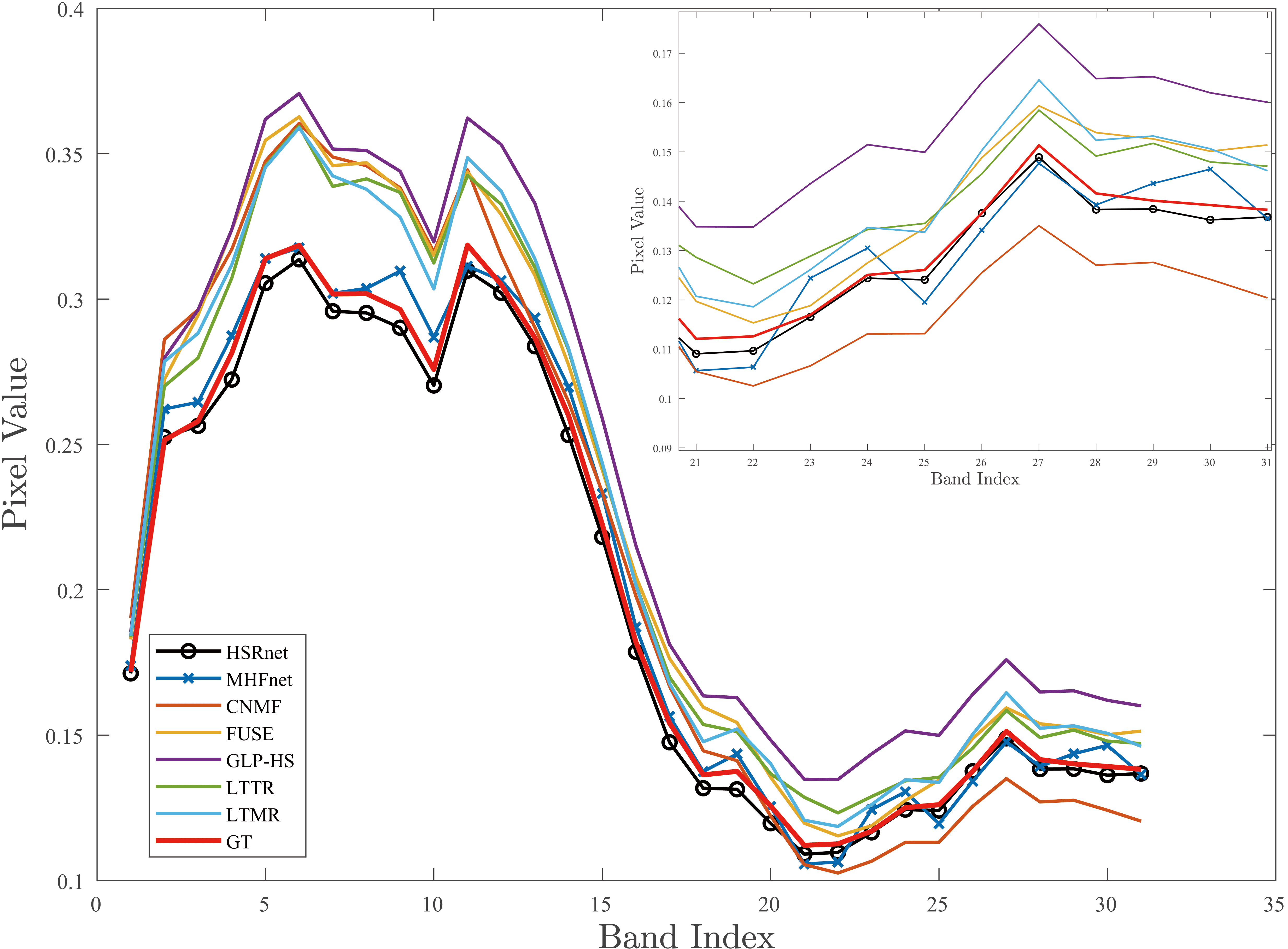}\\
		(a) \textit{balloons} $(276,277)$\\\\
		\includegraphics[height= 0.5\linewidth,width=0.85\linewidth]{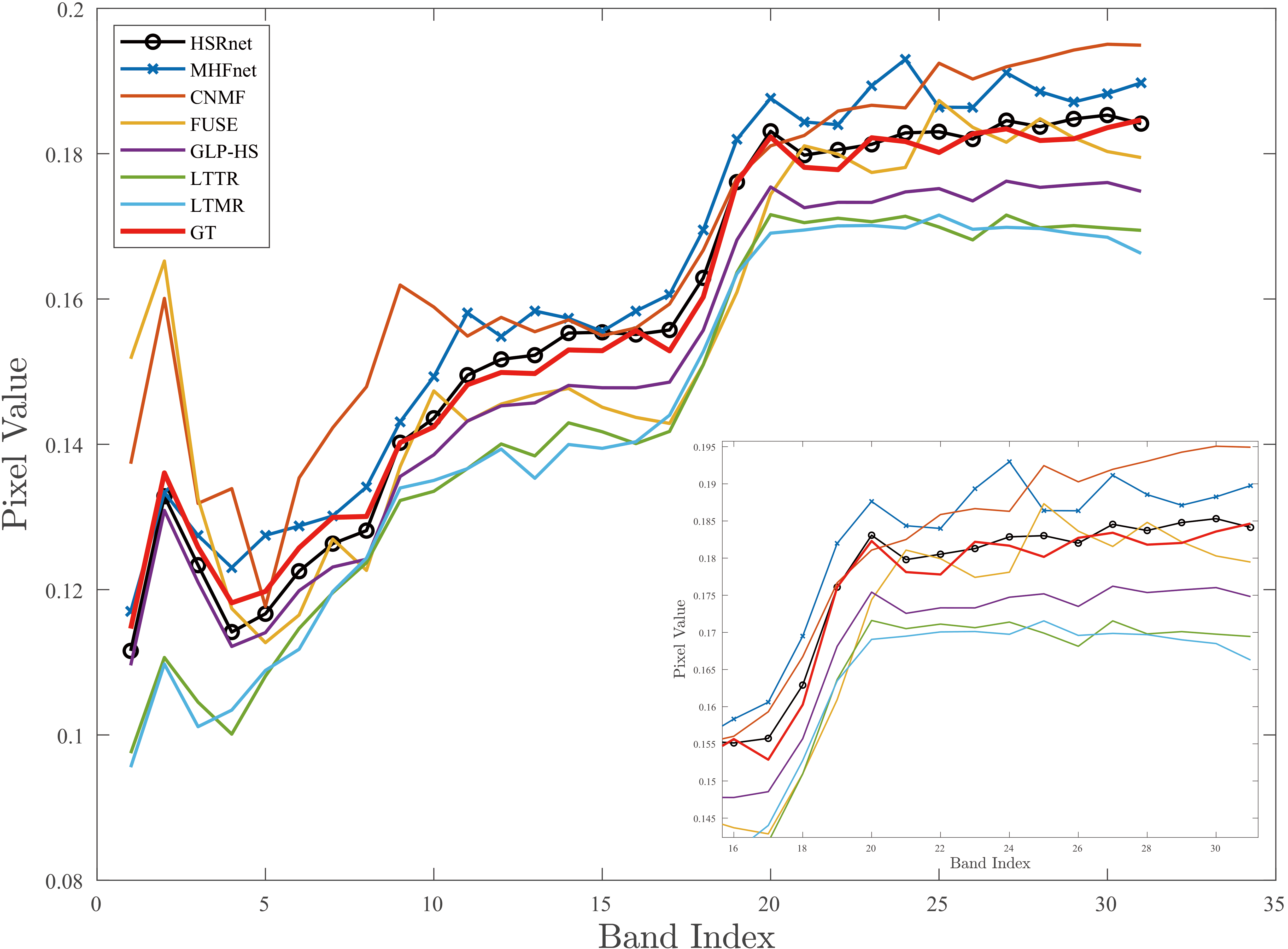}\\
		(b) \textit{fake and real beers} $(272,19)$
	\end{tabular}
	\caption{Selected spectral vectors for the outcomes coming from the different fusion methods and the ground-truth (GT). The indications of the specific dataset and the location of the pixel under analysis are also provided.
	}
	\label{spetral-rCAVE}\vspace{-4mm}
\end{figure}


Afterwards, we conduct the experiments on the whole 11 testing images.
Table \ref{cave11-ave} presents the average QIs on the 11 testing images. 
To ease the readers' burden, we only show the visual results on \textit{balloons}, \textit{clay}, and \textit{fake and real bears}.
Table \ref{qresult-4CAVE} lists the specific QIs of the results on these two images for the different methods.
The proposed method outperforms the compared approaches. Furthermore, the running time of the HSRnet is also the lowest one.
In Fig. \ref{F:cave}, we display the pseudo-color images of the fusion results and the corresponding error maps on three images.
From the error maps in Fig. \ref{F:cave}, it can be observed that the proposed HSRnet approach has a better reconstruction of the high resolution details with respect to the compared methods, thus clearly reducing the errors in the corresponding error maps.

The spectral fidelity is of crucial importance when the fusion of hyperspectral images is considered. In order to illustrate the spectral reconstruction provided by the different methods, we plot the spectral vectors for two exemplary cases, see Fig. \ref{spetral-rCAVE}. It is worth to be remarked that the spectral vectors estimated by our method and the ground-truth ones are very close to each other.


\subsection{Results on Harvard Dataset}\label{exp:general}

\begin{figure}[t]
	\begin{center}
		\begin{minipage}{ 0.98\linewidth}
			\begin{minipage}{ 0.19\linewidth}
				{\includegraphics[width=1\linewidth]{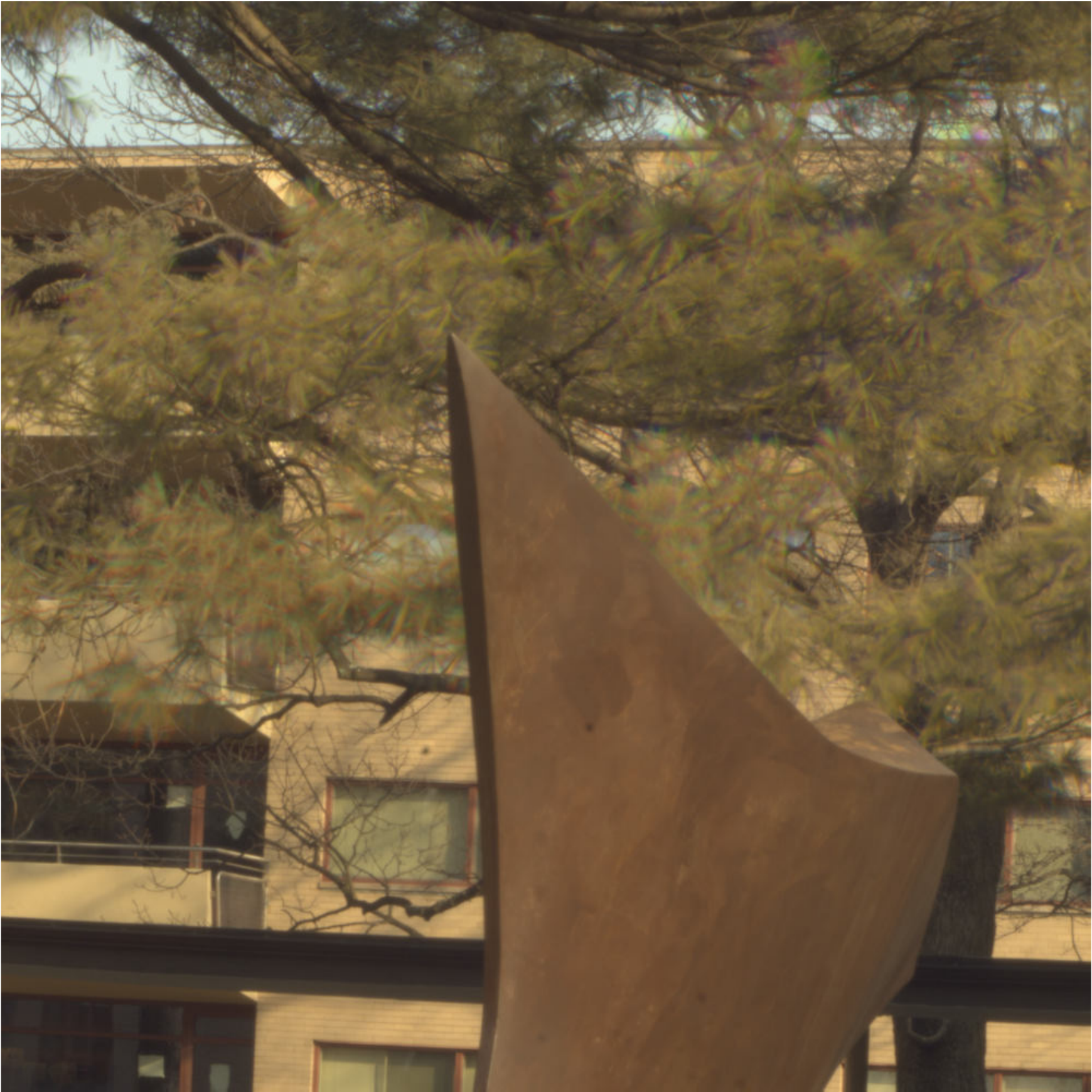}}
				\centering
				{(a)}
			\end{minipage}
			\begin{minipage}{ 0.19\linewidth}
				{\includegraphics[width=1\linewidth]{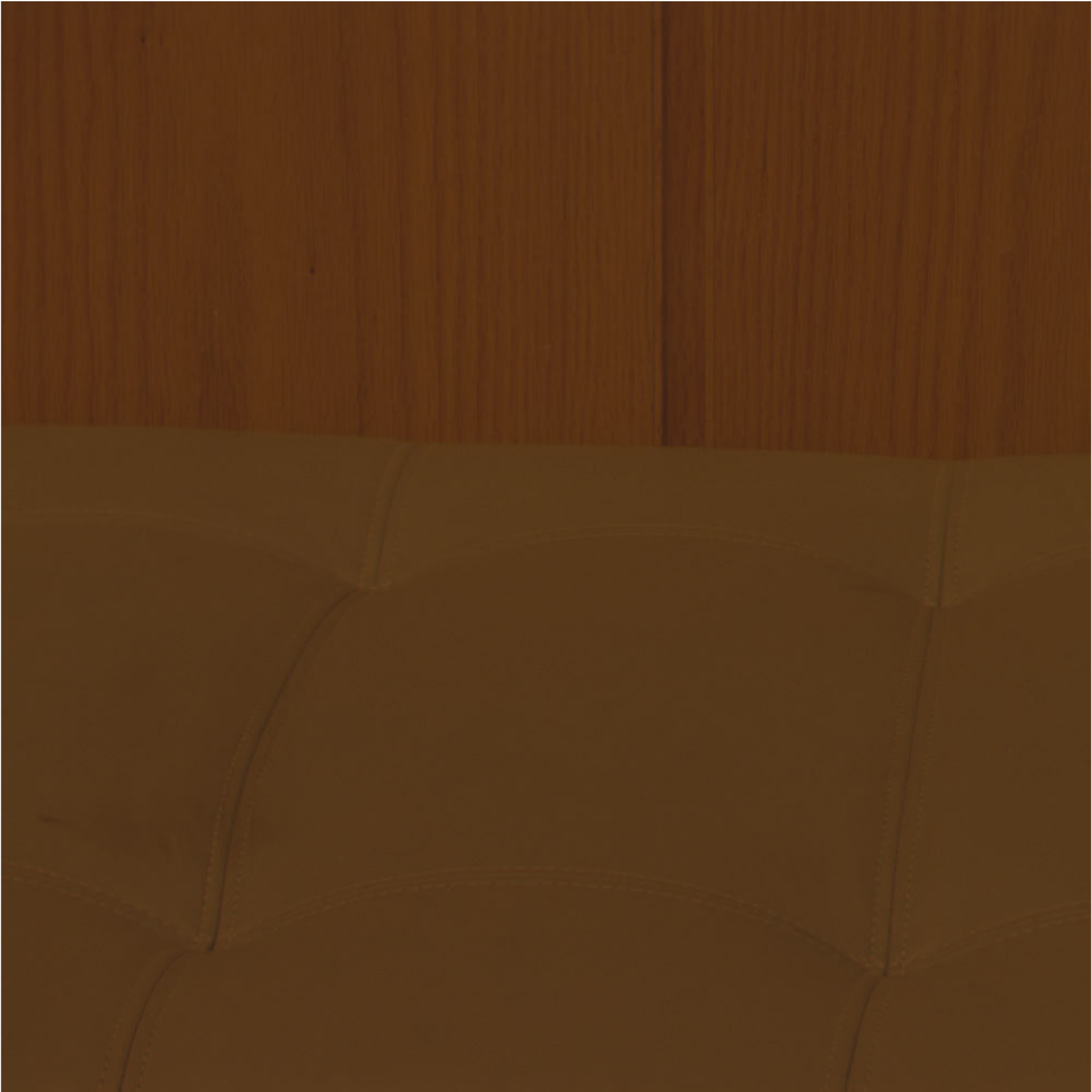}}
				\centering
				{(b)}
			\end{minipage}
			\begin{minipage}{ 0.19\linewidth}
				{\includegraphics[width=1\linewidth]{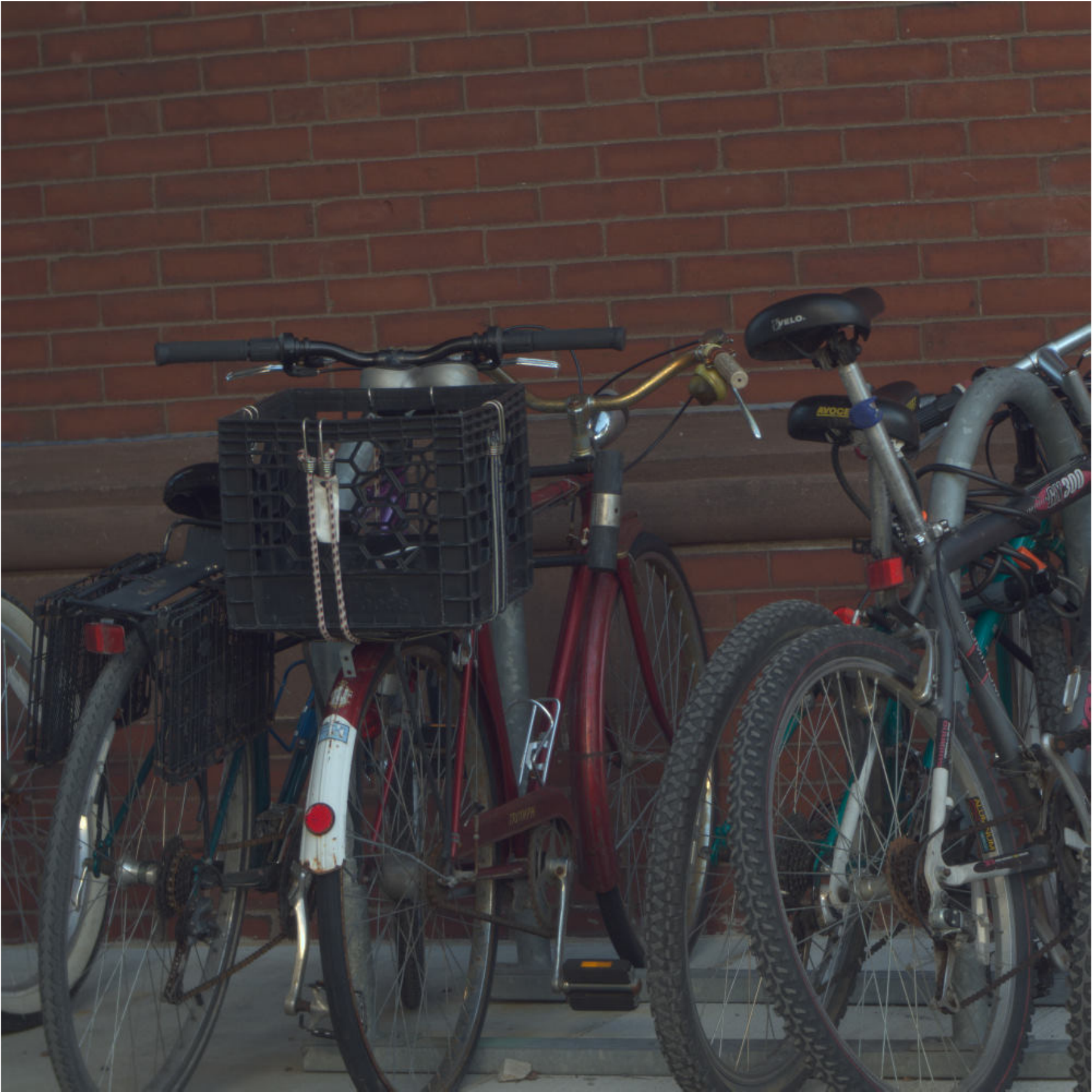}}
				\centering
				{(c)}
			\end{minipage}
			\begin{minipage}{ 0.19\linewidth}
				{\includegraphics[width=1\linewidth]{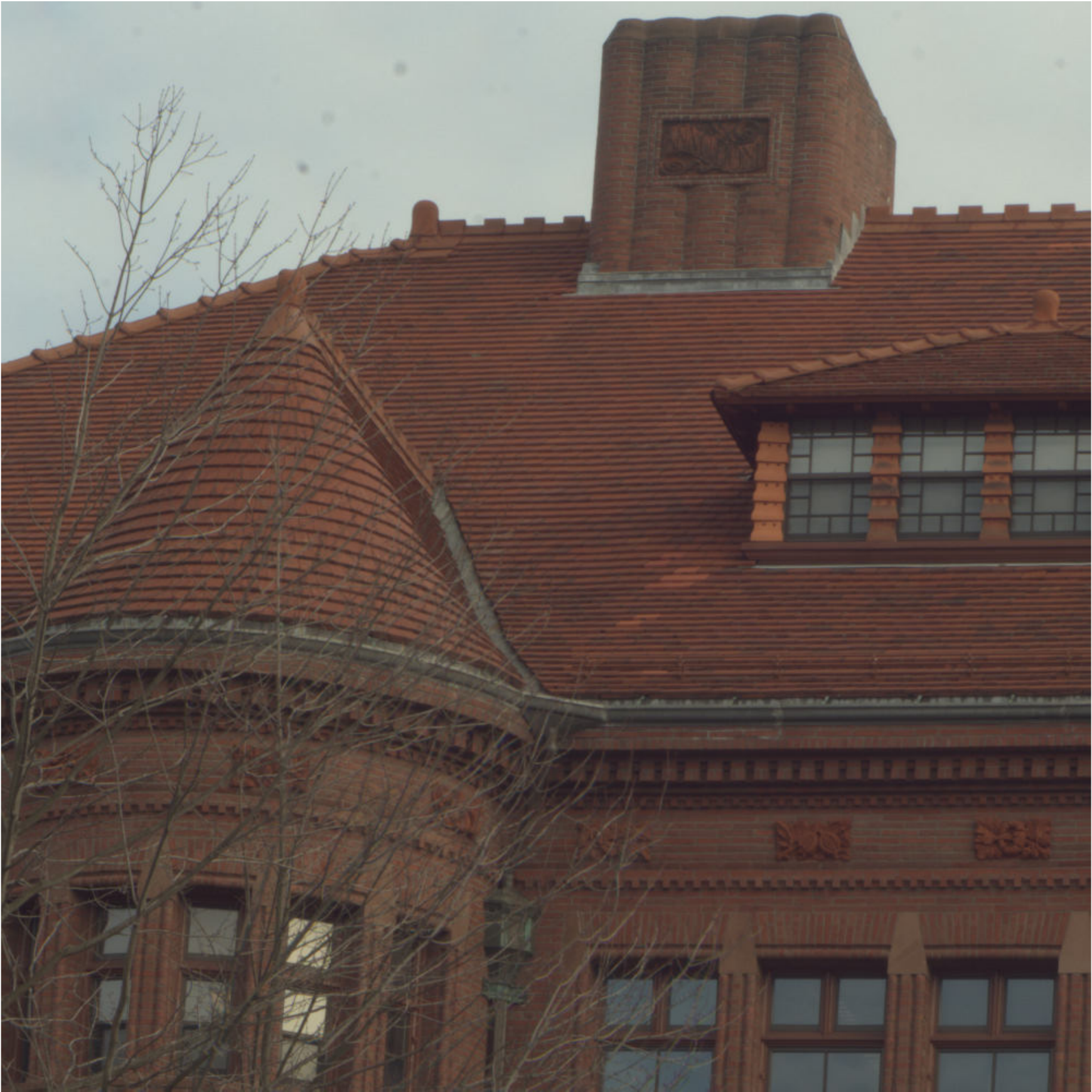}}
				\centering
				{(d)}
			\end{minipage}
			\begin{minipage}{ 0.19\linewidth}
				{\includegraphics[width=1\linewidth]{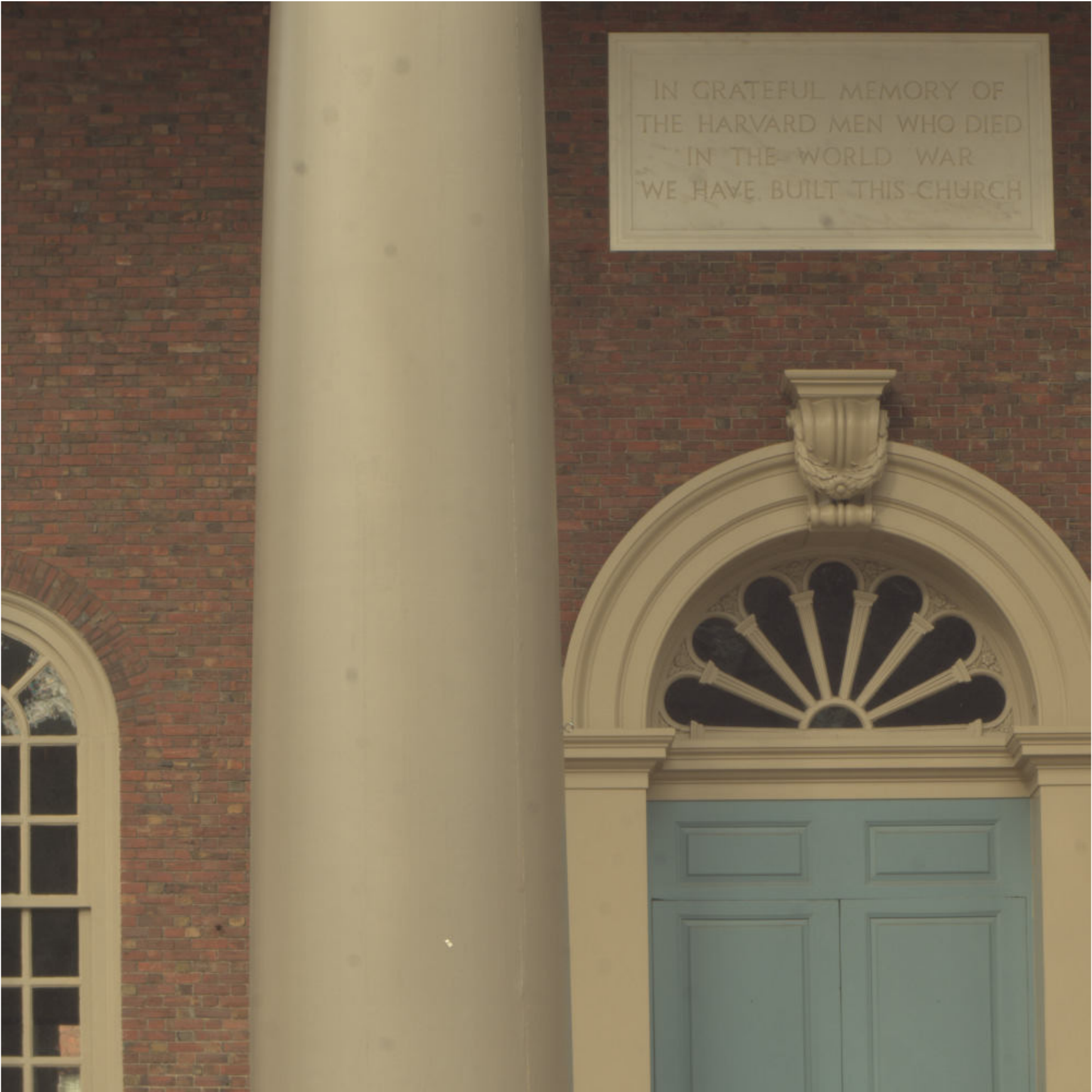}}
				\centering
				{(e)}
			\end{minipage}
			
			\begin{minipage}{ 0.19\linewidth}
				{\includegraphics[width=1\linewidth]{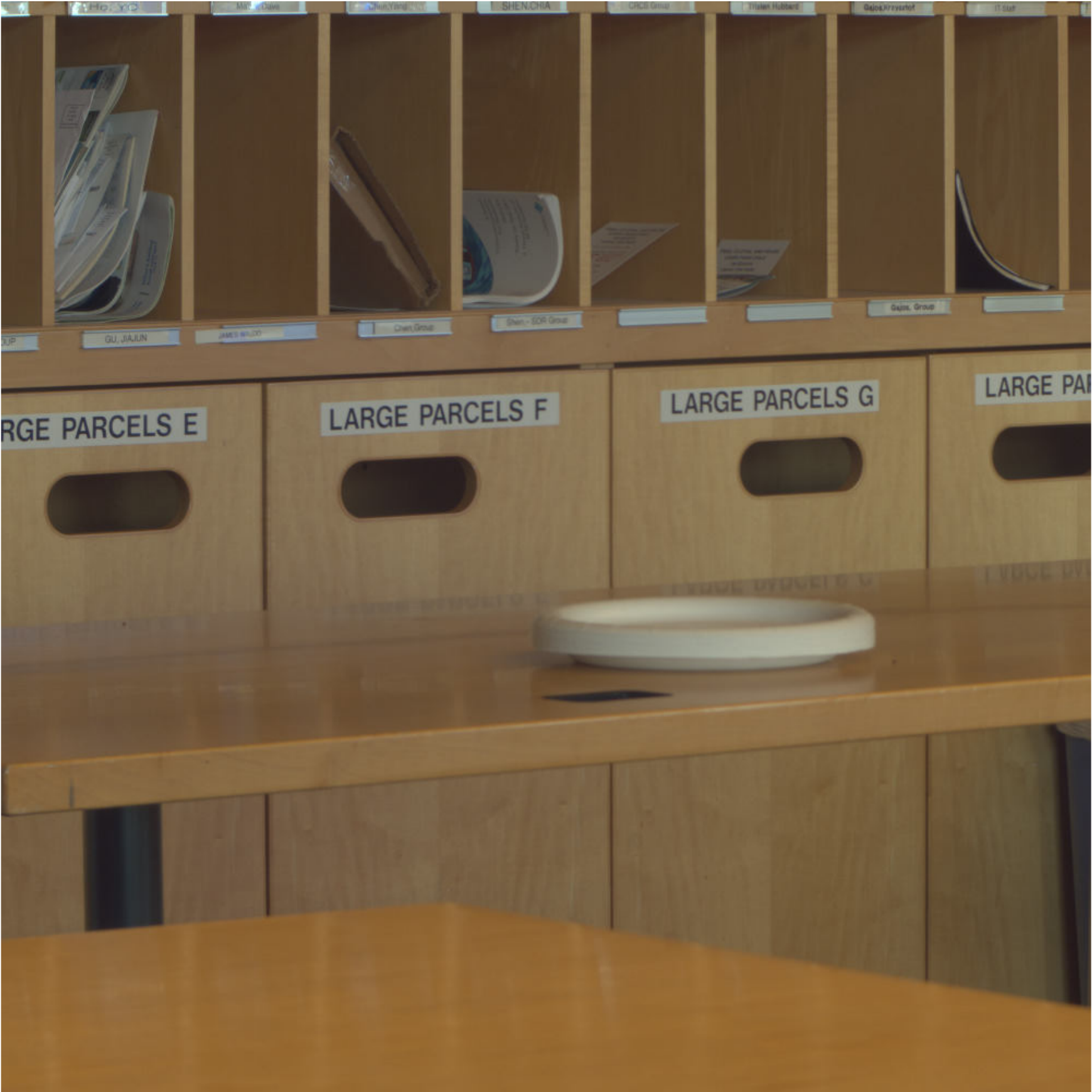}}
				\centering
				{(f)}
			\end{minipage}
			\begin{minipage}{ 0.19\linewidth}
				{\includegraphics[width=1\linewidth]{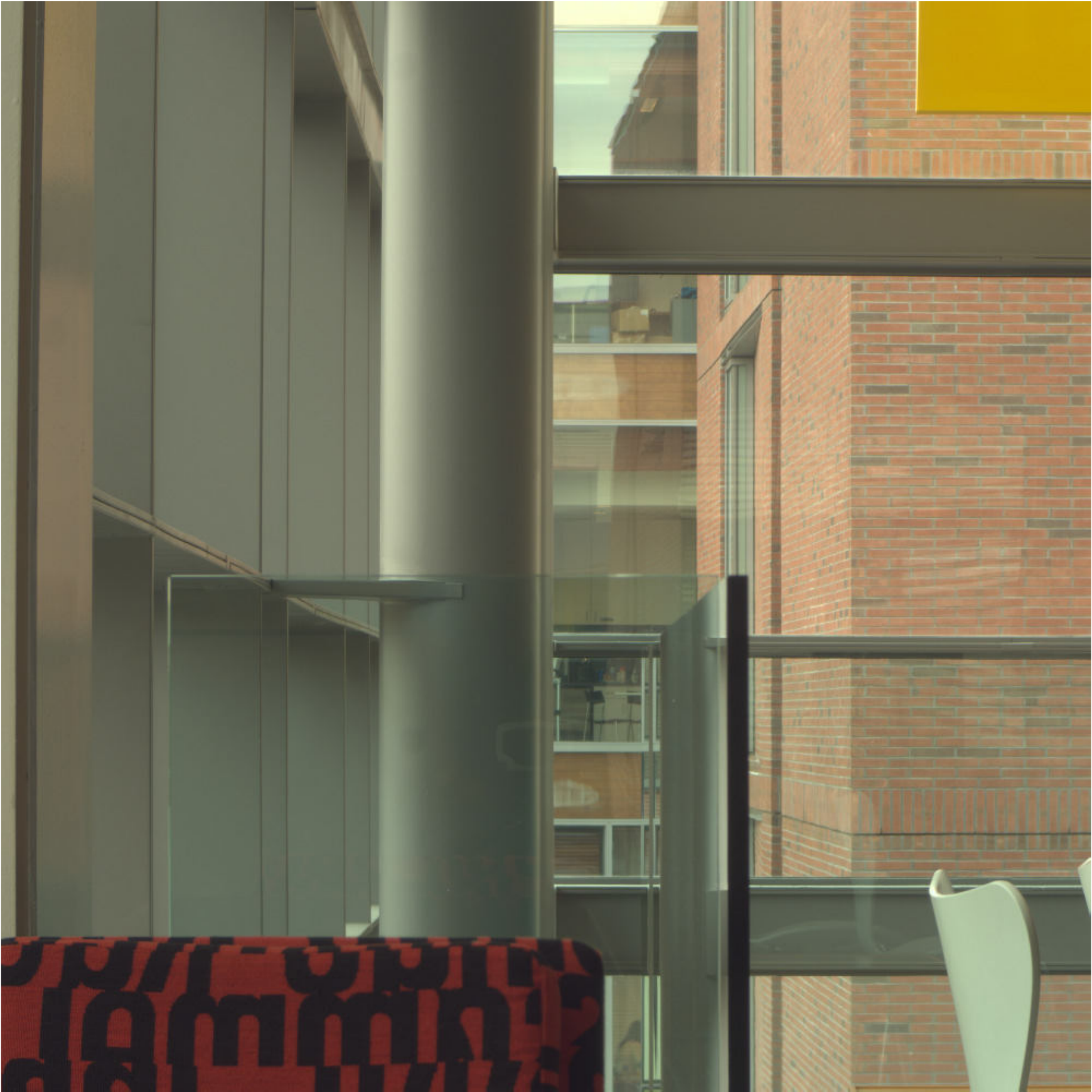}}
				\centering
				{(g)}
			\end{minipage}
			\begin{minipage}{ 0.19\linewidth}
				{\includegraphics[width=1\linewidth]{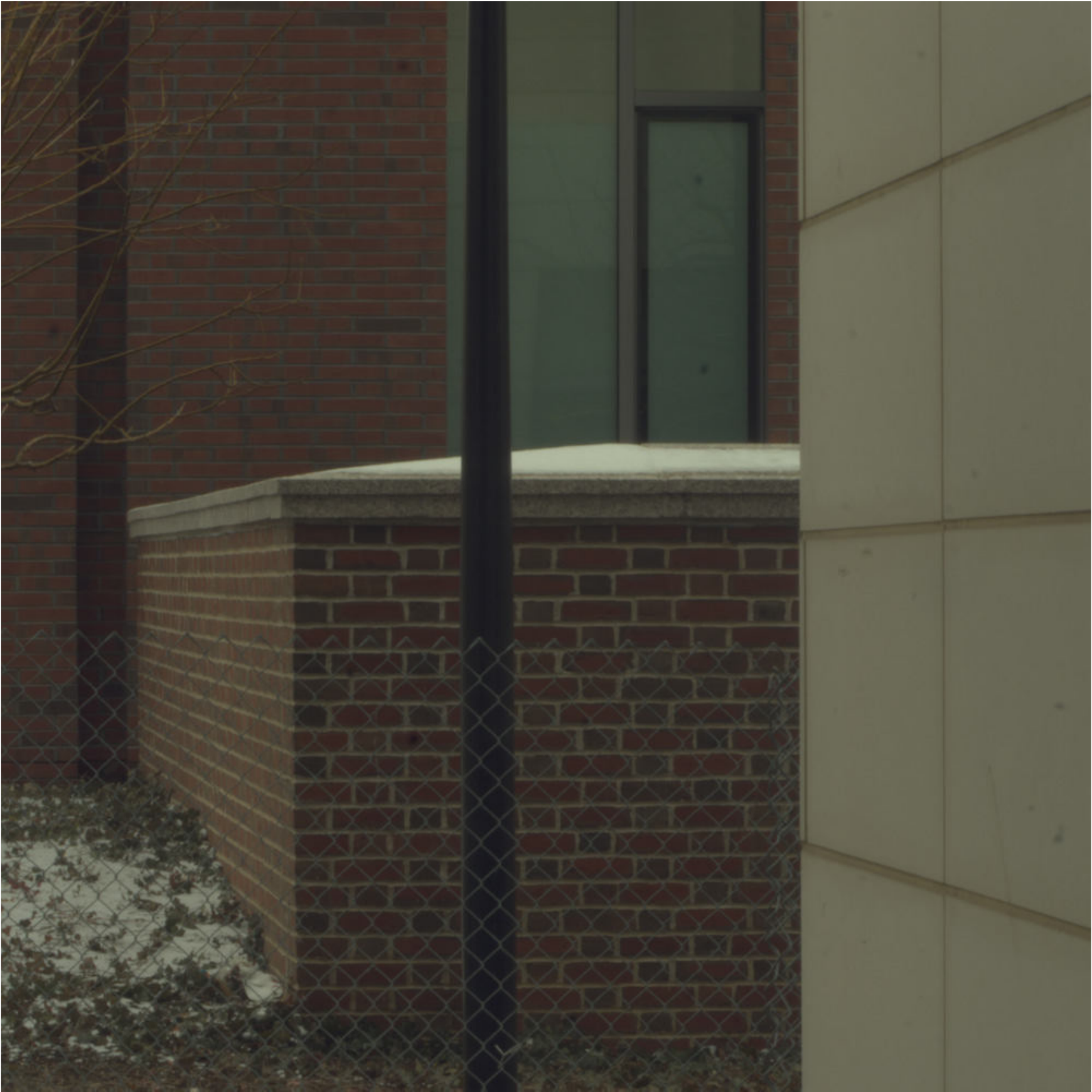}}
				\centering
				{(h)}
			\end{minipage}
			\begin{minipage}{ 0.19\linewidth}
				{\includegraphics[width=1\linewidth]{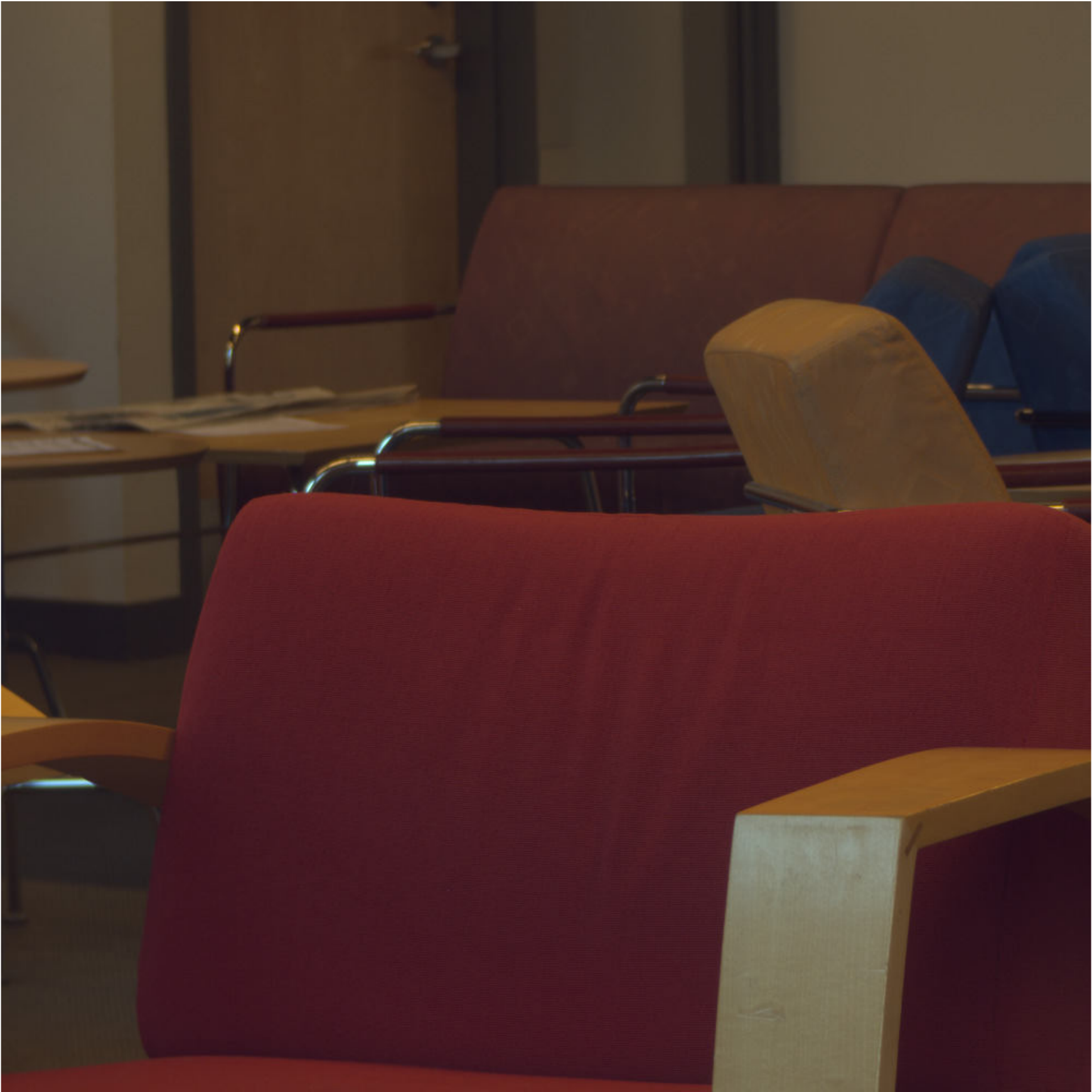}}
				\centering
				{(i)}
			\end{minipage}
			\begin{minipage}{ 0.19\linewidth}
				{\includegraphics[width=1\linewidth]{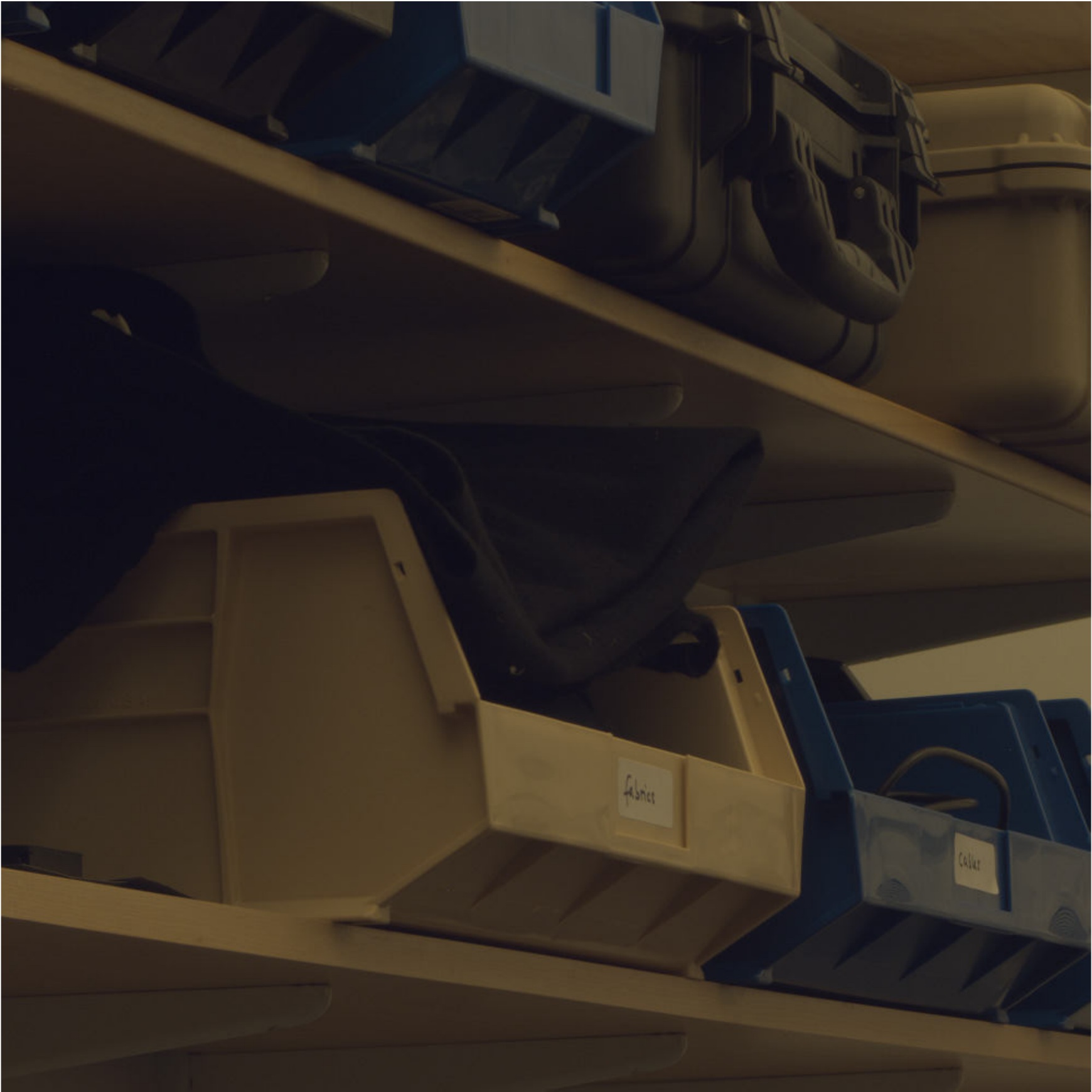}}
				\centering
				{(j)}
			\end{minipage}
			\centering
		\end{minipage}
	\end{center}
	\caption{The 10 testing images from the Harvard dataset. (a) \textit{tree}, (b) \textit{cushion}, (c)\textit{bikes}, (d) \textit{roof}, (e) \textit{door}, (f) \textit{cabinet}, (g) \textit{window}, (h) \textit{wall}, (i) \textit{chairs}, (j) \textit{baskets}.} \label{harvard_test}
\end{figure}

The Harvard dataset is a public dataset that has 77 HSIs of indoor and outdoor scenes including different kinds of objects and buildings.
Every HSI has a spatial size of 1392$\times$1040 with 31 spectral bands, and the spectral bands are acquired at an interval of 10nm in the range of 420-720nm. 10 images are randomly selected for testing. The test images are shown in Fig. \ref{harvard_test}.

As in the previous settings, the original data is regarded as the ground-truth HR-HSI.
The LR-HSI data is simulated as in Sec. \ref{Sec-Train}.
Instead, the HR-MSI (not already available for this dataset) is obtained by applying the method provided by \cite{cie2006fundamental}, where the spectral response functions are obtained from CIE\footnote {http://www.cvrl.org}.

We would like to remark that both our method and the MHFnet are trained on the CAVE dataset, and we directly test them on the Harvard dataset without any retraining or fine-tuning. Thus, the performance on the Harvard dataset of these two methods could reflect their generalization abilities.

\begin{table}[t]
	\centering\renewcommand\arraystretch{1}\setlength{\tabcolsep}{6pt}\footnotesize
	\caption{Average QIs and related standard deviations of the results on 100 patches extracted from the images on the Harvard dataset. The best values are highlighted in boldface.}
	\begin{tabular}{l|c|c|c|c}
		\Xhline{1.2pt}
		Method & PSNR & SAM & ERGAS & SSIM \\ \hline
		CNMF 	&27.6$\pm$3.7 & 3.62$\pm$2.2 & 3.86$\pm$4.1 	& 0.95$\pm$0.05 \\ 
		FUSE 	&26.7$\pm$3.7 	& 5.40$\pm$4.1	& 4.07$\pm$4.0 	& 0.94$\pm$0.06 \\ 
		GLP-HS 	&26.0$\pm$3.4 & 4.74$\pm$3.3 & 4.26$\pm$3.3 	& 0.93$\pm$0.06 \\ 
		LTTR	& 27.4$\pm$3.5& 4.65$\pm$2.5 & 4.87$\pm$3.1 	& 0.94$\pm$0.06 \\
		LTMR 	&26.9$\pm$3.7 & 6.06$\pm$3.0 & 4.29$\pm$3.3 	& 0.92$\pm$0.07 \\ 
		MHFnet 	&26.6$\pm$5.2 & 8.09$\pm$4.6 & 62.18$\pm$178.2 	& 0.88$\pm$0.11 \\ 
		HSRnet 	&\textbf{29.3}$\pm$4.4 & \textbf{3.44}$\pm$2.0 & \textbf{3.5}$\pm$2.2		& \textbf{0.97}$\pm$0.03 \\ \hline
		Best value& +$ \infty $ & 0 & 0 & 1 \\ \Xhline{1.2pt}
	\end{tabular}
	\label{harvard-ave}
\end{table}

\begin{table}[t]
	\centering\renewcommand\arraystretch{1}\setlength{\tabcolsep}{6pt}\footnotesize
	\caption{Average QIs and related standard deviations of the results for 10 testing images on the Harvard dataset. The best values are highlighted in boldface.}
	\begin{tabular}{l|c|c|c|c}
		\Xhline{1.2pt}
		Method 	& PSNR & SAM & ERGAS & SSIM \\ \hline
		CNMF & 34.3$\pm$3.8 & 4.72$\pm$2.3 & 4.37$\pm$2.4 & 0.94$\pm$0.02  \\ 
		FUSE & 32.9$\pm$3.8 & 7.48$\pm$3.5 & 4.79$\pm$2.0 & 0.93$\pm$0.03  \\ 
		GLP-HS & 35.0$\pm$4.8 & 4.87$\pm$2.2 & 4.26$\pm$1.6 & 0.93$\pm$0.04  \\ 
		LTTR & 36.1$\pm$5.4 & 6.06$\pm$2.3 & 6.19$\pm$2.2 & 0.90$\pm$0.07  \\ 
		LTMR & 37.2$\pm$4.5 & 6.13$\pm$2.3 & 4.82$\pm$3.1 & 0.93$\pm$0.05  \\ 
		MHFnet & 36.4$\pm$5.5 & 7.03$\pm$4.0 & 16.57$\pm$14.6 & 0.91$\pm$0.08  \\ 
		HSRnet & \textbf{39.5}$\pm$\textbf{4.7} & \textbf{3.38}$\pm$1.1 & \textbf{3.27}$\pm$1.5 & \textbf{0.97}$\pm$0.02  \\ \hline
		Best value& +$ \infty $ & 0 & 0 & 1 \\ \Xhline{1.2pt}
	\end{tabular}
	\label{harvard10-ave}
\end{table}

\begin{table}[t]
	\setlength{\tabcolsep}{1.47pt}
	\caption{QIs of the results for the different methods and the running times on (a) \emph{trees} ,(c) \emph{bikes}, and (h) \emph{wall} for the Harvard dataset. G indicates that the method is running on the GPU device, while C denotes the use of the CPU. The best values are highlighted in boldface.}
	\centering{
		\begin{tabular}{l|ccccccc}
			\Xhline{1.2pt}
			\multicolumn{8}{c}{(a) 1000 $\times$ 1000} \\ \hline
			Method & CNMF & FUSE & GLPHS & LTTR & LTMR & MHFnet & HSRnet\\
			PSNR & 32.54 & 31.27 & 34.03 & 31.63 & 32.99 & 35.25 & \textbf{37.54} \\ 
			SAM & 5.21 & 7.95 & 5.41 & 7.73 & 7.30 & 5.00 & \textbf{3.01} \\ 
			ERGAS & 3.91 & 5.39 & 4.10 & 8.68 & 4.79 & 29.05 & \textbf{3.1} \\ 
			SSIM & 0.912 & 0.882 & 0.911 & 0.829 & 0.867 & 0.917 & \textbf{0.961} \\ 
			\Xhline{1.2pt}
			\multicolumn{8}{c}{(c) 1000 $\times$ 1000} \\ \hline
			Method & CNMF & FUSE & GLPHS & LTTR & LTMR & MHFnet & HSRnet\\
			PSNR & 33.74 & 31.67 & 33.52 & 34.26 & 36.77 & 38.24 & \textbf{39.25} \\ 
			SAM & 4.15 & 7.75 & 5.12 & 6.19 & 6.06 & 5.00 & \textbf{3.56} \\ 
			ERGAS & 3.28 & 3.80 & 3.87 & 4.56 & 2.90 & 8.05 & \textbf{2.38} \\ 
			SSIM & 0.938 & 0.924 & 0.879 & 0.908 & 0.938 & 0.957 & \textbf{0.974} \\ 
			\Xhline{1.2pt}
			\multicolumn{8}{c}{(h) 1000 $\times$ 1000} \\ \hline
			Method & CNMF & FUSE & GLPHS & LTTR & LTMR & MHFnet & HSRnet\\
			PSNR & 39.69 & 3.007 & 39.33 & 42.55 & 41.90 & 43.97 & \textbf{44.76} \\ 
			SAM & 5.04 & 9.11 & 5.83 & 5.94 & 6.65 & 5.27 & \textbf{3.91} \\ 
			ERGAS & 7.56 & 7.26 & 7.01 & 7.44 & 6.61 & 14.33 &\textbf{ 3.77} \\ 
			SSIM & 0.921 & 0.942 & 0.959 & 0.974 & 0.972 & 0.977 &\textbf{ 0.989} \\ \hline
			\makecell[c]{Average \\ time(s)}&\footnotesize{102.1(C)}&\footnotesize	{7.3(C)}&\footnotesize	{16.1(C)}&\footnotesize {2049.5(C)}&\footnotesize {864.3(C)}&\footnotesize	{6.8(G)}&\footnotesize {\textbf{2.4}(G)}\\
			
			\Xhline{1.2pt}
	\end{tabular}}
	\label{qresult-4}
\end{table}

\begin{figure*}[t]
	\centering
	\begin{minipage}[t]{0.94\linewidth}
		\begin{minipage}[t]{0.12\linewidth}
			{\includegraphics[width=1\linewidth]{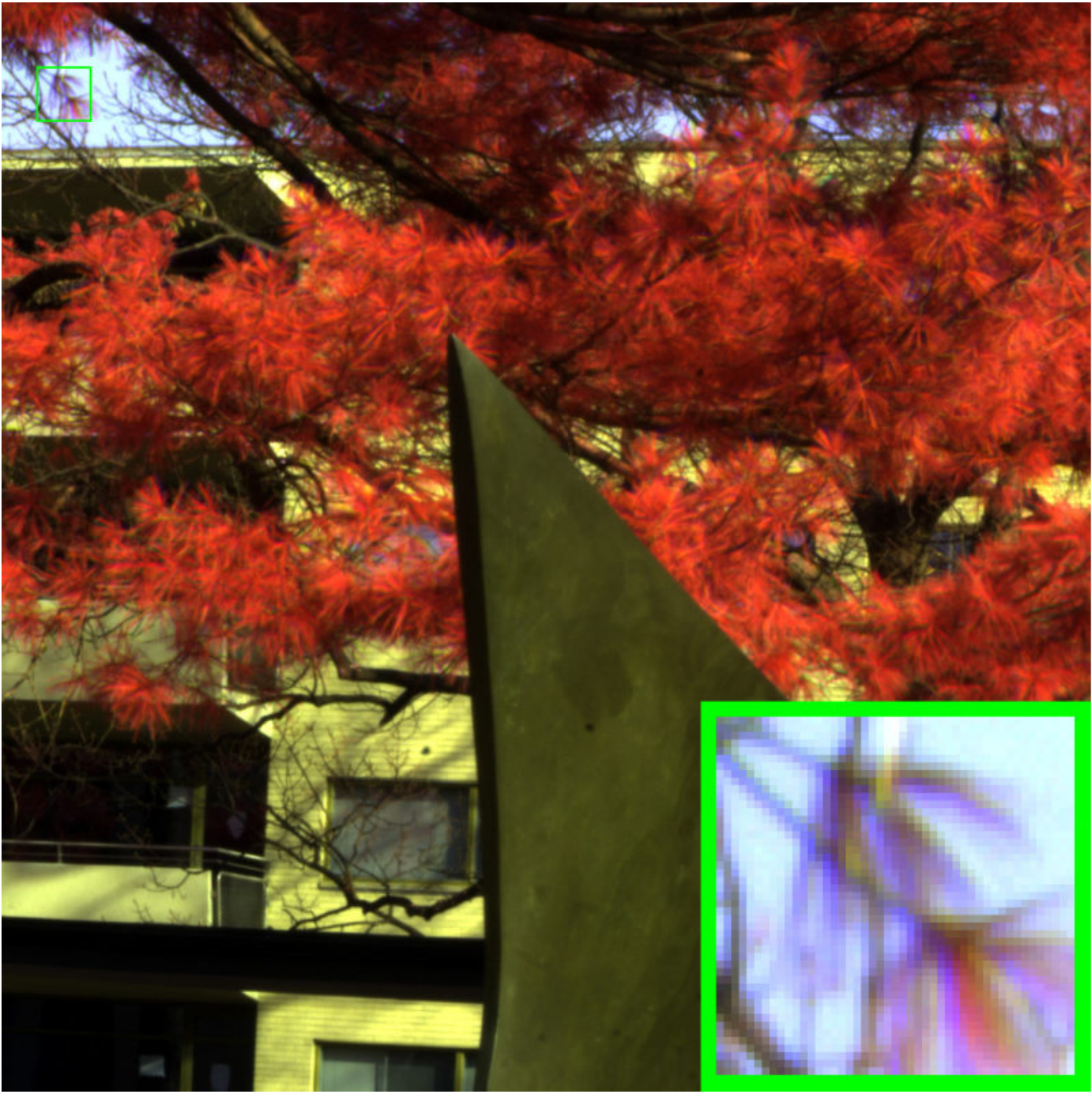}}
			{\includegraphics[width=1\linewidth]{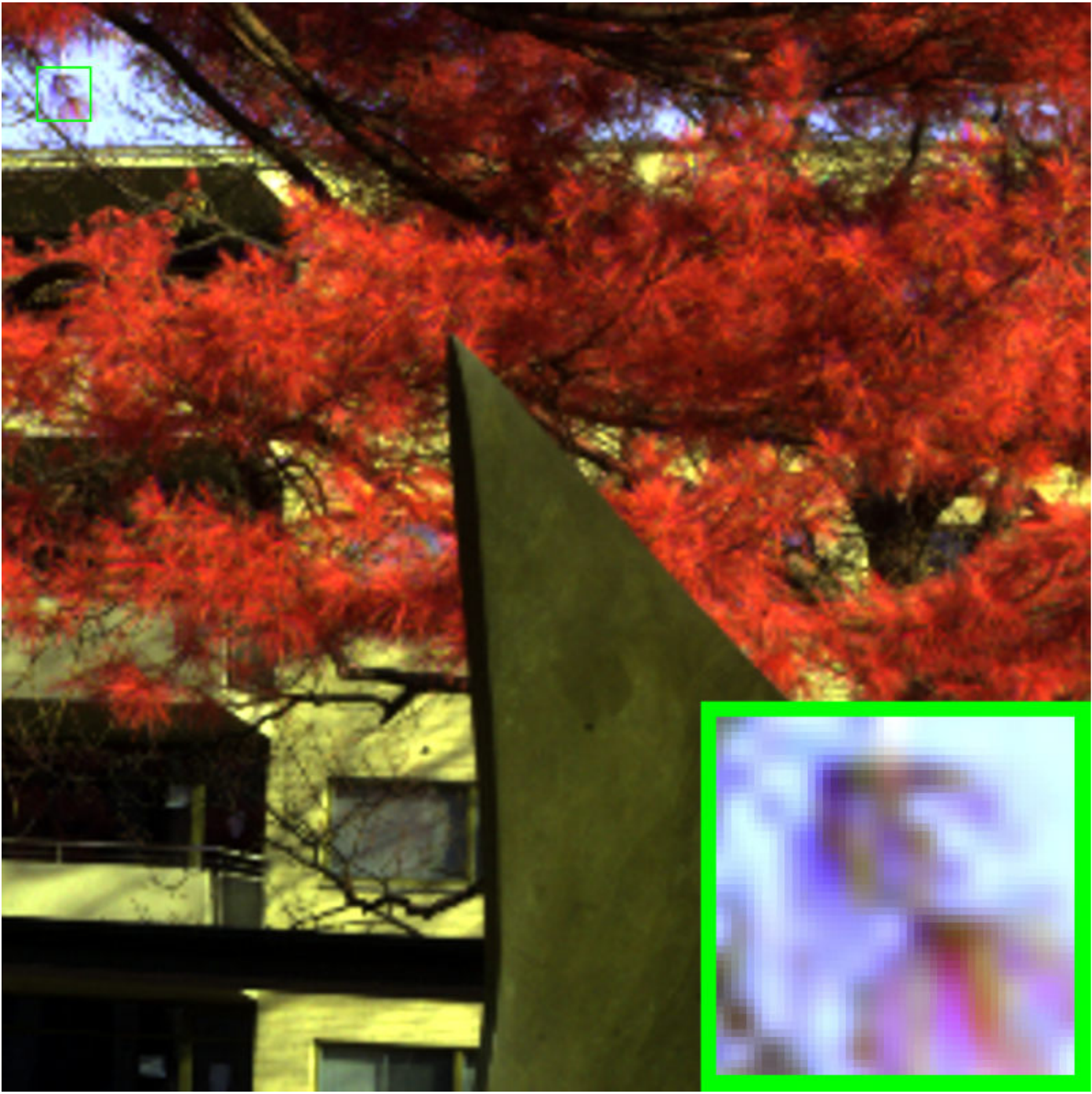}}
			\centering
		\end{minipage}
		\begin{minipage}[t]{0.12\linewidth}
			{\includegraphics[width=1\linewidth]{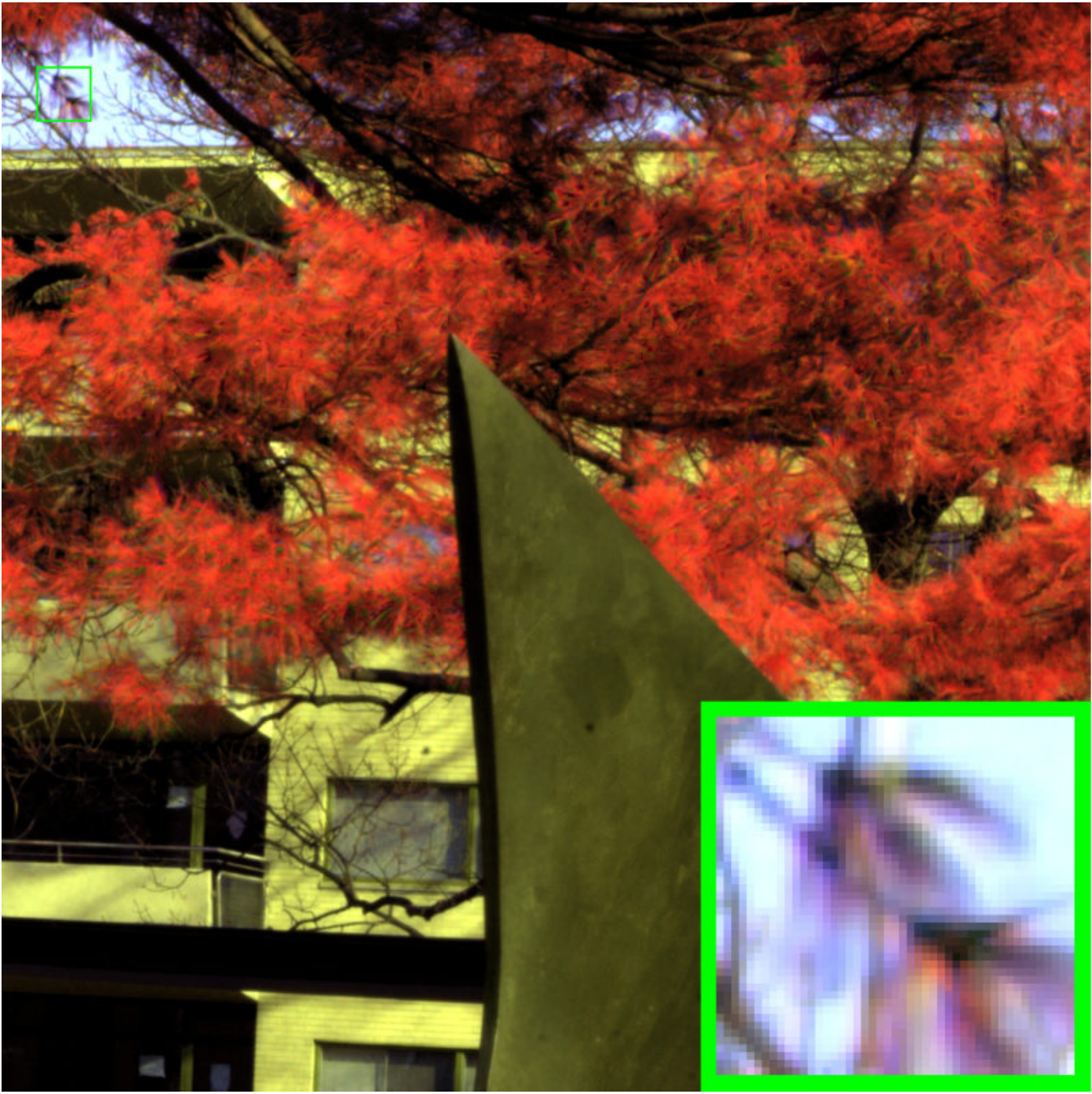}}
			{\includegraphics[width=1\linewidth]{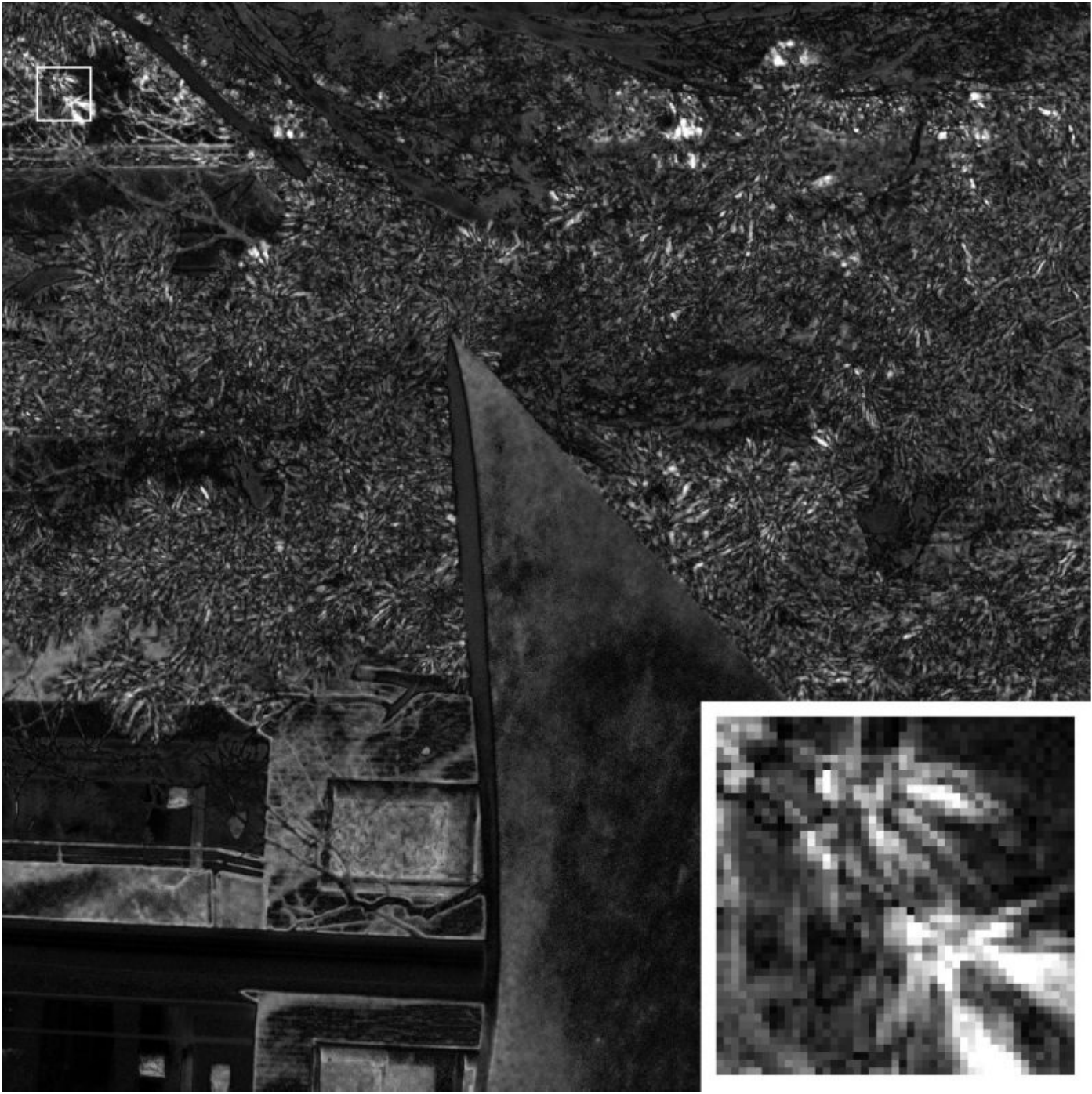}}
			\centering
		\end{minipage}
		\begin{minipage}[t]{0.12\linewidth}
			{\includegraphics[width=1\linewidth]{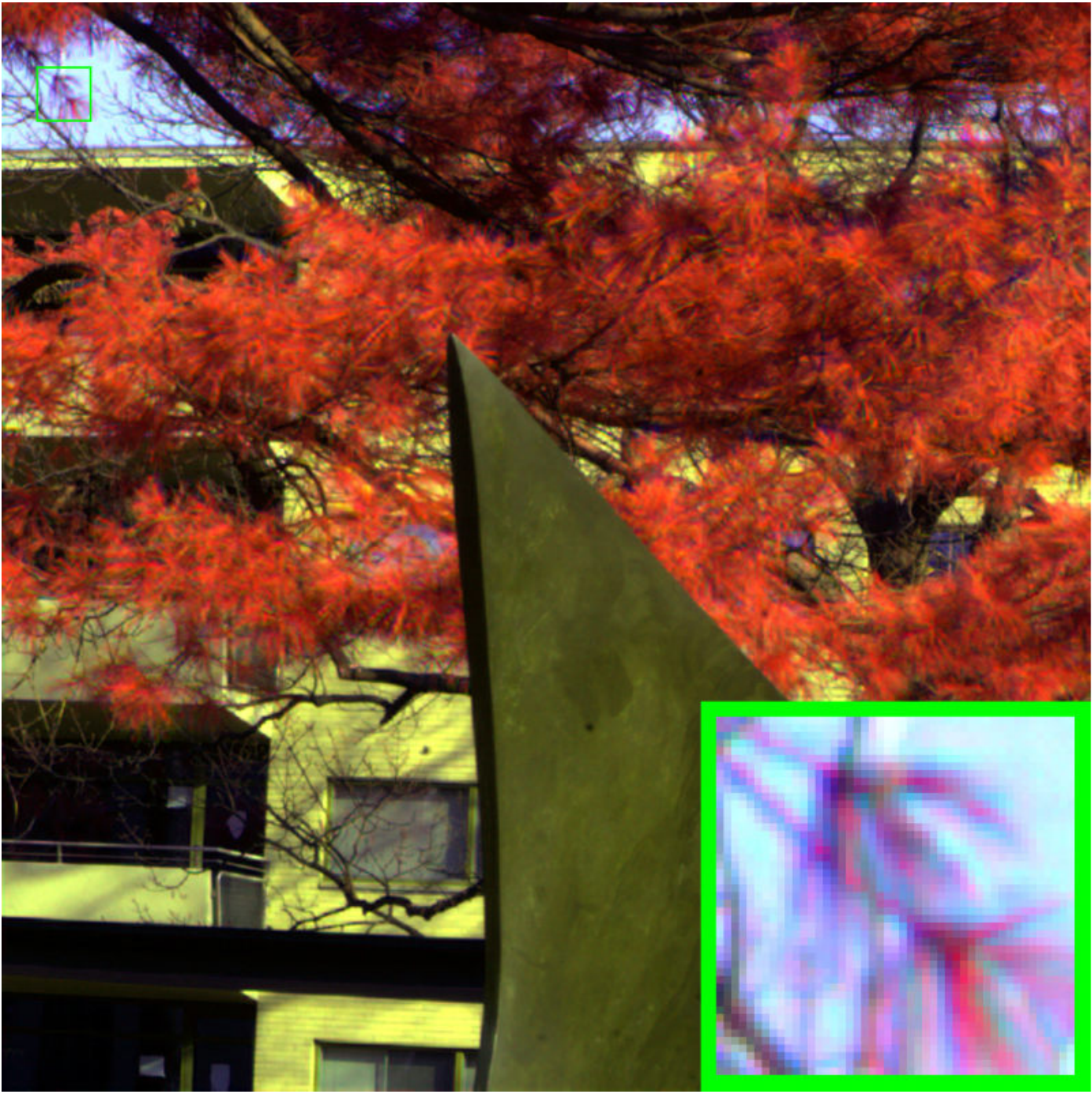}}
			{\includegraphics[width=1\linewidth]{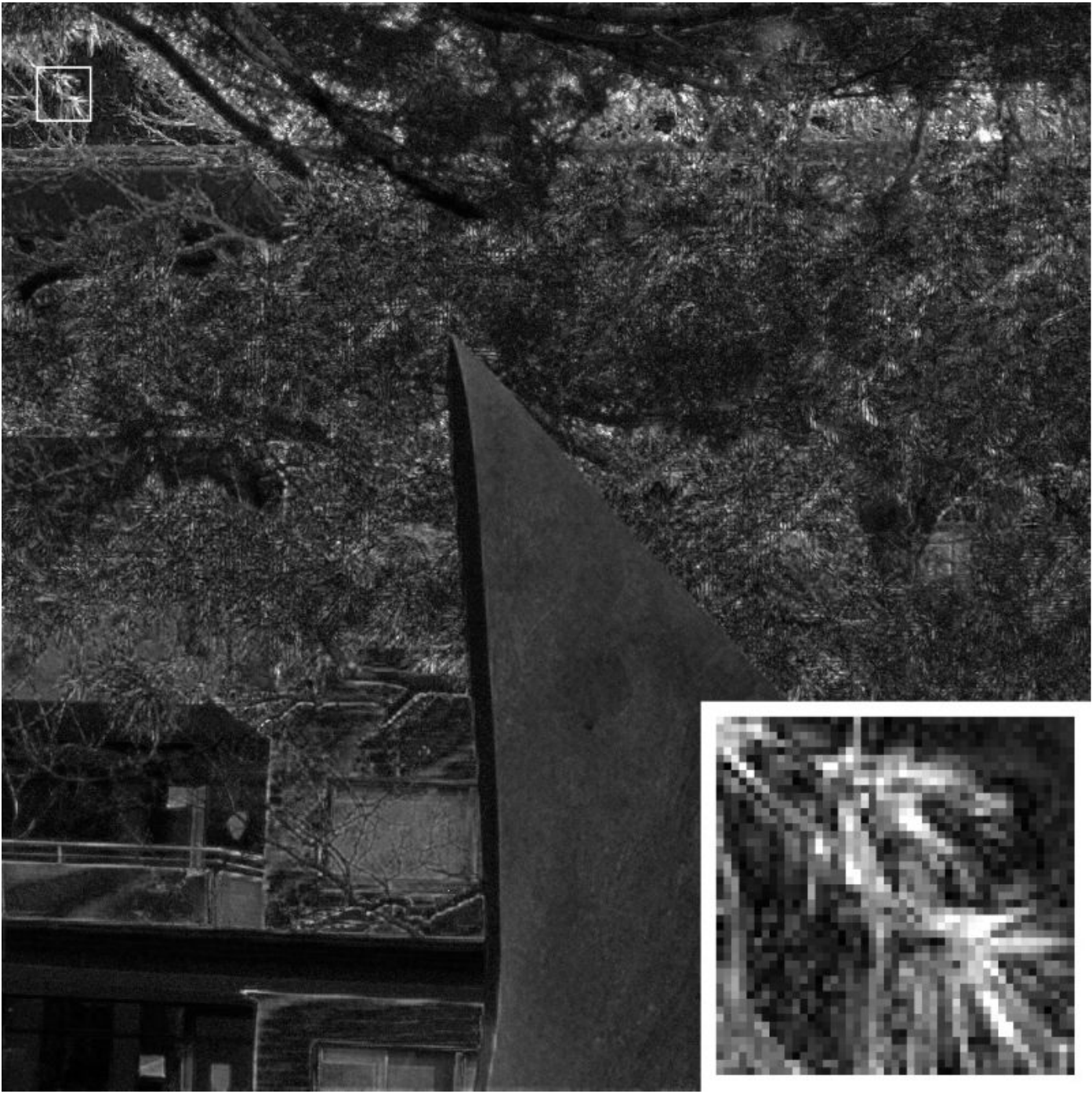}}
			\centering
		\end{minipage}
		\begin{minipage}[t]{0.12\linewidth}
			{\includegraphics[width=1\linewidth]{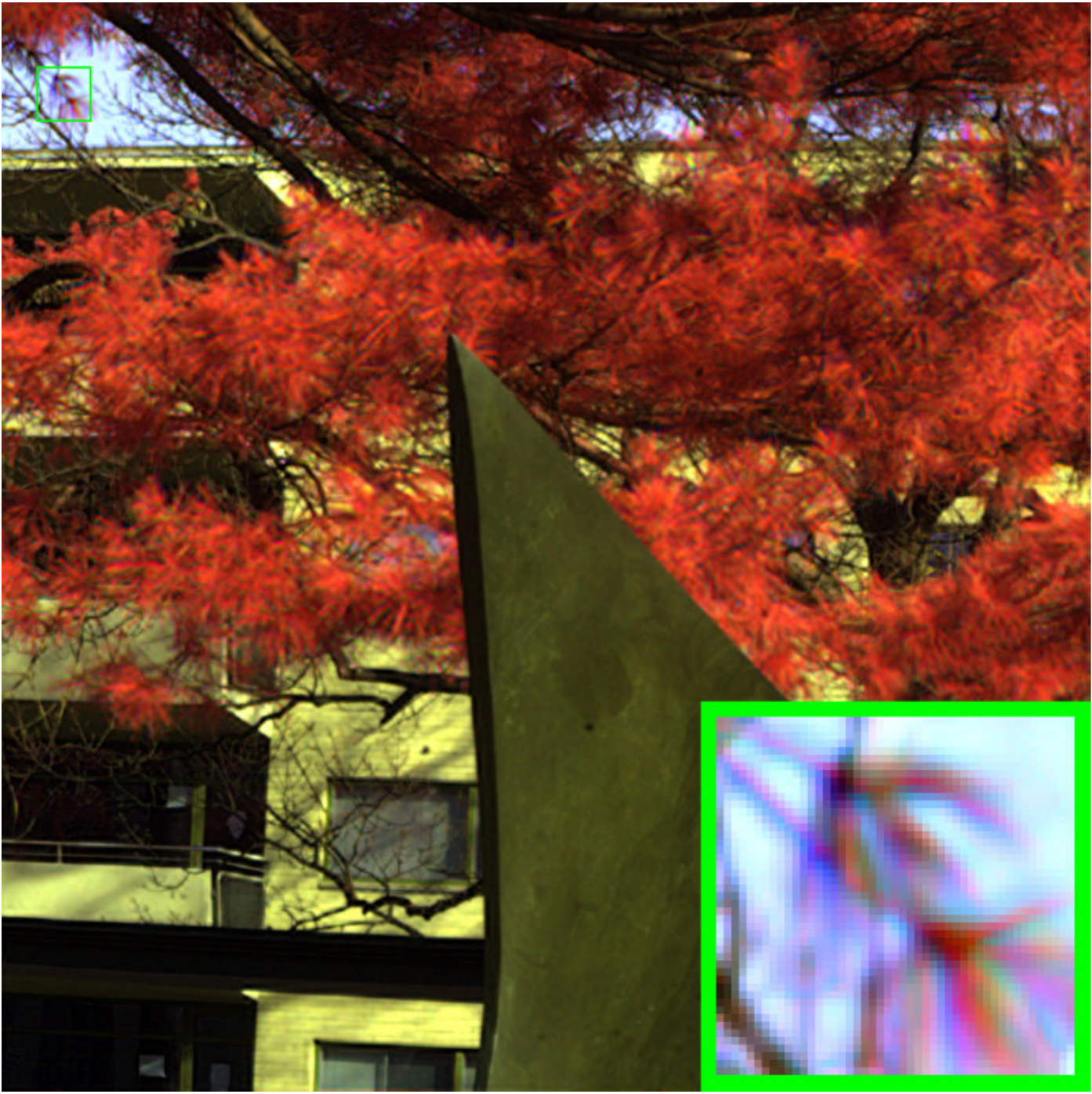}}
			{\includegraphics[width=1\linewidth]{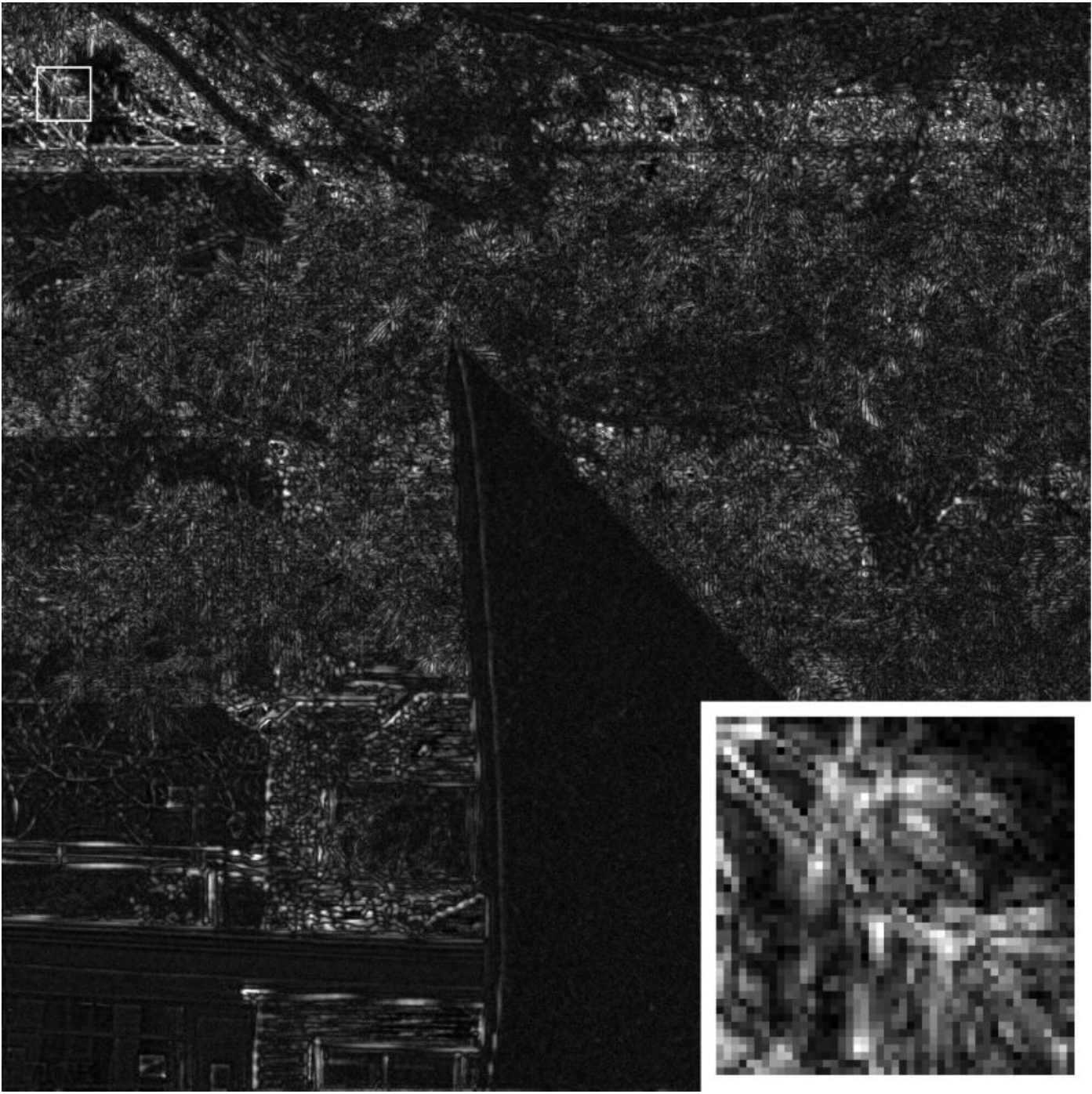}}
			\centering
		\end{minipage}
		\begin{minipage}[t]{0.12\linewidth}
			{\includegraphics[width=1\linewidth]{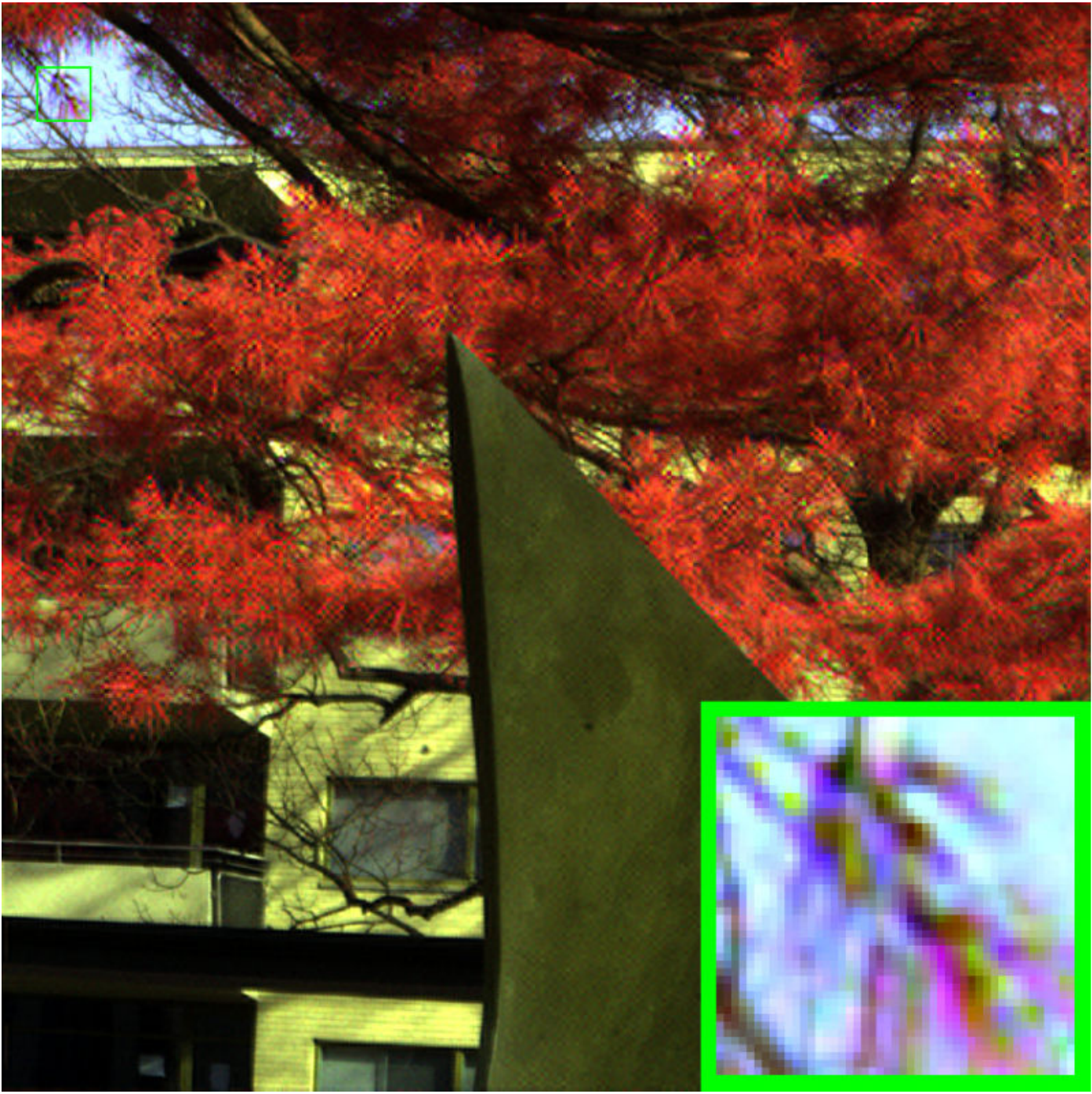}}
			{\includegraphics[width=1\linewidth]{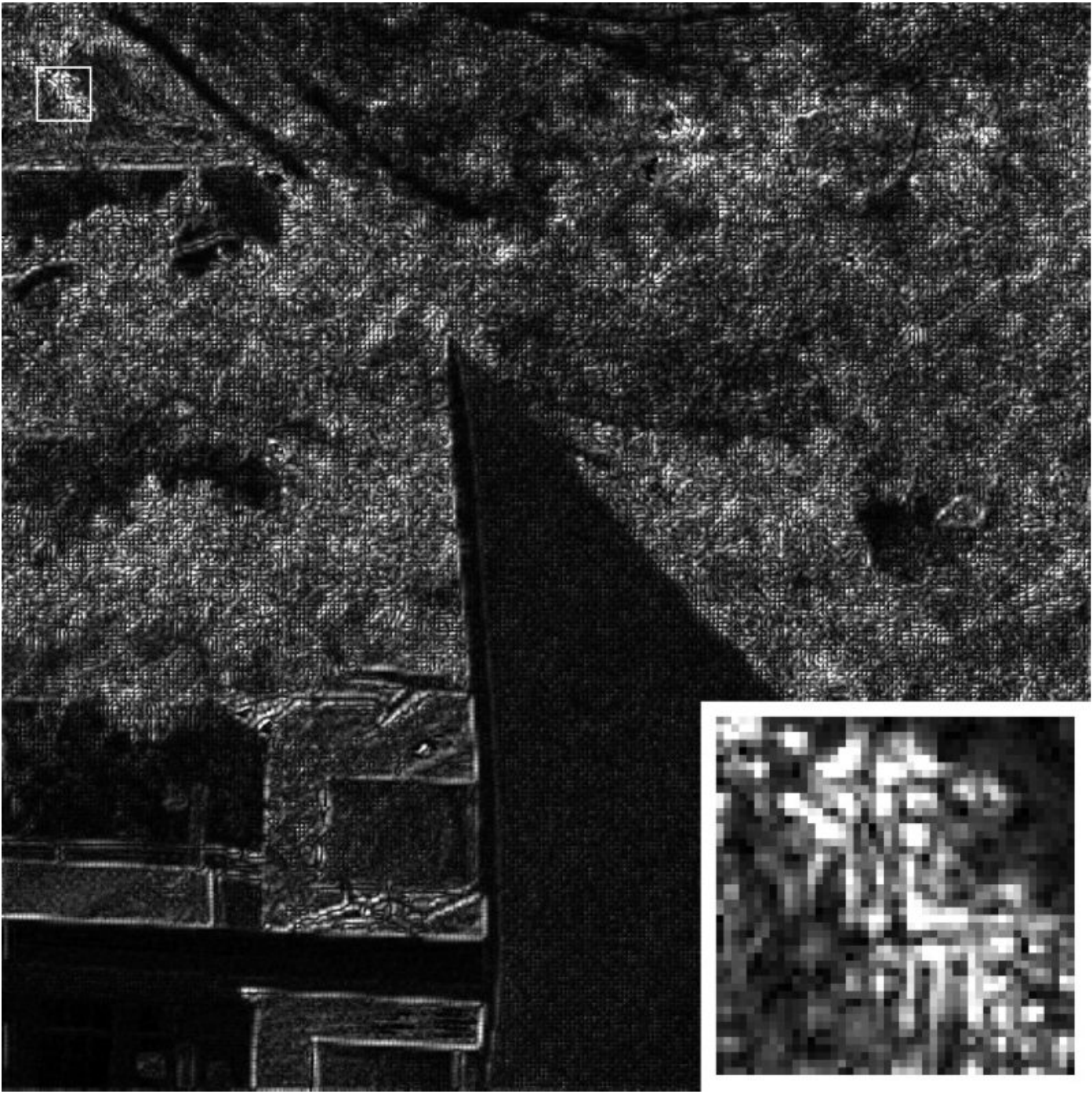}}
			\centering
		\end{minipage}
		\begin{minipage}[t]{0.12\linewidth}
			{\includegraphics[width=1\linewidth]{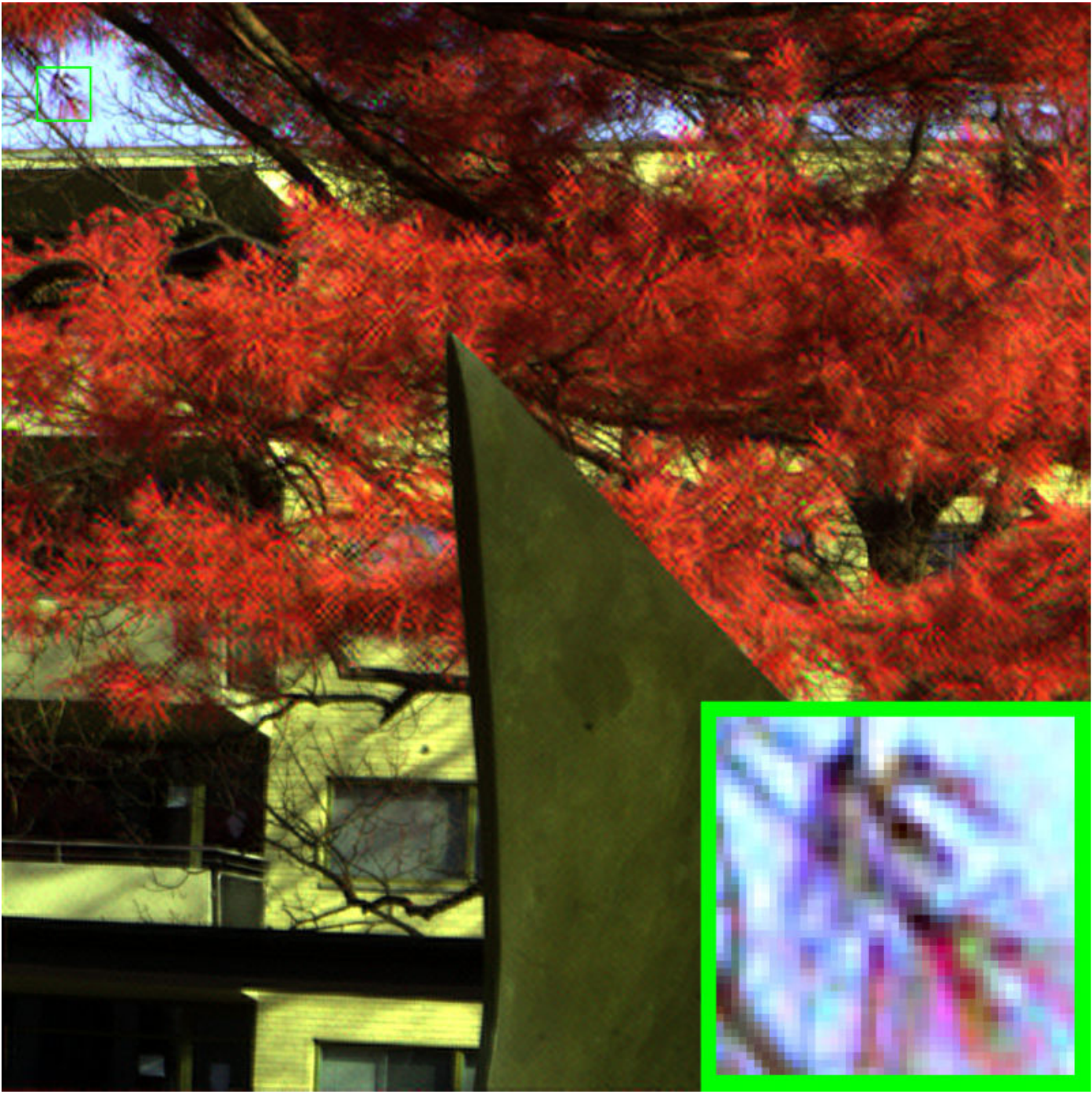}}
			{\includegraphics[width=1\linewidth]{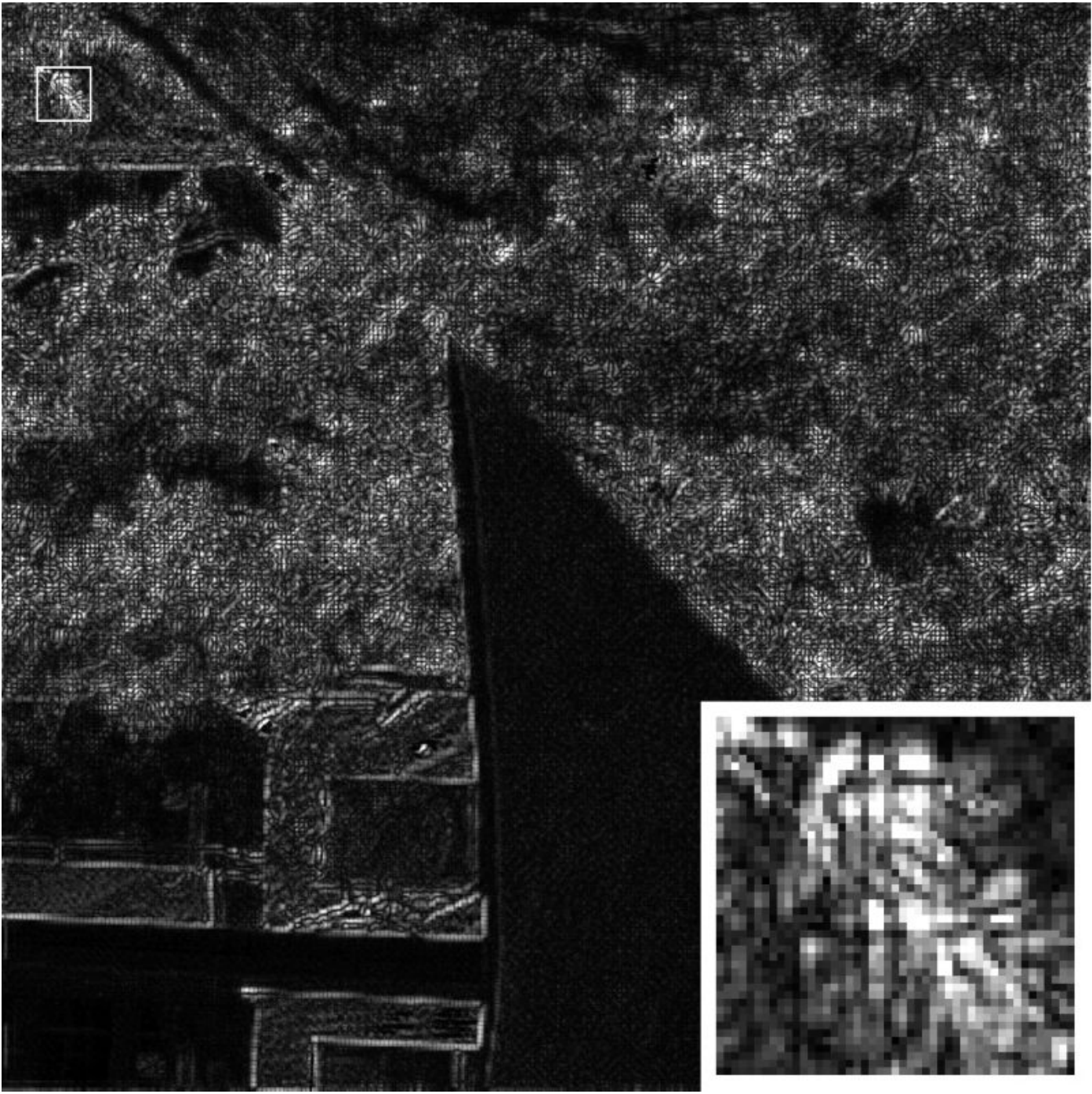}}
			\centering
		\end{minipage}
		\begin{minipage}[t]{0.12\linewidth}
			{\includegraphics[width=1\linewidth]{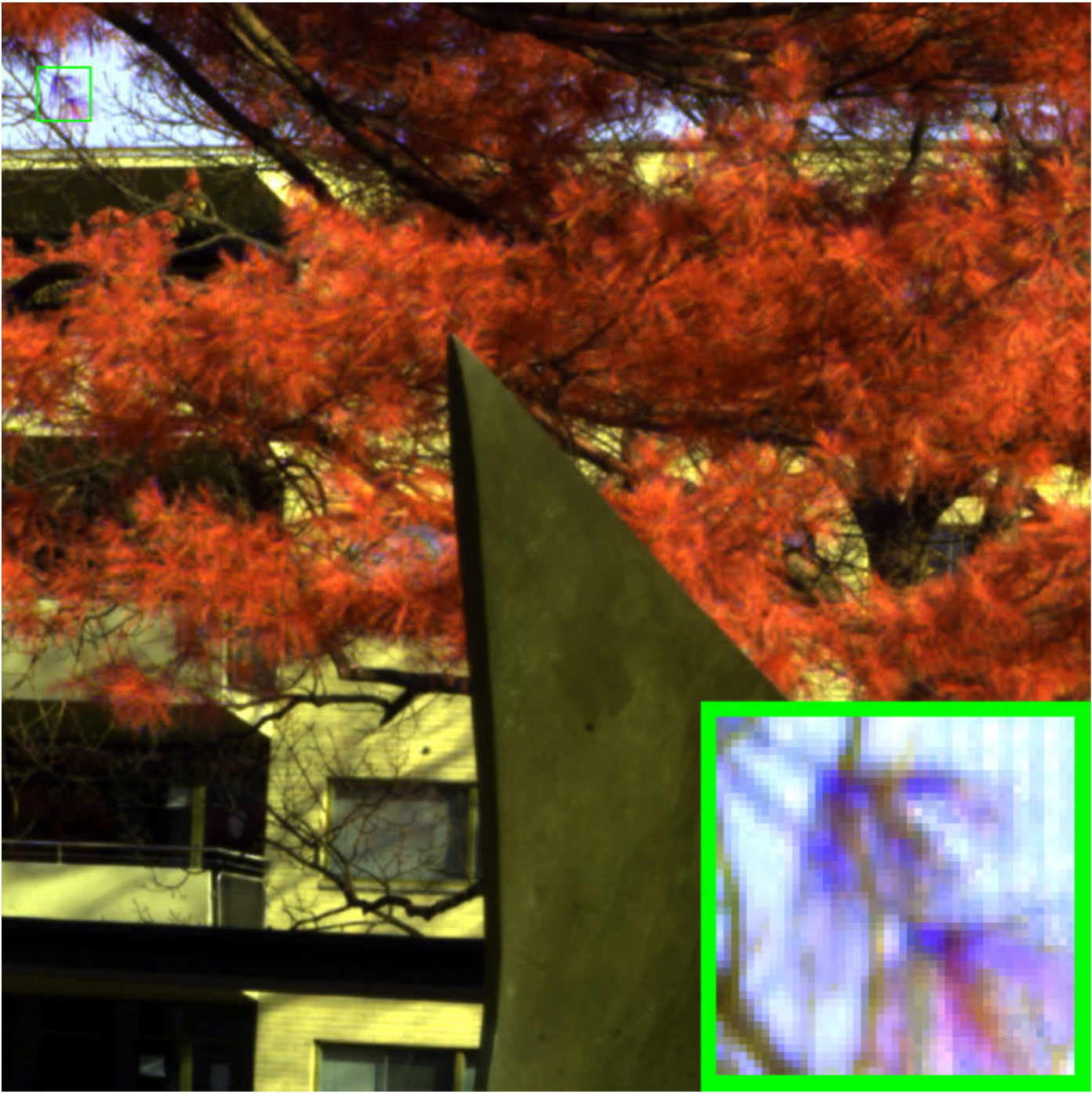}}
			{\includegraphics[width=1\linewidth]{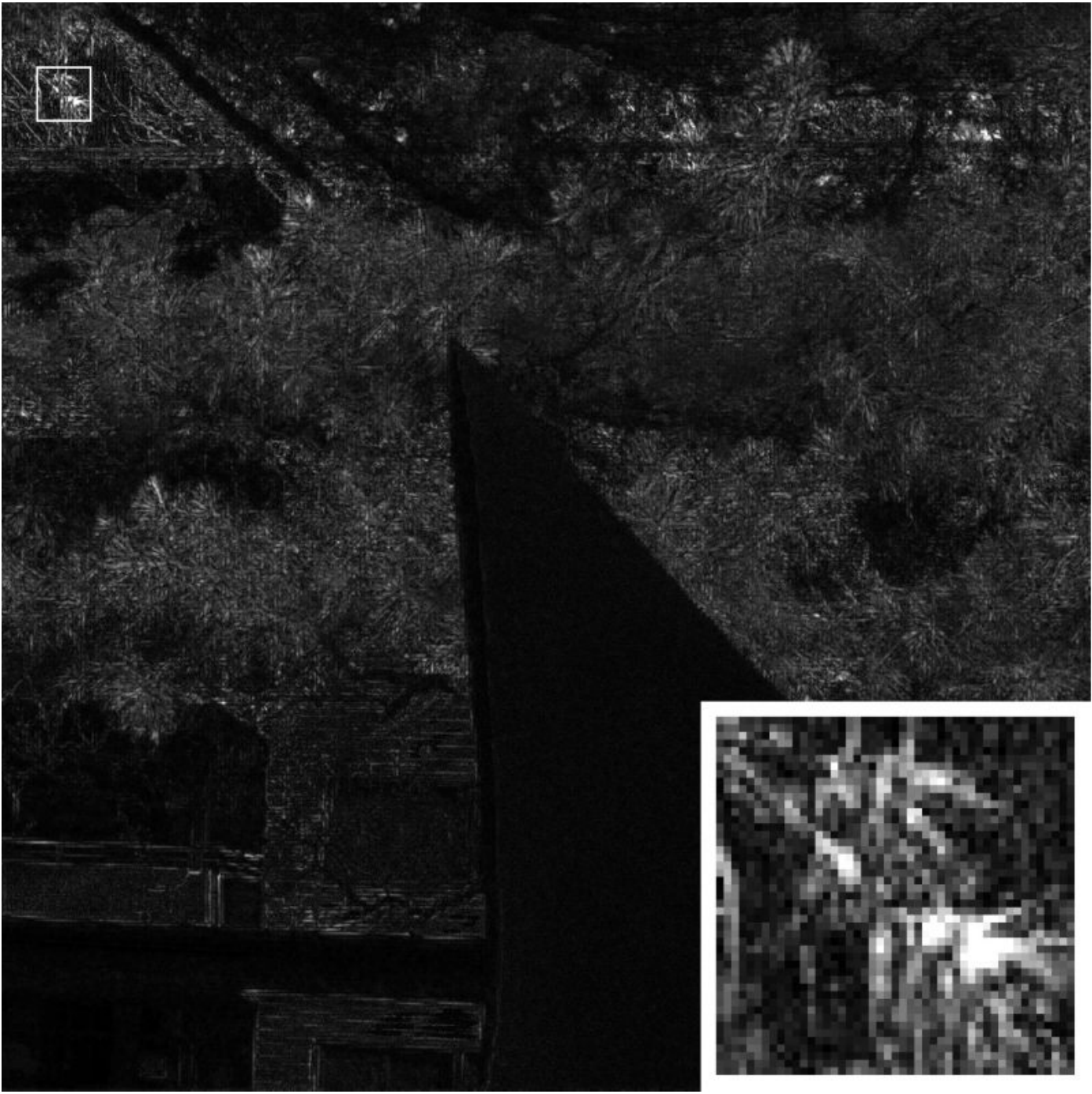}}
			\centering
		\end{minipage}
		\begin{minipage}[t]{0.12\linewidth}
			{\includegraphics[width=1\linewidth]{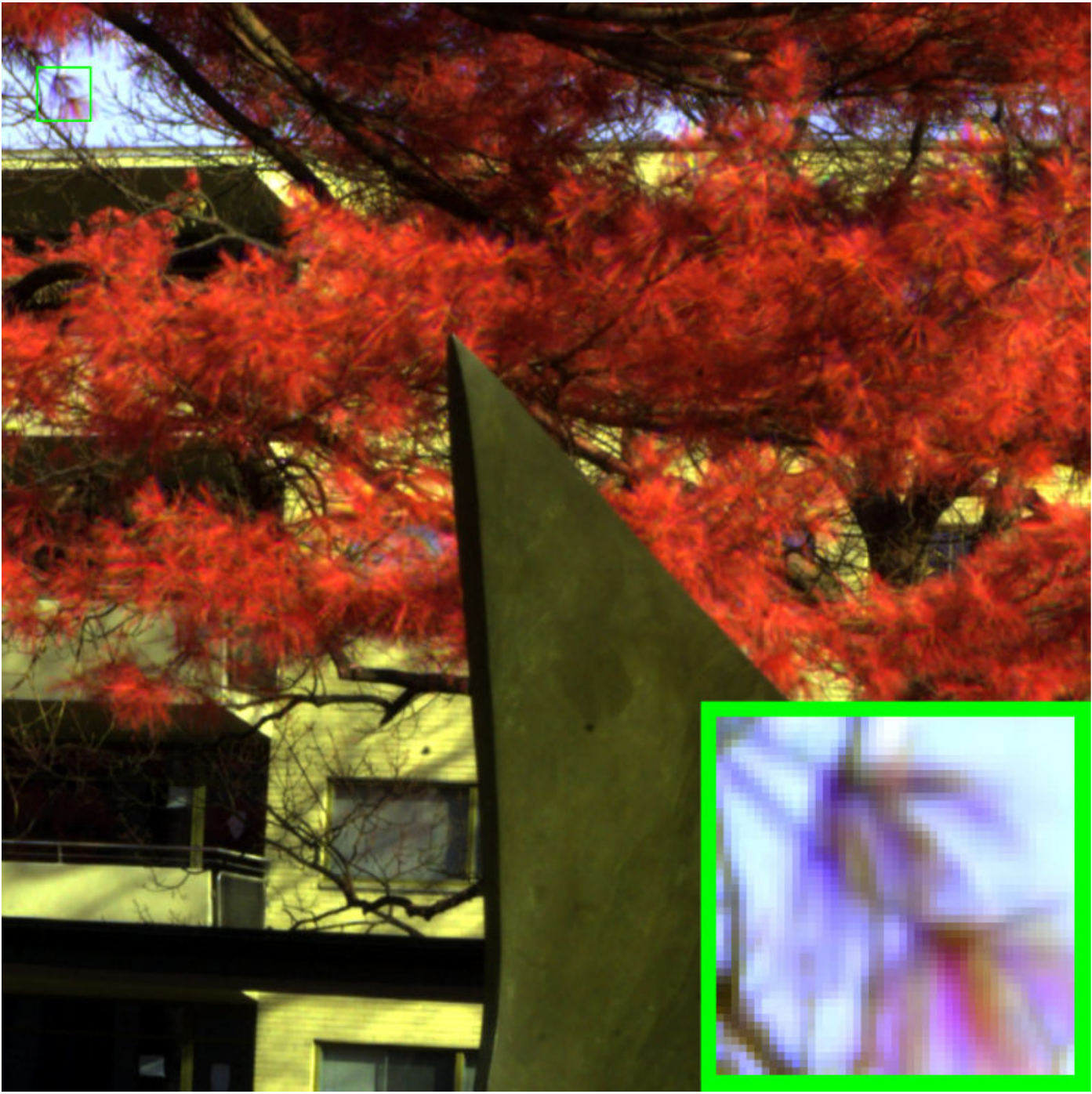}}
			{\includegraphics[width=1\linewidth]{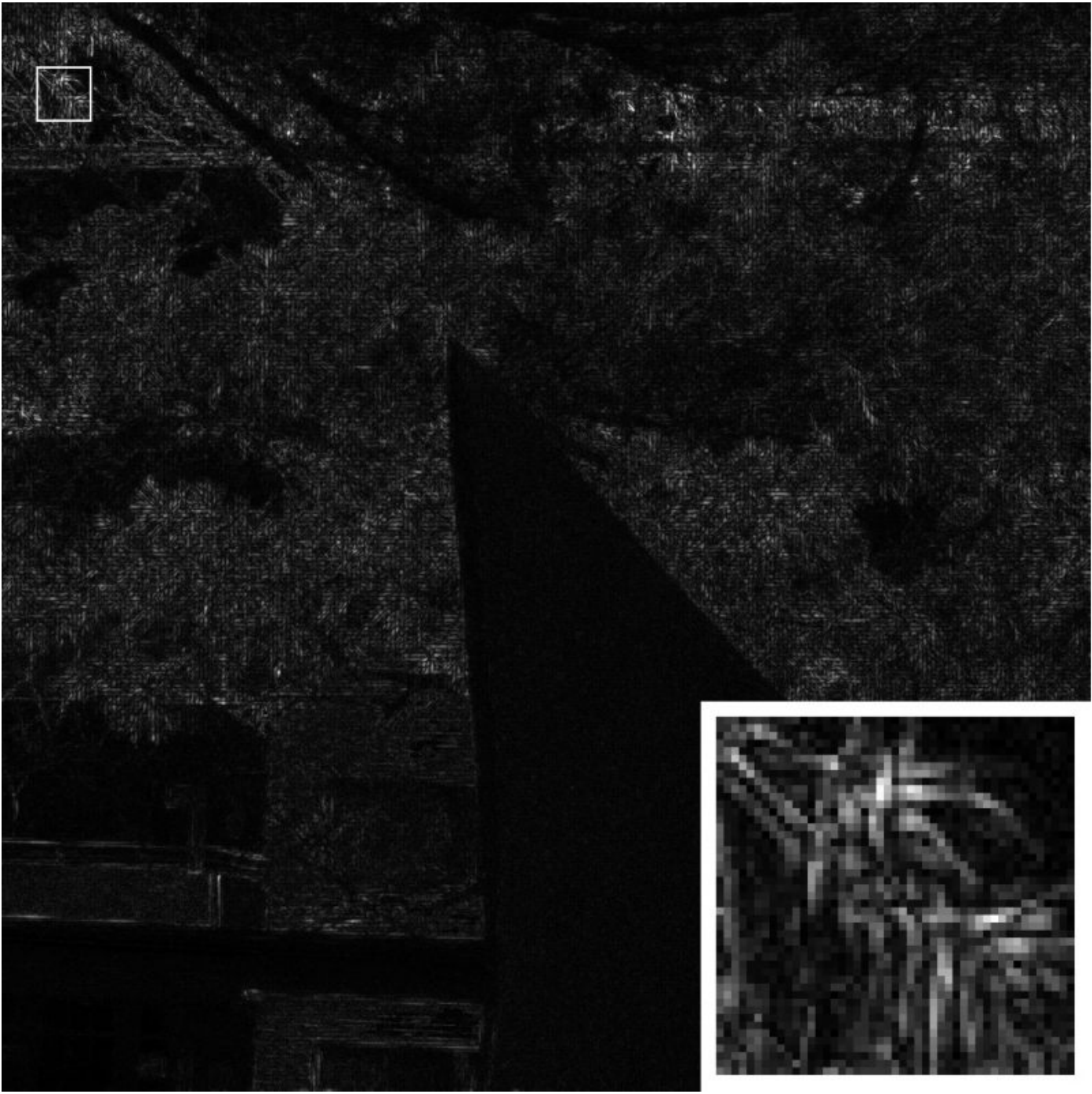}}
			\centering
		\end{minipage}
		
		\vspace{5pt}
		
		\begin{minipage}[t]{0.12\linewidth}
			{\includegraphics[width=1\linewidth]{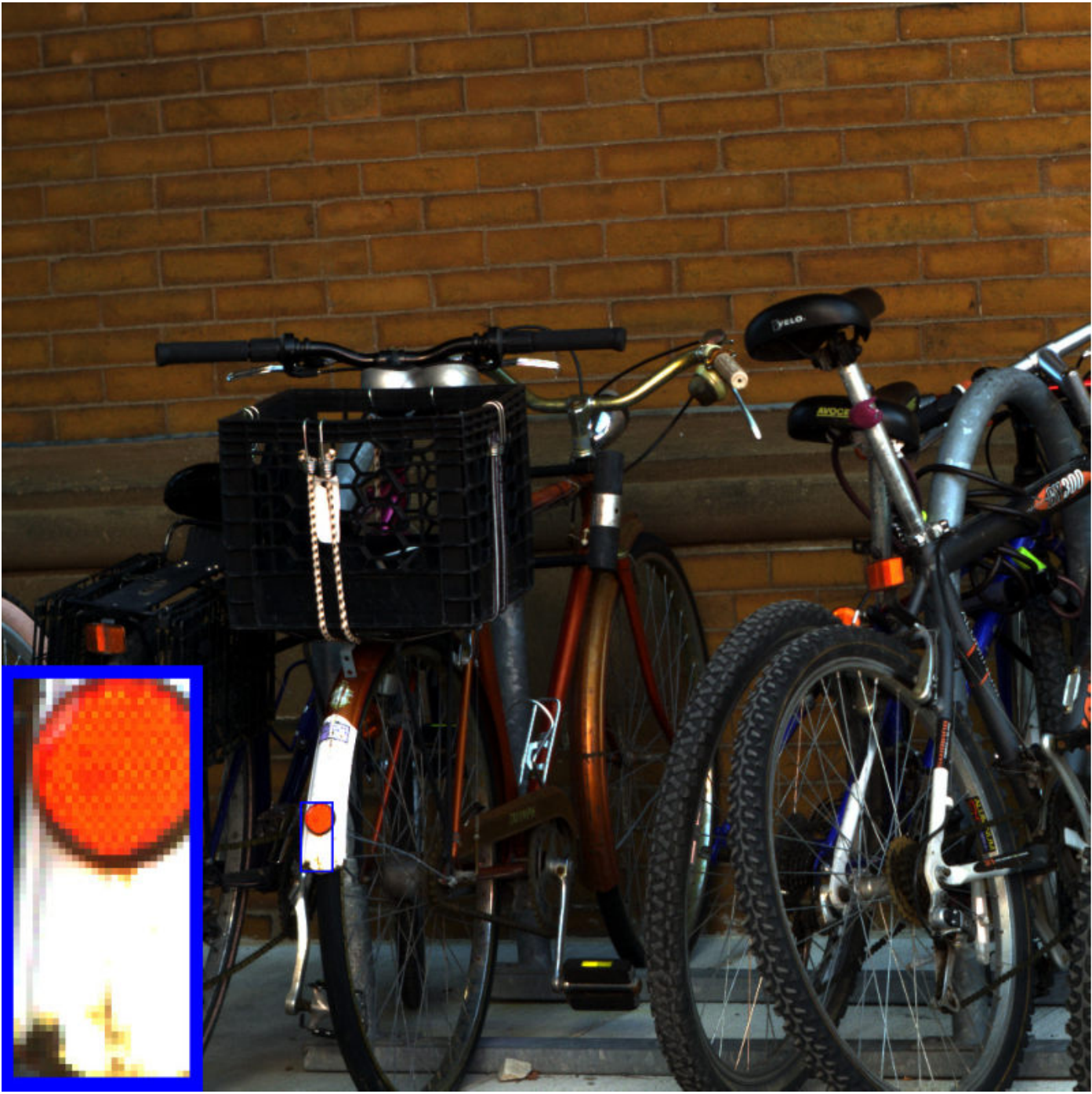}}
			{\includegraphics[width=1\linewidth]{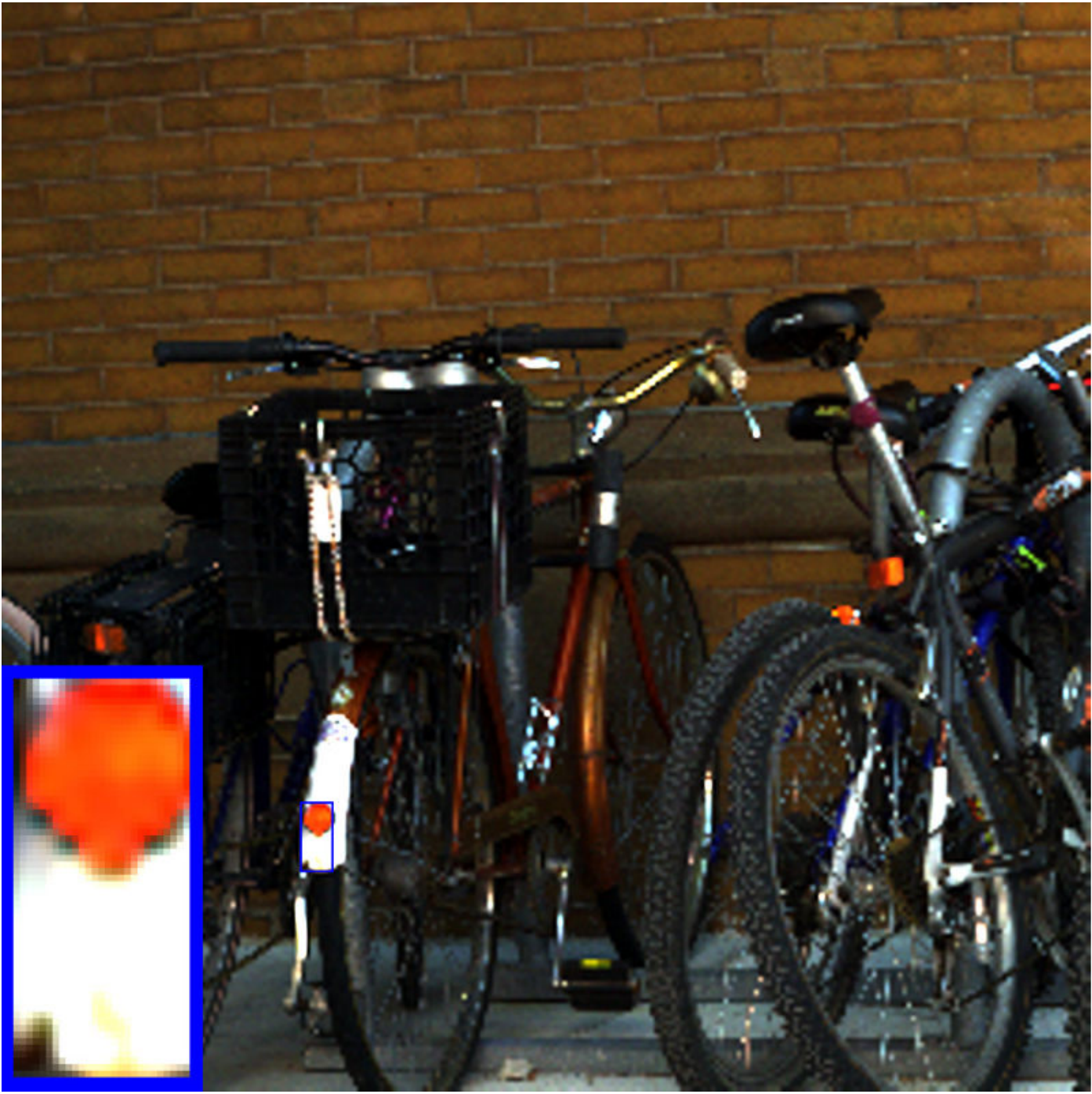}}
			\centering

		\end{minipage}
		\begin{minipage}[t]{0.12\linewidth}
			{\includegraphics[width=1\linewidth]{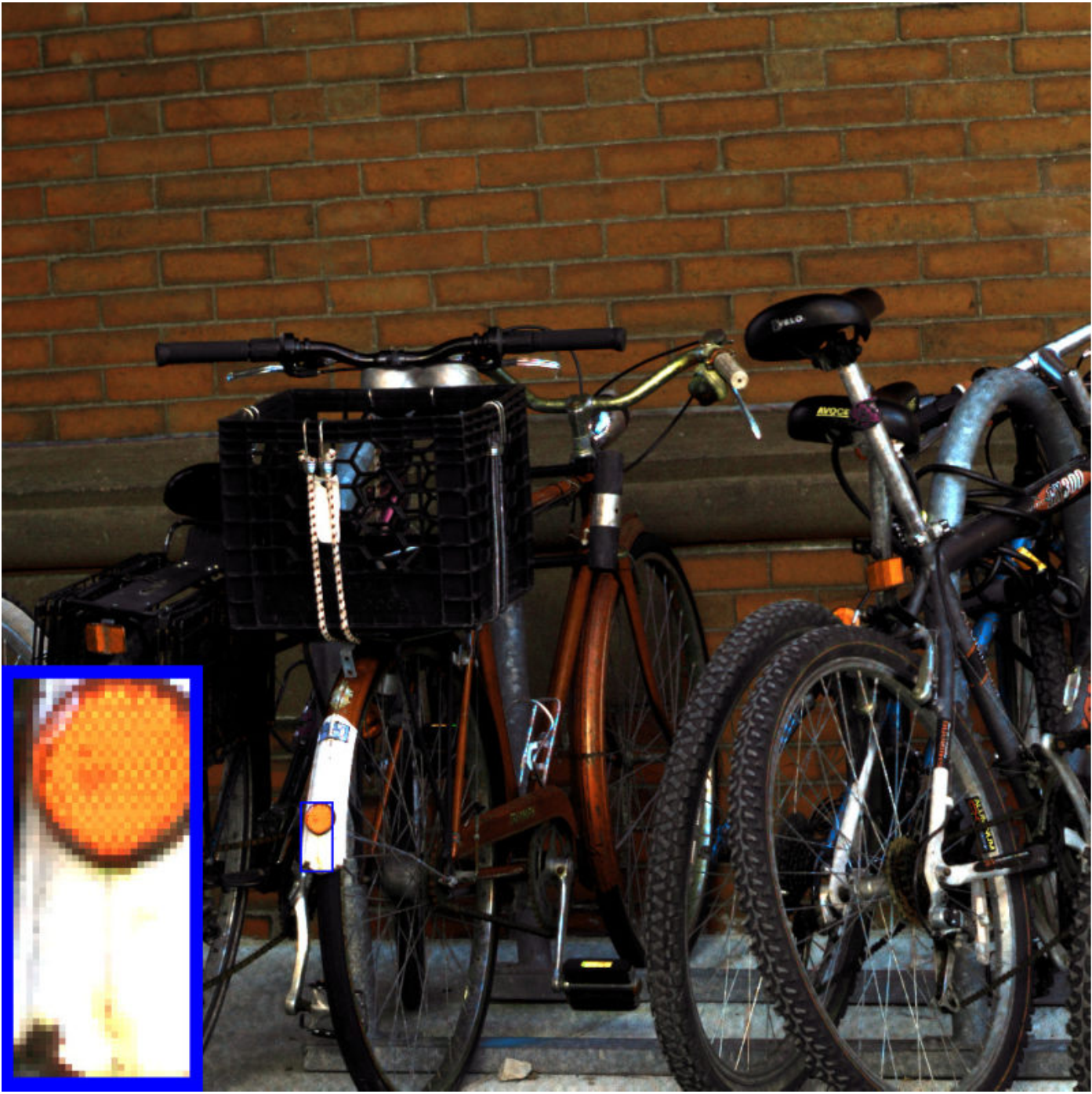}}
			{\includegraphics[width=1\linewidth]{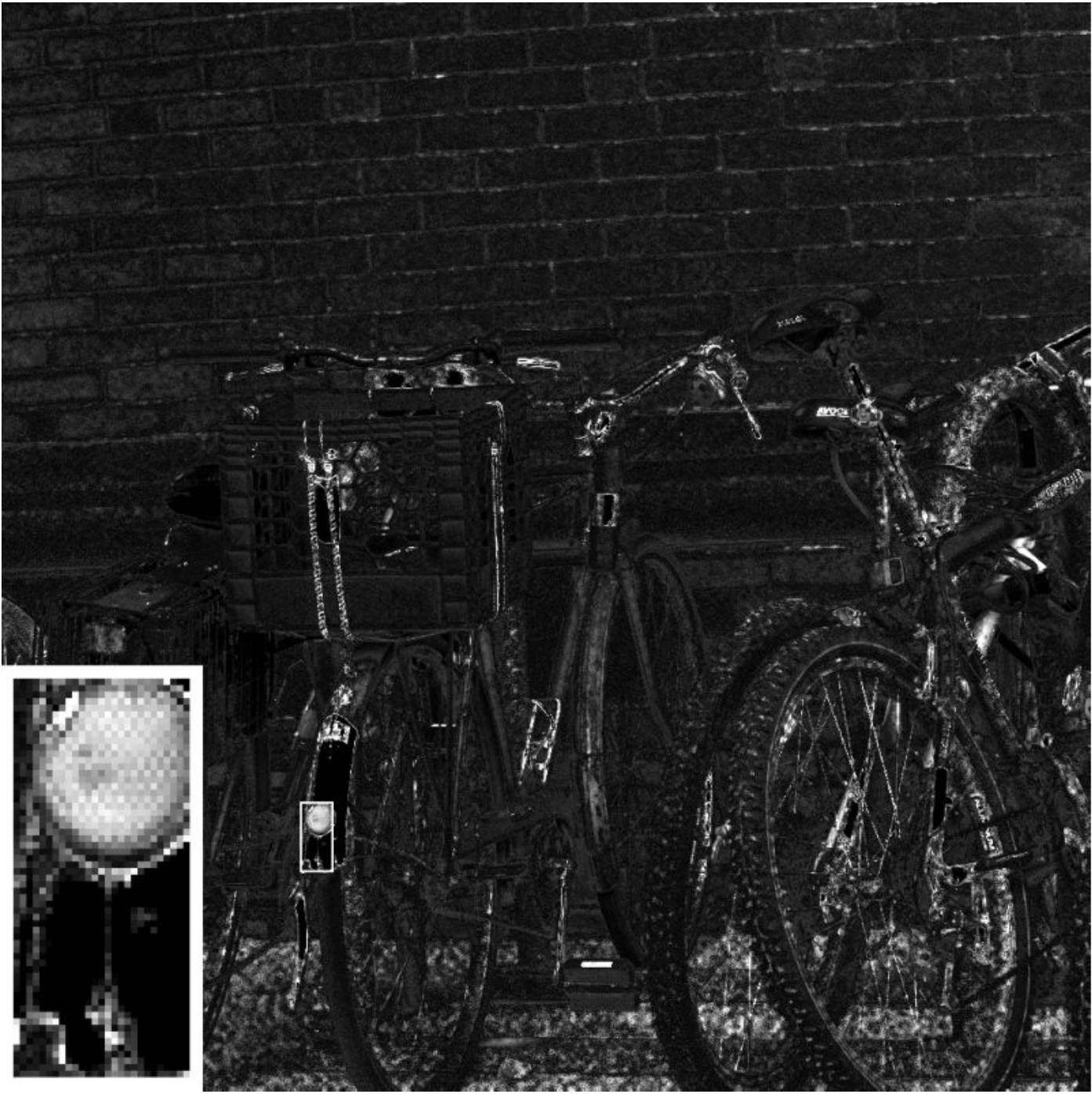}}
			\centering

		\end{minipage}
		\begin{minipage}[t]{0.12\linewidth}
			{\includegraphics[width=1\linewidth]{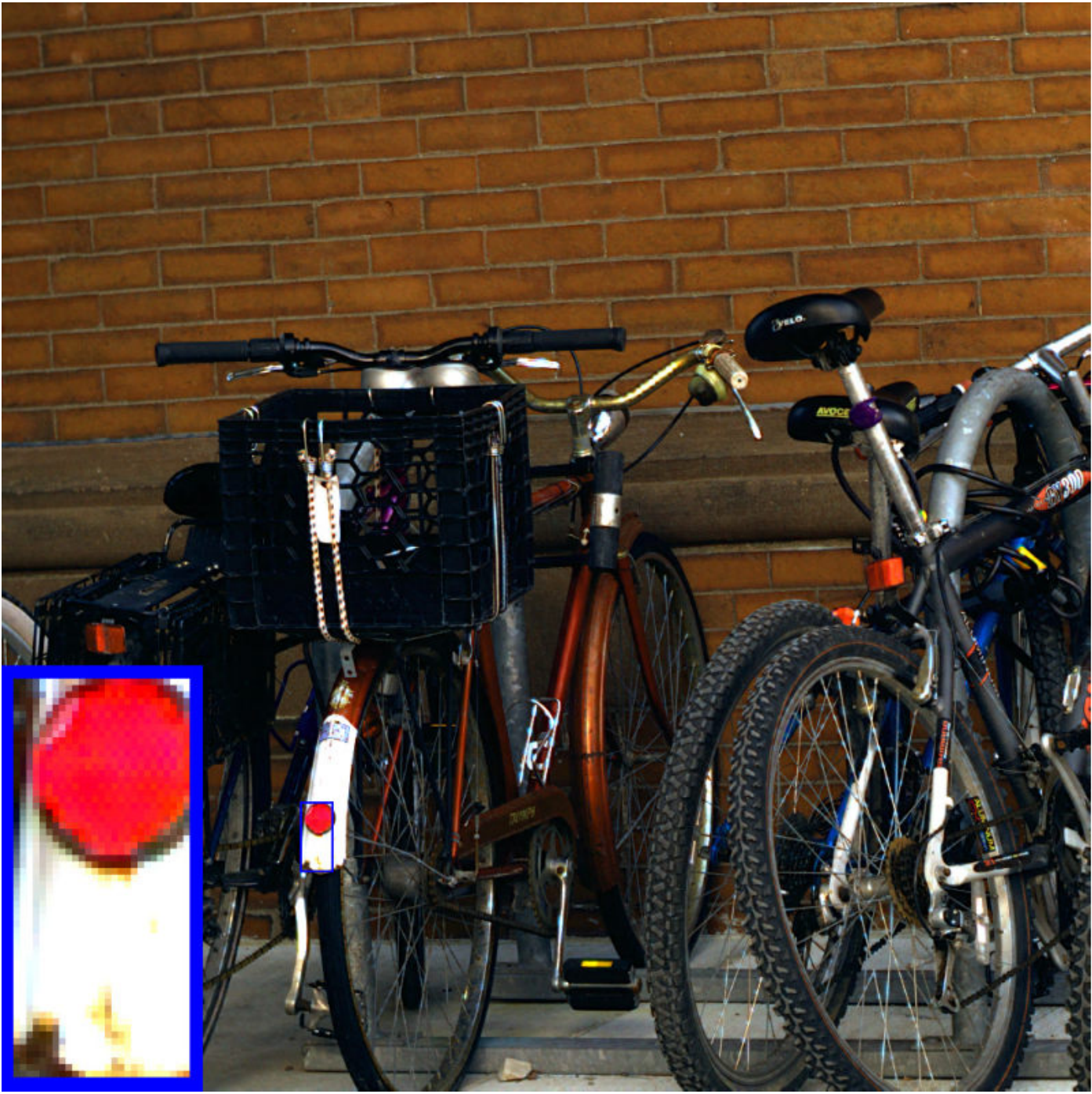}}
			{\includegraphics[width=1\linewidth]{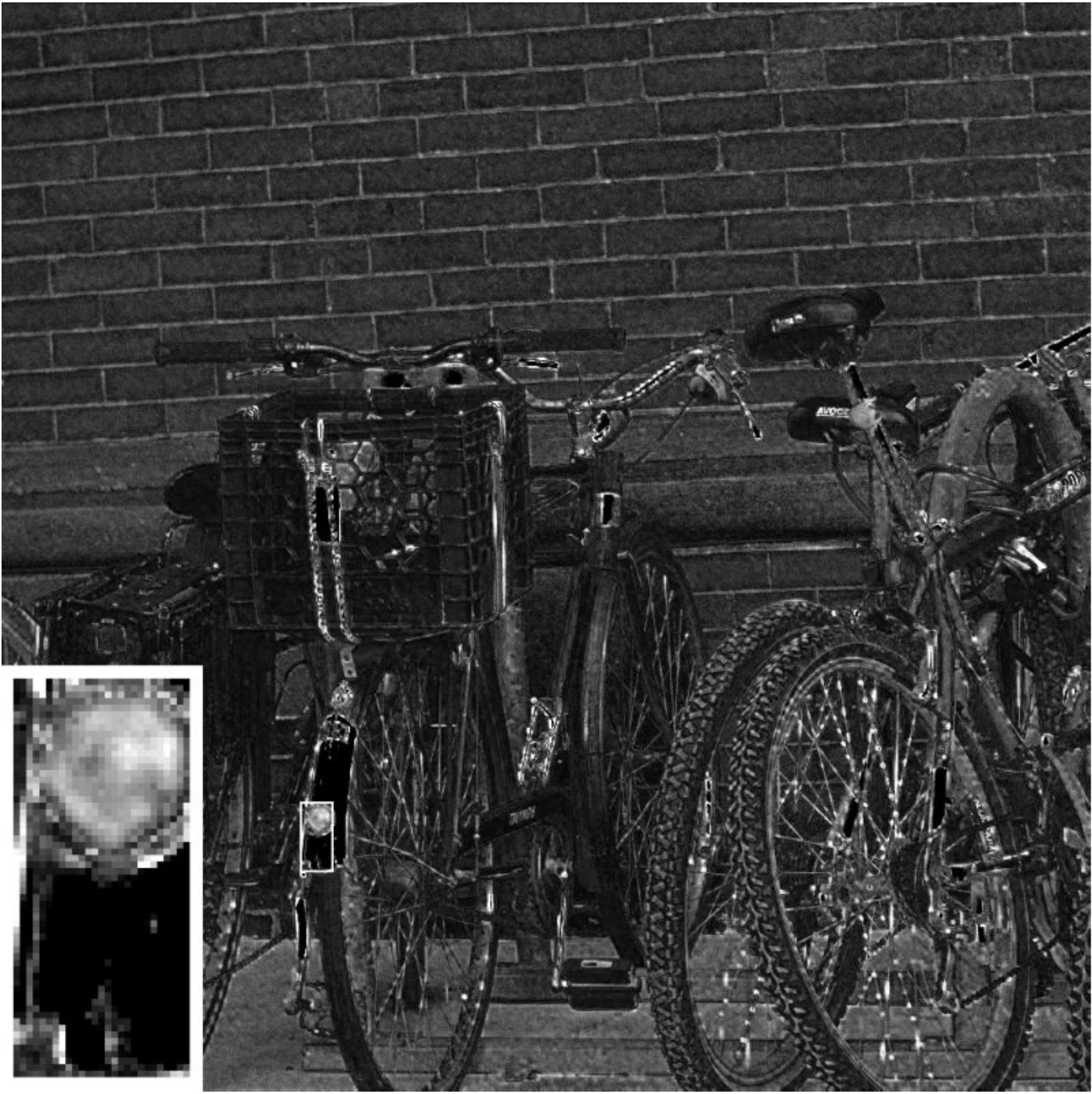}}
			\centering

		\end{minipage}
		\begin{minipage}[t]{0.12\linewidth}
			{\includegraphics[width=1\linewidth]{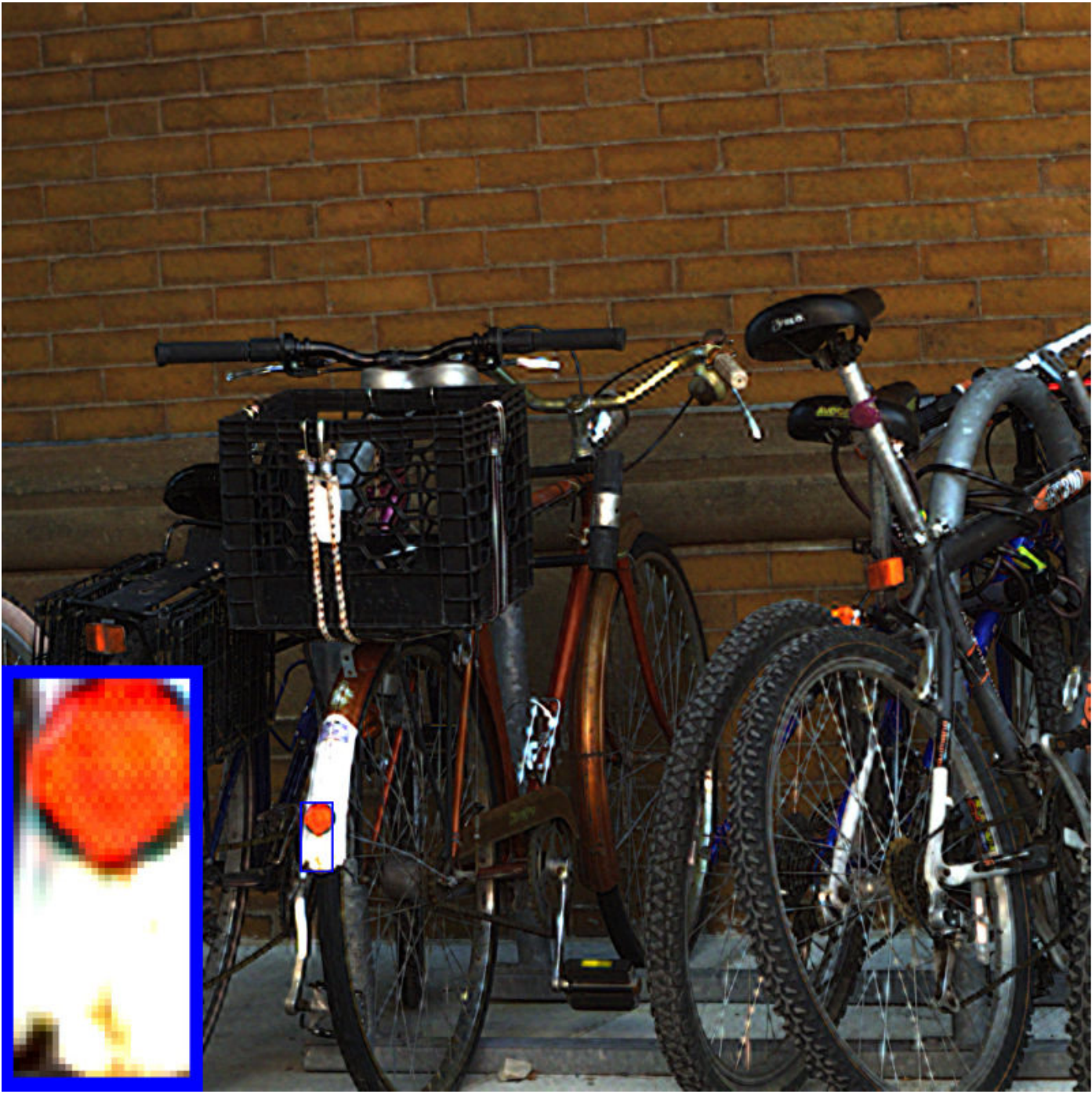}}
			{\includegraphics[width=1\linewidth]{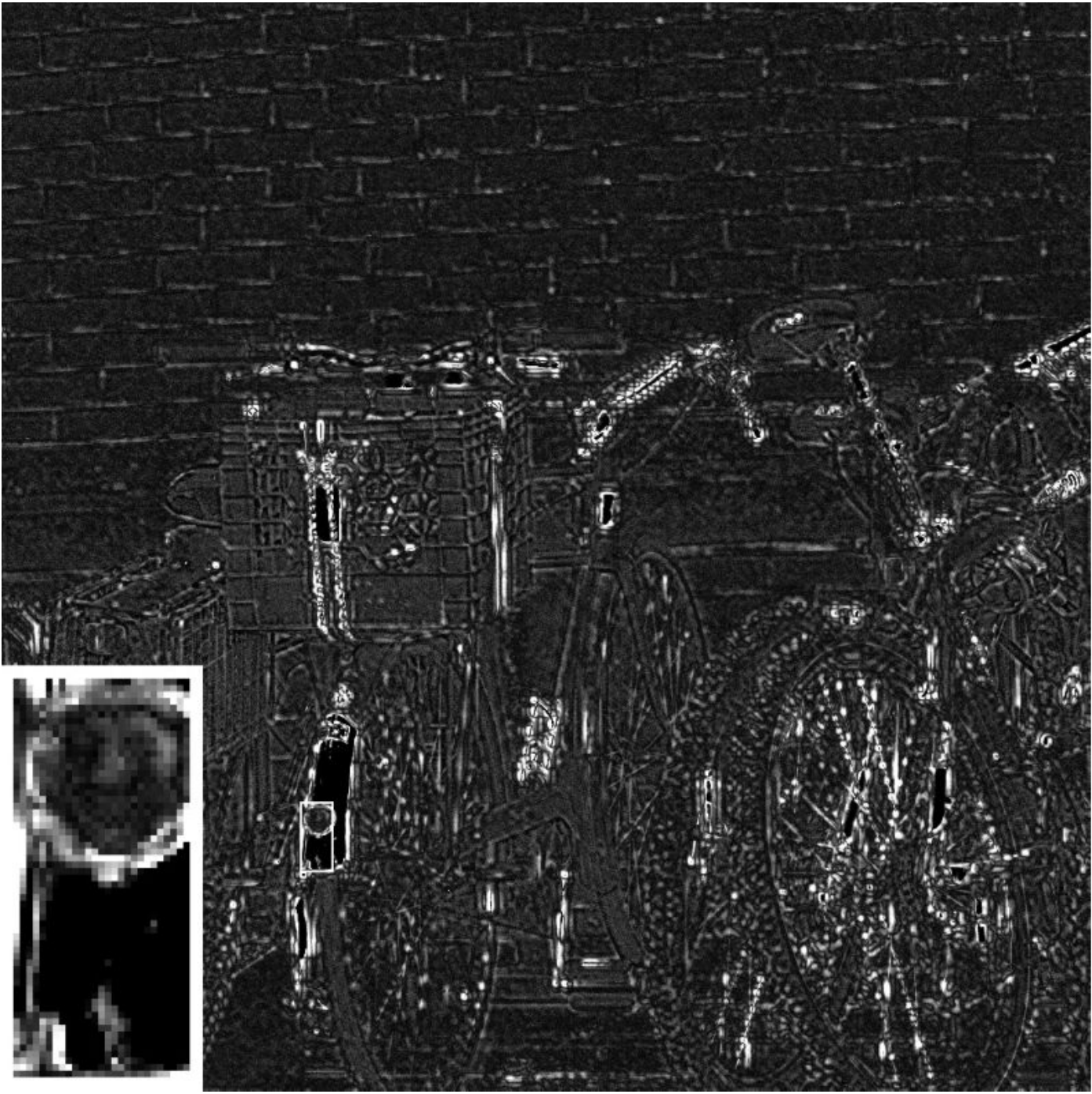}}
			\centering

		\end{minipage}
		\begin{minipage}[t]{0.12\linewidth}
			{\includegraphics[width=1\linewidth]{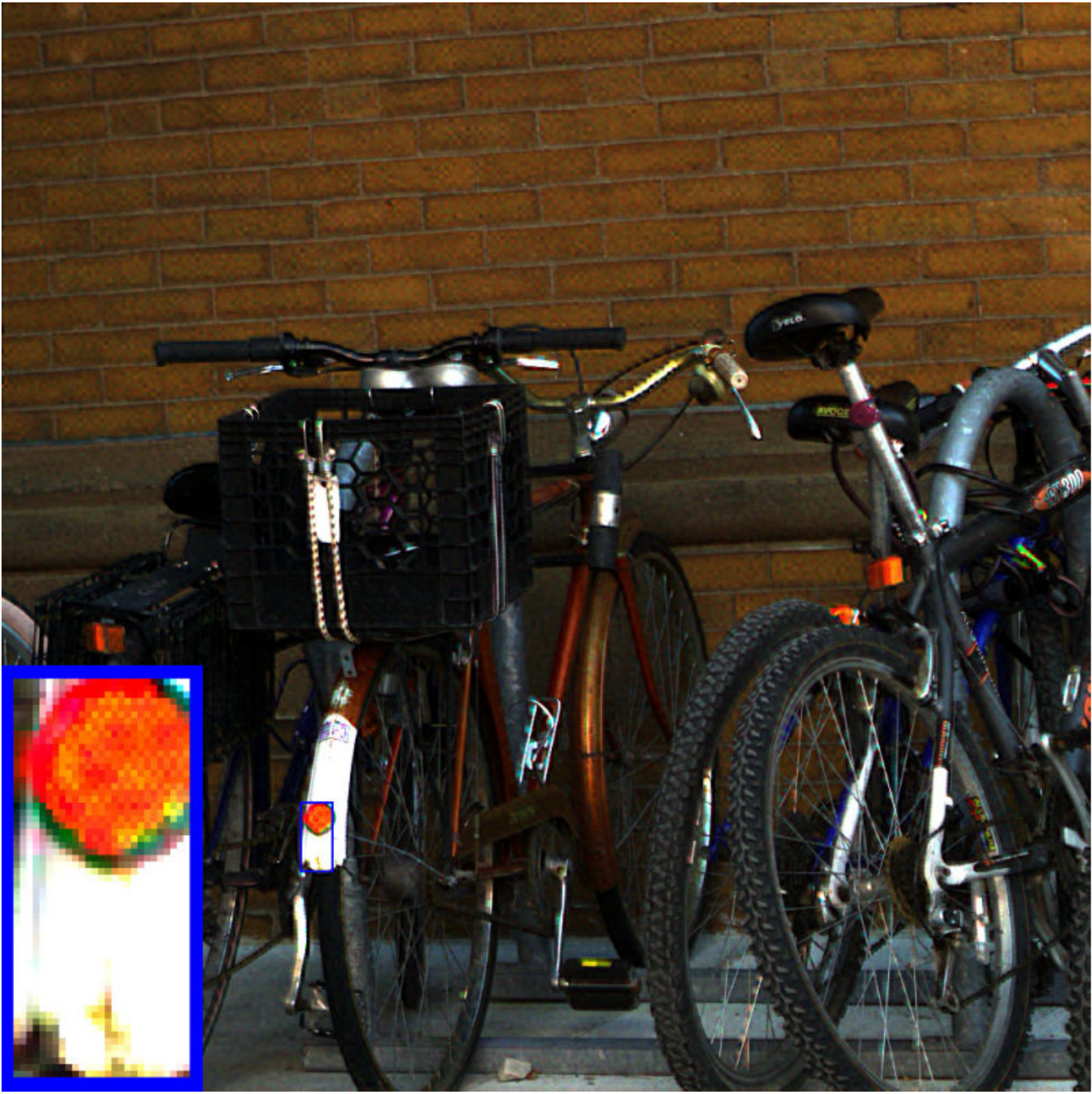}}
			{\includegraphics[width=1\linewidth]{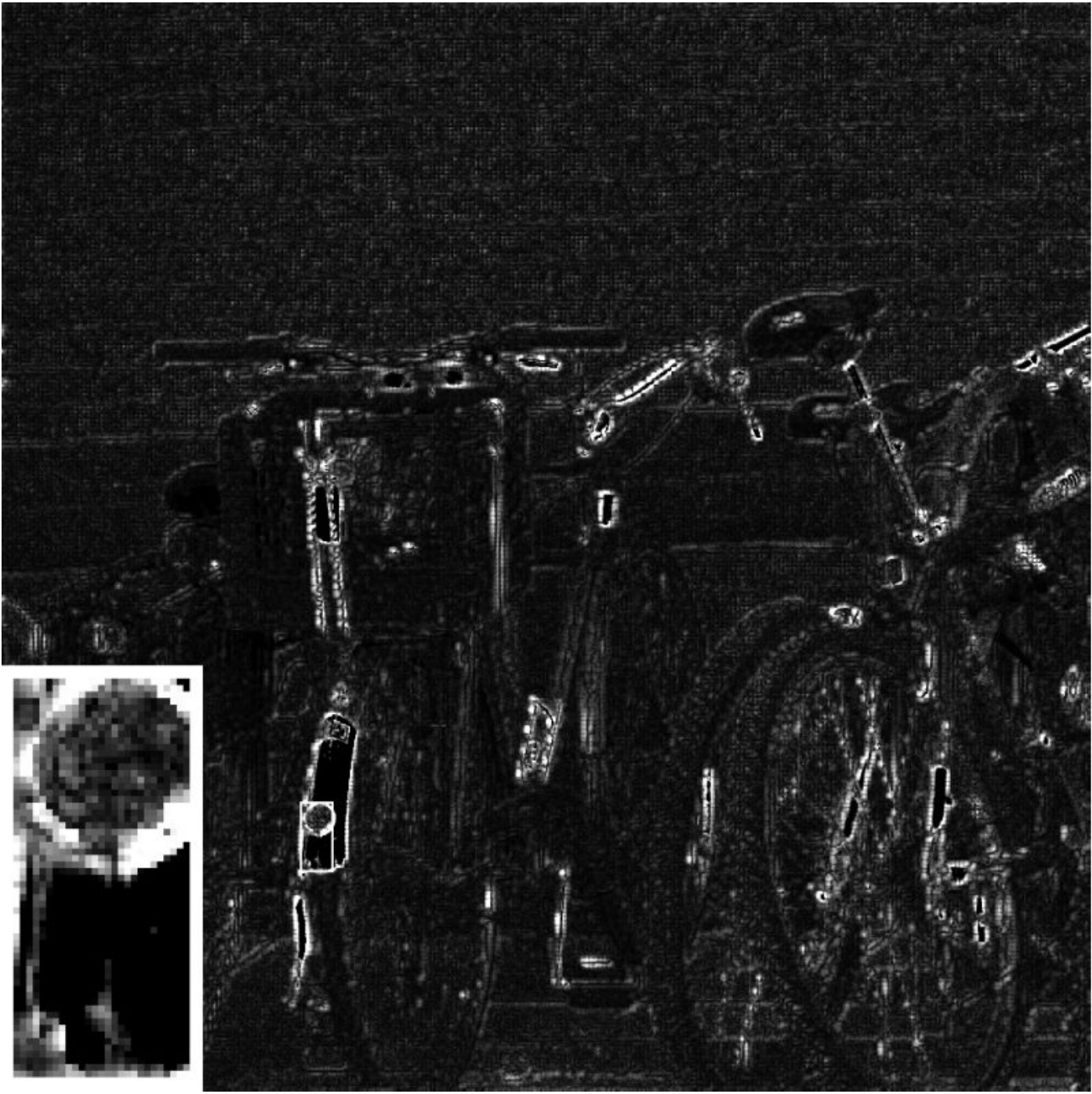}}
			\centering

		\end{minipage}
		\begin{minipage}[t]{0.12\linewidth}
			{\includegraphics[width=1\linewidth]{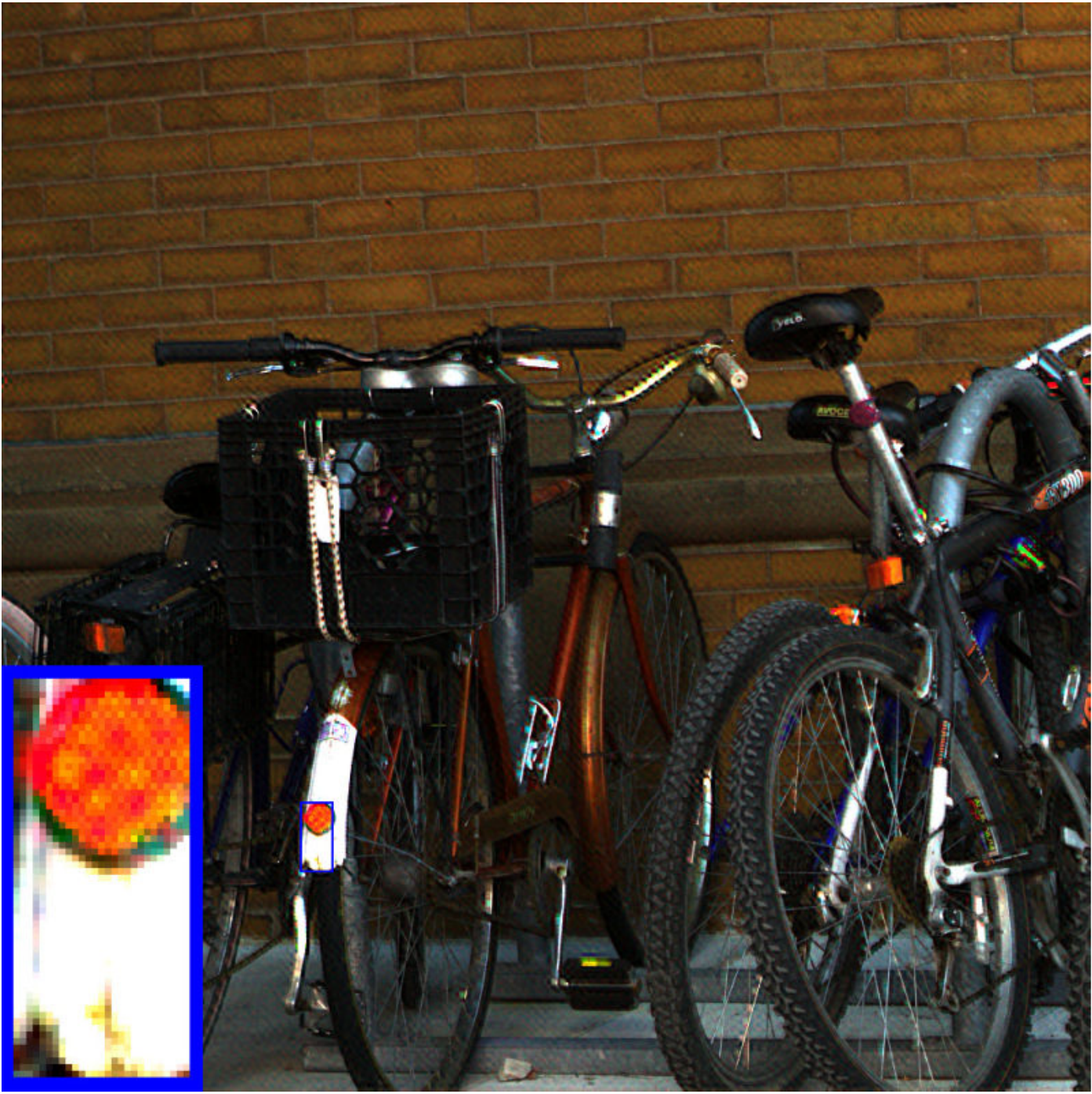}}
			{\includegraphics[width=1\linewidth]{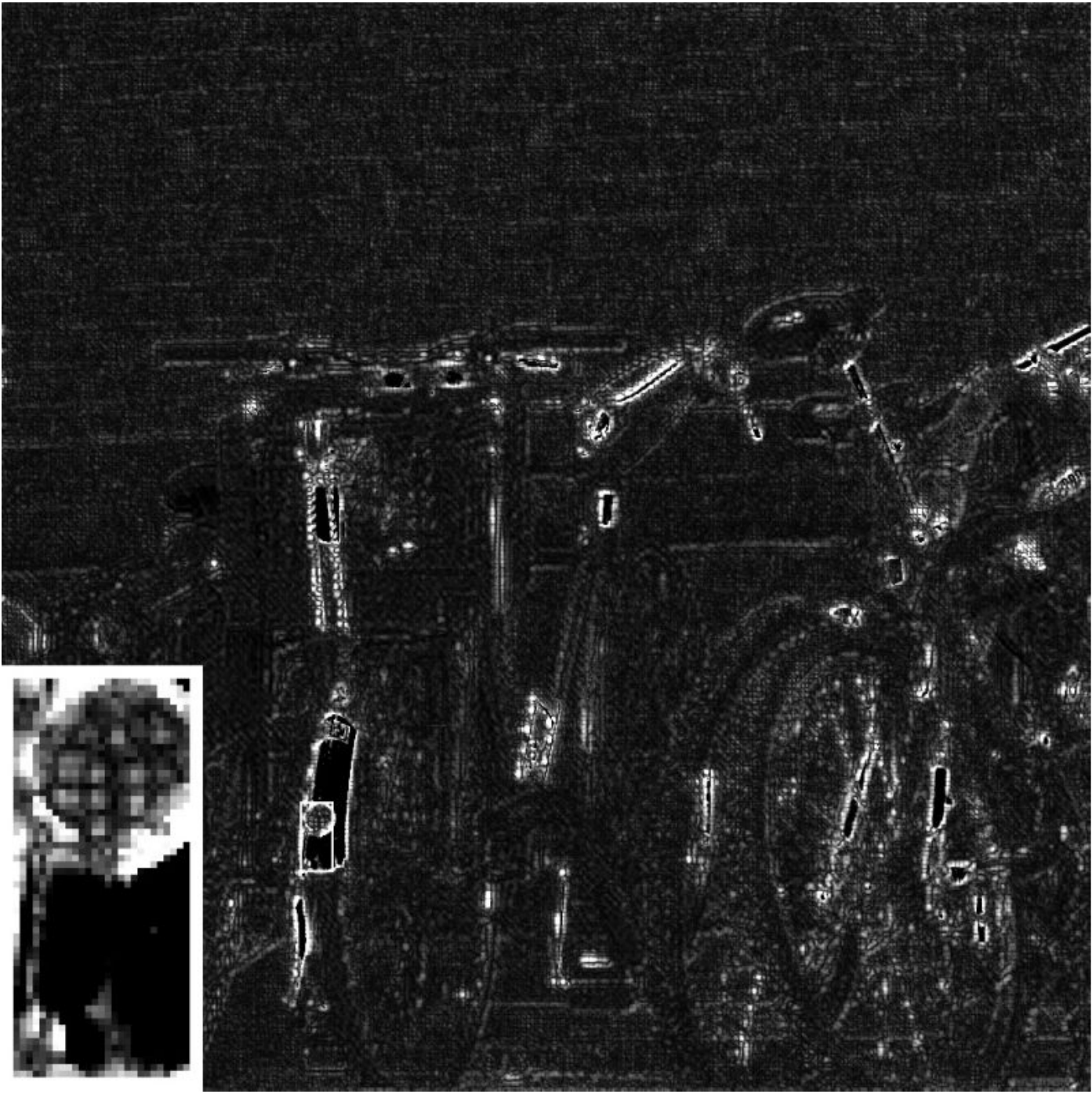}}
			\centering

		\end{minipage}
		\begin{minipage}[t]{0.12\linewidth}
			{\includegraphics[width=1\linewidth]{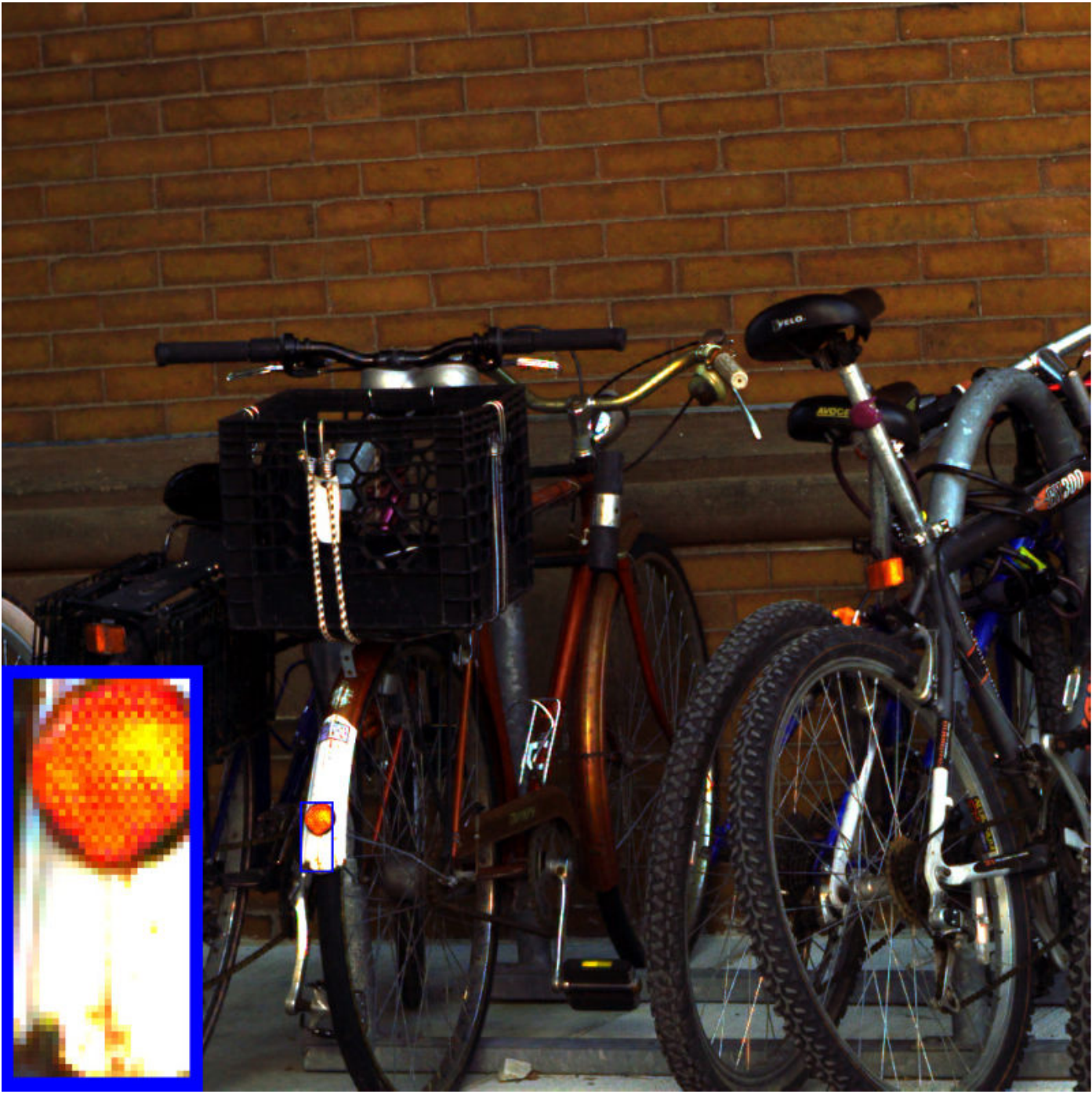}}
			{\includegraphics[width=1\linewidth]{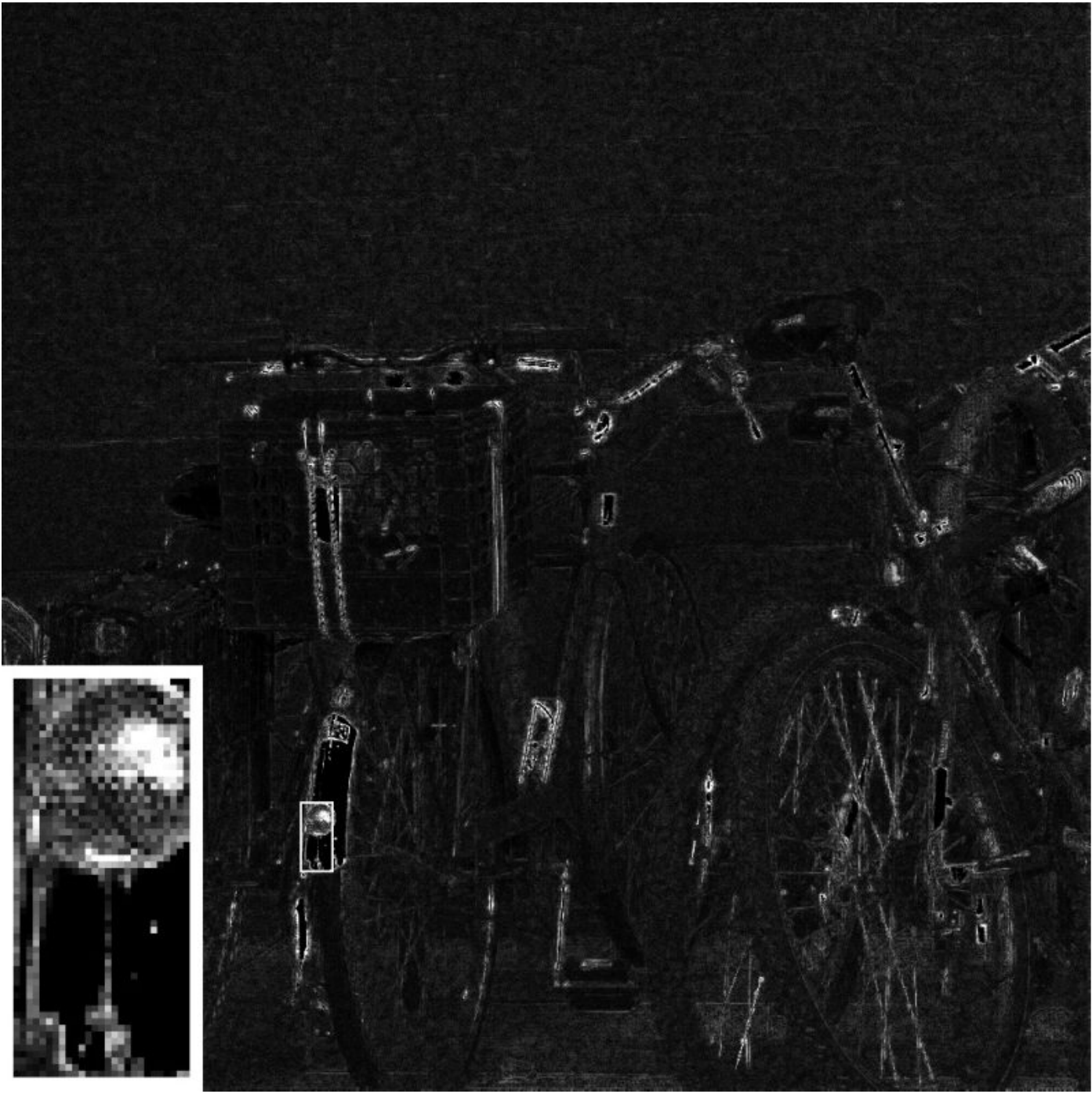}}
			\centering

		\end{minipage}
		\begin{minipage}[t]{0.12\linewidth}
			{\includegraphics[width=1\linewidth]{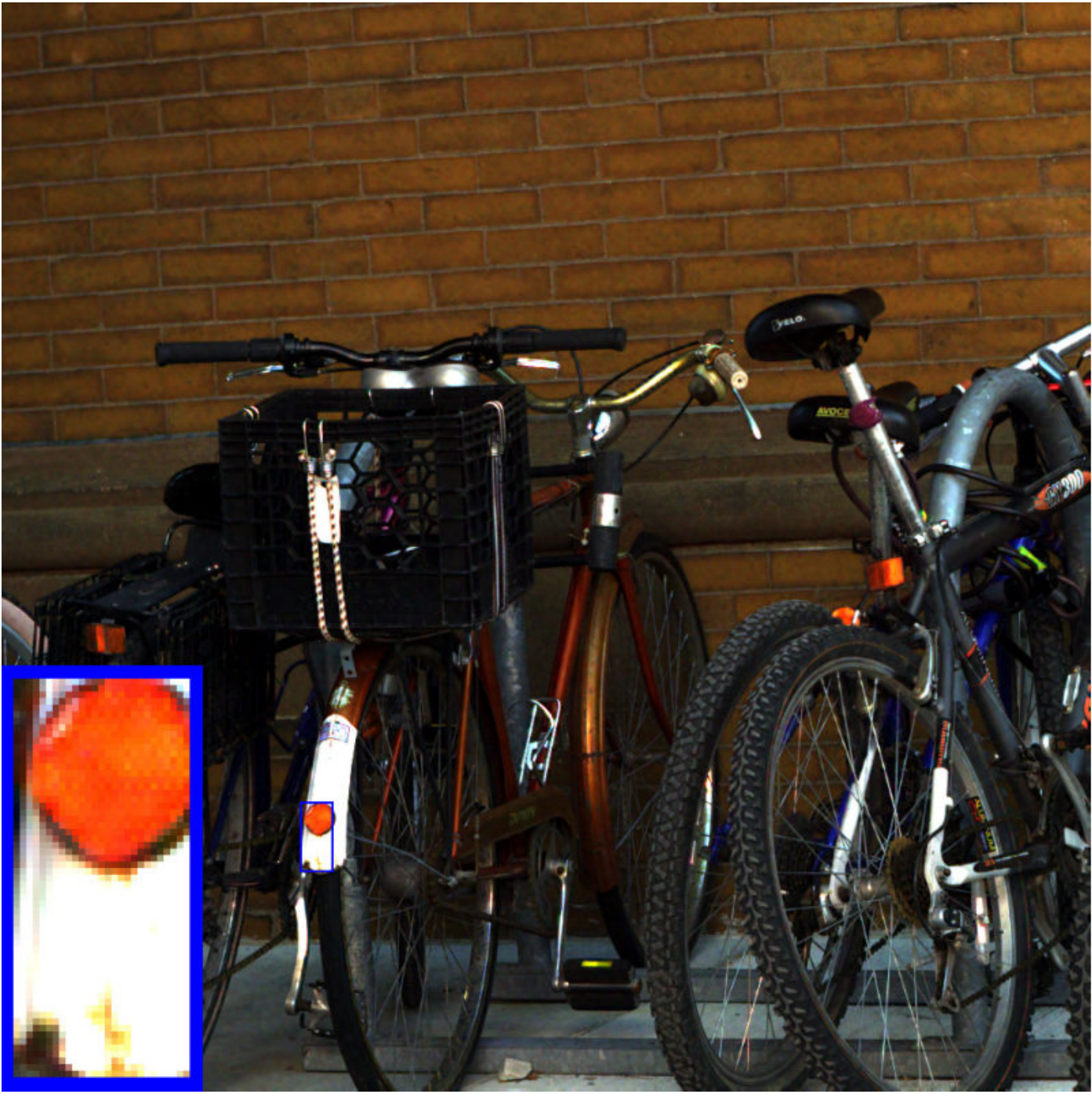}}
			{\includegraphics[width=1\linewidth]{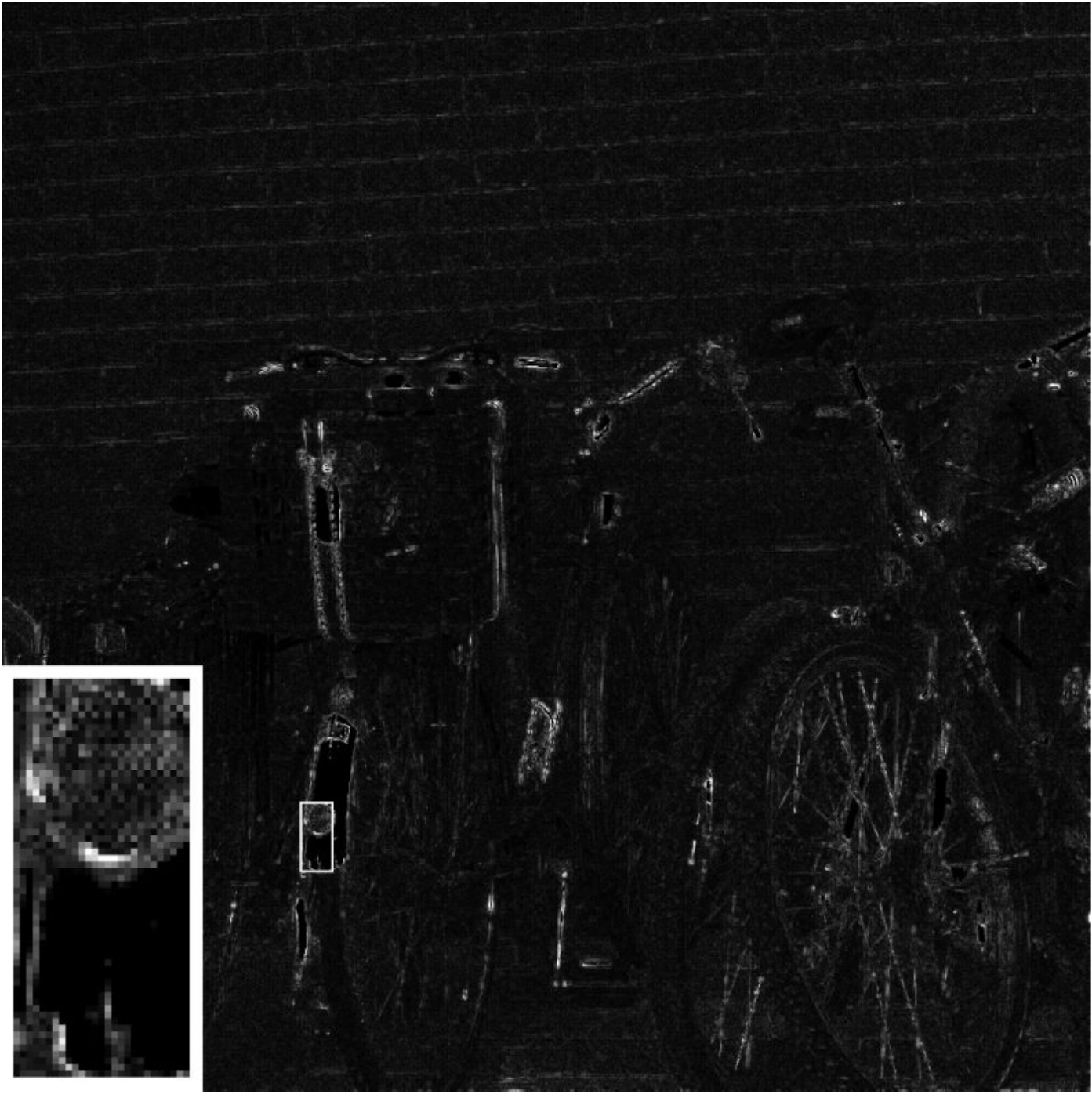}}
			\centering

		\end{minipage}
	
			\vspace{5pt}
	
	\begin{minipage}[t]{0.12\linewidth}
		{\includegraphics[width=1\linewidth]{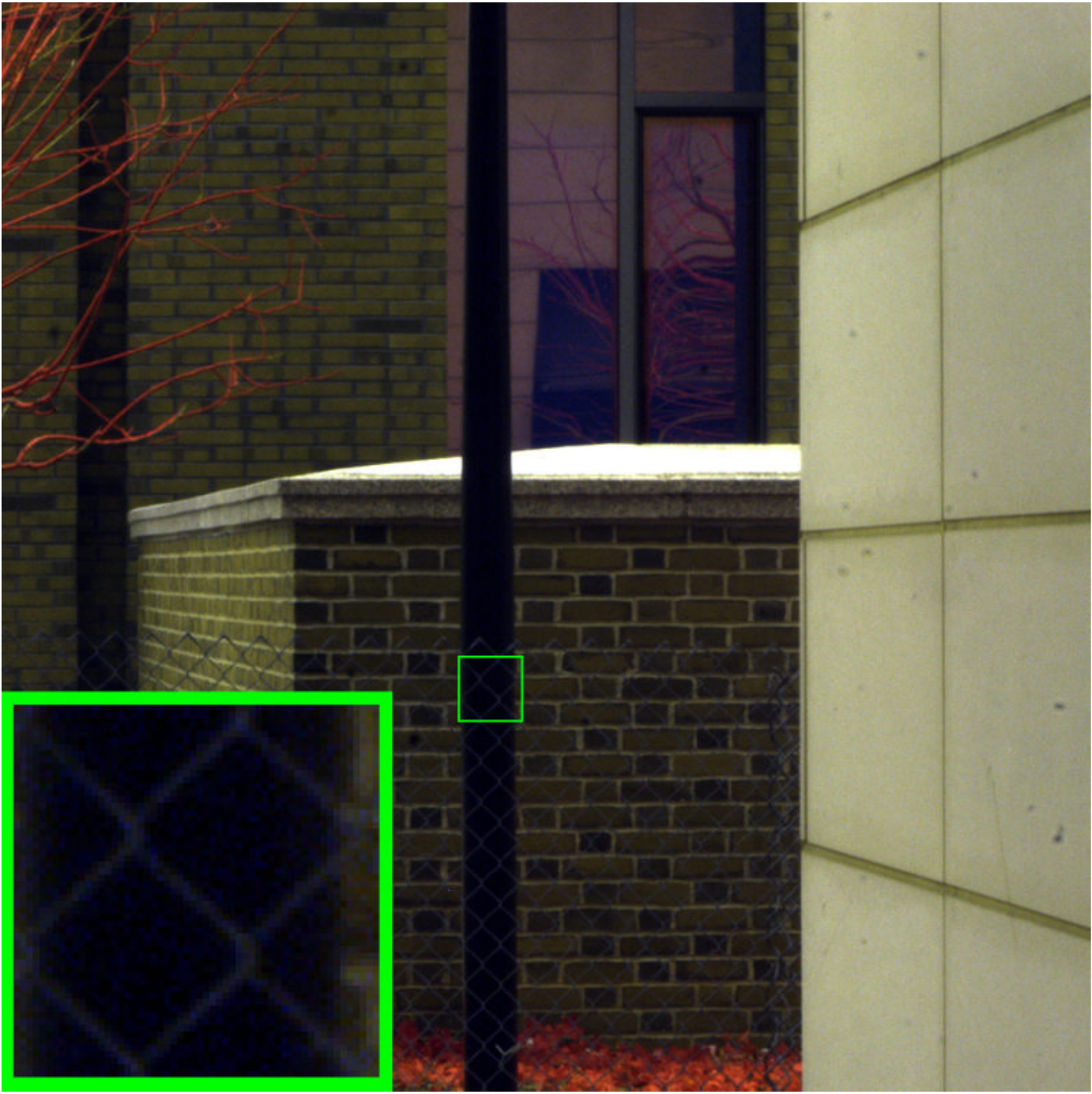}}
		{\includegraphics[width=1\linewidth]{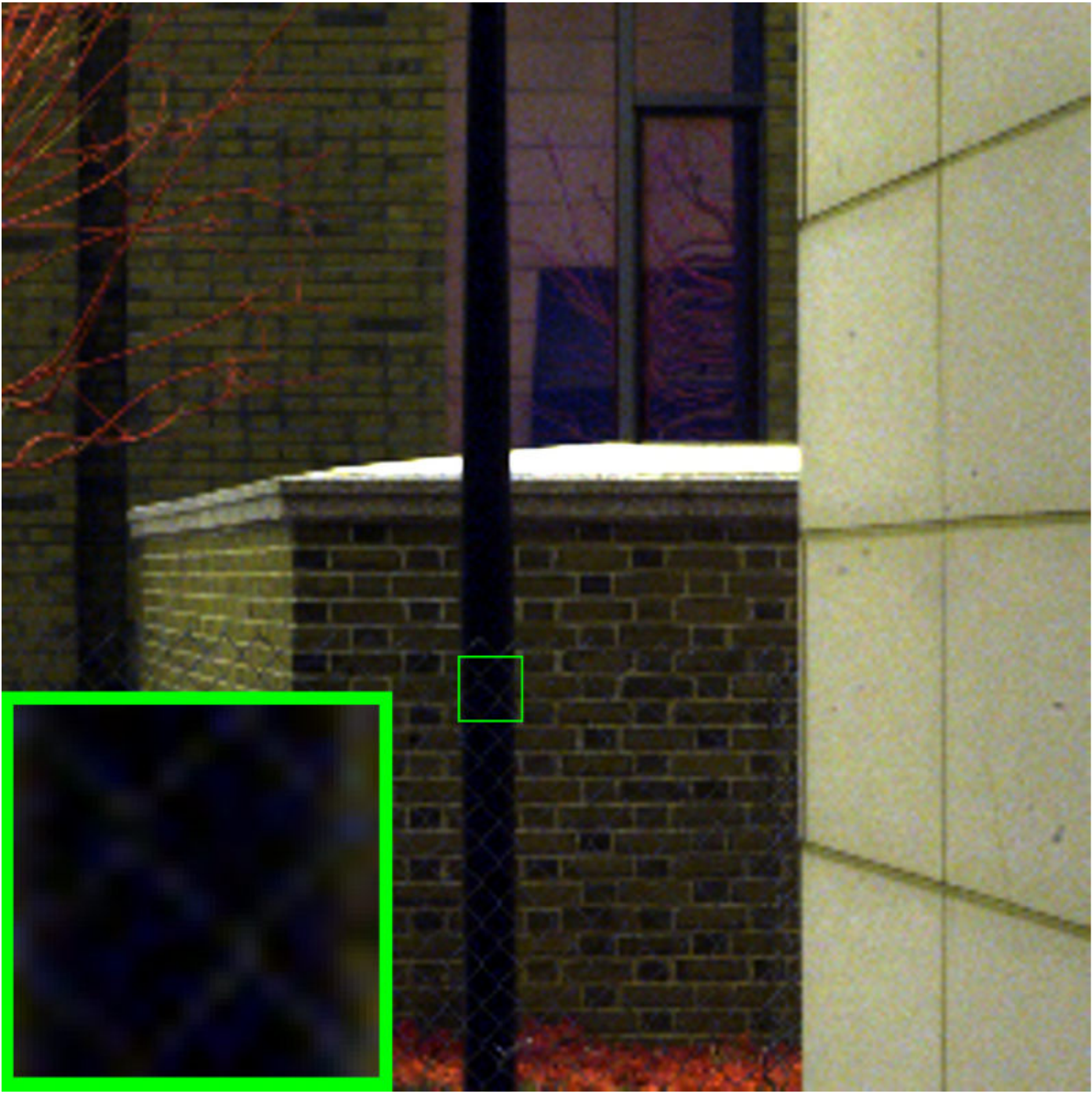}}
		\centering
		{GT}
	\end{minipage}
	\begin{minipage}[t]{0.12\linewidth}
		{\includegraphics[width=1\linewidth]{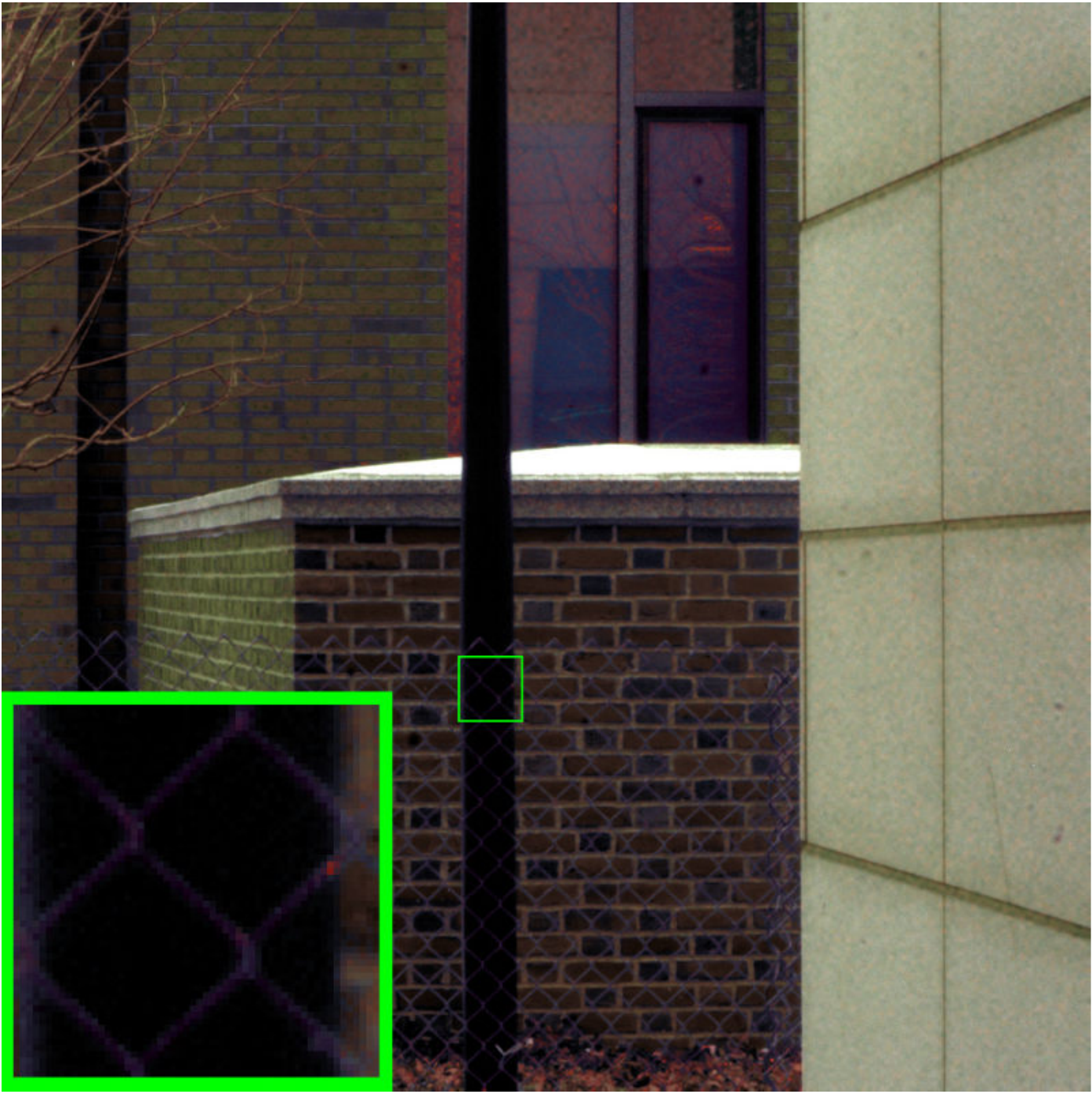}}
		{\includegraphics[width=1\linewidth]{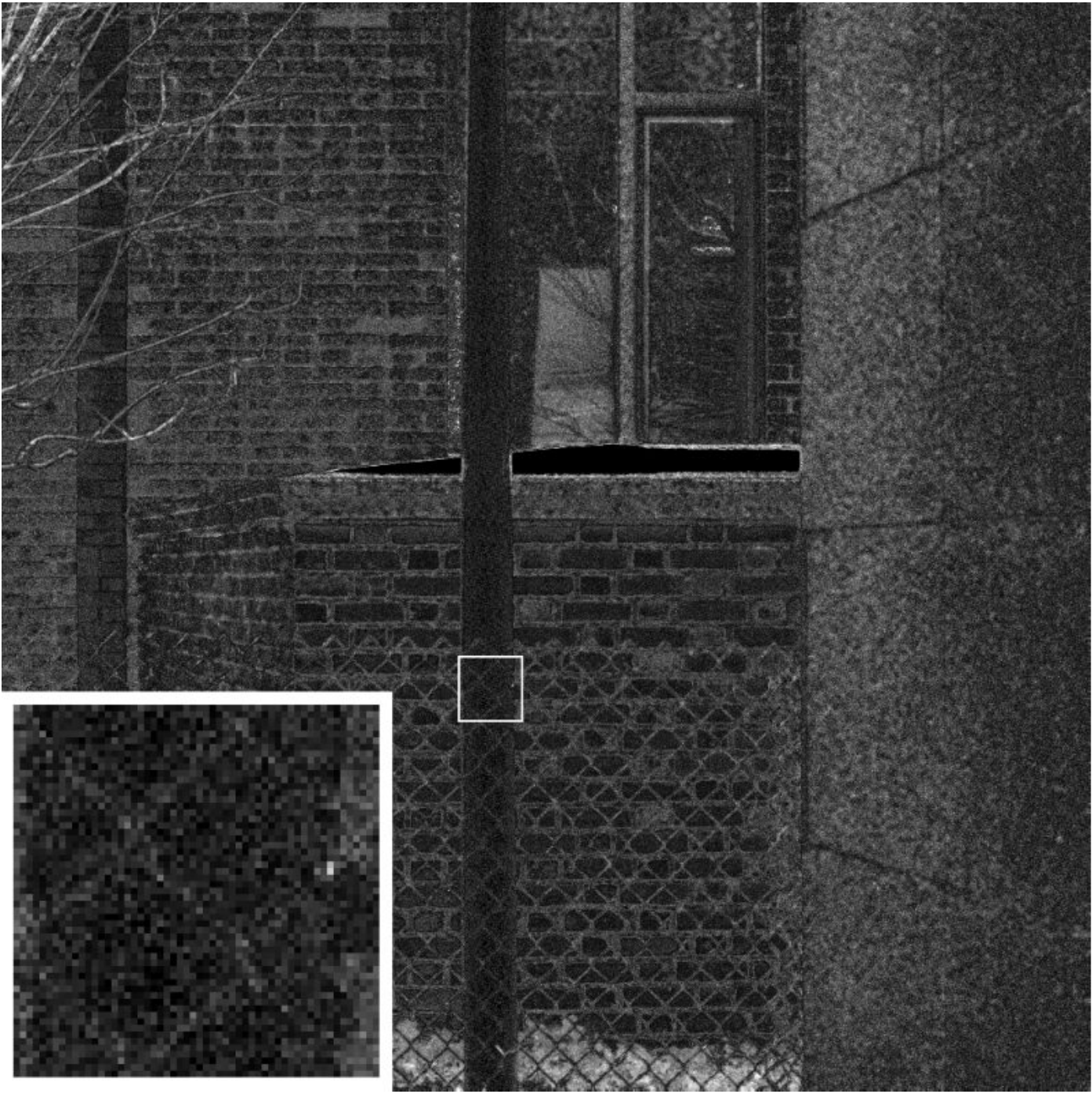}}
		\centering
		{CNMF \cite{CNMF}}
	\end{minipage}
	\begin{minipage}[t]{0.12\linewidth}
		{\includegraphics[width=1\linewidth]{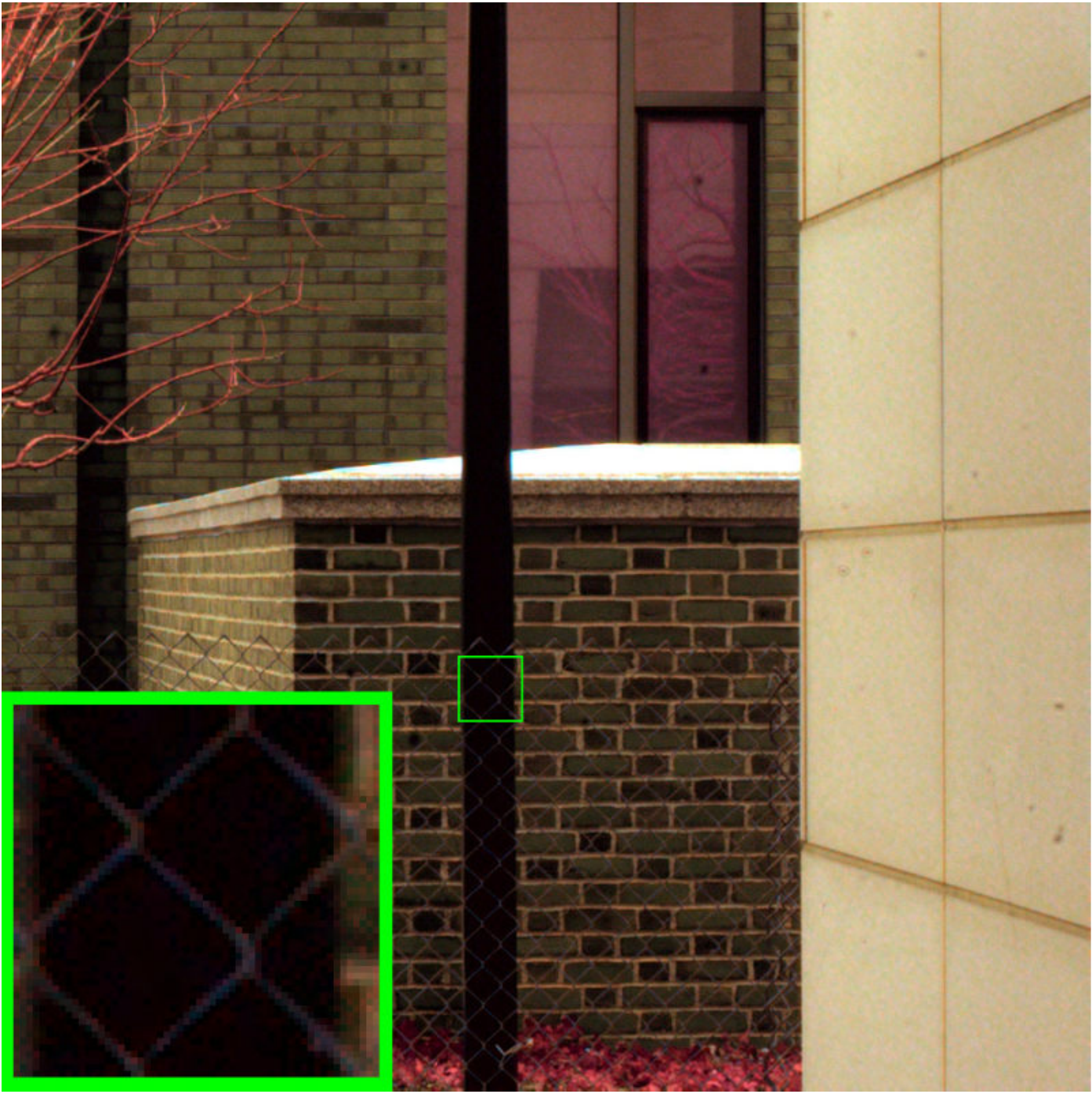}}
		{\includegraphics[width=1\linewidth]{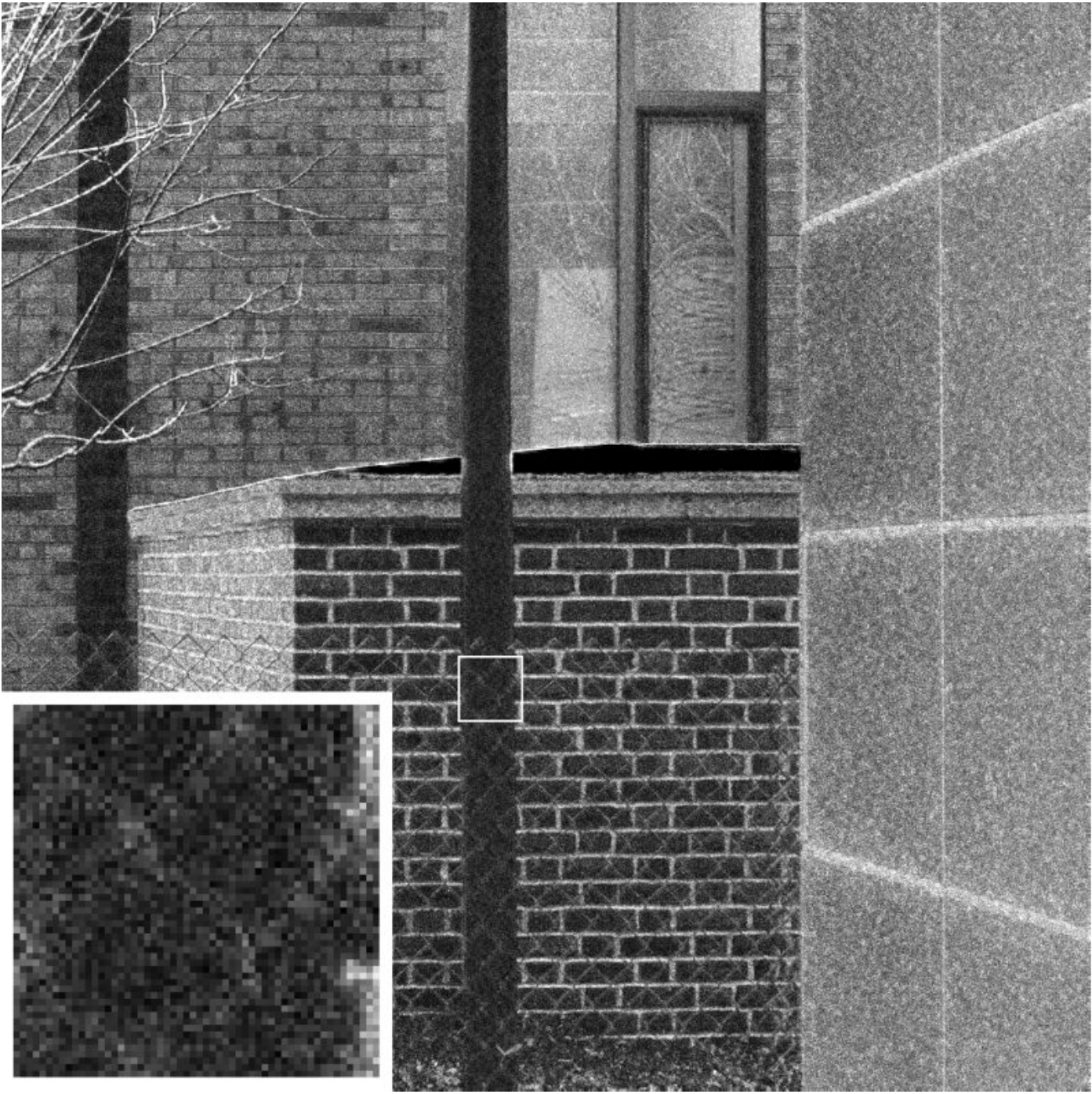}}
		\centering
		{FUSE \cite{FUSE}}
	\end{minipage}
	\begin{minipage}[t]{0.12\linewidth}
		{\includegraphics[width=1\linewidth]{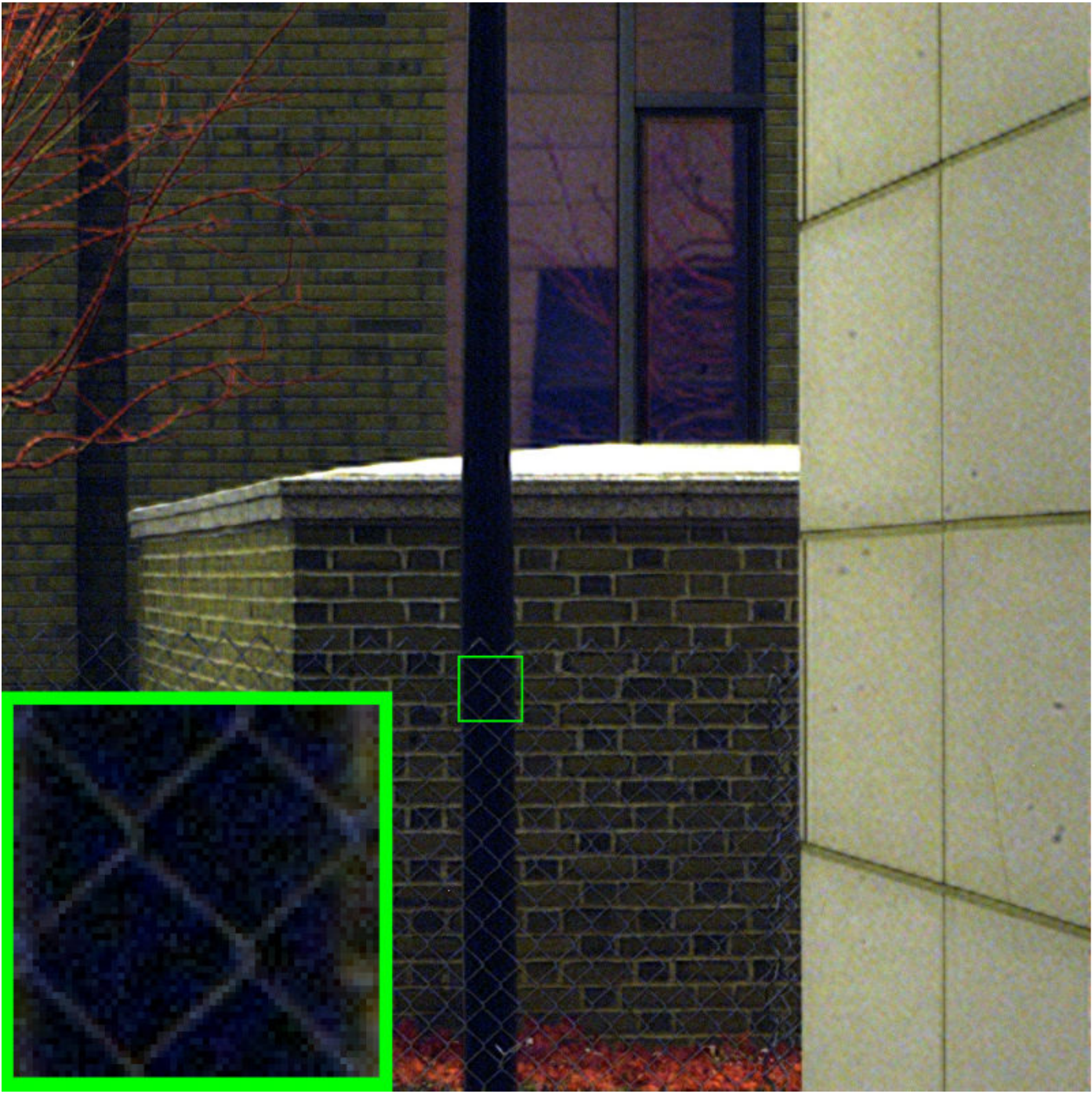}}
		{\includegraphics[width=1\linewidth]{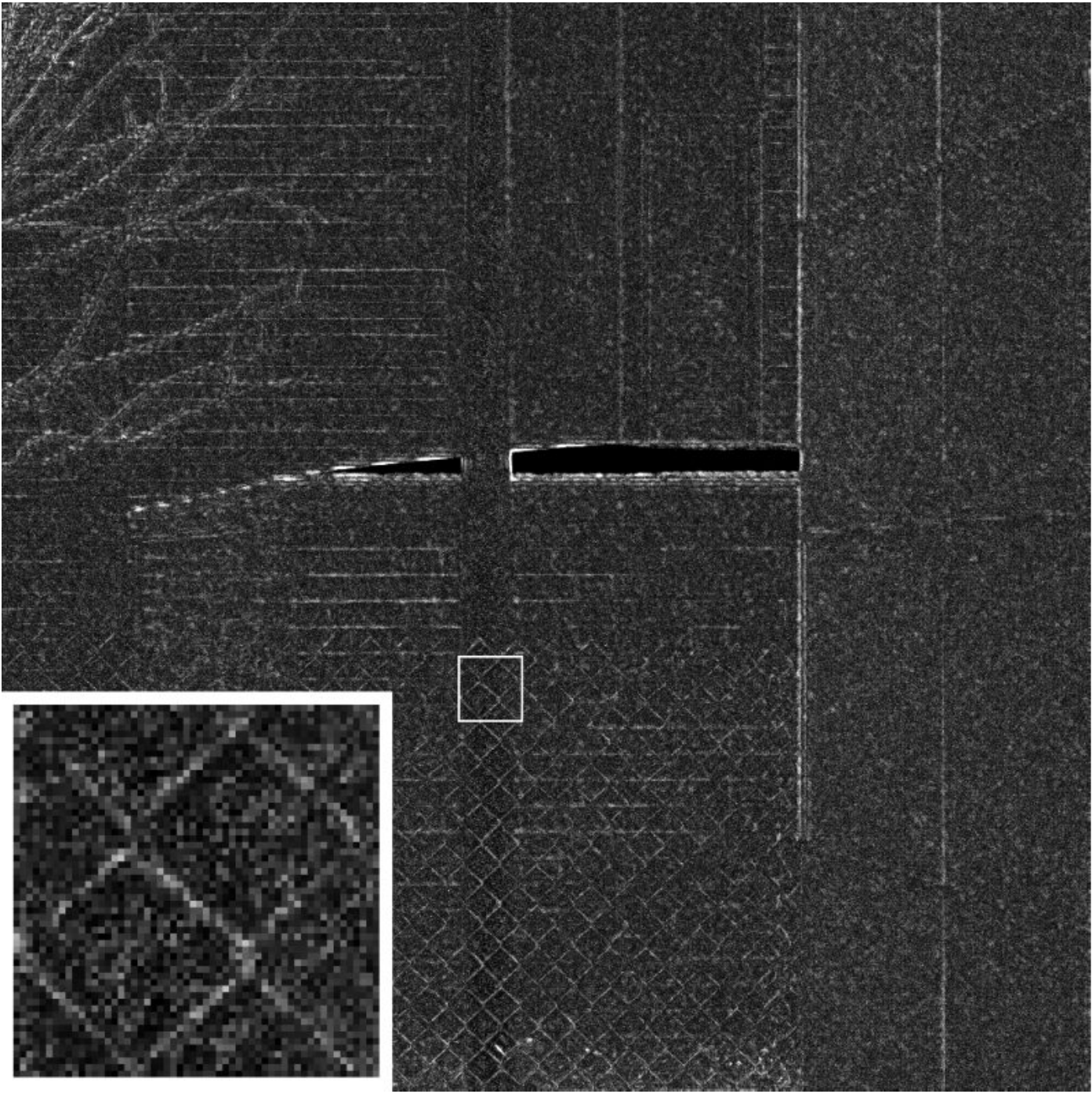}}
		\centering
		{GLP-HS \cite{GLP-HS}}
	\end{minipage}
	\begin{minipage}[t]{0.12\linewidth}
		{\includegraphics[width=1\linewidth]{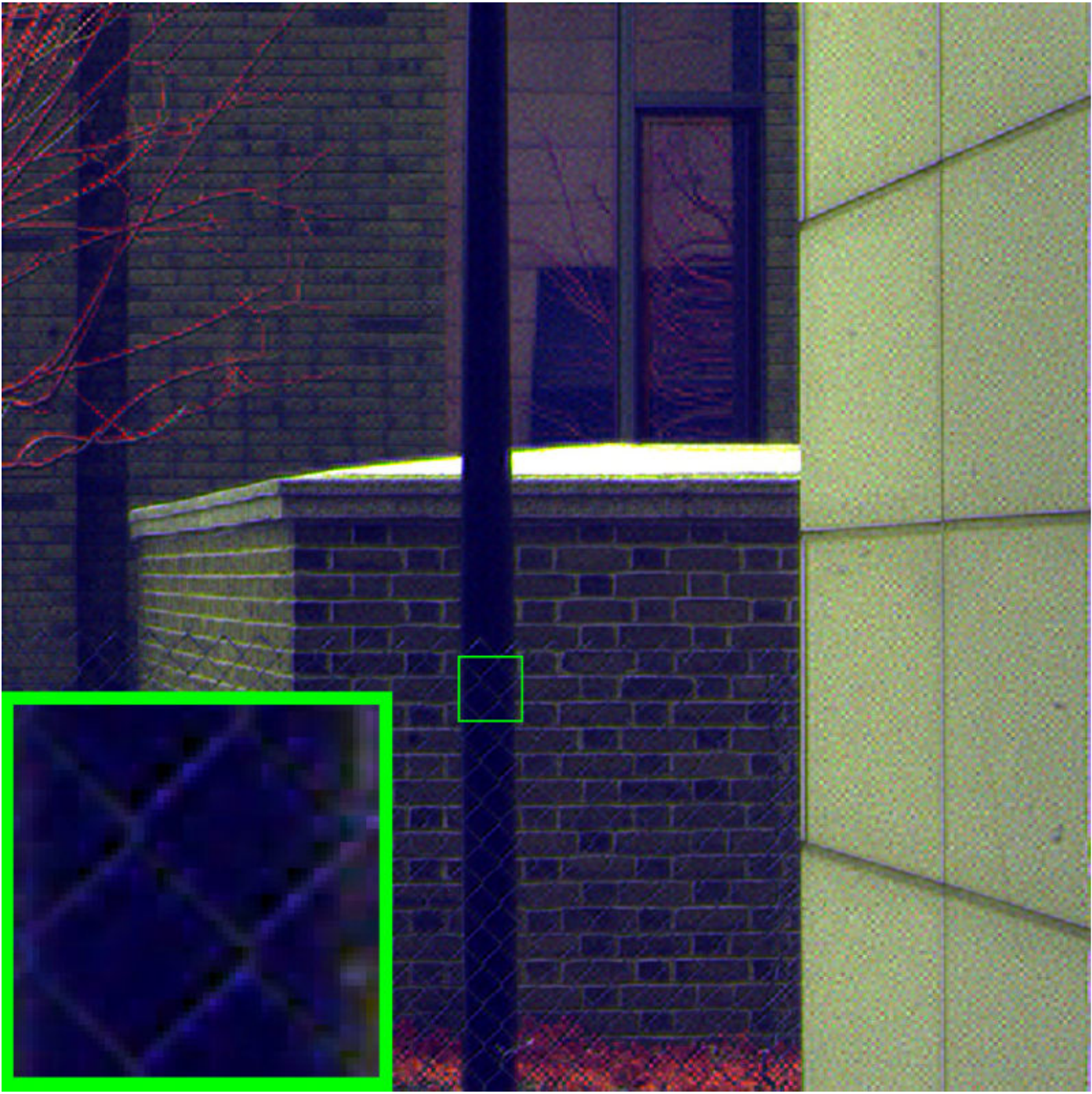}}
		{\includegraphics[width=1\linewidth]{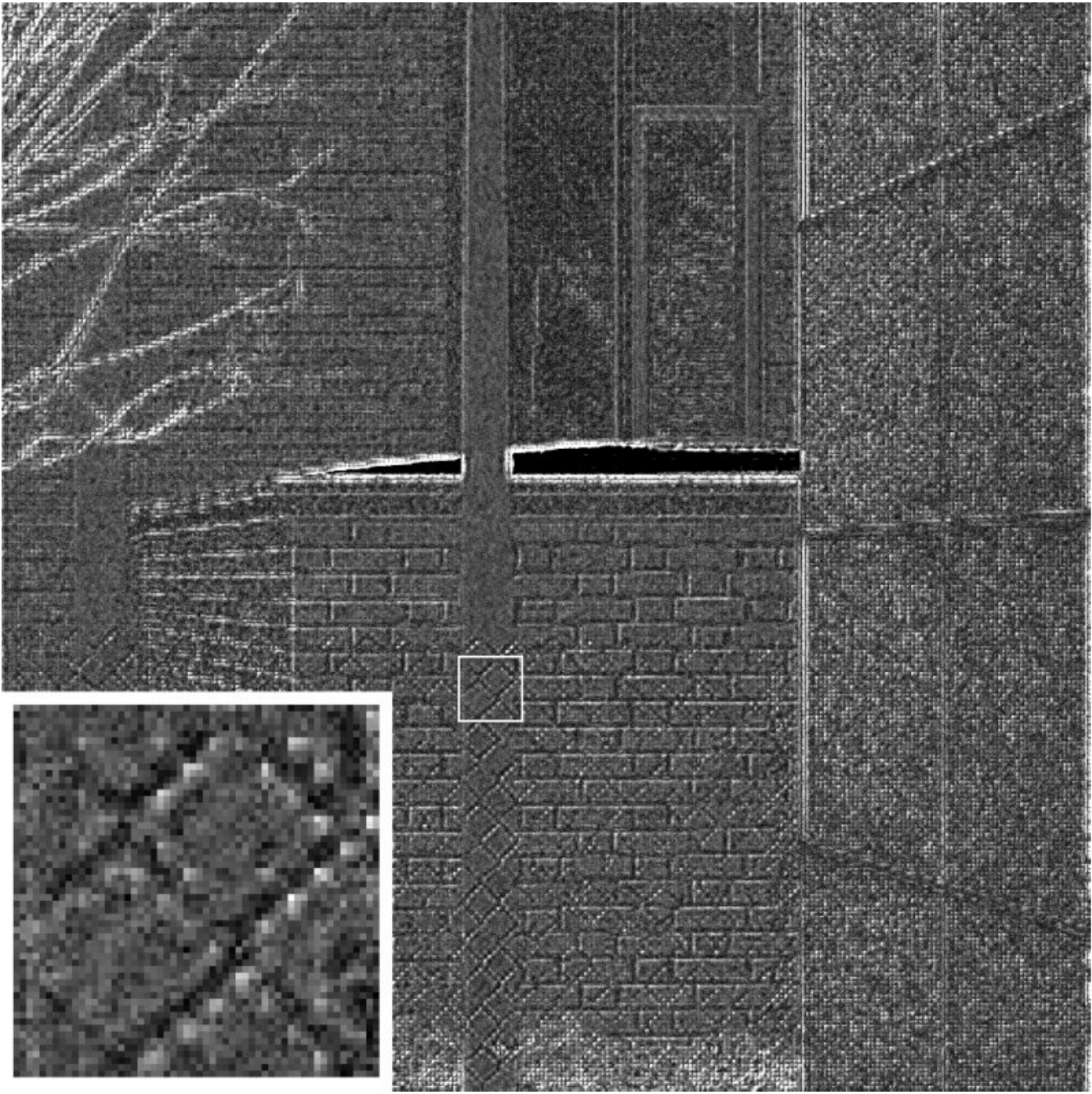}}
		\centering
		{LTTR \cite{LTTR}}
	\end{minipage}
	\begin{minipage}[t]{0.12\linewidth}
		{\includegraphics[width=1\linewidth]{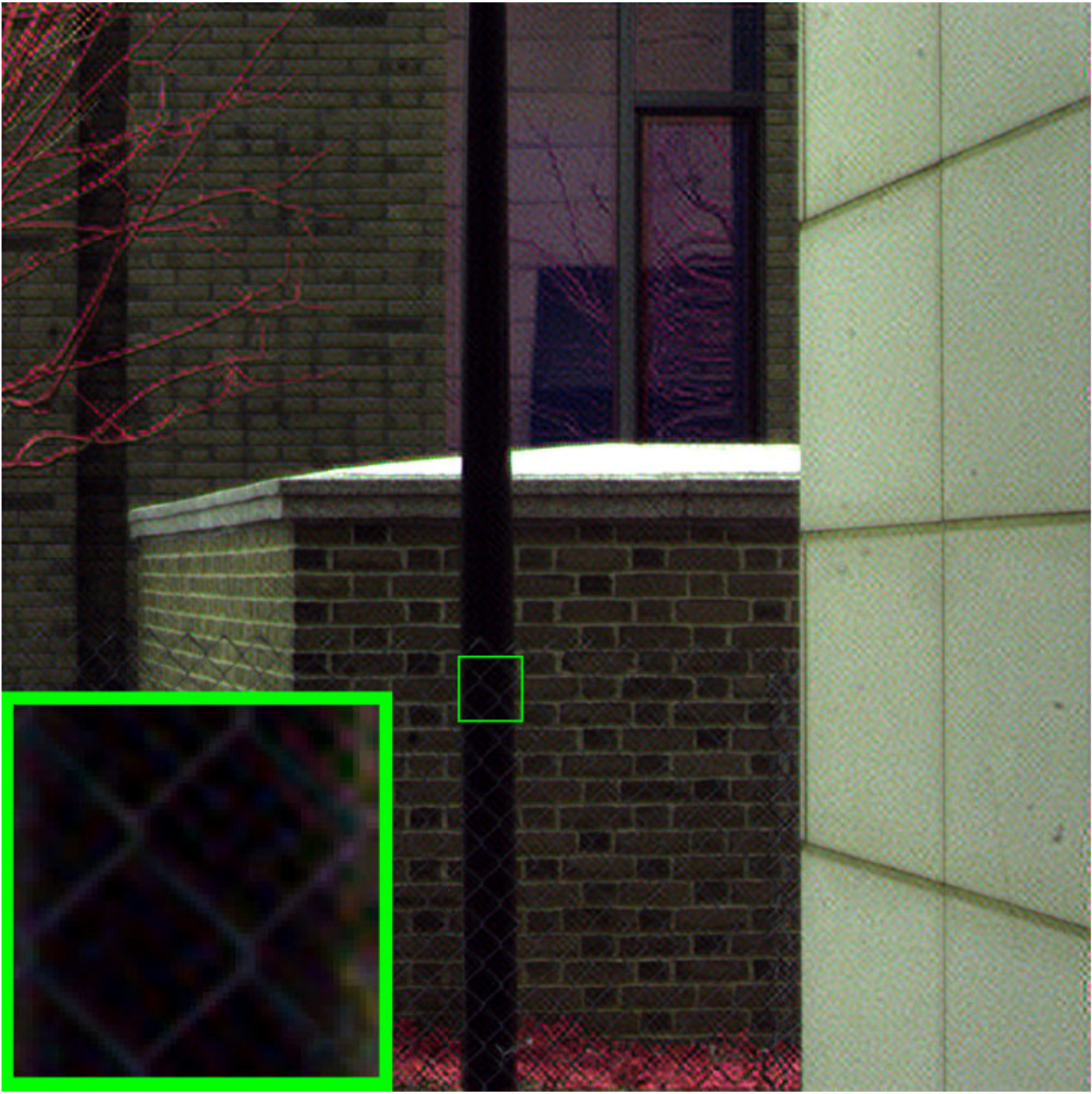}}
		{\includegraphics[width=1\linewidth]{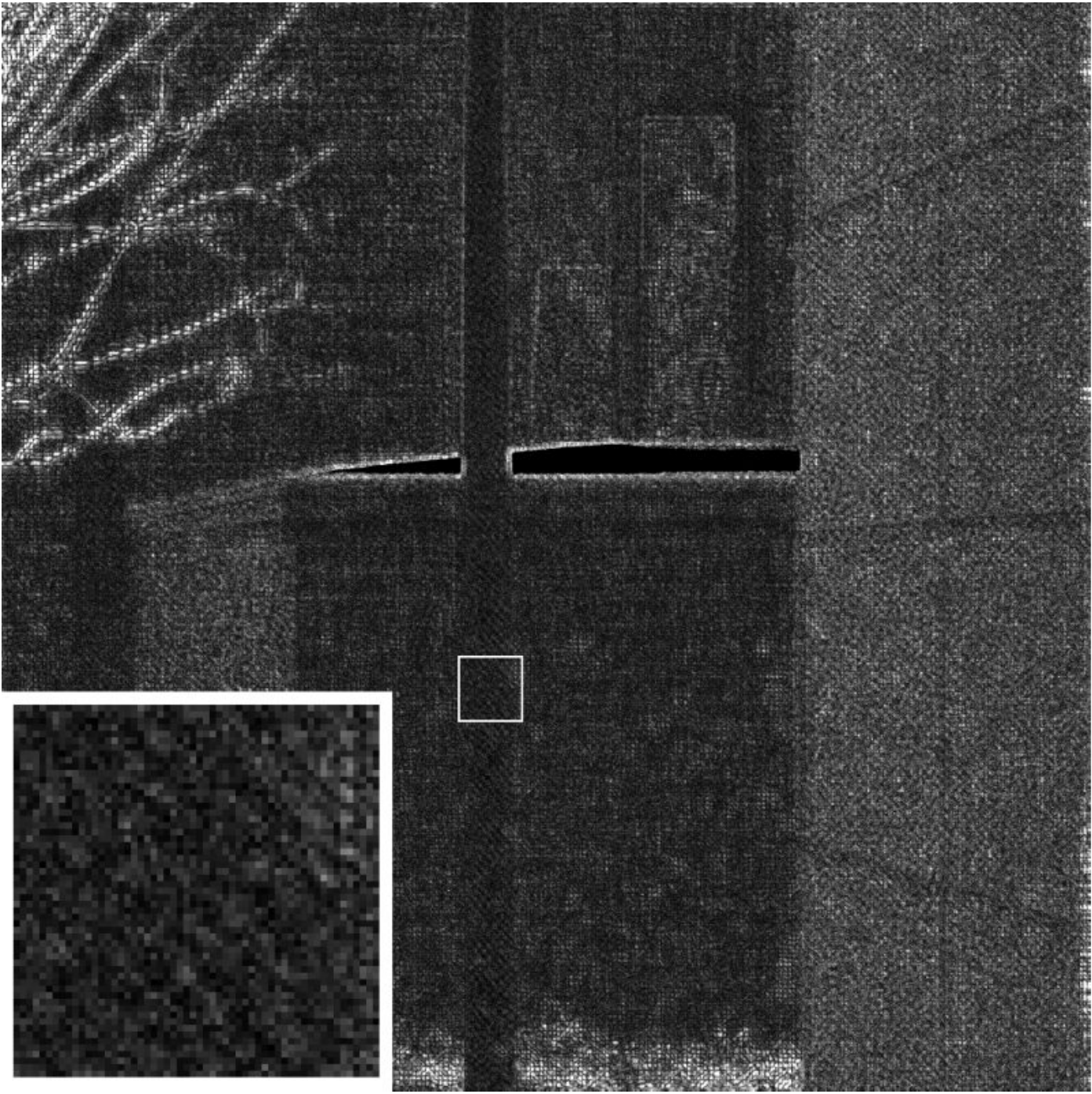}}
		\centering
		{LTMR \cite{LTMR}}
	\end{minipage}
	\begin{minipage}[t]{0.12\linewidth}
		{\includegraphics[width=1\linewidth]{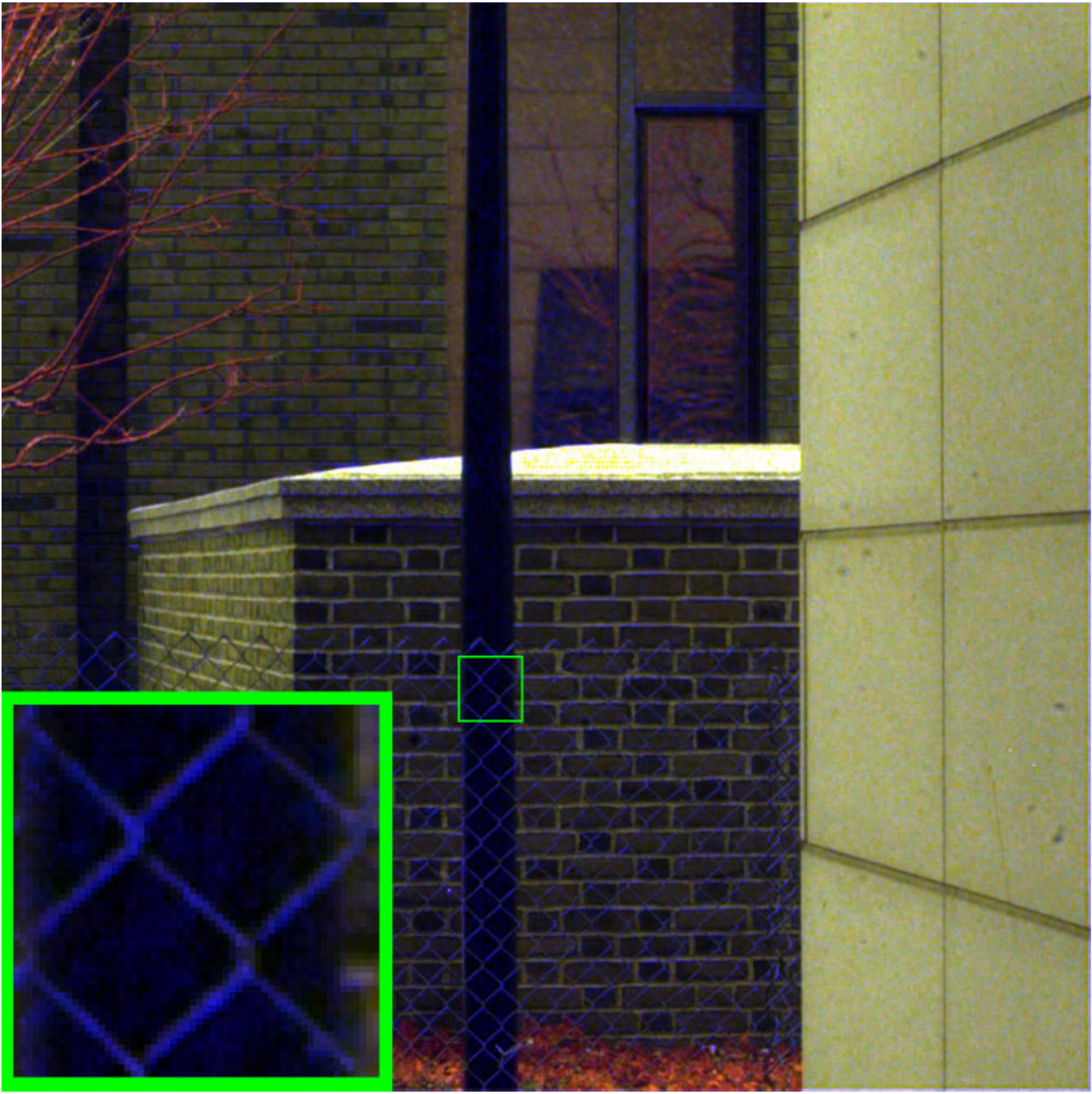}}
		{\includegraphics[width=1\linewidth]{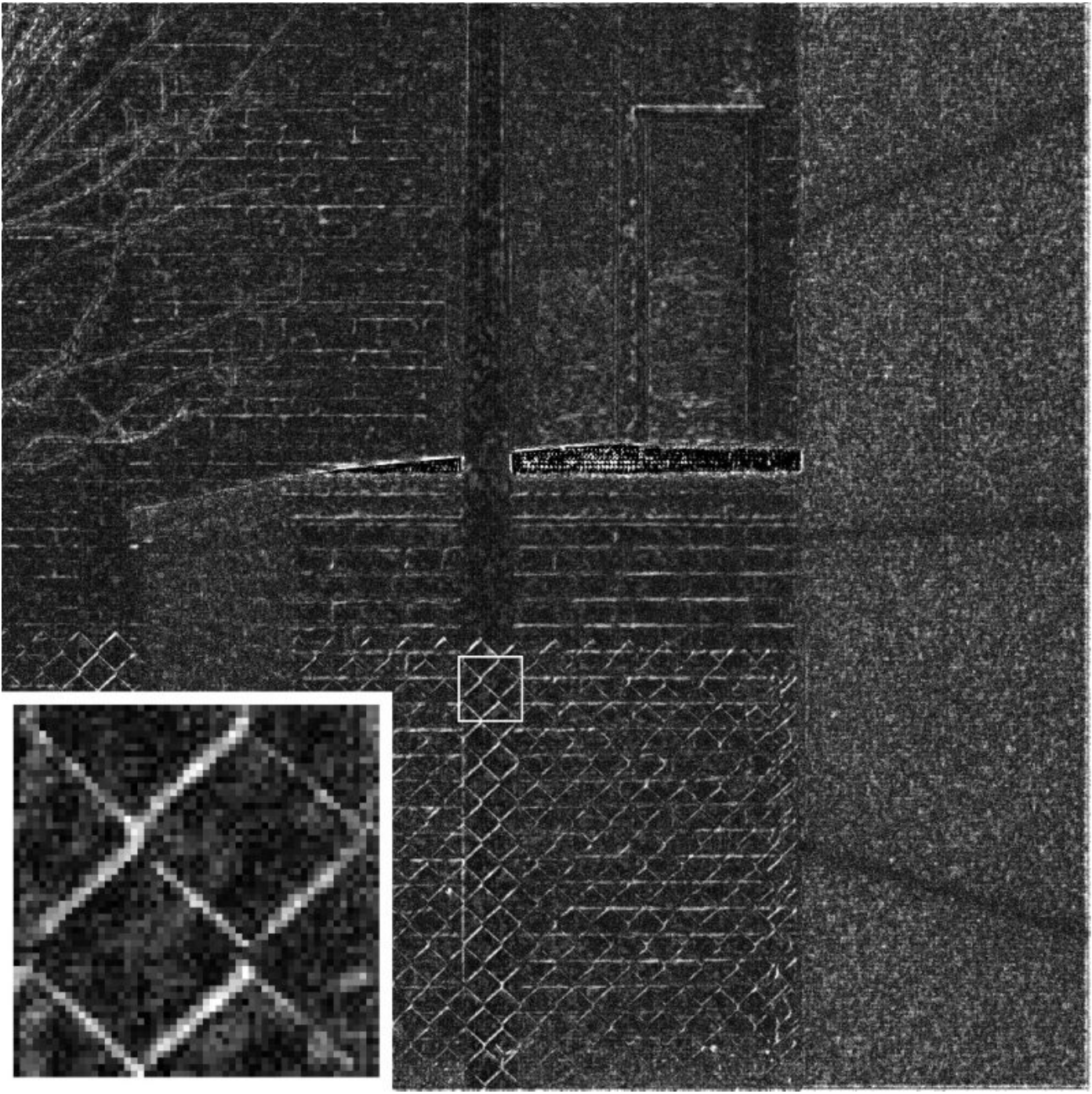}}
		\centering
		{MHFnet \cite{xie2019multispectral}}
	\end{minipage}
	\begin{minipage}[t]{0.12\linewidth}
		{\includegraphics[width=1\linewidth]{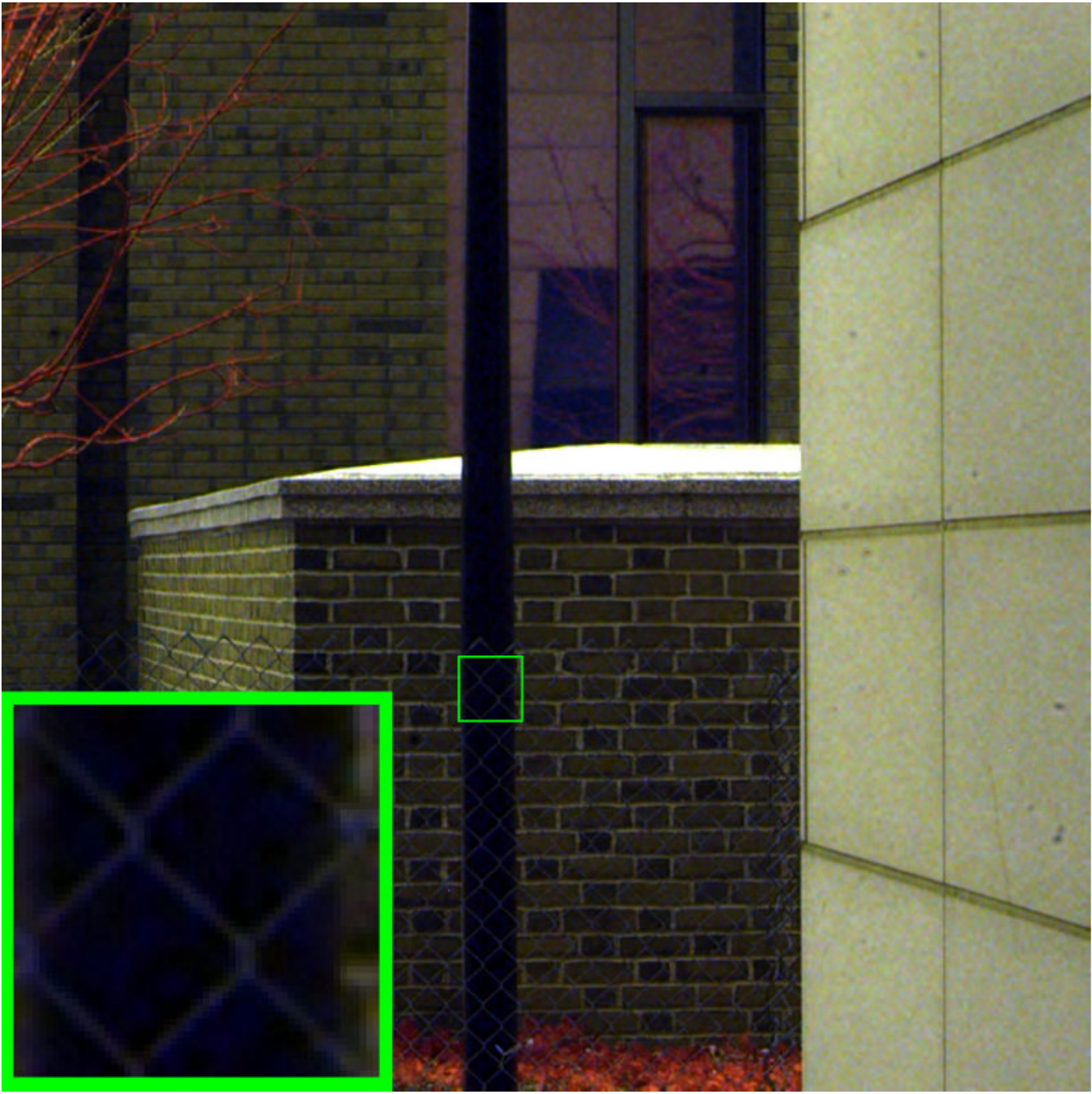}}
		{\includegraphics[width=1\linewidth]{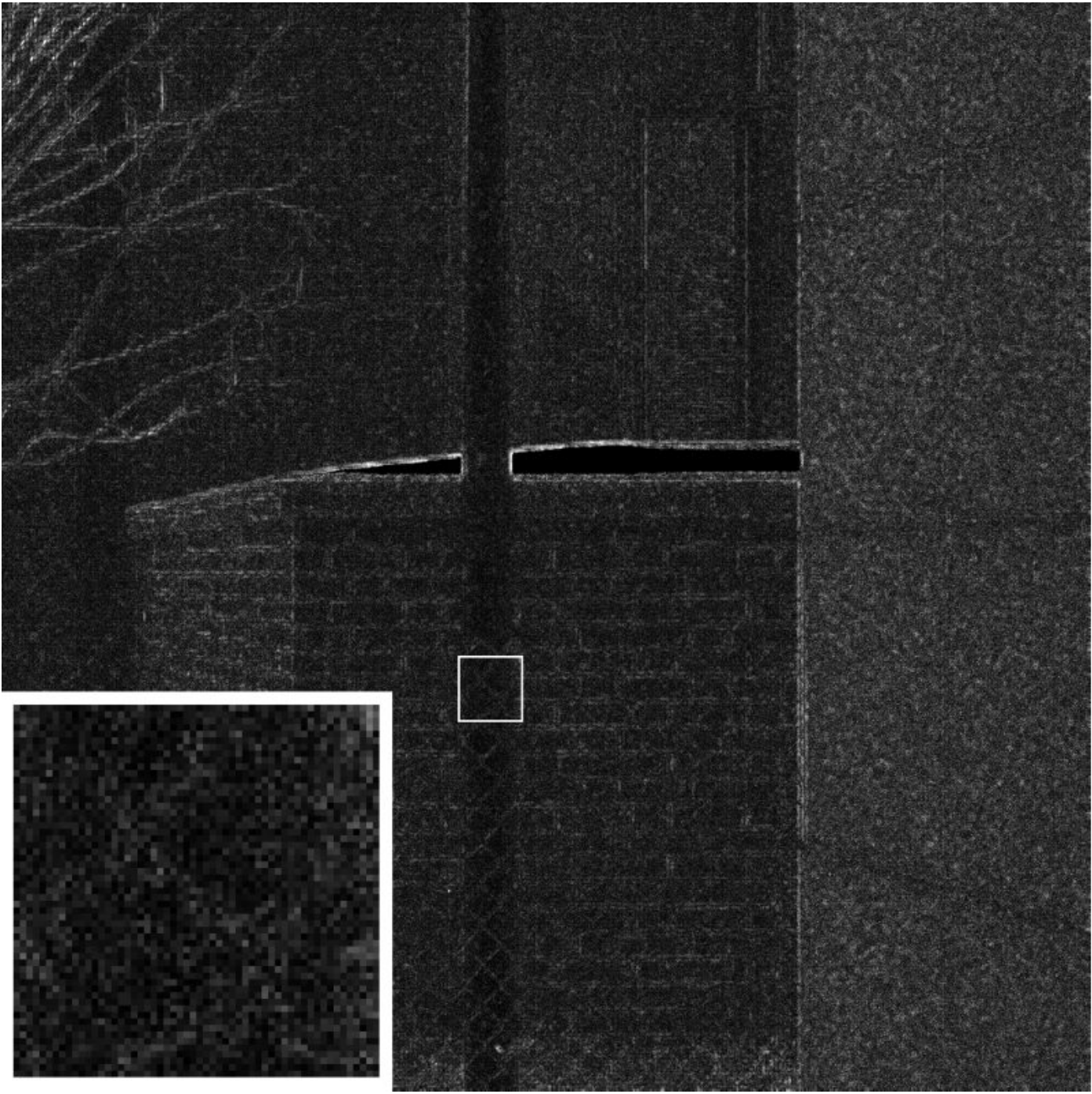}}
		\centering
		{HSRnet}
	\end{minipage}

	\end{minipage}
	\begin{minipage}[t]{0.04\linewidth}
		\vspace{0.7cm}
		{\includegraphics[height=10\linewidth,width=1\linewidth]{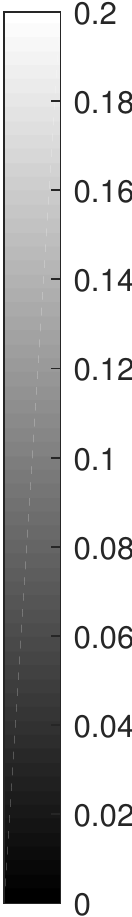}}
	\end{minipage}
	\caption{The first column: the true pseudo-color images from the original Harvard dataset and the corresponding LR-HSI images of \textit{tree} (R-30, G-27, B-7) (1st-2nd rows), \textit{bikes} (R-31, G-18, B-9) (3rd-4th rows), and (h) \textit{window} (R-31, G-28, B-1) (5th-6th rows). 2nd-8th columns: the true pseudo-color fused products and the corresponding residuals for the different methods in the benchmark pointing out some close-ups to facilitate the visual analysis.
	}
	\label{F:harvard-1}
\end{figure*}

Moreover, we firstly divide these 10 testing images into patches of size $128 \times 128$ randomly selecting 100 patches.
Table \ref{harvard-ave} shows the QIs of the results for the different methods on these 100 patches.
We can observe that our method is still the best method for all the different QIs, while the margins between our method and the MHFnet become larger than those in Table \ref{cave-ave}.
Particularly, the ERGAS value of the MHFnet ranks last place.
Thus, this test corroborates that the proposed approach has a better generalization ability than the compared deep learning-based method (\textit{i.e.}, the MHFnet).

Table \ref{harvard10-ave} records the average QIs and the corresponding standard deviations for the different methods using the 10 testing images. 
Table \ref{qresult-4} gives the QIs and the running times for three specific datasets of the Harvard dataset. The proposed method ranks first with the lowest running time.
Finally, considering the details in the pseudo-color images in Fig. \ref{F:harvard-1}, we can see that the results of our method get the highest qualitative performance, thus obtaining error maps that are very dark (\textit{i.e.}, with errors that tend to zero everywhere).

\subsection{Ablation Study}\label{newstruct}
\subsubsection{High-pass filters} 
In order to investigate the effects of the use of high-pass filters, we compare our HSRnet with its variant that is similar to the original HSRnet but without any high-pass filtering.
After removing the high-pass filters, the data cube $\mathcal{C}_{0}$ in Fig. \ref{structure} is obtained by concatenating the LR-HSI $\mathcal{Y}$ and the downsampled version of the HR-MSI, \textit{i.e.}, $\mathcal{Z}^{D}$. The network is trained on the same training data of the HSRnet with the same training settings.
Table \ref{nohp} presents the average QIs of these two networks on the 11 testing images for the CAVE dataset and the 10 testing images for the Harvard dataset.
As we can see from Table \ref{nohp}, the mean values and standard deviations of the proposed network are much better than that of the one without the high-pass filters.
This demonstrates that the use of high-pass filters lead to better and more stable performance.
In particular, the QIs of the Harvard testing images prove that the filters significantly provide better generalization ability.
Thus, the high-pass filters are of crucial importance for competitive performance of the proposed HSRnet. 


\begin{figure}[t]
	\begin{center}
		{\includegraphics[width=0.9\linewidth]{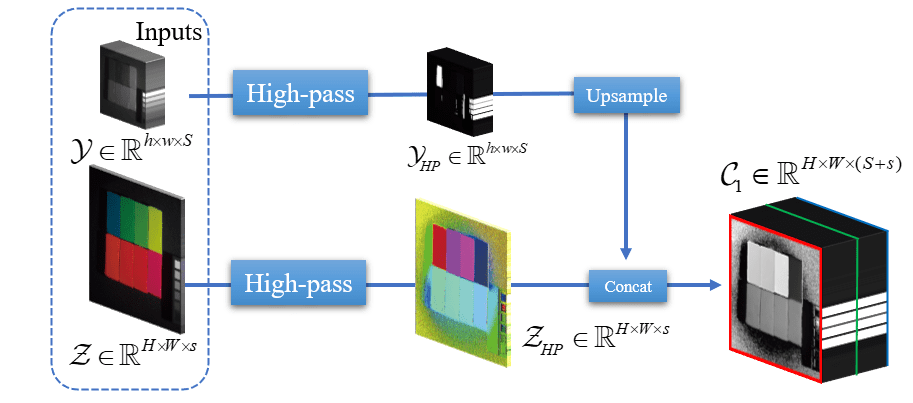}}
		\caption{Concatenation strategy with single scale. If we use this simple and single scale structure to replace the multiscale concatenation  $ \mathcal{C}_{1}$ of our HSRnet in Fig. \ref{structure}, it will get worse outcome than our HSRnet, which validates the importance of our multi-scale concatenation.
	}\label{single_scale_structure}
	\end{center}
\end{figure}

\begin{table}[t]
	\centering\small
	\renewcommand\arraystretch{0.9}\setlength{\tabcolsep}{6pt}
	\caption{Average QIs and related standard deviations of the results on the CAVE and the Harvard datasets using the proposed method with and without the high-pass (HP) filters. The best values are highlighted in boldface.}\label{nohp}
	\begin{tabular}{l|c|c|c|c}
		\Xhline{1.2pt}
		\multicolumn{5}{c}{CAVE}  \\ \hline
		Method 		& PSNR 	& SAM 	& ERGAS & SSIM \\ \hline
		
		w/o HP 	& 39.4$\pm$3.3 & 3.88$\pm$1.3 & 3.60$\pm$2.4 & 0.98$\pm$0.01  \\
		with HP 	  	& \textbf{44.0$\pm$2.9} & \textbf{3.09$\pm$1.0} & \textbf{1.93$\pm$1.0} & \textbf{0.99$\pm$0.00}\\ \Xhline{1.2pt}
		\multicolumn{5}{c}{Harvard}  \\ \hline
		Method 		& PSNR 	& SAM 	& ERGAS & SSIM \\ \hline
		
		w/o HP 	& 32.8$\pm$5.6 & 4.96$\pm$2.6 & 8.12$\pm$6.6 & 0.90$\pm$0.07 \\
		with HP 		& \textbf{39.5$\pm$4.7} & \textbf{3.38$\pm$1.3} & \textbf{3.27$\pm$1.5} & \textbf{0.97$\pm$0.02} \\ \Xhline{1.2pt}
	\end{tabular}	
\end{table}

\subsubsection{Multi-scale module} 
Concatenating multi-scale images is a key part of our network architecture. This leads to the extraction of several details at two different scales, which represent useful information for the super-resolution processing.
To prove the strength of this module, we compare our original HSRnet and the simpler architecture that only uses the main scale, $ \mathcal{C}_{1} $ in proposed Network is replaced by the one in Fig. \ref{single_scale_structure}.
The results of the two compared approaches are reported in Table \ref{compare_single}. The QI values show the necessity of the multi-scale module in our HSRnet representing a part of the proposed architecture that is less important than the high-pass filtering, but relevant in order to improve the performance measured by some QIs, see \textit{e.g.} the SAM and the ERGAS.

\begin{table}[t]
	\centering\small
	\renewcommand\arraystretch{0.9}\setlength{\tabcolsep}{6pt}
	\caption{Average QIs and related standard deviations of the results on the CAVE and the Harvard datasets using the proposed method with a different number of scales. The best values are highlighted in boldface.}\label{compare_single}
	\begin{tabular}{l|c|c|c|c}
		\Xhline{1.2pt}
		\multicolumn{5}{c}{CAVE}  \\ \hline
		Method 		& PSNR 	& SAM 	& ERGAS & SSIM \\ \hline
		
		one scale 	& 42.9$\pm$3.3 & 3.20$\pm$1.1 & 2.18$\pm$1.2 & \textbf{0.99}$\pm$0.00 \\
		HSRnet 	  	& \textbf{44.0}$\pm$2.9 & \textbf{3.09}$\pm$1.0 & \textbf{1.93}$\pm$1.0 & \textbf{0.99}$\pm$0.00\\ \Xhline{1.2pt}
		\multicolumn{5}{c}{Harvard}  \\ \hline
		Method 		& PSNR 	& SAM 	& ERGAS & SSIM \\ \hline
		
		one scale 	& 38.8$\pm$4.4 & 3.66$\pm$1.9 & 3.64$\pm$1.8 & \textbf{0.97}$\pm$0.02 \\ 
		HSRnet 		& \textbf{39.5}$\pm$4.7 & \textbf{3.38}$\pm$1.3 & \textbf{3.27}$\pm$1.5 & \textbf{0.97}$\pm$0.02 \\ \Xhline{1.2pt}
	\end{tabular}	
\end{table}

\subsection{Comparison with MHFnet}\label{vs}
To our knowledge, the MHFnet developed by Xie \etal \cite{xie2019multispectral} outperforms the state-of-the-art of the model-based and the deep learning-based methods, actually representing the best way to address the HSI super-resolution problem.
Due to the fact that the MHFnet and our HSRnet are both deep learning-based methods, in this subsection, we keep on discussing about the HSRnet comparing it with the MHFnet. 

\subsubsection{Sensitivity to the number of training samples} We train the MHFnet and our HSRnet with different numbers of training samples to illustrate their sensitivity with respect to this parameter. 
We randomly select 500, 1000, 2000, and 3136 samples from the training data.
Testing data consists of 7 testing images on the CAVE dataset and 10 testing images on the Harvard dataset.
Table \ref{diffrernt-num-t} reports the average QIs of the results obtained by the MHFnet and by our HSRnet varying the number of the training samples.
From the results on the CAVE dataset in Table \ref{diffrernt-num-t}, we can note that the MHFnet performs well when the training samples are less. This can be attributed to its elaborately designed network structure. Our method steadily outperforms the MHFnet in the cases of 2000 and 3196 training samples.
Instead, from the results on the Harvard dataset, we can remark that the generalization ability of our method is robust with respect to changes of the numbers of the training samples (due to the use of the high-pass filters in the architecture). Whereas the MHFnet shows poor performance due to its manual predefined parameters that are sensitive to scene changes.

\begin{table}[t]
	\centering\renewcommand\arraystretch{1}\setlength{\tabcolsep}{3pt}\footnotesize
	\caption{Results of the two deep learning-based methods varying the number of the training samples. The best values are highlighted in boldface.}\label{diffrernt-num-t}
	\begin{tabular}{c|c|l|c|c|c|c}
		\Xhline{1.2pt}
		
		Datasets & \# training data & Methods & PSNR & SAM & ERGAS & SSIM \\ \hline\hline
		\multirow{8}*{CAVE} & \multirow{2}*{3136}
		& MHFnet & 43.27 & 4.34 & 2.33 & 0.989\\ 
		&&HSRnet  & \textbf{44.00} & \textbf{3.09} & \textbf{1.93} & \textbf{0.992}  \\  \cline{2-7}
		&\multirow{2}*{2000}
		& MHFnet & 43.37 & 4.50 & 2.39 & 0.988 \\ 
		&&HSRnet  &\textbf{43.91} & \textbf{3.03} & \textbf{1.96} &\textbf{ 0.992} \\ 
		\cline{2-7}
		&\multirow{2}*{1000}
		& MHFnet & \textbf{43.42} & 4.47 & 2.34 & 0.988 \\
		&&HSRnet  &43.40 & \textbf{3.16} & \textbf{2.08} & \textbf{0.991}  \\  \cline{2-7}
		&\multirow{2}*{500}
		& MHFnet &   \textbf{42.74} & 4.77 &\textbf{ 2.50} & 0.987 \\ 
		&&HSRnet  & 40.99 & \textbf{3.65} & 2.89 & 0.987 \\  \hline\hline
		\multirow{8}*{Harvard} & \multirow{2}*{3136}
		& MHFnet & 36.41 & 7.03 & 16.57 & 0.915 \\ 
		&& HSRnet & \textbf{39.53} & \textbf{3.38} & \textbf{3.27} & \textbf{0.970} \\ \cline{2-7}
		&\multirow{2}*{2000}
		& MHFnet & 36.54 & 6.93 & 18.42 & 0.912 \\
		&& HSRnet &\textbf{39.87} & \textbf{3.40} & \textbf{3.33} & \textbf{0.970} \\ \cline{2-7}
		&\multirow{2}*{1000}
		& MHFnet & 36.16 & 6.99 & 26.49 & 0.916 \\
		&&HSRnet  & \textbf{39.44} & \textbf{3.47} &\textbf{ 3.54} & \textbf{0.968}  \\ \cline{2-7}
		&\multirow{2}*{500}
		& MHFnet & 36.18 & 7.41 & 25.95 & 0.903 \\
		&&HSRnet  & \textbf{38.69} & \textbf{3.55} & \textbf{3.81} &\textbf{ 0.966} \\ \Xhline{1.2pt}
	\end{tabular}
\end{table}

\begin{figure}[t]
	\begin{center}
		
		{\includegraphics[width=0.6\linewidth]{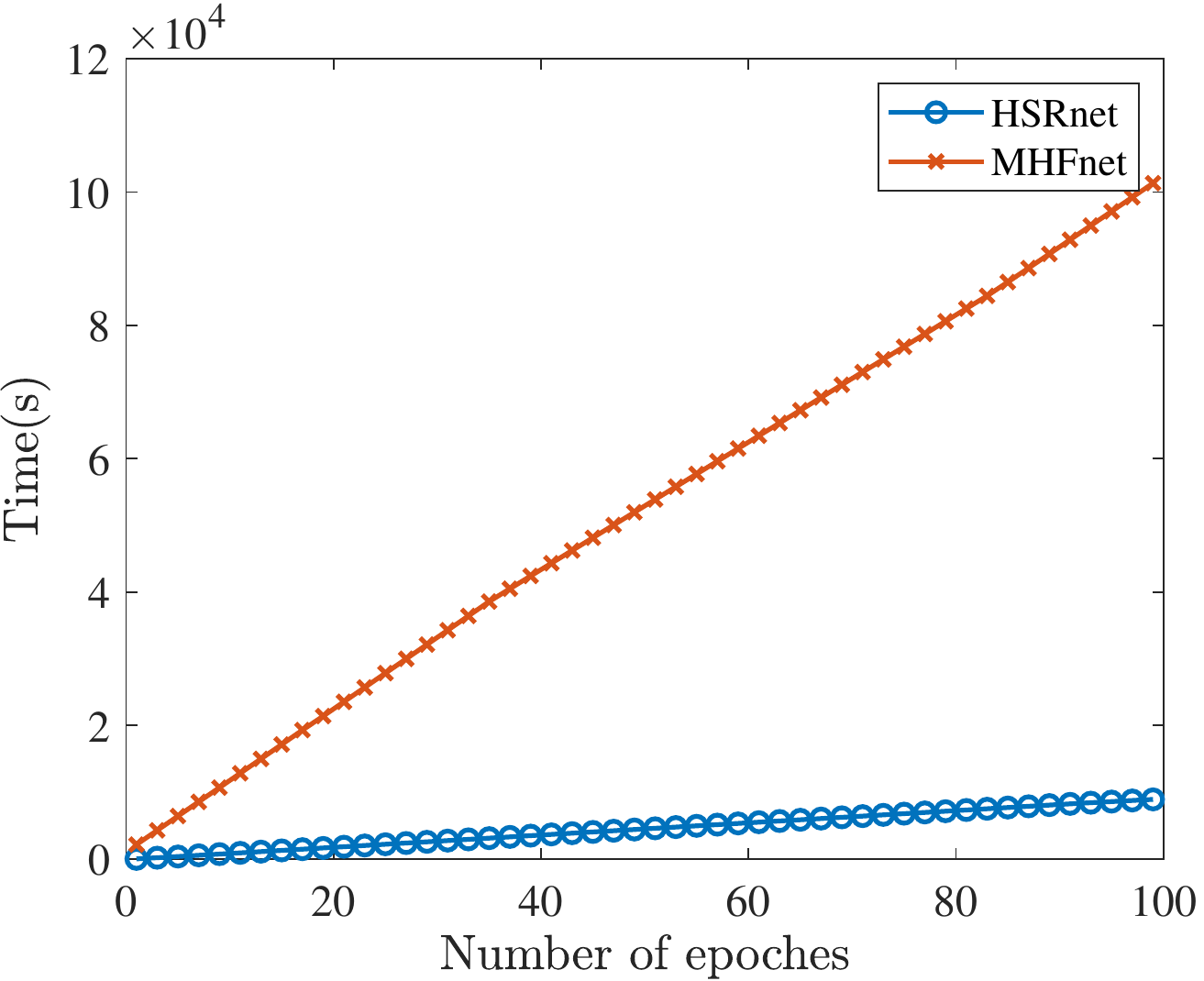}}
		\caption{The comparison of the training times for the MHFnet and the proposed HSRnet.}\label{timec}
	\end{center}

\end{figure}

\subsubsection{Network generalization}
In the above content, MHFnet and our HSRnet are both trained with CAVE data. We can find that our HSRnet outperforms the MHFnet in all the experiments on the testing data provided by the Harvard dataset. This shows the remarkable generalization ability of our network. To further corroborate it, we retrain these two networks on training samples provided by the Harvard dataset. Namely, we extract from the Harvard dataset 3763 training samples, in which the HR-MSI is of size $64 \times 64$ and the LR-HSI is of size $16 \times 16$. 
As previously done, we select the same 11 images from the CAVE dataset and the same 10 images from the Harvard dataset to build the testing set. 
We show the QIs of the results for these two networks trained on the Harvard dataset in Table \ref{table-trained-harvard}.
It can be seen that the generalization ability of the MHFnet is still limited. Instead, the proposed approach still shows an excellent generalization ability when it is used on CAVE data but trained on the Harvard samples.

\begin{table}[hpt]
	\centering\small
	\renewcommand\arraystretch{0.9}\setlength{\tabcolsep}{6pt}
	
	\caption{Average QIs and related standard deviations of the results for the networks trained on the Harvard dataset. The best values are highlighted in boldface.}\label{table-trained-harvard}
	\begin{tabular}{l|c|c|c|c}
		\Xhline{1.2pt}
		\multicolumn{5}{c}{CAVE}  \\ \hline
		Method & PSNR & SAM & ERGAS & SSIM \\ \hline
		
		MHFnet &  34.9$\pm$2.5 & 13.15$\pm$4.2 & 5.73$\pm$2.4 & 0.93$\pm$0.02 \\
		HSRnet  & \textbf{40.5}$\pm$2.8 & \textbf{4.21}$\pm$1.6 & \textbf{3.20}$\pm$1.6 &\textbf{ 0.98}$\pm$0.01 \\ \Xhline{1.2pt}
		\multicolumn{5}{c}{Harvard}  \\ \hline
		Method & PSNR & SAM & ERGAS & SSIM \\ \hline
		MHFnet &  \textbf{41.0}$\pm$5.3 & 3.36$\pm$1.6 & 3.33$\pm$1.9 & 0.97$\pm$0.02\\ 
		HSRnet  & 40.1$\pm$5.5 & \textbf{3.06}$\pm$1.1 & \textbf{2.49}$\pm$1.0 & \textbf{0.98}$\pm$0.02 \\ \Xhline{1.2pt}
	\end{tabular}
\end{table}

\subsubsection{Parameters and training time}
MHFnet contains 3.6 million parameters, instead, 2.1 million parameters have to be learned by our HSRnet.
In Fig. \ref{timec}, we plot the training time with respect to the epochs. We can find that our network needs much less training time than MHFnet. Actually, from Tables \ref{qresult-4CAVE} and \ref{qresult-4}, the testing time of our HSRnet is also less than that of the MHFnet. Indeed, fewer parameters result in less training and testing times, making our method more practical.

\section{Conclusions}\label{conclusion}

In this paper, a simple and efficient deep network architecture has been proposed for addressing the hyperspectral image super-resolution issue. The network architecture consists of two parts: $i$) a spectral preservation module and $ii$) a spatial preservation module that has the goal to reconstruct image spatial details starting from multi-resolution versions of input data. The combination of these two parts is performed to get the final network output. This latter is compared with the reference (ground-truth) image under the Frobenius norm based loss function. This is done with the aim of estimating the network parameters during the training phase. 

Extensive experiments demonstrated the superiority of our HSRnet with respect to recent state-of-the-art hyperspectral image super-resolution approaches. Additionally, advantages of our HSRnet have been reported also from other points of view, such as, the network generalization, the limited computational burden, and the robustness with respect to the number of training samples.



%


\bibliographystyle{IEEEtran}
\bibliography{references}

%
%
%




\end{document}